\documentclass{aa}

\pdfoutput =1
\usepackage{natbib}
\usepackage{amsmath}
\usepackage{amssymb}
\usepackage{mathtools}
\usepackage{graphicx}
\graphicspath{ {./figures/} }
\usepackage[multidot]{grffile}
\usepackage{txfonts}
\usepackage{color}
\usepackage{threeparttable}
\usepackage{array}
\usepackage{booktabs}
\usepackage{rotating}
\usepackage{verbatim}
\usepackage{appendix}
\usepackage{float}
\usepackage{pdflscape}

\usepackage{multicol}
\usepackage{morefloats}
\usepackage{comment}
\usepackage{bm}
\usepackage[utf8]{inputenc}
\usepackage{hyperref} 
\hypersetup{colorlinks, citecolor=blue,  linkcolor=blue,  urlcolor=blue}

\newcommand{\msun}{M_{\rm \odot}}
\newcommand{\mbh}{M_{\rm BH}}

\newcommand{\sgra}{Sgr~A*~}

\newcommand{\ltfrac}[2]{\mbox{\large$\frac{#1}{#2}$}} 
\newcommand{\code}[1]{\mbox{\texttt{#1}}}
\newcommand{\fig}{\rm Fig.~}
\newcommand{\rhi}{R_{high}}
\newcommand{\rlo}{R_{low}}
\newcommand{\chisq}{\chi ^2}

\begin{document}

\title{The Core-shift of Sagittarius~A* as a Discriminant between Disk and Jet Emission Models with millimeter-VLBI} 
\titlerunning{Core-shift of Sagittarius~A* Emission Models with mm-VLBI}
\authorrunning{Fraga-Encinas, R. et al. }
%
\author{R. Fraga-Encinas\inst{\ref{inst1}}
        \and
        M. Mo{\'s}cibrodzka\inst{\ref{inst1}}
        \and
        H. Falcke\inst{\ref{inst1}, \ref{inst2}}}      
\institute{Department of Astrophysics / IMAPP, Radboud University, P.O. Box 9010, 6500 GL Nijmegen, The Netherlands.\label{inst1}
            \and
           ASTRON, Oude Hoogeveensedijk 4, 7991 PD Dwingeloo, The Netherlands.\label{inst2} \\          
\email{r.fraga@astro.ru.nl}
            }

\date{Submitted to A$\&$A.  Under revision.  Accepted xxxx}

\abstract
{The nature of the emission region around Sagittarius~A* (\sgra), the supermassive black hole at the Galactic Center, remains under debate. A prediction of jet models is that a frequency-dependent shift in the position of the radio core (core-shift) of active galactic nucleii (AGN) occurs when observing emission dominated by a highly collimated relativistic outflow.} 
{We use millimeter Very Long Baseline Interferometry to study the frequency-dependent position of \sgra 's radio core, estimate the core-shift for different emission models, investigate the core-shift evolution as a function of viewing angle \& orientation, and study its behaviour in the presence of interstellar scattering.}
{We simulate images of the emission around \sgra for accretion inflow models (disks) and relativistic outflow models (jets). They are based on three-dimensional general relativistic magnetohydrodynamic simulations. We create flux density maps at 22, 43 \& 86~GHz sampling different viewing angles and orientations of \sgra , and examine the effects of scattering.} 
{Jet-dominated models show significantly larger core-shifts - in some cases by a factor of 16 - than disk-dominated models, intermediate viewing angles ($i=30\degr,45\degr$) show the largest core-shifts. Our jet models follow a power-law relation for the frequency-dependent position of \sgra 's core. Their core-shifts decrease as the position angle increases from $0\degr$ to $90\degr$. Disk models do not fit well a power-law relation and their core-shifts are insensitive to changes in viewing angle. We place an upper limit of $241.65 \pm 1.93~ \mu  as~cm^{-1}$ for the core-shift of jet models including refractive scattering.}
{The core-shift can serve as a tool to discriminate between jet \& disk-dominated models and to constrain the geometry of \sgra. Our jet models agree with earlier predictions of AGN with conical jets, and the core-shift is retrievable even in the presence of interstellar scattering.}

\maketitle


\section{Introduction}\label{sec:introduction}
Almost five decades ago a compact radio source was discovered at the Galactic Center \citep{balick:1974}. This well-studied radio source, Sagittarius A* (Sgr~A*), is associated with a supermassive black hole (SMBH) with a mass of $\sim~4.1 \times 10^6 \msun$ \citep{reid:2004, gillessen:2009, gravity:2019a} and it is located at a distance of $\sim$~8.1~kpc \citep{ghez:2008, gravity:2019a}. \sgra has a compact radio core size (120~$\mu$as at 86~GHz, \citealt{issaoun:2019_size}) and observations indicate that the millimeter emission is being generated from within a few Schwarzschild radii ($R_s=2GM_{BH}/c^2 \sim 10~ \mu$as) of the black hole \citep{lo:1985, falcke:1998_specsgrA, krichbaum:1998, doeleman:2001, bower:2004, shen:2005, doeleman:2008, fish:2011, doeleman:2012, fish:2016, akiyama:2013, lu:2018}. Given its mass and proximity to Earth makes \sgra the black hole (BH) with the largest angular scale on the sky, therefore being an excellent object to study the processes of radiative emission around a SMBH and to probe its structure at near event-horizon scales (for reviews see \citealt{melia:2001_review}; \citealt{genzel:2010}; \citealt{falcke:2013}; \citealt{goddi:2017}). 

A photon ring and an associated black hole "shadow" with a diameter of $\sim$~50~$\mu$as against the background emission is predicted by General Relativity (Bardeen 1973; Falcke et al. 2000; Takahashi 2004). Being able to resolve this region allows us to study the structure of \sgra and plasma dynamics at near-horizon scales. Observing radio emission coming from this event-horizon-scales and imaging the BH shadow is now possible using Very Long Baseline Interferometry (VLBI) as recently shown in the core of the galaxy M87 \citep{EHT:2019_abs, EHT:2019_ins, EHT:2019_cal, EHT:2019_img, EHT:2019_theo, EHT:2019_feat}.

Thus far it remains to be established whether the observed radio emission in \sgra is generated by radiatively inefficient accretion flows, RIAFs \citep{narayan:1995_spectrum, yuan:2014}, mildly-relativistic down-sized jets coupled to accretion flows \citep{falcke_markoff:2000, markoff:2007}, relativistic jets coupled to RIAFs \citep{moscibrodzka:2013, moscibrodzka:2014}, hot-spots in accretion disks \citep{broderick:2006, doeleman:2009} or various other jet-disk combination models \citep{yuan:2002, broderick:2009_horizon, dexter:2013, chan:2015a, ball:2016, gold:2017, davelaar:2018, chael:2019, ressler:2019}. These models remain unconstrained because there is a thin screen of plasma located between the Earth and the Galactic Center \citep{vanlangevelde:1992, lo:1998, bower:2006}. This scattering screen distorts observations up to millimeter (mm) wavelengths. 

One of the characteristics of \sgra is that its measured size is related to the observing wavelength, and it follows a $\lambda ^2$ dependence (\citealt{davies:1976}). Its intrinsic size is "blurred" by the scattering screen. Electron-density inhomogeneities in this plasma screen cause 
scattering at radio frequencies \citep[e.g.,][]{narayan:1992, thompson:2017_book, johnson:2018}, which produces angular broadening of
the \sgra image and scintillation (i.e., changes in intensity). The latter leading to the appearance of substructure \citep{lee:1975,narayan:1989, gwinn:2014}. 

\begin{figure}[ht]
\centering
\includegraphics[width=0.95\columnwidth]{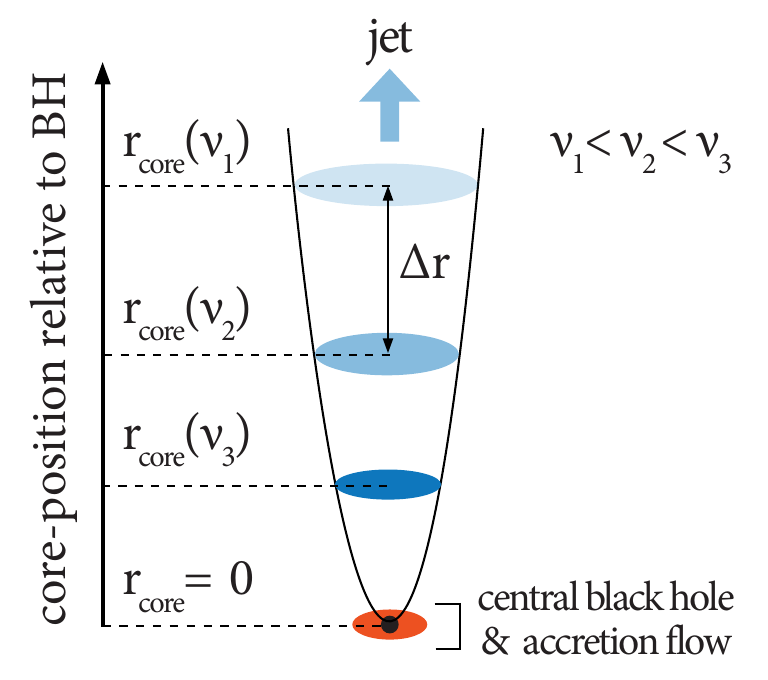}
\caption{\textbf{Geometry of the core-position of Sagittarius~A*} as a function of observing frequency. The central supermassive black hole (black dot) launches a jet which is perpendicular to the accretion flow (orange ellipse). The surfaces with optical depth $\tau =1$ at a given frequency are represented by the blue ellipses. The \textit{core-shift} between two frequencies is indicated by $\Delta r$. Adapted from Figure~4.5 in \citet{lobanov:1996}.}\label{fig:coreshift_geometry}
\end{figure}

VLBI observations of relativistic jets of many active galactic nucleii (AGN) show that a bright compact "core" of radio emission can be pinpointed at the base of their jets. This core exhibits a flat-to-inverted spectrum at radio through sub-millimeter (sub-mm) wavelengths. It has a peak at a frequency of around 350~GHz - the sub-mm "bump" - beyond which the spectrum falls steeply into the infrared. Emission at the sub-mm bump comes from regions nearest to the BH \citep{falcke:1998_specsgrA, shen:2005, bower:2006, doeleman:2008}. Compact relativistic jet models predict that at a given frequency, the radio core will be located at the surface along the jet axis where the optical depth $\tau _{\nu} \approx 1$. Since optical depth for synchrotron emission is frequency-dependent, this implies that the core location changes with observing frequency \citep{blandford_konigl:1979, konigl:1981, lobanov:1998_jets, falcke_biermann:1999a}. This is commonly known as the \textit{core-shift} as illustrated in \fig\ref{fig:coreshift_geometry}. The core-shift is a robust prediction of jet models but not disk models, and it has not yet been carefully tested against observational constraints. This is one of the reasons that motivates our work.

All emission processes have absorption counterparts. In the case of synchrotron emission the counterpart is synchrotron self-absorption (SSA), in which a photon is absorbed due to its interaction with a charged particle in a magnetic field ($B$). SSA is the mechanism responsible for the changes in $\tau _{\nu}$ along the jet, and is the reason why not all of the synchrotron photons propagating through the jet can escape the source \citep{ginzburg:1969}. Low-frequency synchrotron photons (i.e., 22~GHz) suffer stronger absorption by the relativistic electrons in the plasma than high-frequency photons (i.e., 86~GHz). For non-thermal electrons that follow a power-law distribution of particles, the absorption coefficient ($\alpha _{\nu}$) has a strong dependence upon frequency ($\nu$). For full details see Eq. 6.53 of \citet{rybicki_lightman:1979}. The absorption coefficient $\alpha _{\nu}$ is expressed as:
\begin{equation}\label{eq:absorption_coeff}
\alpha _{\nu} = C ~ B ~^{(P+2)/2} \nu ~^{-(P+4)/2}   ~.
\end{equation}
where $C$ is a constant, $B$ is the magnetic field and $P$ is the index of the of the particle distribution law of relativistic electrons. At high observing frequencies the absorption is small, therefore the plasma is optically-thin and we can see "deeper" into the source than at low frequencies. A description for non-thermal electrons is used in this paragraph to illustrate the concept of core-shift, we note that for our work we assume a thermal, relativistic distribution of radiating electrons (as described in Sect.~\ref{sec:emission_models}).

%

Because the displacement of the absolute position of the core with observing frequency is a strong indicator of the presence of a conical or parabolic jet, studying core-shifts is an excellent tool to probe the emission mechanisms around a SMBH. The distance from the central engine to the observed core can be written as: 
\begin{equation}\label{eq:core_position_konigl}
r_{core}(\nu) \propto ~\nu  ~^{-1/k_r} ~, 
\end{equation}
where $r_{core}$ is the core position relative to the BH as a function of observing frequency $\nu$. The index $k_r$ depends on jet properties such as the particle density distribution, electron energy spectrum, jet opening angle and magnetic field. The factor $k_r =1$ is predicted for SSA with equipartition between the magnetic field energy and the energy density of the relativistic particles \citep{blandford_konigl:1979}. The core-shift $\Delta r$ (see \fig\ref{fig:coreshift_geometry}) between two observing frequencies 
$\nu _1$ and $\nu _2$ can be expressed as:
\begin{equation}\label{eq:core_shift_freq}
\Delta r = r_{core}(\nu _1) - r_{core}(\nu _2) ~. 
\end{equation}

In previous work using high-precision astrometry, core-shifts have been measured along the jets of many AGN 
\citep[e.g.,][]{lara:1994, lobanov:1998_jets, kadler:2004, hada:2011, haga:2015, kovalev:2008, voitsik:2018}. In the case of M81, whose nuclear radio source M81* shares similarities with \sgra , core-shift studies have placed constraints on the size of the emission region around M81*, with he central BH located within a region of $\pm 0.2$~mas, and confirmed that M81* is a compact core-jet source \citep{bietenholz:2004}.

For \sgra the presence of a core-shift has been investigated using the Galactic Center pulsar 
\citep{bower:2015_motion}. The pulsed radio emission of the magnetar was phase-referenced 
to the core-position of \sgra using two band-pairs (15-8~GHz \& 43-22~GHz). This work places an 3$\sigma$ upper-limit on 
the core-shift of 0.3~$\rm mas~cm^{-1}$ in right ascension and 0.2~$\rm mas~cm^{-1}$ in declination. Observations of \sgra at 43~GHz \& 86~GHz with the Korean VLBI Network (KVN) report a core-shift of 0.3~$\rm mas~cm^{-1}$ \citep{zhao:2017}. 
However, the direction of the core-shift is inconsistent with that measured by \cite{bower:2015_motion}. The authors consider the contribution of the calibrators own core-shift plus blending effects 
due to the large KVN beamsize \citep{rioja:2014} to be the reason for the discrepancy.

The goal of this work is to predict the frequency-dependent position of \sgra 's radio core and to estimate the core-shift for different jet-plus-disk models. Our predictions are made using three-dimensional general relativistic magnetohydrodynamical (3D GRMHD) simulations, ray-tracing and scattering screen models. We have chosen to investigate combined emission models of jets plus accretion inflows \citep{moscibrodzka:2014, shiokawa:2013} because they reproduce well the observed spectrum and intrinsic size of \sgra. We also investigate the evolution of the core-shift as a function of viewing angle \& orientation on the sky, and study the core-shift behaviour in the presence of interstellar scattering.

This paper is organized as follows: Sect.~\ref{sec:emission_models} describes the emission models that can explain the observed radiation, Sect.~\ref{sec:scattering_background} provides context on the effects that interstellar scattering has on \sgra observations, Sect.~\ref{sec:results} presents our results, discusses the modeling of core-shifts, investigates the effects of an interstellar scattering screen and studies the behaviour of cores-shift as a function of \sgra orientation, Sect.~\ref{sec:discussion} discusses constraints from observations, future prospects and Sect.~\ref{sec:conclusions} summarizes our conclusions. Lastly, the appendices contain a library of model images created for this work. 

%
\section{Disk \& Jet Emission Models for Sagittarius A*}\label{sec:emission_models}
 
Our underlying, dynamical model of plasma flow onto \sgra is produced in 3D GRMHD simulations. We assume that the observed \sgra emission
is produced by a two component (disk-plus-jet) model. The model consists of a radiatively inefficient accretion flow (RIAF) onto a spinning black hole, i.e. a turbulent disk,  \textit{plus} a strongly magnetized, low density, relativistic outflow, i.e. a jet \citep{blandford_znajek:1977}. Because the spin of \sgra is unconstrained we use a fiducial spin value of $a_*=0.94$. Particle heating determines which part of the jet-disk system dominates the appearance. The turbulent plasma inflows and outflows have different amounts of magnetization, plasma is weakly magnetized in the accretion disk,
whereas is strongly magnetized in the jet outflows. These simulations are described in detail in \cite{shiokawa:2013} and \cite{moscibrodzka:2014}. 

Next, we use the general relativistic radiative transfer scheme \code{IPOLE} \citep{moscibrodzka:2018} to generate synthetic radio maps 
of the model. The radiative transfer simulations are carried out assuming the following fixed parameters for the \sgra system: 
black hole mass $\mbh = 4.5 \times 10^6~\msun$ and distance $D = 8.6 \times 10^3$ pc from the observer as used in \cite{moscibrodzka:2014}. These values are slightly different than the latest results from \cite{gravity:2019a}. The distance $D$ is necessary for 
the conversion from absolute luminosity to observed flux density. All models are normalized to produce a total flux of 2~Jansky at a 
frequency of 86~GHz to be consistent with observed fluxes of \sgra \citep{issaoun:2019_size}. The normalization is done by changing 
the value of the mass unit $\mathcal{M}$ which translates the mass accretion rate $\dot{M}$ into physical units of $\msun$/year. In practice this means that 
the matter densities in an entire model are multiplied by a constant scaling density factor. $\mathcal{M}$ also changes the magnetic field strength. 

We run \code{IPOLE} to integrate the radiative transfer equation including synchrotron emission and synchrotron self-absorption. 
We assume a thermal, relativistic Maxwell-J\"{u}ttner electron distribution function (eDF) for the radiating electrons in both the disk and jet regions. 
The ratio of the proton-to-electron temperature (${T_p}/{T_e}$) is not followed in the considered GRMHD simulation.
Hence in what follows we parameterize in terms of the plasma parameter $\beta$ (i.e. the ratio of the gas pressure to magnetic pressure $\beta={P_{gas}}/{P_{mag}}$~), and in terms of some coupling constants $R_{high}$ and $R_{low}$ so that 
\begin{equation}\label{eq:temp_ratio}
\frac{T_p}{T_e}= R_{high} \frac{\beta^2}{1+\beta^2} + R_{low} \frac{1}{1+\beta^2} ~.
\end{equation}
The coupling constants $R_{high}$ and $R_{low}$ describe the proton-to-electron coupling in weakly and strongly magnetized
plasma, respectively. For weakly magnetized plasma $\beta \gg 1$ (e.g., inside a turbulent accretion disk), $T_p/T_e \to R_{high}$. 
For strongly magnetized plasma $\beta\ll1$ (e.g., along a jet), $T_p/T_e \to R_{low}$. 
This approach is adopted from \citet{moscibrodzka:2016_spin} and it is motivated by recent studies of electron-proton couplings 
in collisionless plasma \citep{howes:2010, ressler:2015, chael:2018, kawazura:2018}. By increasing the value of $R_{high}$ we can 
move smoothly from models where the source appearance is dominated by disk emission to models where the source appearance is dominated by jet emission . In this work we do not scan the whole parameter space of $\rhi$ and $\rlo$, we chose to investigate two models: a bright disk and a bright jet. Emission models with $( \rhi~,~\rlo ) = (3, 3)$ result in bright disks. 
For these disk models we assume that electrons are strongly coupled to protons both in the disk and in the jet, and that the 
temperature ratio is $T_p/T_e=3$ everywhere \citep{moscibrodzka:2009}. In the equatorial plane of the accretion disk 
the plasma density is highest, therefore emission from the disk dominates. In contrast, models with $( \rhi~,~\rlo )=(20, 1)$ result 
in bright jets. For these jet models, electrons are weakly coupled to protons 
in the accretion disk ($T_p/T_e=20$), but they remain strongly coupled to protons in the jet ($T_p/T_e=1$); hence synchrotron emission 
from the jets overpowers the disk emission.

\begin{figure}[h]
\centering
\includegraphics[width=0.7\columnwidth]{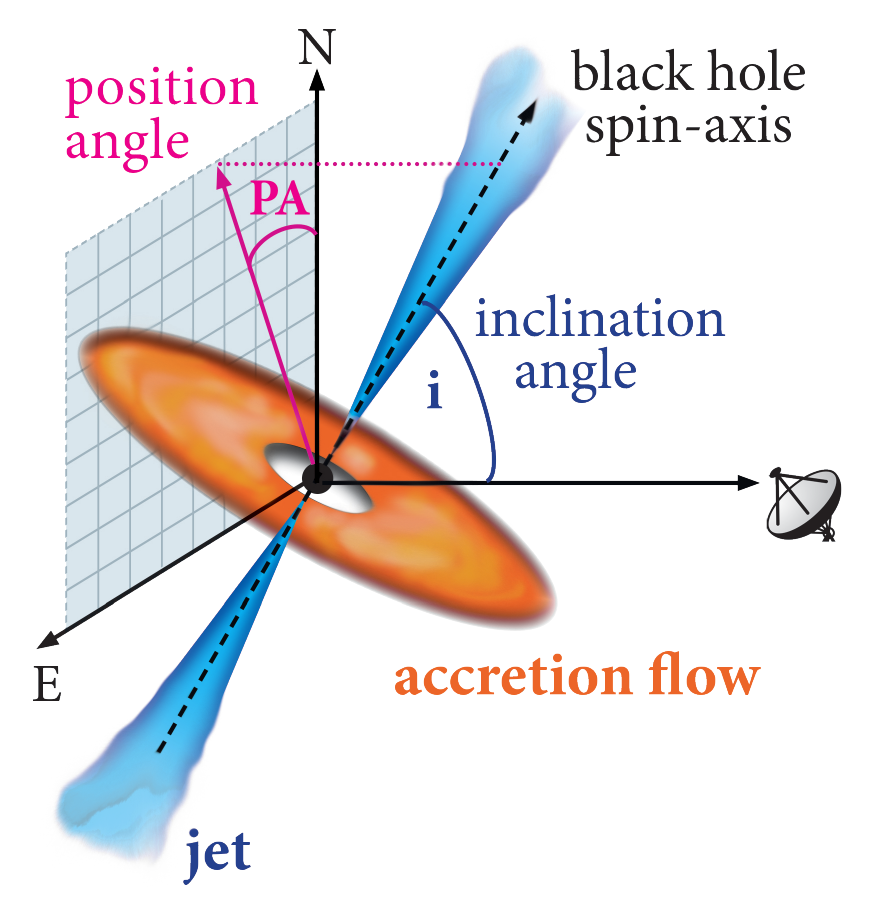}
\caption{\textbf{Geometry of the emission models}. The angle between the observer's line-of-sight and the black hole spin-axis is
defined as the inclination angle ($i$). The orientation of the spin-axis on the plane of the sky is the position angle ($PA$). 
A completely face-on accretion disk occurs at $i=0 \degr$, an edge-on disk is seen at $i=90 \degr$. The spin-axis points due North 
when $PA=0 \degr$, and due East when $PA=90 \degr$, orientation changes East of North (ie. counter-clockwise).}\label{fig:BH_geometry}
\end{figure}

Another parameter that we change to generate different models is the inclination (i.e. viewing) angle $i$. This is the angle between the observer's line-of-sight and the BH spin-axis as illustrated in \fig\ref{fig:BH_geometry}. The morphology of the emission models is strongly dependent on $i$. Recent findings have shown that $i$ can be partially constrained for the jet models presented here using observations at 86~GHz \citep{issaoun:2019_size}. Furthermore, we also change the position angle ($PA$), which is the orientation of the BH spin-axis on the plane of the sky.

%
\section{Interstellar Scattering Model for Sagittarius A*}\label{sec:scattering_background}
 Studies of \sgra at 43~GHz and 86~GHz \citep[e.g.,][]{krichbaum:1993, lu:2011} have shown that its intrinsic structure starts to be 
unveiled as the effects of the scatter broadening become less dominant at frequencies 
$\gtrsim$~43~GHz \citep{doeleman:2001, bower:2004, shen:2005, krichbaum:2006}. 
Recent work reports sub-structure at micro-arcsecond scales in the emission from \sgra at 86~GHz \citep{brinkerink:2019}. Additionally, \citet{issaoun:2019_size} have obtained the first image of \sgra at 86~GHz where the intrinsic source structure has been deconvolved from the effects the interstellar scattering. Using observation of very scattered pulsars within the Galactic Center suggest that at least two scattering screens are required to explain the pulsar observations, with one component being within $\leq $700 pc of the GC and another faraway at $\sim$2~kpc from Earth \citep{dexter:2017}.

\begin{figure}[ht]
\centering
\includegraphics[width=0.9\columnwidth, trim= 0cm 0cm 0cm 0.5cm, clip=true]{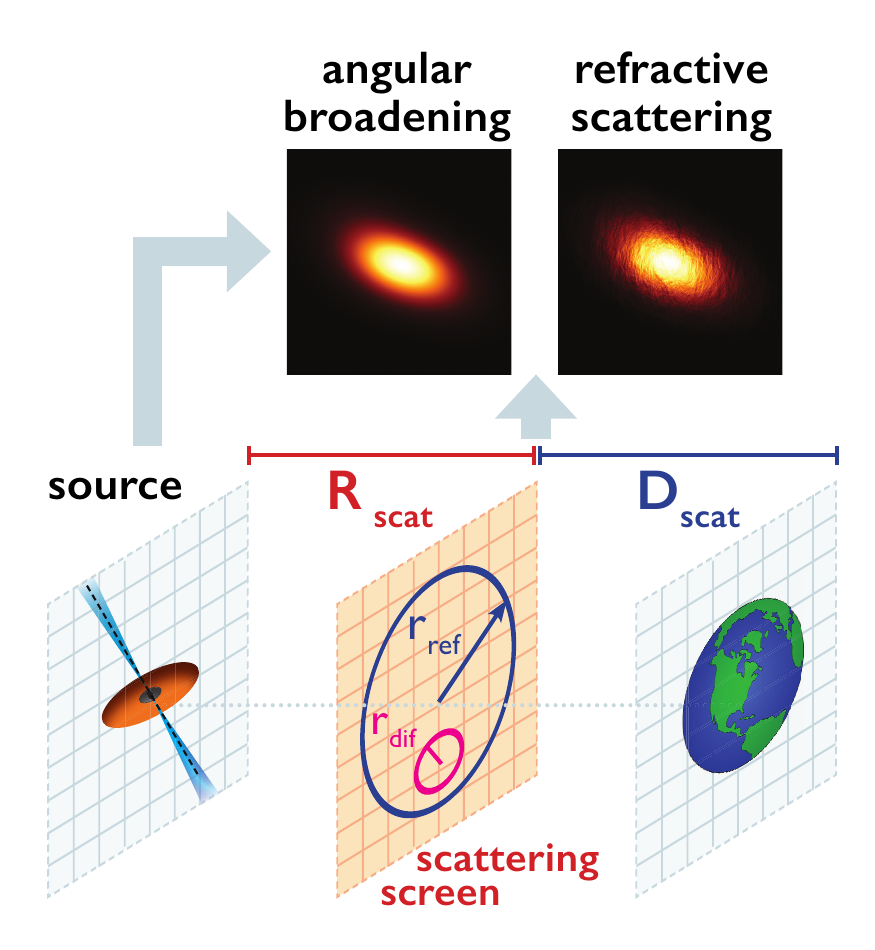}
\caption{\textbf{Geometry of the Scattering Screen}. This diagram shows the diffractive $(r_{dif})$ and refractive $(r_{ref})$ scales, screen-Earth distance $(D_{scat})$, screen-\sgra distance $(R_{scat})$. Modified from Fig. 1 in \citet{psaltis:2018}.}\label{fig:scatter_screen}
\end{figure}

When investigating the effects of scattering, there are two important spatial scales to consider: the \textit{diffractive} scale ($r_{dif}$) and
the \textit{refractive} scale ($r_{ref}$). \fig\ref{fig:scatter_screen} illustrates the scattering screen geometry. The smallest scale on which
flux variations due to turbulence happen is $r_{dif}$. Diffractive scattering leads to scatter-broadening of the source. The image size projected onto the scattering screen is defined as $r_{ref}$ \citep{narayan:1989, psaltis:2015}. Refractive scattering causes substructure in the image and it has already been observed in \sgra at 13~mm \citep{gwinn:2014}. 

These scales can be written in terms of the observing wavelength $\lambda$ as follows \citep{psaltis:2018}:
\begin{equation}\label{eq:diffractive_scale}
r_{dif} \sim \frac{\lambda}{(1+D_{scat}/R_{scat})~ \theta_{scat}} ~ \sim \lambda ^{-1}, 
\end{equation}
\begin{equation}\label{eq:refractive_scale}
r_{ref} \sim \theta_{scat}~D_{scat} ~ \sim \lambda ^{2}, 
\end{equation}
where $\theta_{scat} \sim \lambda^2$ gives the angular size of the scattered image of \sgra assuming a Gaussian scattering kernel. In the regime where $r_{ref} > r_{dif}$ strong scattering occurs. When $r_{dif}$ and $r_{ref}$ become comparable, refractive noise (which introduces stochastic changes) needs to be considered when imaging \sgra \citep{johnson:2018}.

In this work we use the \code{Python} library \code{eht-imaging} \citep{chael:2016} and its module \code{stochastic-optics} \citep{johnson:2016} 
to simulate the effects of the scattering screen using a scattering model that is physically motivated by observations of \sgra (\citealt{johnson:2018, psaltis:2018, issaoun:2019_size}).

%
\section{Results}~\label{sec:results}
Here we present flux density maps of \sgra for our disk and jet emission models (Sect.~\ref{sec:model_flux_maps}), we make core-shift predictions for different models without scattering (Sect.~\ref{sec:coreshift_calculations}) and then include the effects of interstellar scattering on core-shift modelling (Sect.~ \ref{sec:adding_scattering}). In addition,
we investigate the effects of changing the inclination angle and the position angle of \sgra (Sect.~\ref{sec:position_angle_changes}).

\subsection{Model Radio Maps of Sagittarius~A*}~\label{sec:model_flux_maps}
We study 14 radiative transfer models with various combinations of parameters as summarized 
in Table~\ref{tab:model_params}. We change the inclination angle ($i$) with respect to the observer's line-of-sight 
and the proton-to-electron coupling constants ($\rhi$ \& $\rlo$). We generate maps of the flux density ($S_\nu$) 
for each model at three observing frequencies (22~GHz, 43~GHz \& 86~GHz) and at seven inclination 
angles ($i=1 \degr$, 15$\degr$, 30$\degr$, 45$\degr$, 60$\degr$, 75$\degr$ \& 90$\degr$). 
We keep the position angle on the sky of the BH spin-axis at $PA=0\degr$. 
Later in this paper, we create maps where the orientation angle is rotated to $PA=45\degr$ 
and to $PA=90\degr$ East of North (see Section~\ref{sec:position_angle_changes}).

\begin{table*}
\centering
\tiny
\begin{tabular}{c c c c c c c c c c}
\toprule
\toprule
Model & ID & $i$ & $R_{low}$ & $R_{high}$ & $\mathcal{M}$  & $\dot{M}$  	&$S_{86GHz}$  & $S_{43GHz}$  & $S_{22GHz}$ \\
...         & ...  & [$\degr$]   & ...   & ...  & [g] & [$\msun$~/yr] & [Jy] & [Jy] & [Jy] \\  \addlinespace[2pt]      
\midrule
Jet  & J1    & 1   & 1  & 20   & $1.44\times 10^{20}$ 	    & $1.62\times 10^{-7}$  & 2.00  & 1.80	& 1.84   \\ \addlinespace[2pt]
Jet  & J15   & 15  & 1  & 20   & $1.89\times 10^{20}$		& $2.12\times 10^{-7}$  & 2.00  & 1.51	& 1.28   \\ \addlinespace[2pt]
Jet  & J30   & 30  & 1  & 20   & $2.34\times 10^{20}$		& $2.63\times 10^{-7}$  & 2.00  & 1.35	& 0.87   \\ \addlinespace[2pt]
Jet  & J45   & 45  & 1  & 20   & $2.82\times 10^{20}$		& $3.17\times 10^{-7}$  & 2.00  & 1.26	& 0.82   \\ \addlinespace[2pt]
Jet  & J60   & 60  & 1  & 20   & $3.51\times 10^{20}$		& $3.94\times 10^{-7}$  & 2.00  & 1.23	& 0.70   \\ \addlinespace[2pt]
Jet  & J75   & 75  & 1  & 20   & $4.44\times 10^{20}$		& $4.99\times 10^{-7}$  & 2.00  & 1.31	& 0.77   \\ \addlinespace[2pt]
Jet  & J90   & 90  & 1  & 20   & $4.98\times 10^{20}$		& $5.60\times 10^{-7}$  & 2.00  & 1.35	& 0.74   \\ 
\midrule
Disk  & D1    & 1   & 3  & 3   & $1.54\times 10^{18}$		& $1.73\times 10^{-9}$  & 2.00  & 1.65	& 0.73   \\ \addlinespace[2pt]
Disk  & D15   & 15  & 3  & 3   & $1.55\times 10^{18}$		& $1.74\times 10^{-9}$  & 2.00  & 1.66	& 0.71   \\ \addlinespace[2pt]
Disk  & D30   & 30  & 3  & 3   & $1.55\times 10^{18}$		& $1.74\times 10^{-9}$  & 2.00  & 1.56	& 0.66   \\ \addlinespace[2pt]
Disk  & D45   & 45  & 3  & 3   & $1.57\times 10^{18}$		& $1.76\times 10^{-9}$  & 2.00  & 1.39	& 0.59   \\ \addlinespace[2pt]
Disk  & D60   & 60  & 3  & 3   & $1.66\times 10^{18}$		& $1.87\times 10^{-9}$  & 2.00  & 1.18	& 0.49   \\ \addlinespace[2pt]
Disk  & D75   & 75  & 3  & 3   & $1.94\times 10^{18}$		& $2.18\times 10^{-9}$  & 2.00  & 0.93	& 0.39   \\ \addlinespace[2pt]
Disk  & D90   & 90  & 3  & 3   & $2.51\times 10^{18}$		& $2.83\times 10^{-9}$  & 2.00  & 0.78	& 0.35   \\ 
\bottomrule
\addlinespace[5pt]
\end{tabular}
\caption{ \textbf{Model parameters}. The observer's inclination angle in degrees ($i$) 
is the angle between the observer's line-of-sight and the black hole spin axis; 
the coupling constants $R_{low}$ and $R_{high}$ describe the proton-to-electron coupling in strongly
and weekly magnetized plasma; the mass unit ($\mathcal{M}$) and mass accretion rate ($\dot{M}$) 
scale the total flux density to observed values at 86~GHz. The total integrated flux densities at each frequency ($S_{86GHz}$, $S_{43GHz}$ and $S_{22GHz}$) are given in Jansky.}\label{tab:model_params}
\end{table*}

\begin{figure*}
\caption{\textbf{Flux Density ($S_\nu$) maps of disk models}. Rows of panels show disk models with
 different inclinations ($i=1 \degr$, 30$\degr$, 60$\degr$ \& 90$\degr$, 
 \textit{top to bottom}). The position angle on the sky of the BH spin-axis is $PA=0 \degr$ for all models. 
 Columns show the maps at different frequencies (22~GHz, 43~GHz \& 86 GHz, \textit{left to right}). 
 The field of view for each panel is 120~$\times$~120~$GM/c^2$ (645~$\times$~645~$\mu$as). The bottom colorbar 
 indicates the flux density in square-root scale, the y-axis shows the distance from the BH  
 (located at 0,0) in units of $GM/c^2$ and the top x-axis shows that same distance in $\mu$as. The green dot represents the position of 
 the intensity weighted centroid.}\label{fig:flux_density_disk_models}
\centering
\includegraphics[width=0.27\textwidth]{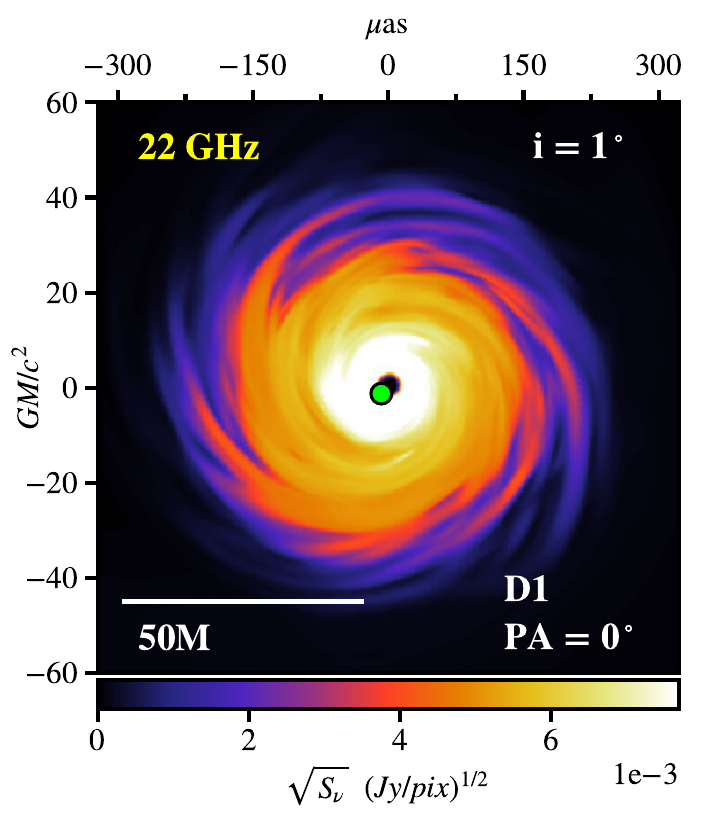}
\includegraphics[width=0.27\textwidth]{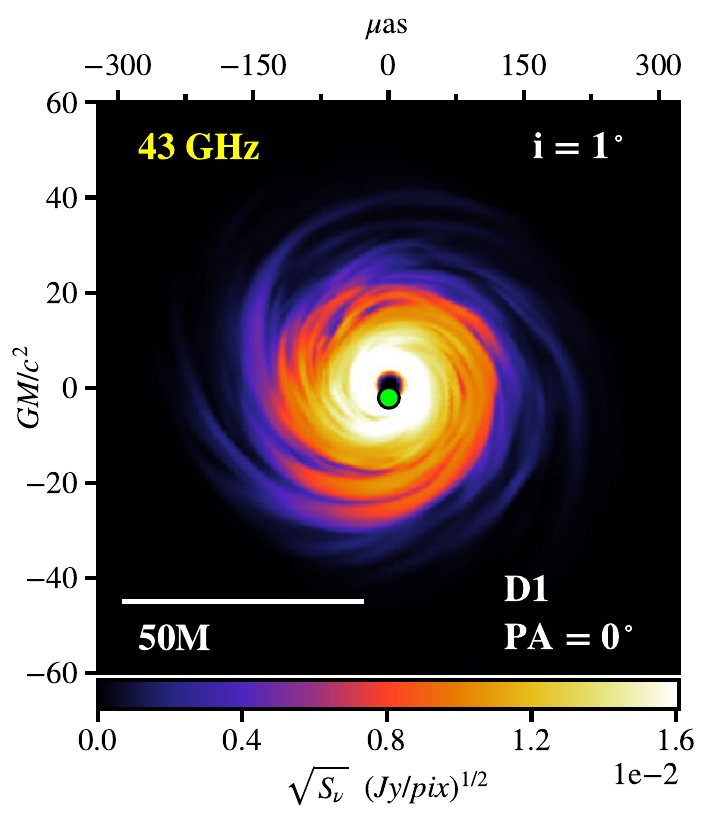}
\includegraphics[width=0.27\textwidth]{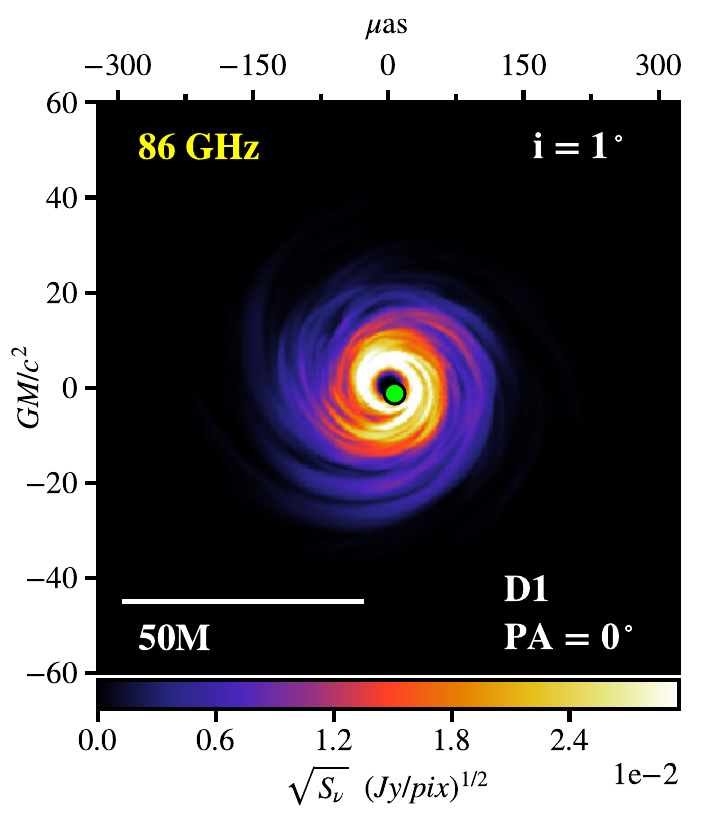}\\
\includegraphics[width=0.27\textwidth]{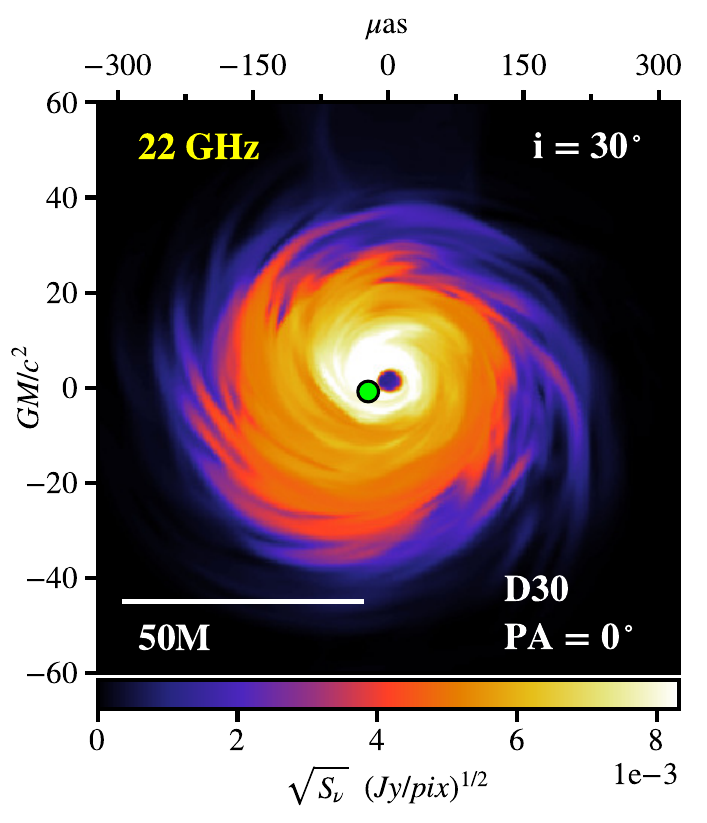}
\includegraphics[width=0.27\textwidth]{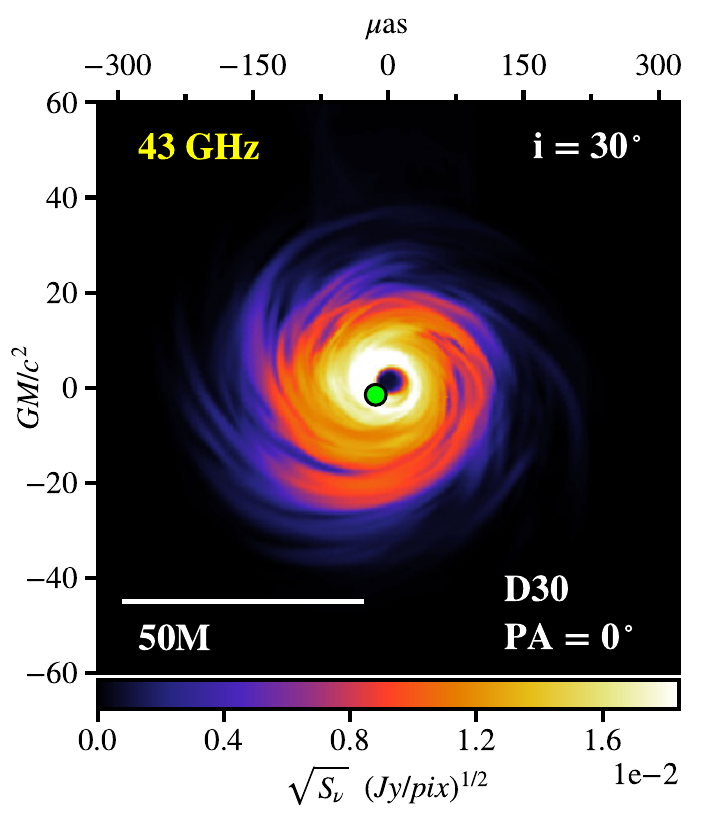}
\includegraphics[width=0.27\textwidth]{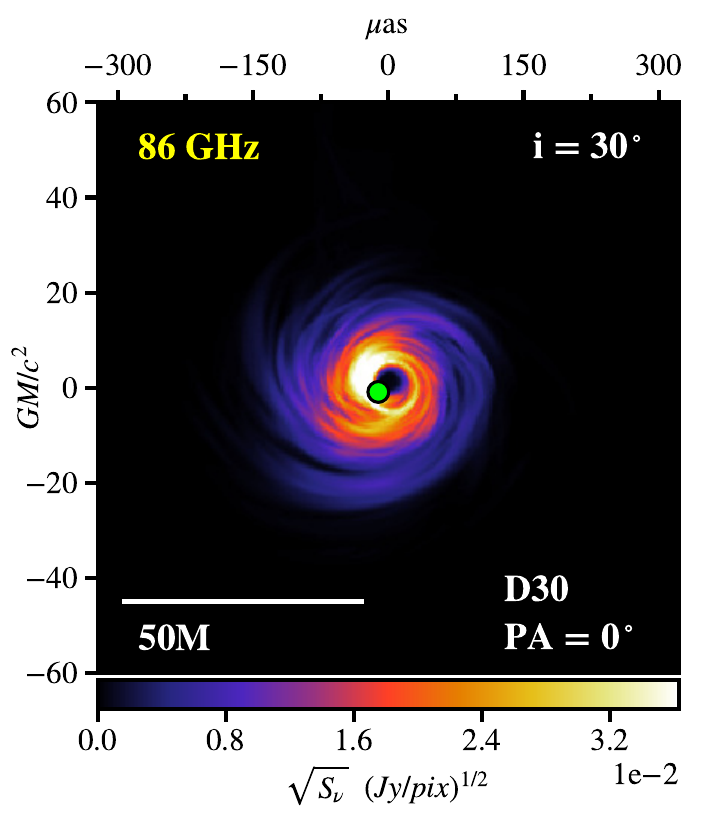}\\
\includegraphics[width=0.27\textwidth]{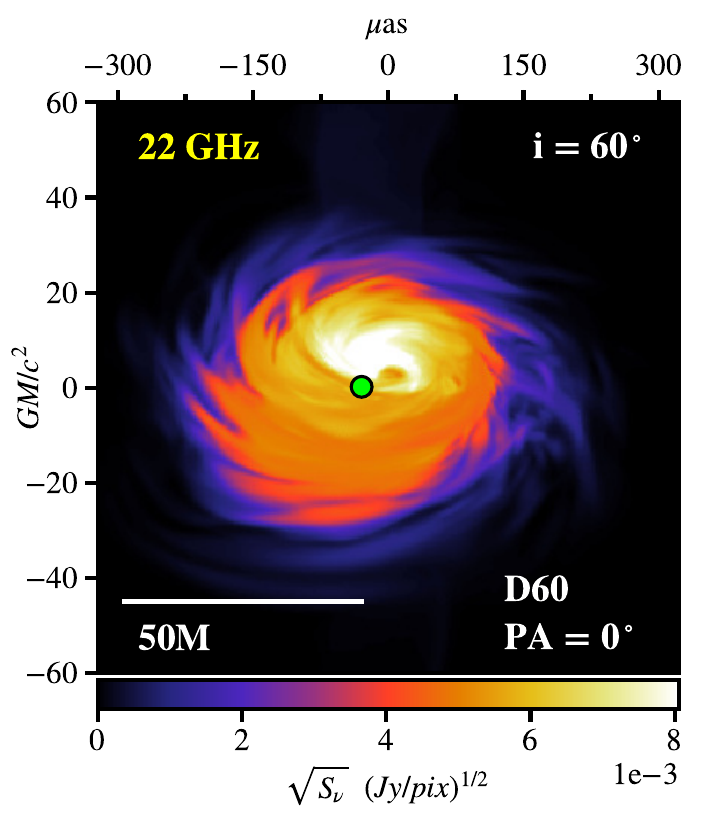}
\includegraphics[width=0.27\textwidth]{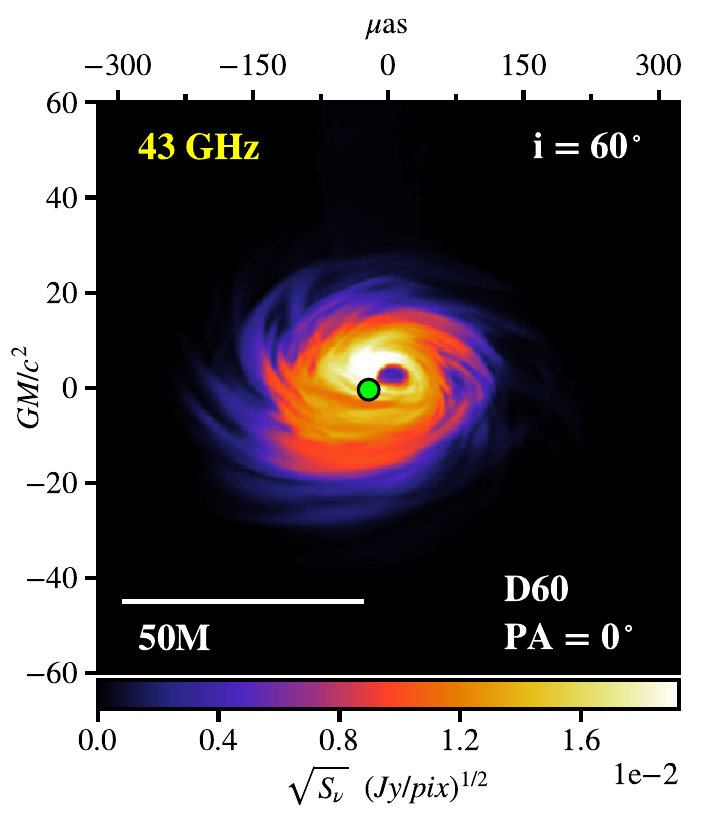}
\includegraphics[width=0.27\textwidth]{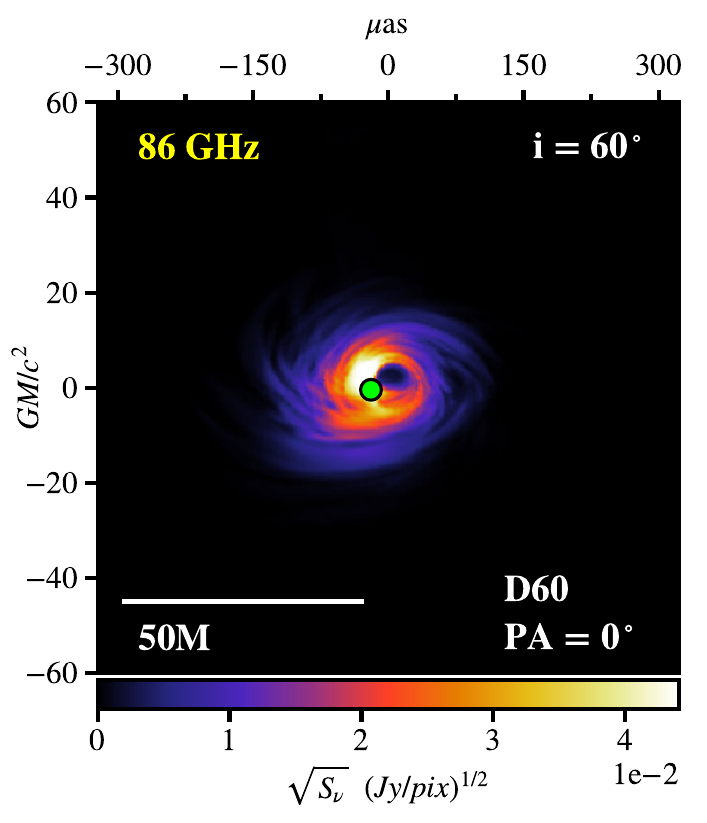}\\
\includegraphics[width=0.27\textwidth]{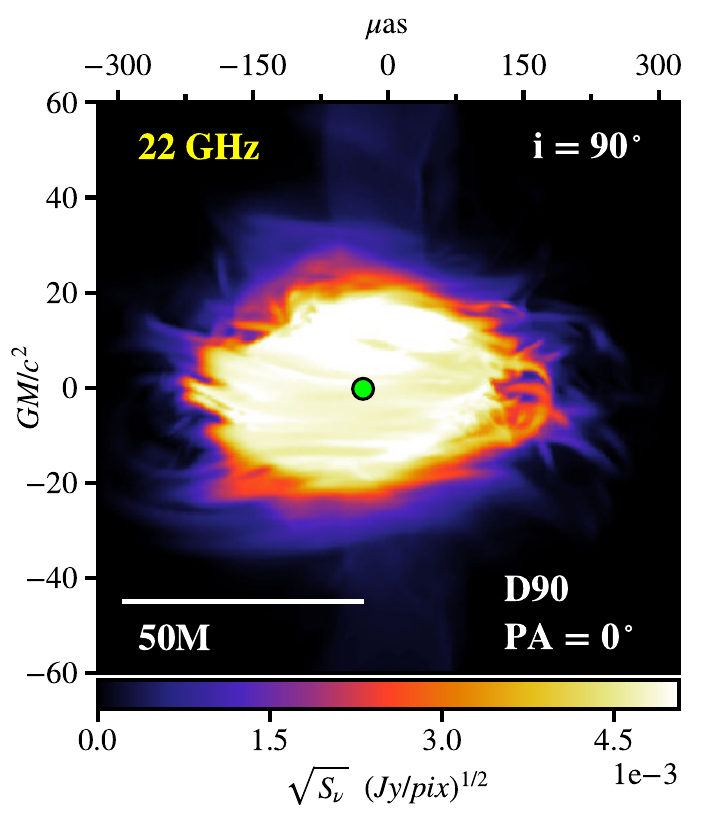}
\includegraphics[width=0.27\textwidth]{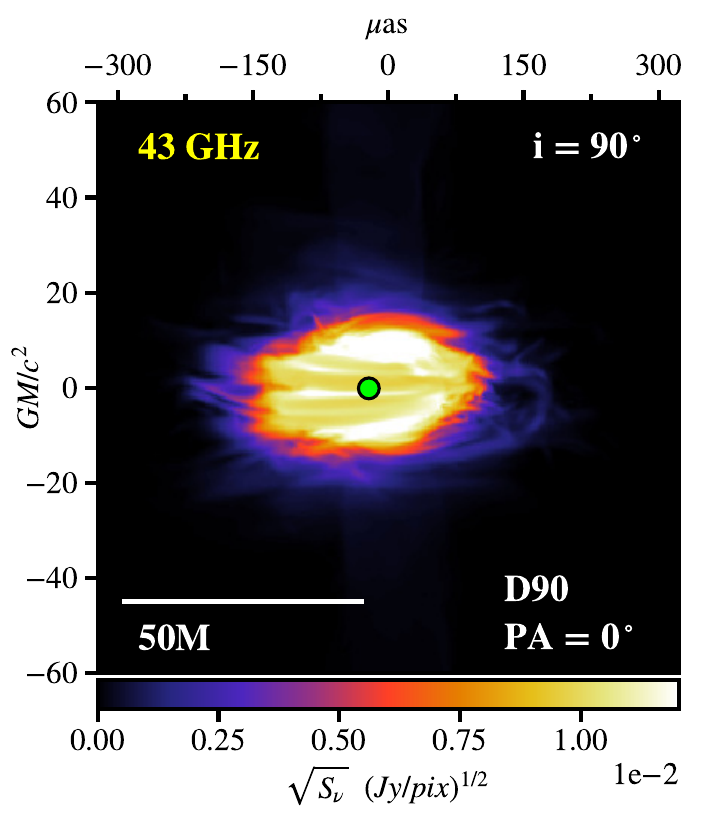}
\includegraphics[width=0.27\textwidth]{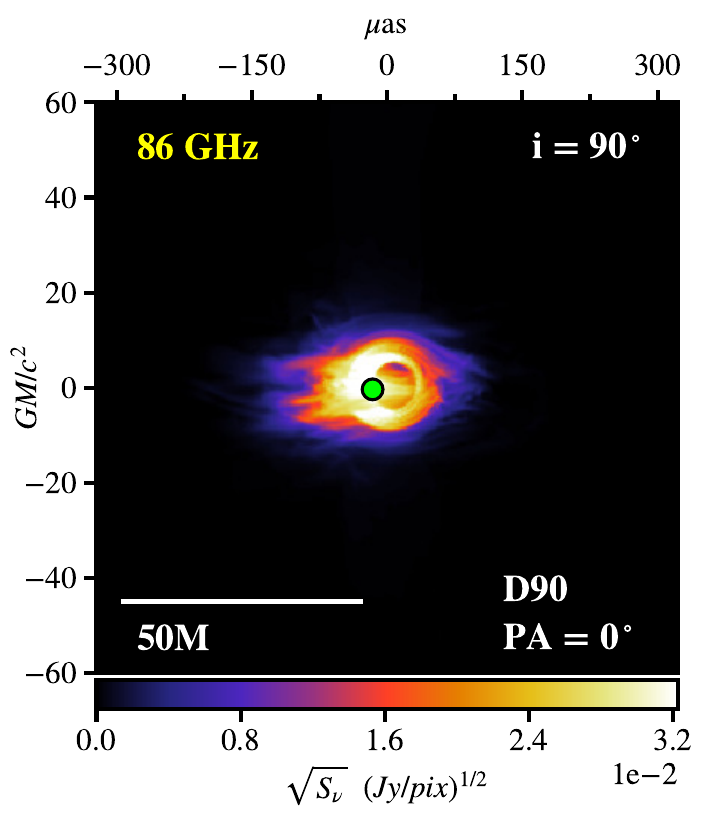}\\
\end{figure*}

\begin{figure*}
\caption{\textbf{Flux Density ($S_\nu$) maps of jet models}. Rows show jet models with
 different inclinations with respect to the observer's line of sight ($i=1 \degr$, 30$\degr$, 60$\degr$ \& 90$\degr$, 
 \textit{top to bottom}). The position angle on the sky of the BH spin-axis is $PA=0 \degr$ for all models. 
 Columns show the maps at different frequencies (22~GHz, 43~GHz \& 86 GHz, 
 \textit{left to right}). The field of view for each panel is 300~$\times$~300~$GM/c^2$ (1600~$\times$~1600~$\mu$as). 
 The bottom colorbar indicates the flux density in square-root scale, the y-axis shows the distance 
 from the BH (located at 0,0) in units of $GM/c^2$ and the top x-axis that same distance in $\mu$as. The green dot represents the position 
 of the intensity weighted centroid.}\label{fig:flux_density_jet_models}
\centering
\includegraphics[width=0.27\textwidth]{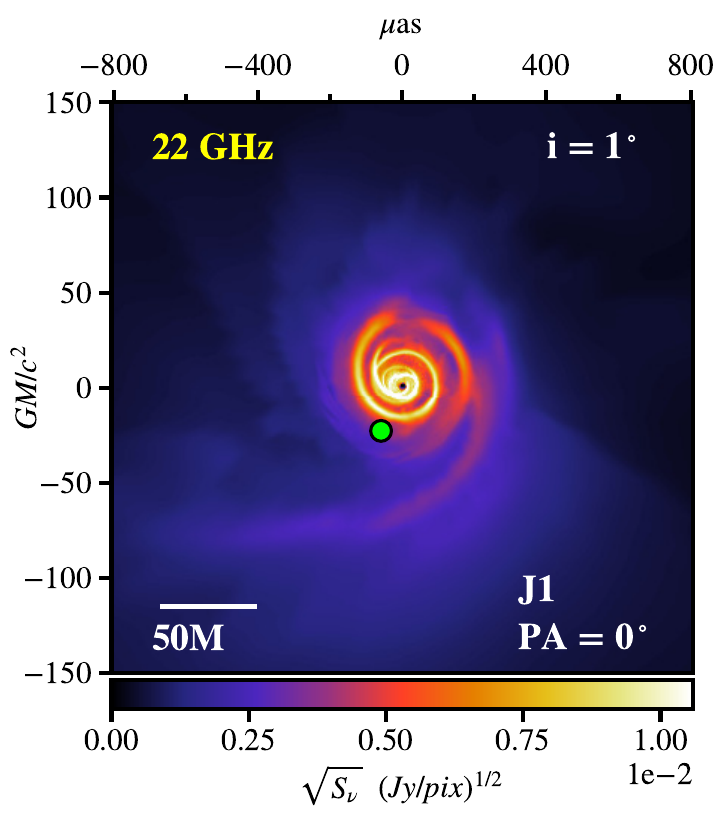}
\includegraphics[width=0.27\textwidth]{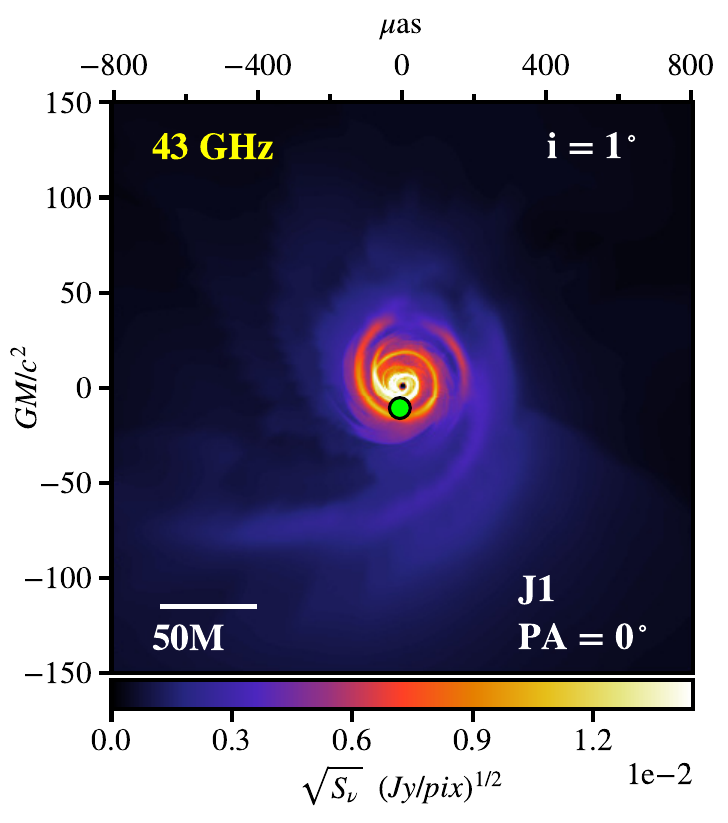}
\includegraphics[width=0.27\textwidth]{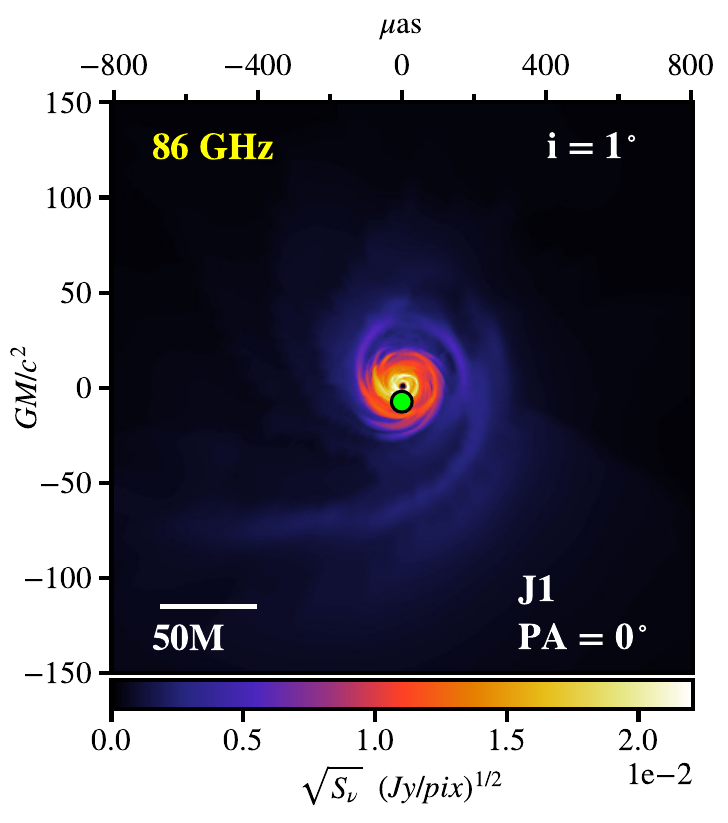}\\
\includegraphics[width=0.27\textwidth]{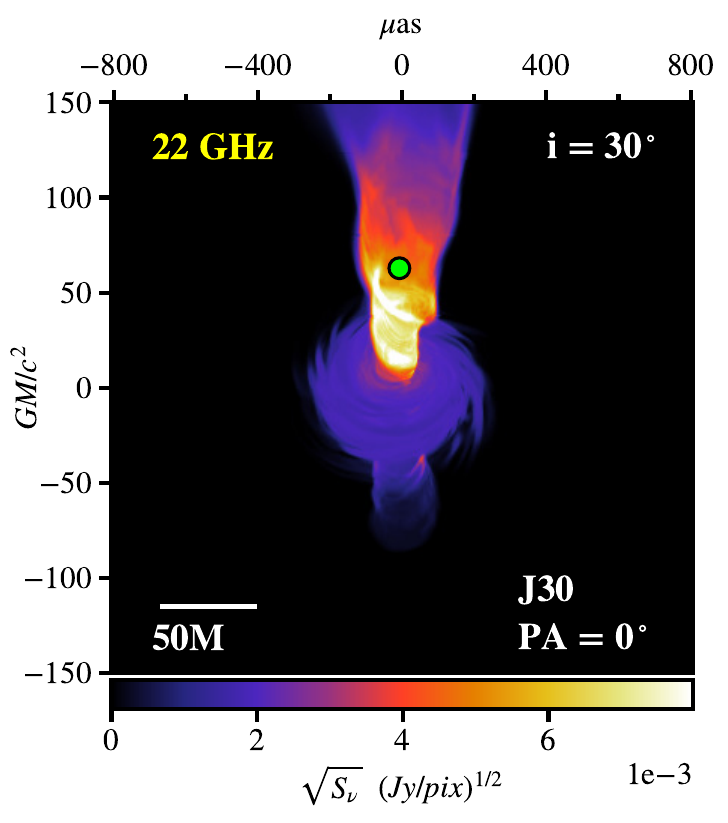}
\includegraphics[width=0.27\textwidth]{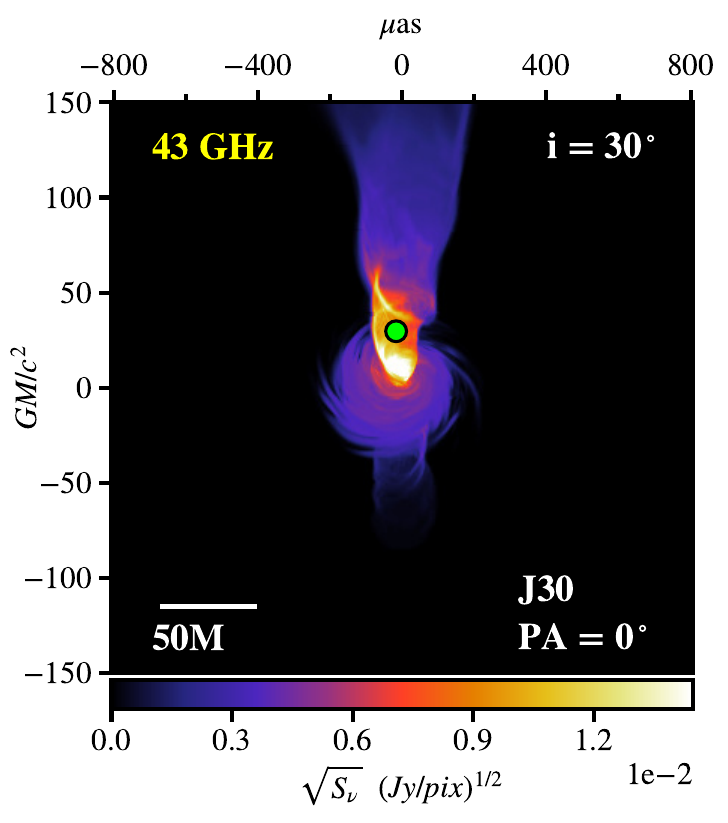}
\includegraphics[width=0.27\textwidth]{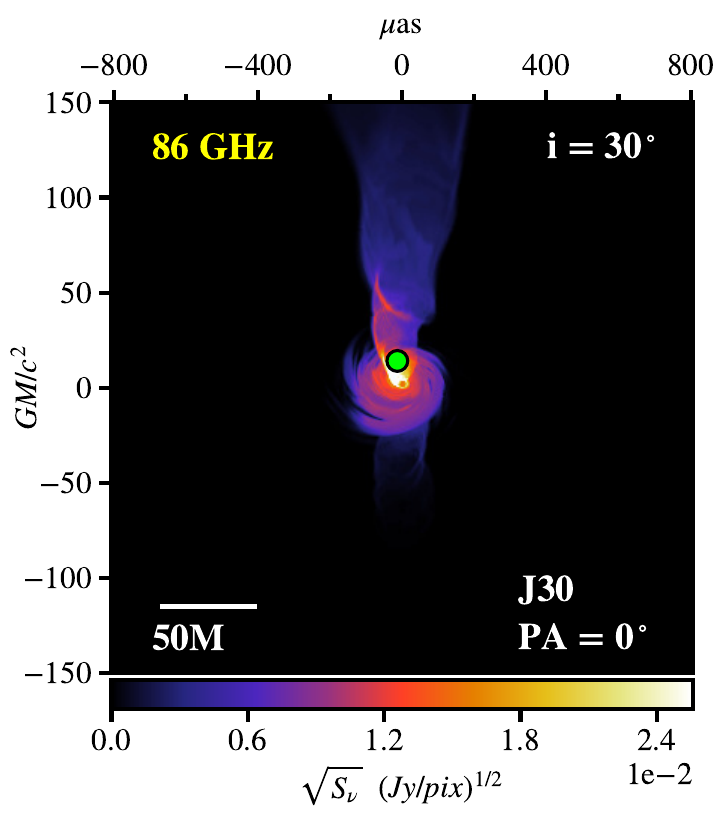}\\
\includegraphics[width=0.27\textwidth]{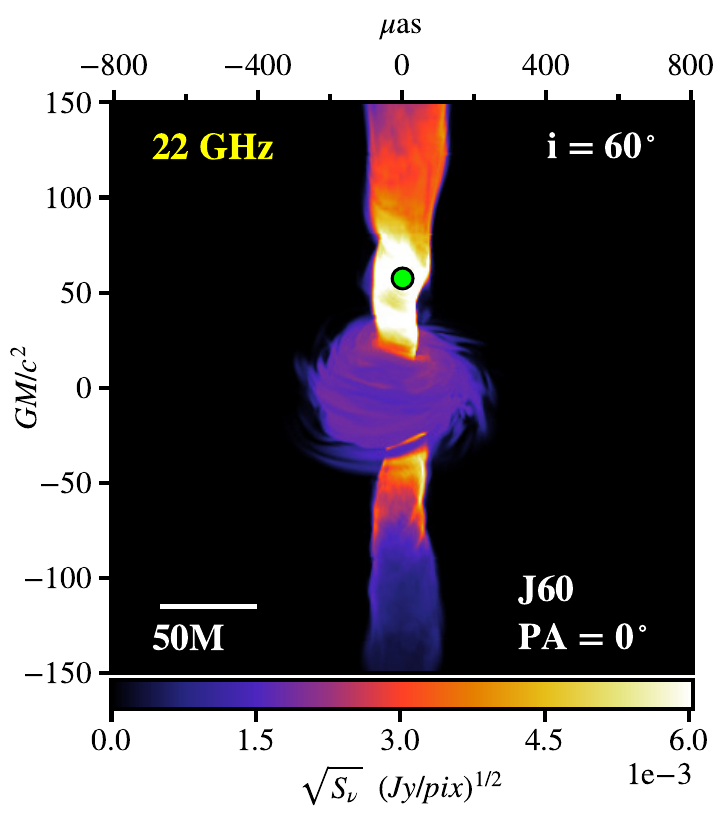}
\includegraphics[width=0.27\textwidth]{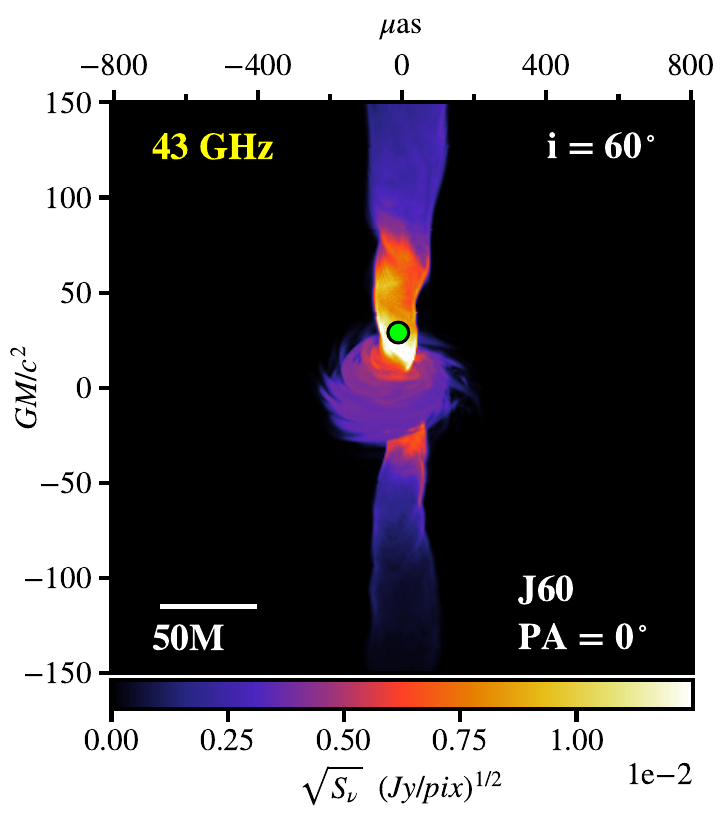}
\includegraphics[width=0.27\textwidth]{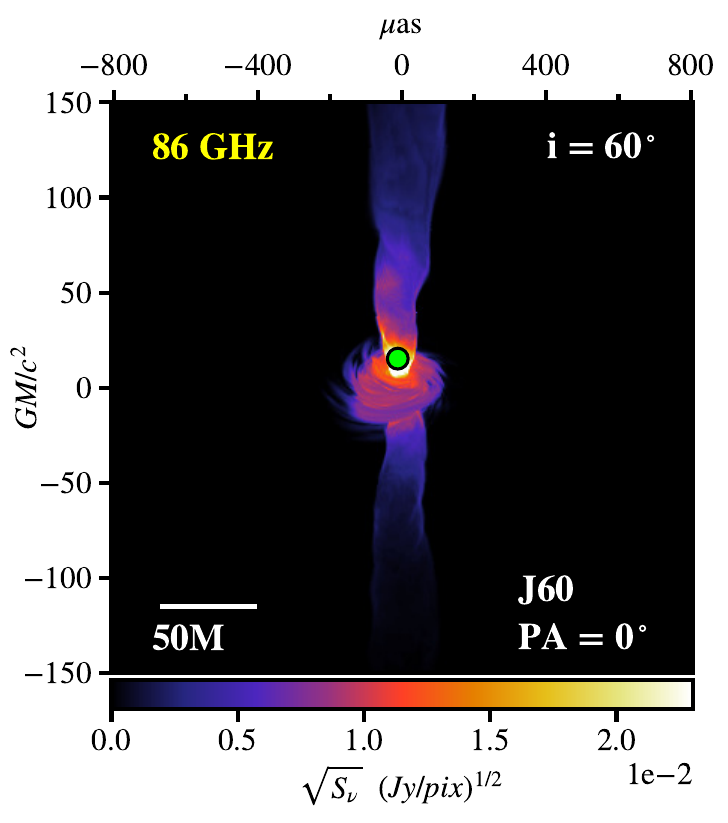}\\
\includegraphics[width=0.27\textwidth]{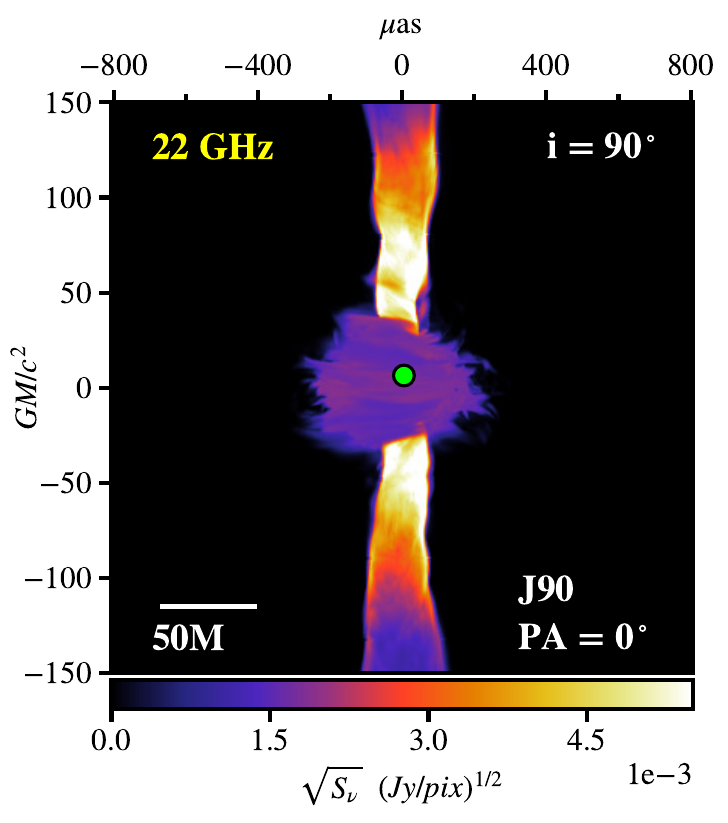}
\includegraphics[width=0.27\textwidth]{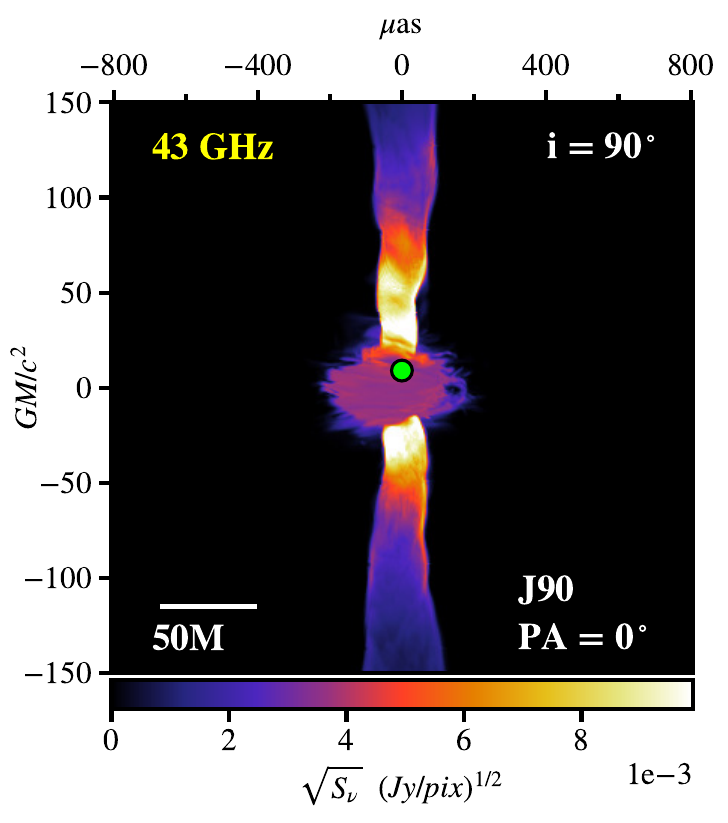}
\includegraphics[width=0.27\textwidth]{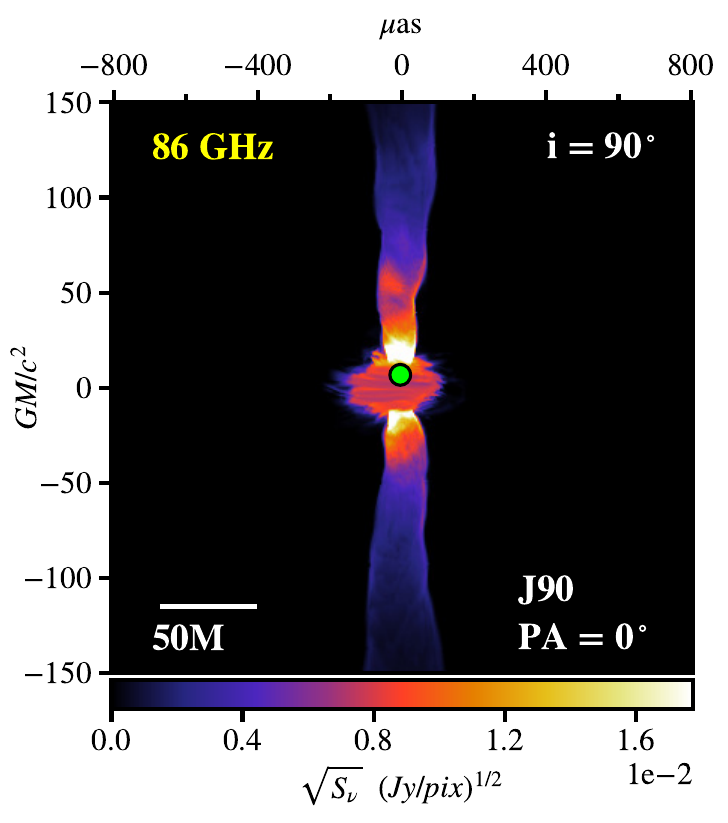}\\
\end{figure*}

The resulting flux density maps are presented in \fig\ref{fig:flux_density_disk_models} for disk models, and 
in \fig\ref{fig:flux_density_jet_models} for jet models. The inclinations shown are in
steps of $\Delta i =$~30$\degr$. Nevertheless, for the sake of completeness, additional figures for intermediate 
inclinations ($i=$15$\degr$, 45$\degr$ \& 75$\degr$) are included in Appendix~\ref{sec:unscattered_models_library}. 
The radio maps in \fig\ref{fig:flux_density_disk_models} and \fig\ref{fig:flux_density_jet_models} 
show the \textit{intrinsic} image of \sgra, angular broadening due to interstellar scattering is thus far not included. We choose to show the square-root of the flux density rather than the flux density in a linear scale so that low flux, faint regions can be clearly discerned.

For each emission model at 43~GHz \& 86~GHz we create a snapshot of an area in the Galactic Center with dimensions $300\times 300 ~R_g$, where the gravitational radius corresponds to a size of $R_g=6.6 \times 10^{11}~ {\rm cm}$ for \sgra. The field of view (FOV) is doubled to $600 \times 600~R_g$ for models at 22~GHz so extended emission at large scales in the jet models can be captured in the maps. In addition, to study the effects of scattering at 22~GHz large FOVs are required due to the large angular broadening produced by the scattering kernel at this frequency. FOV size and resolution are summarized in Table~\ref{tab:fields_view}.

\begin{table}
\centering
\tiny
\setlength\tabcolsep{5pt}
\begin{tabular}{c c c c c}
\toprule
\toprule
$FOV$	                         &$FOV$                          &$FOV$	                            &$\theta$	         &$\nu$  \\ 
$[Gm/c^2]$                 &[pix]                                &[$\mu$as]                      &[$\mu$as/pix]  &[GHz] \\  \addlinespace[2pt]
\midrule
$600 \times 600$		 &$3070 \times 3070$     &$6442 \times 6442$   	&2.1                    &22 \\  \addlinespace[2pt]
$300 \times 300$		 &$1535 \times 1535$     &$3221 \times 3221$   	&2.1                    &43/86 \\
\bottomrule
\addlinespace[5pt]
\end{tabular}
\caption{ \textbf{Fields of View}. This table summarizes the Fields of View ($FOV$) and resolution ($\theta$) of the emission model maps at different frequencies ($\nu$) generated by the radiative transfer scheme \code{IPOLE}.}\label{tab:fields_view}
\end{table}

As expected, the appearance of the emission models depends strongly on the assumed electron temperature. In \fig\ref{fig:flux_density_disk_models} the emission from \sgra is dominated by the disk due to the strong electron-proton
coupling in all regions, while in \fig\ref{fig:flux_density_jet_models} emission is dominated by the jets due to 
strong electron-proton coupling occurring in the jet regions.  

As we move from top row to bottom row in \fig\ref{fig:flux_density_disk_models} and \fig\ref{fig:flux_density_jet_models}, the inclination angle $i$ changes from almost completely face-on ($i=1\degr$) to completely edge-on ($i=90\degr$). An interesting feature starts to emerge for disk models with high inclination when observed at 86~GHz 
(\fig\ref{fig:flux_density_disk_models}, bottom right). Although the disk is viewed edge-on, its appearance reveals a distinct ring-like structure. The extreme gravitational field generated by the black hole bends the path followed by the photons radiated by the disk, so light coming from the back side of the accretion disk can be seen by the observer. \sgra acts as a gravitational lens that disturbs the geometry of the emission. The noticeable ring is the photon orbit and the central darker area is the shadow of the black hole. The left side of the disk is brighter, as indicated by the intensity centroid (green dot) being displaced from the geometrical center of the map. See Sect.~\ref{sec:coreshift_calculations} for a full description of centroid calculations. As matter swirls around the SMBH at relativistic speeds, the plasma on that side of the disk moves towards the observer and the radio emission is Doppler boosted thus appearing brighter. 

Jet models (\fig\ref{fig:flux_density_jet_models}) show a different morphology. The accretion disk appears faint compared to emission coming from the polar jets. At intermediate inclinations, emission from the jet pointing towards the observer (which is doppler boosted) dominates over the emission coming from the counter-jet (pointing away) and emission originating from the disk. At $i=90\degr$, the emission is bright for both jets above and below the dim disk. In these jet models we cannot discern the boundary of the event horizon even at the highest observing frequency we calculate here because they have a higher optical depth.

%
\subsection{Core-shift Modelling without Scattering}~\label{sec:coreshift_calculations}
We obtain the radio core positions at each frequency using the following method. First, we calculate the intensity-weighted centroid of each map (i.e., the image "first moment") as indicated by the green dots in Fig.~\ref{fig:flux_density_disk_models} and Fig.~\ref{fig:flux_density_jet_models}. The centroid coordinates $(x_{c}, y_{c})$ are given by:
\begin{equation}\label{eq:weighted_centroid}
\begin{array}{lcl} 
x_{c} = \ltfrac{\sum_{i=1}^{i=i_{max}} \sum_{j=1}^{j=j_{max}} x_i ~ S_{i,j}}{\sum_{i=1}^{i=i_{max}} \sum_{j=1}^{j=j_{max}}  S_{i,j}}   
&  ~ , ~ ~  
y_{c} = \ltfrac{\sum_{i=1}^{i=i_{max}} \sum_{j=1}^{j=j_{max}} y_j ~ S_{i,j}}{\sum_{i=1}^{i=i_{max}} \sum_{j=1}^{j=j_{max}}  S_{i,j}}   
\end{array}
\end{equation}
where $(x_i, y_j)$ is the location of a given pixel, $S_{i,j}$ is the value of the flux density in that pixel, and the dimensions of the map in pixels are $(i_{max} \times  j_{max})$ . The errors in the intensity-weighted centroid position ($x_c, y_c$) can be expressed as:
\begin{equation} \label{eq:sigma_xy_cent}
\begin{array}{lcl} 
\sigma _{x_c} =\frac{1}{\sqrt{\sum_{i=1}^{i_{max}} \sum_{j=1}^{j_{max}}   S_{i,j}}}
&  ~ , ~ ~  
\sigma _{y_c} =\frac{1}{\sqrt{\sum_{i=1}^{i_{max}} \sum_{j=1}^{j_{max}}   S_{i,j}}}
\end{array}
\end{equation}
\begin{equation} \label{eq:sigma_centroid}
\sigma _{x_c} = \sigma _{y_c} = \frac{1}{\sqrt{S_{tot}}}
\end{equation}
where $S_{tot}$ is the total flux density of the map.

Next, we calculate the core position relative to the black hole ($r_{core}$), as illustrated in Fig.~\ref{fig:coreshift_geometry}. 
This is the offset between the centroid $(x_{c}, y_{c})$ and the geometrical center of the image $(x_0, y_0)$ 
where the black hole is located: 
\begin{equation}\label{eq:shift}
r_{core} = \sqrt{(x_{c} - x_0 )^2 + (y_{c} - y_0 )^2  } ~. 
\end{equation}

\begin{figure}[ht]
\centering
\includegraphics[width=1.0\columnwidth]{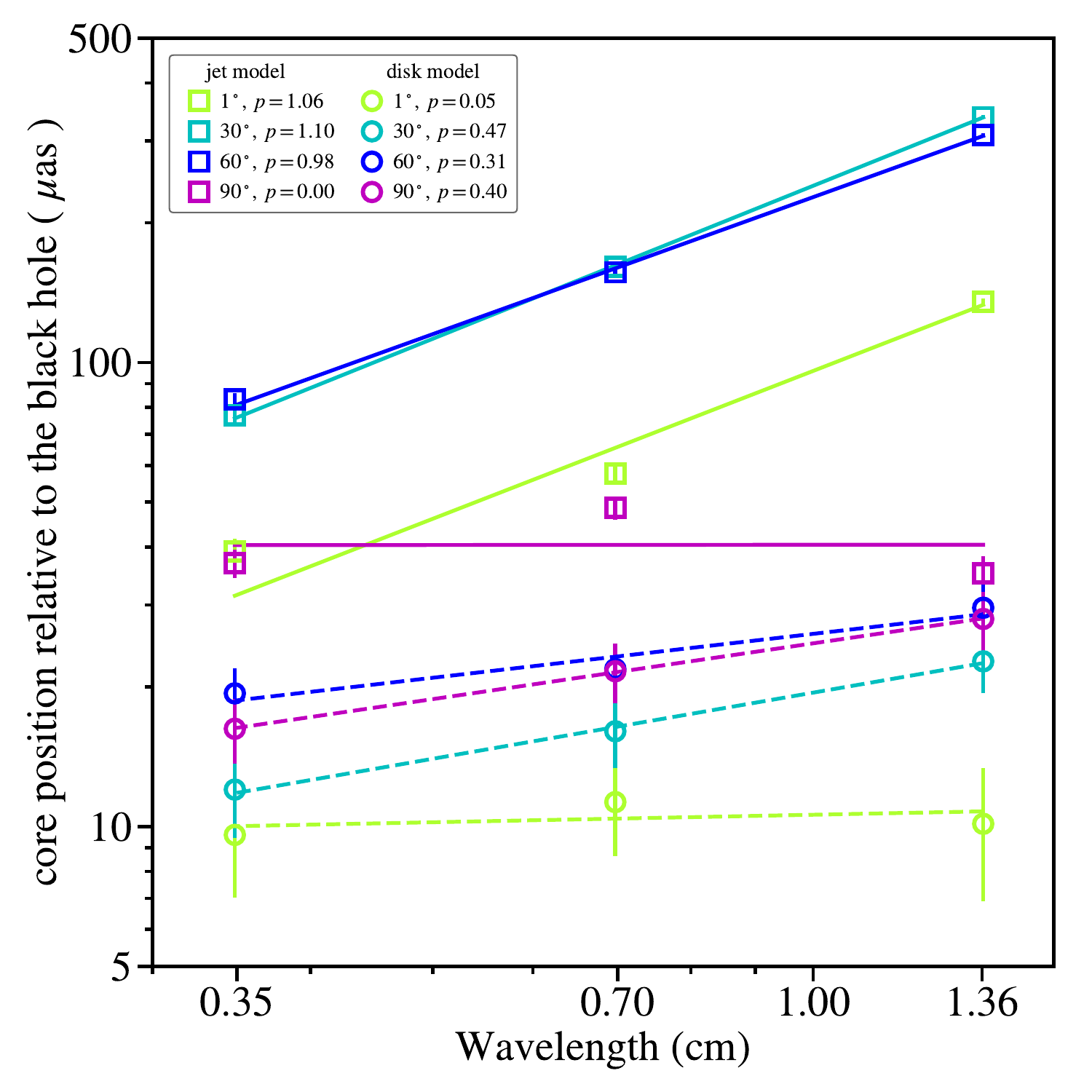}
\caption{\textbf{Core position as a function of wavelength for intrinsic jet \& disk models}. Squares indicate the core
 position relative to the central black hole at 13.6~mm, 7~mm \& 3.5~mm (22/43/86~GHz) for jet models. Colors represent different 
 inclination angles. Circles indicate the core positions for disk models. All models have an orientation of $PA=0\degr$. 
 Solid lines (jets) and dashed lines (disks) show the best-fit power-law using a least-squares fitting method. Error bars
 represent 1$\sigma$ errors in the core position data. Parameters, fitting errors and $\chisq$ are summarized in
 Table~\ref{tab:plaw_params_PA0_models_unscattered}. The legend indicates the inclination ($i$) and power-law index ($p$) 
 for a given model.}\label{fig:plaw_lambda_PA0_models_unscattered}
\end{figure}

The error in the core position $r_{core}$ is:
\begin{equation} \label{eq:sigma_rcore}
\sigma _{r_{core}}=\frac{\Big( (x_c~\sigma _{x_c})^2 + (y_c~\sigma _{y_c})^2 \Big)^{1/2}}{\sqrt{(x_c - x_0 )^2 + (y_c - y_0 )^2  }} ~. 
\end{equation}
Starting with Equation~\ref{eq:core_position_konigl}, we can write the frequency-dependent core position relation to 
be a power-law function in terms of the wavelength $\lambda$ as is commonly used:
\begin{equation}\label{eq:core_position_lambda}
r_{core}~(\lambda)= A~~ \lambda ^{p} ~~,
\end{equation}
where the power-law index is $p=1/k_{r}$~, and the coefficient $A$ indicates the amplitude (i.e., amount) of core-shift. 
For each emission model, we fit a power law to $r_{core}$ at the three 
observing wavelengths 13.6~mm, 7~mm \& 3.5~mm (22~GHz, 43~GHz \& 86~GHz) and calculate $p$, $A$, and their errors using the non-linear least squares fitting method \code{optimize.curvefit} from \code{Python}. These models do not include scattering. Table~\ref{tab:plaw_params_PA0_models_unscattered} summarizes these results. 
Figure~\ref{fig:plaw_lambda_PA0_models_unscattered} shows the best power-law fits for the intrinsic  disk and jet models presented in \fig\ref{fig:flux_density_disk_models} and \fig\ref{fig:flux_density_jet_models}. 

\begin{table}
\tiny
\centering
\setlength\tabcolsep{4pt}
\begin{tabular}{c c c c c c c c}
\toprule
\toprule
Model	 & ID	     &$i$  	     &index $p$  &$\sigma _p$  	&$A$   	           &$\sigma _A$       &$\chi ^2$          \\
...            &...        &[$\degr$] & ...                 & ...              &[$\mu$as/cm]  &[$\mu$as/cm]         & ...  \\
\midrule
Jet		 & J1     & 1			& 1.06		&$\pm$ 0.05  	& 95.99 			&$\pm$ 1.73    			& 2.88 \\ \addlinespace[2pt]
Jet		 & J15	& 15			& 1.13		&$\pm$ 0.02  	& 220.69   		&$\pm$ 1.85  			& 0.19\\ \addlinespace[2pt]
Jet		 & J30	& 30			& 1.10		&$\pm$ 0.02  	& 240.35 			&$\pm$ 1.92 	  		& 0.03\\ \addlinespace[2pt]
Jet		 & J45	& 45			& 1.08		&$\pm$ 0.02  	& 244.51   		&$\pm$ 1.97  			& 0.10\\ \addlinespace[2pt]
Jet		 & J60	& 60			& 0.98		&$\pm$ 0.02 	& 227.20  			&$\pm$ 1.97    			& 0.14\\ \addlinespace[2pt]
Jet		 & J75	& 75			& 0.79		&$\pm$ 0.03  	& 166.37   		&$\pm$ 1.95  			& 0.22\\ \addlinespace[2pt]
Jet		 & J90	& 90			& 0.00		&$\pm$ 0.07  	& 40.45		    	&$\pm$ 2.12    		 	& 2.68\\				
\midrule
Disk	 & D1 	& 1			& 0.05		&$\pm$ 0.29  	& 10.60 			&$\pm$ 2.12   			& 0.13\\ \addlinespace[2pt]
Disk	 & D15	& 15			& 0.60		&$\pm$ 0.29  	& 13.45			&$\pm$ 1.99   	     	& 0.30\\ \addlinespace[2pt]
Disk	 & D30	& 30			& 0.47		&$\pm$ 0.19  	& 19.42   			&$\pm$ 2.05    			& 0.01\\ \addlinespace[2pt]
Disk	 & D45	& 45			& 0.36		&$\pm$ 0.15  	& 23.89 			&$\pm$ 2.15 				& 0.13\\ \addlinespace[2pt]
Disk	 & D60	& 60			& 0.31		&$\pm$ 0.14 	& 26.00  			&$\pm$ 2.28    			& 0.14\\ \addlinespace[2pt]
Disk	 & D75	& 75			& 0.35		&$\pm$ 0.14  	& 26.17 			&$\pm$ 2.44  			& 0.14\\ \addlinespace[2pt]
Disk	 & D90	& 90			& 0.40		&$\pm$ 0.16 	& 24.81   			&$\pm$ 2.51    			& 0.00\\
\bottomrule
\addlinespace[5pt]
\end{tabular}
\caption{ \textbf{Power-law parameters for jet \& disk models}. Results from power-law fits using a least-squares fitting method. 
Values include the power-law index ($p$), its error ($\sigma _p$), the power-law coefficient ($A$) which indicates the amount of core-shift, its error ($\sigma _A$), and the chi-squared value of the fit (~$\chi ^2$). Models are labeled with an identification (ID) according to 
their inclination angle ($i$).}\label{tab:plaw_params_PA0_models_unscattered}
\end{table}

\begin{table}
\centering
\tiny
\setlength\tabcolsep{5pt}
\begin{tabular}{c c c c c c c c}
\toprule
\toprule
$i$~[$\degr$]	                  &1           &15	         &30	        &45  	       &60          &75		     &90 \\
\midrule
 $A_{jet}/A_{disk}$			&9.06     &16.36   	&12.38     &10.23	  &8.74		 &6.38 		&1.63\\
\bottomrule
\addlinespace[5pt]
\end{tabular}
\caption{ \textbf{Ratio of core-shifts}. $A_{jet}/A_{disk}$ indicates the ratio of the power-law coefficient ($A$) between jet and disk models at a given inclination ($i$). Values for the amount of core-shift $A$ are shown in Table~\ref{tab:plaw_params_PA0_models_unscattered}.}\label{tab:A_ratio_mod_PA0}
\end{table}

We can see in \fig\ref{fig:plaw_lambda_PA0_models_unscattered} that the intrinsic core-shifts of \sgra for models with intermediate inclinations ($i=30\degr, 60\degr$) are larger than core-shifts for $i=1\degr$ models or for completely edge-on models ($i=90\degr$). Jet models show a larger amount of core-shift ($A$) and steeper slope ($p$) in the power-law relation than disk models, the former follow quite well a power-law function as seen from the values in Table~\ref{tab:plaw_params_PA0_models_unscattered}. This result agrees with predictions of a displacement in the observed cores of AGN with conical jets \citep{konigl:1981}. \cite{falcke_biermann:1999a} predict values of $p=0.9-1.1$ increasing with inclination. Our results are within the same $p$ range but the slight trend with inclination occurs in the opposite direction, as our $p$ decreases with inclination. This could be due to the jet collimation shape in their analytical work being slightly different from the collimation shape in our 3D GRMHD models. A purely conical jet would yield $p=1$. 

For jet models, we obtain core positions relative to the BH ranging from $r_{core} \sim 30-400~\mu$as. In contrast, disk models show smaller values ranging from $r_{core} \sim 9-25~\mu$as and have much flatter slopes.  If we compare the core-shift of jet models versus disk models, as indicated by the ratio $A_{jet} / A_{disk}$ (see Table~\ref{tab:A_ratio_mod_PA0}), we can see that it is $\sim$2-16 times larger. The core-shift is a clear discriminant between jet-dominated and disk-dominated emission models for \sgra. Therefore, it can be used as a tool to constrain the values of $R_{high}$ and $R_{low}$, which describe the proton-to-electron coupling on the plasma surrounding \sgra.

We should note that even at an inclination of $i=1\degr$ the core-shift is detectable in disk models. At $i=0\degr$ (completely face-on disk) the core-shift should be 0 due to symmetry in the flux density maps, the bright face-on disk would have similar flux densities to the right and left of the disk. This also explains why the value in jet models at $i=90\degr$ (solid magenta line) is so much lower than at intermediate inclinations. At this edge-on inclination the polar jets have similar flux densities above and below the faint accretion disk, which results in the intensity-weighted centroids at all wavelengths being quite close to the geometrical center of the map. Hence, resulting in a much flatter power-law slope and smaller value of $A$. In the jet model at $i=1\degr$ there is a bright arm appearing in the southwest region (see Fig.~\ref{fig:flux_density_jet_models}, top row). We consider this feature to be a temporal enhancement of emission in the snapshot image used. This leads to the centroid position being “pulled” slightly southward, and it explains why at this inclination the power-law fit is worse than at intermediate inclinations (see Table~\ref{tab:plaw_params_PA0_models_unscattered}). The implications of variability in \sgra are discussed in Sect.~\ref{sec:future}.

\begin{table*}
\centering
\tiny
\setlength\tabcolsep{5pt}
\begin{tabular}{c c c c c c c c}
\toprule
\toprule
$\theta _{maj,0}$	  		&$\theta _{minj,0}$			&$PA_G$					&$D_{scat}$	&$R_{scat}$  &$r_{in}$	&$\alpha$		&Reference\\
 $[$mas$]$							&[mas]								&$[\degr]$				&[kpc]				&[kpc]			&[km]			& ...				&  ...  \\
\midrule
1.380~$\pm$~0.013		& 0.703~$\pm$~0.013 		& 81.9~$\pm$~0.2  	& 5.4  				&2.7				&$800$		&$1.38$ 		&\citet{johnson:2018}\\
\bottomrule
\addlinespace[5pt]
\end{tabular}
\caption{ \textbf{Scattering Kernel Parameters}. The following values reproduce the
scattering screen at a reference wavelength of $\lambda _0 \equiv $~1~cm. The sizes along the major \& minor axes of the
anisotropic Gaussian ($\theta_{maj,0}$, $\theta_{min,0}$), position angle ($PA_G$), screen-\sgra distance ($R_{scat}$), screen-observer 
distance ($D_{scat}$), inner scale ($r_{in}$) for the power spectrum of the electron-density fluctuations in the screen, and the index ($\alpha$) of the power-law for that power spectrum produced by turbulence as given by \citet{johnson:2018}. The scattering screen geometry is shown in 
\fig\ref{fig:scatter_screen}.}\label{tab:scat_screen_parameters}
\end{table*}

%
\subsection{Core-shift Modelling including Scattering}~\label{sec:adding_scattering}
 The scattering model produces an anisotropic Gaussian that scales as $\lambda ^2$. This Gaussian can be parametrized 
in terms of the full-width at half-maximum (FWHM) sizes along the major and minor axes ($\theta _{maj,0}$ and 
$\theta _{min,0}$), and the position angle on the sky of the Gaussian ($PA_G$). They are given at a reference wavelength 
$\lambda _0$ of 1 cm. The parameter values of our scattering screen model are summarized in Table~\ref{tab:scat_screen_parameters}. 
The distances from the scattering screen to \sgra  ($R_{scat}$) and from the scattering screen to the observer ($D_{scat}$), as illustrated in Fig.~\ref{fig:scatter_screen} are taken from observations of the Galactic Center. We use the inner scale ($r_{in}$) for the power spectrum of the energy-density fluctuations in the scattering screen, and a power-law index ($\alpha$) for that power-spectrum generated by turbulence in the screen. All values are adopted from \citet{johnson:2018}.

\begin{figure*}
\centering
\includegraphics[width=0.3\textwidth]{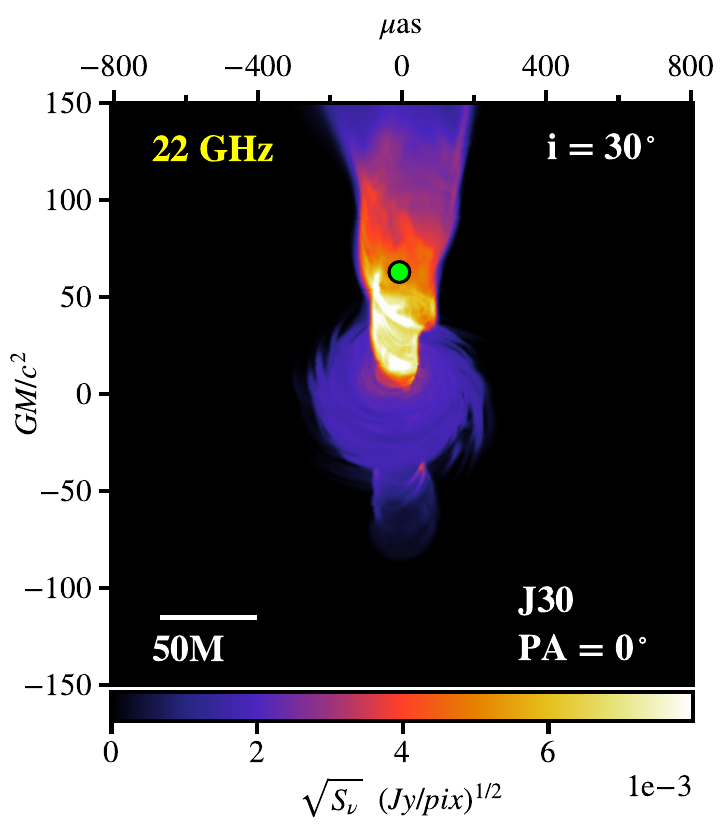}
\includegraphics[width=0.3\textwidth]{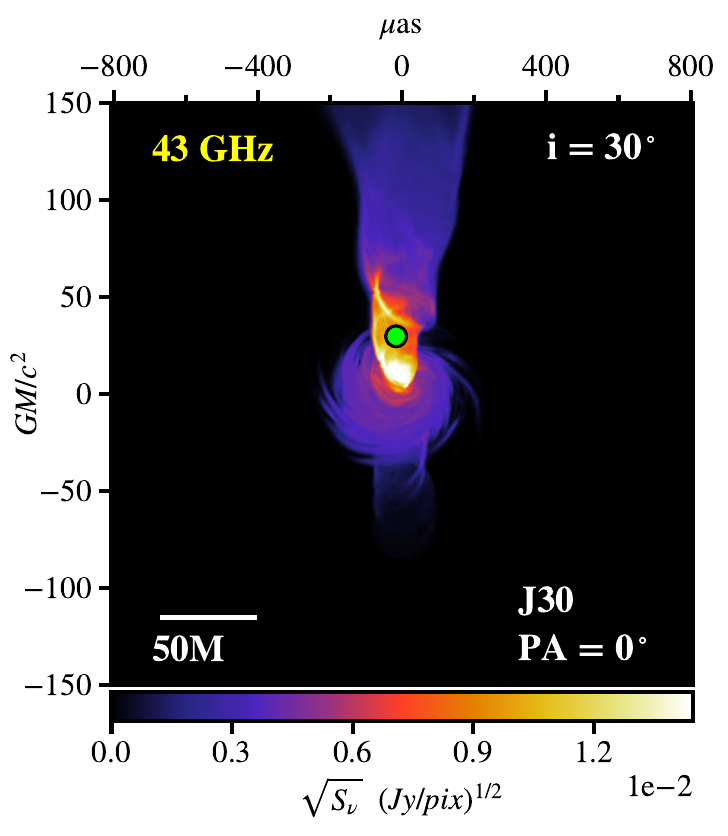}
\includegraphics[width=0.3\textwidth]{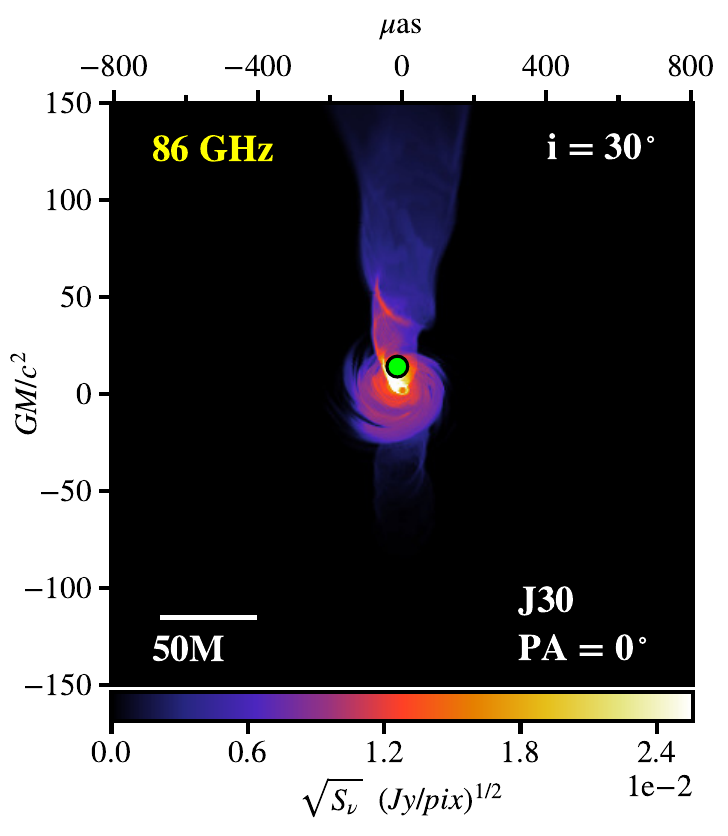}\\
\includegraphics[width=0.3\textwidth]{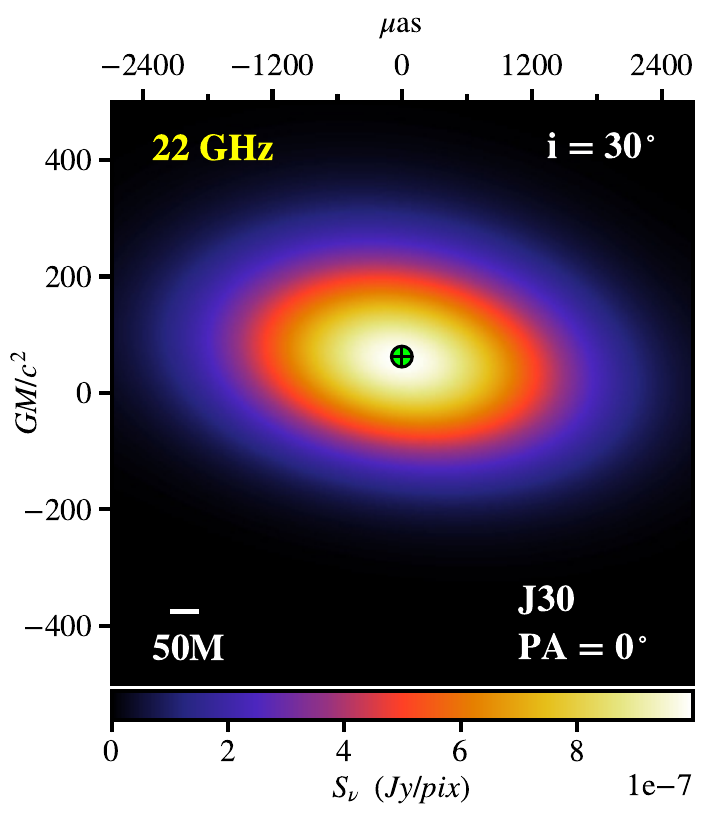}
\includegraphics[width=0.3\textwidth]{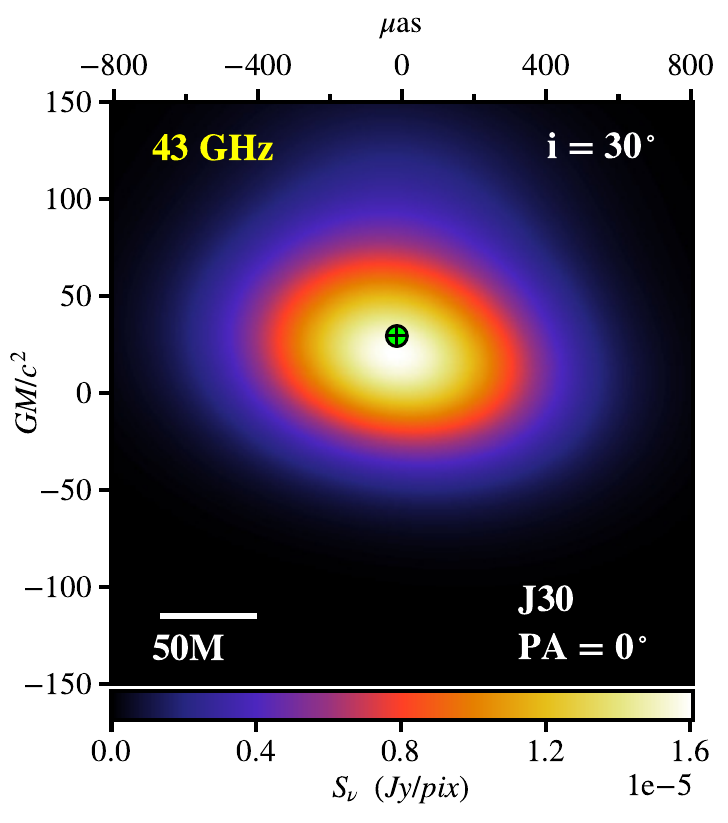}
\includegraphics[width=0.3\textwidth]{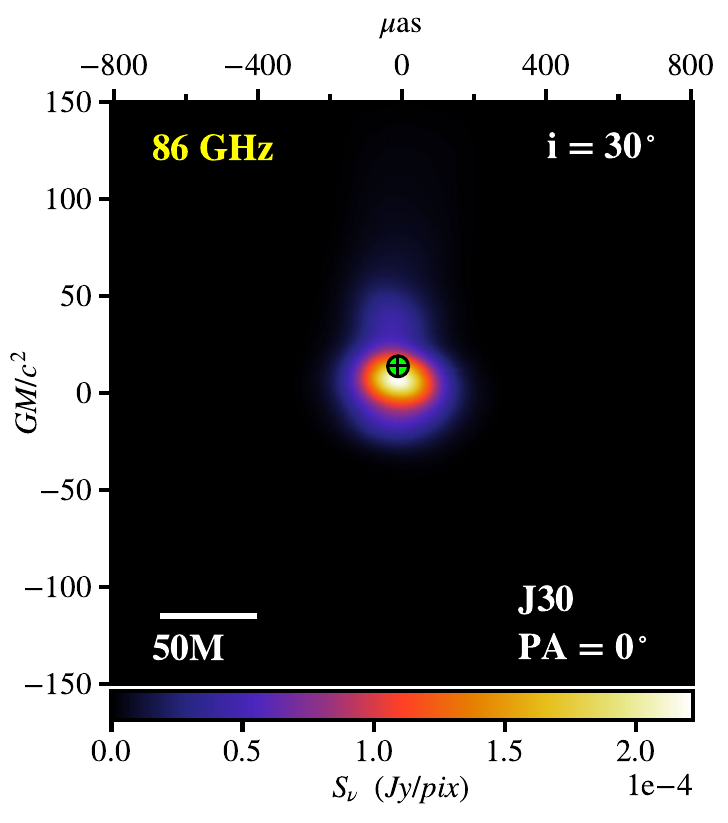}\\
\includegraphics[width=0.3\textwidth]{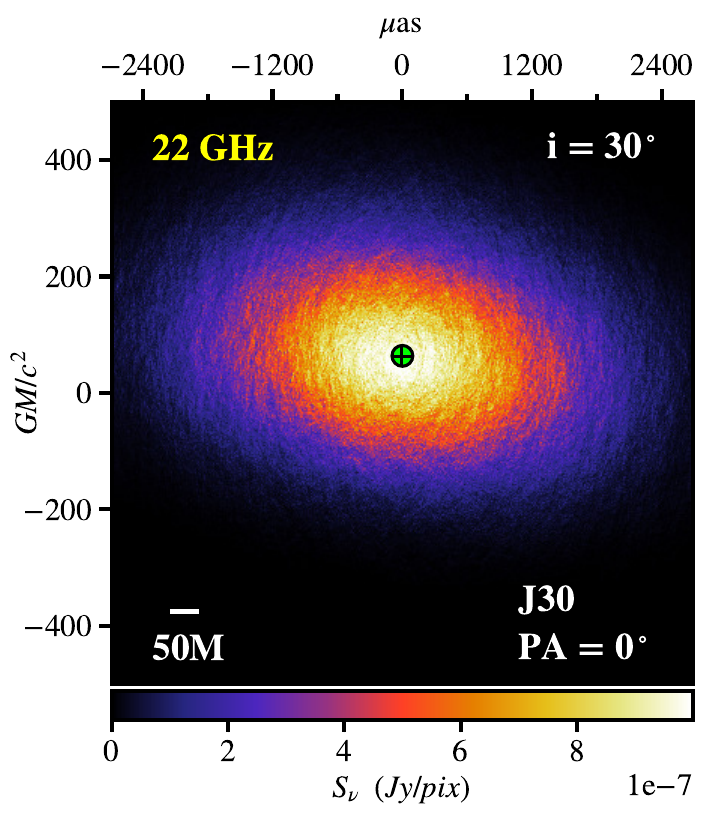}
\includegraphics[width=0.3\textwidth]{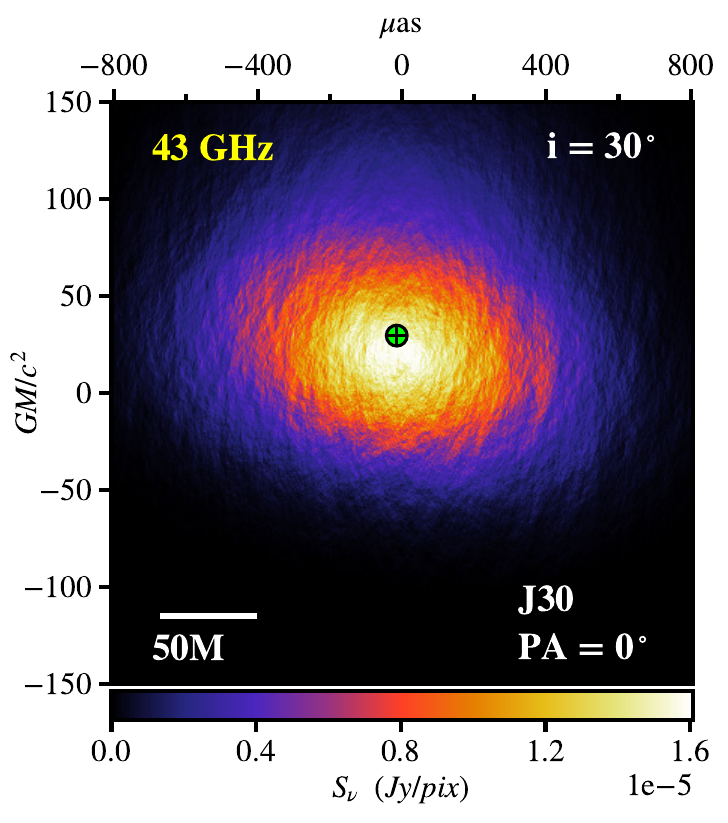}
\includegraphics[width=0.3\textwidth]{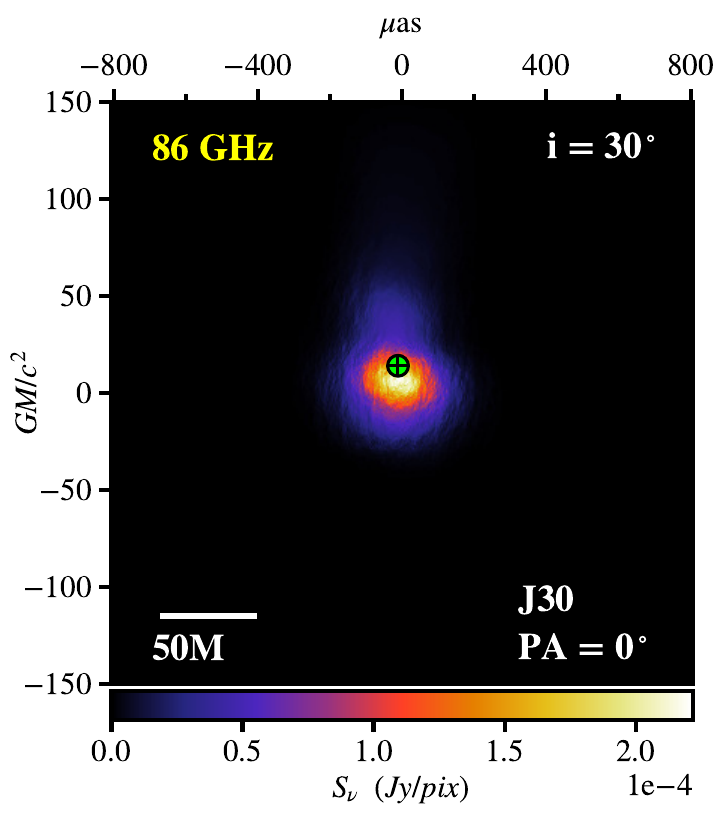}
\caption{\textbf{Jet model (J30) including scattering}. Each column shows a jet model with
 inclination $i=30 \degr$ at 22~GHz, 43~GHz \& 86~GHz 
 (\textit{left to right}). The position angle on the sky is $PA=0 \degr$. 
 Rows from \textit{top to bottom} show the intrinsic 3D-GRMHD jet model, the scatter-broadened image, 
 and the average image including refractive scattering. Note that the FOV for images at 22~GHz including 
 scattering \textit{(left column)} is doubled compared to the FOV at 43~GHz \& 86~GHz. The color stretch is 
 also different. The intrinsic model panels display the square-root of the flux density (\textit{top row}). 
 The scattered maps show a linear scale (\textit{second \& third rows}). The green dot indicates
 the centroid of each image. Black cross-hairs (\textit{second \& third rows}) indicate 
 the location of the centroid of the model \textit{before} scattering.}\label{fig:scattered_jet_model_i30}
\end{figure*}

\begin{figure*}
\centering
\includegraphics[width=0.49\textwidth]{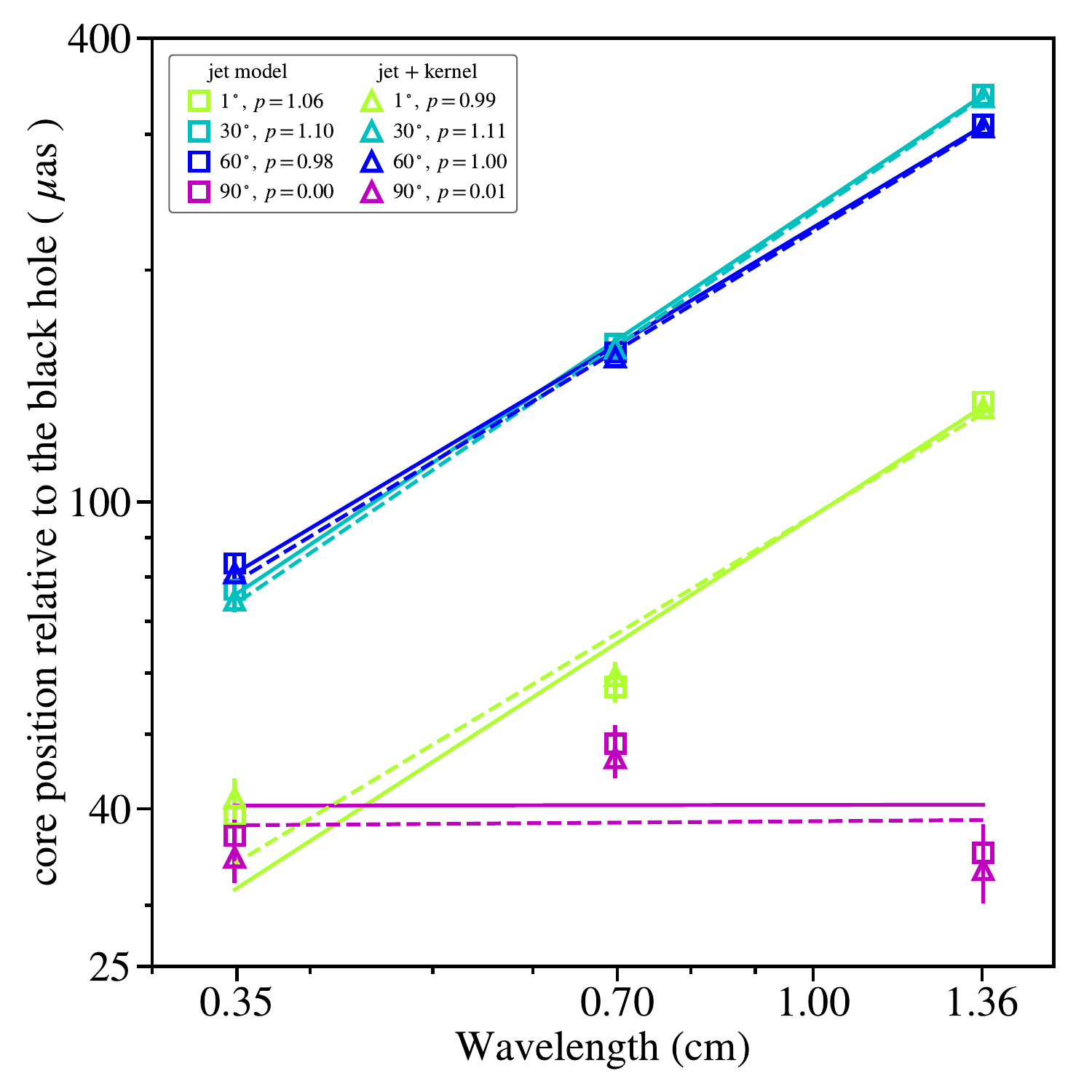}
\includegraphics[width=0.49\textwidth]{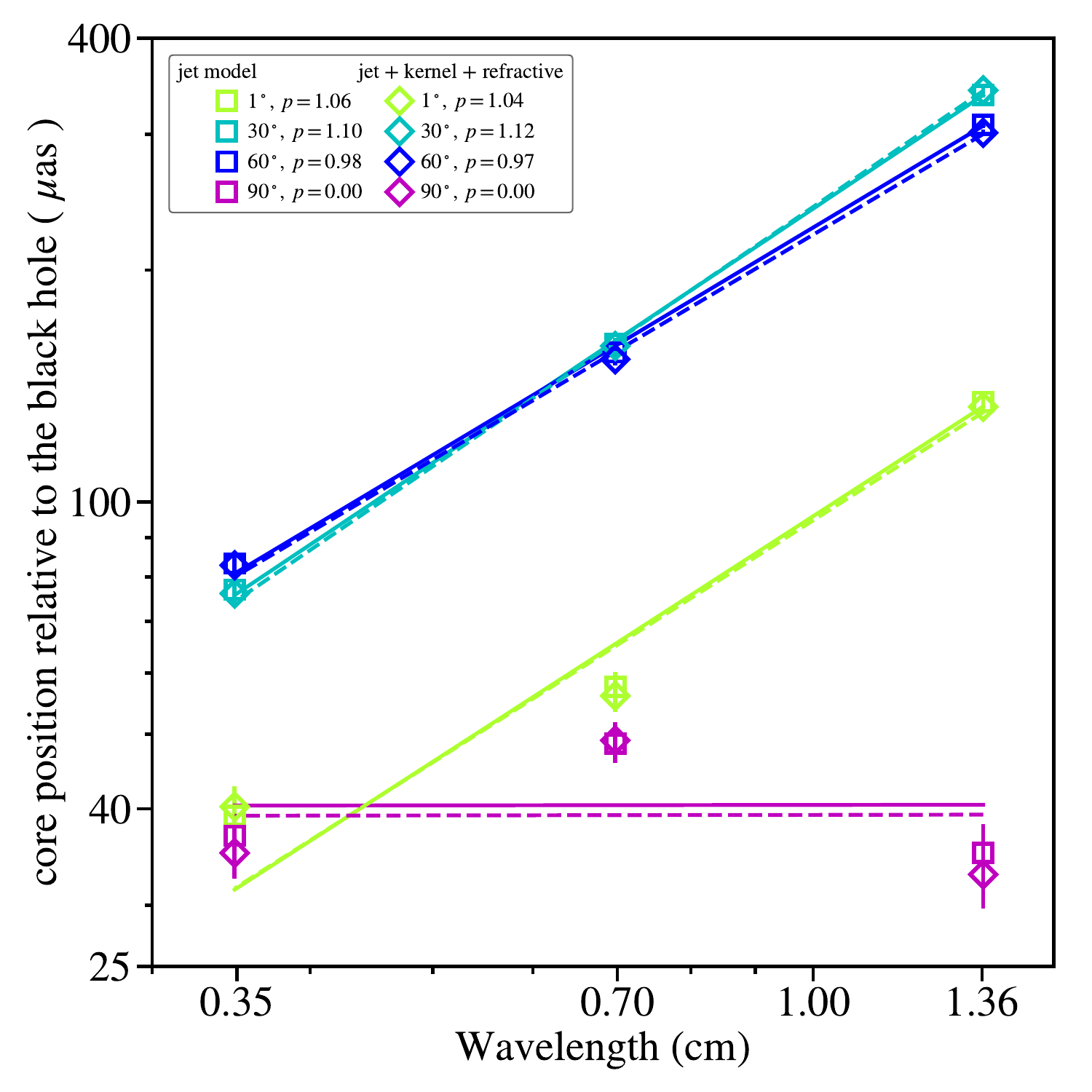}
\caption{\textbf{Core position as a function of wavelength for scattered jet models}. \textit{Left panel}: Squares indicate the core position 
relative to the central black hole at 13.6~mm, 7~mm \& 3.5~mm (22/43/86~GHz) for \textit{unscattered} jet models. Triangles indicate the core position for jet models that have been scattered with the kernel. Colors represent different inclination angles ($i$) with respect to the observer's line of sight. For all models, the orientation on the sky of the spin-axis is $PA=0 \degr$. Solid lines (jet models) and dashed lines (jets scattered with 
the kernel) show the best-fit power-law using a least-squares fitting method. Error bars indicate 1$\sigma$ uncertainty in the core position data points. Errors in the fit parameters and $\chisq$ are summarized in Table~\ref{tab:plaw_params_all_PA_jet_scattered}.
The values for the inclination ($i$) and the power-law index ($p$) are given in the legend.\textit{ Right panel}: Same as the left but showing jet models (squares) scattered with the kernel and also including refractive scattering (diamonds).}\label{fig:power_law_jet_PA0_scattered}
\end{figure*}

Using the Python library \code{eht-imaging} \citep{chael:2016} and \code{stochastic-optics} \citep{johnson:2016} we convolve our \textit{intrinsic} model images of \sgra 
(presented in Fig.~\ref{fig:flux_density_disk_models} and Fig.~\ref{fig:flux_density_jet_models}) with the scattering kernel 
to produce angular broadening. Then, we generate a refractive phase screen 
that introduces stochastic variations in the flux density (i.e. "speckles") using random numbers as Fourier components. 
The final scattered maps of \sgra include refractive scattering and diffractive scattering. 
An example for a jet model, $J30$, is shown in Figure~\ref{fig:scattered_jet_model_i30}. The panels on the first row show the intrinsic \sgra model maps, 
the second row shows the scatter-broadened image due to the anisotropic 
kernel with parameters defined in Table~\ref{tab:scat_screen_parameters}, 
and the last row displays the final scattered image 
of \sgra incorporating stochastic effects due to refractive noise. At the end of this paper we include a library of maps 
showing the effects of scattering for jet models (Appendix~\ref{sec:scattered_jet_models_library}) 
and disk models (Appendix~\ref{sec:scattered_disk_models_library}). 

As seen in Fig.~\ref{fig:scattered_jet_model_i30}, at long wavelengths (13.6~mm/22~GHz) \sgra is heavily broadened (almost engulfing the whole FOV) due to the high level of diffractive scattering (Fig.~\ref{fig:scattered_jet_model_i30}, first column). 
Even with an orientation of $PA=0 \degr$ (i.e. the polar jet runs in the N-S direction and is nearly perpendicular to the scattering Gaussian with $PA_G \sim 82 \degr $), most of the angular broadening occurs along the E-W direction and the jet features get completely blurred out. Interstellar scattering is very dominant at this observing wavelength. At the shortest wavelength (3.5~mm/86~GHz, Fig.~\ref{fig:scattered_jet_model_i30} last column), the plasma starts becoming optically-thin so the jet begins to be revealed as a faint feature in the north direction, the counter-jet remains unseen. 

By inspecting the locations of the centroids for each map (green dots) and comparing them to the centroids \textit{prior} 
to scattering (black cross-hairs), we can see that scattering appears to have no effect on the centroid position. This is due to the effect being very small so that it would require zooming in to the central area of the map extremely closely. There are minor changes in the centroid location that become more evident when plotting the core position as a function of $\lambda$ before and after scattering, as shown in \fig\ref{fig:power_law_jet_PA0_scattered}. 

In \fig\ref{fig:power_law_jet_PA0_scattered} the left panel shows the intrinsic jet models (squares, solid lines) and the jet models after simulating the effects of the scattering kernel (triangles, dashed lines). The right panel shows the same intrinsic jet models compared to scattered models that include scatter broadening \textit{plus} refractive scattering (diamonds, dashed lines). After accounting for scattering these jet models still fit very well a power-law relation. Adding refractive scattering at these wavelengths (\fig\ref{fig:power_law_jet_PA0_scattered}, right panel) has a negligible effect in the core-shift when comparing it to the effect of purely angular broadening due to the kernel (\fig\ref{fig:power_law_jet_PA0_scattered}, left panel). For our study of core-shift as a function of inclination angle and position angle we present the effects of scatter broadening \textit{plus} refractive scattering at these $\lambda$. This decision is also motivated by results of \cite{johnson:2018} suggesting that refractive noise becomes dominant over the intrinsic structure of \sgra for $\lambda > 13$~mm and that it is less dominant than previously thought at 1.3~mm, so we want to explore the effects of refractive noise within that range of $\lambda$. 
We find that even when jet features are completely "blurred out"  by the scattering screen (Fig.~\ref{fig:scattered_jet_model_i30}), a core-shift measurement is still retrievable. Thus, the presence of a scattering screen has negligible effects on core-shift estimates.

%
\subsection{Effects of Changing the Position Angle}~\label{sec:position_angle_changes}
In Sections~\ref{sec:model_flux_maps}- \ref{sec:adding_scattering} we modelled core-shifts assuming a position angle of  $PA=0\degr$.  In this section we investigate how changing the orientation of \sgra to other $PAs$ affects the core-shift.  We change the position angle of the BH spin-axis in the direction East of North (i.e., rotate counter-clockwise) to $PA=0\degr$, $45\degr$, $90\degr$ while keeping the inclination angle $i$ constant and the orientation of the scattering kernel unchanged ($PA_{G}=81.9 \degr$). We repeat the process for all inclinations and emission models.

\begin{figure*}
\centering
\includegraphics[width=0.3\textwidth]{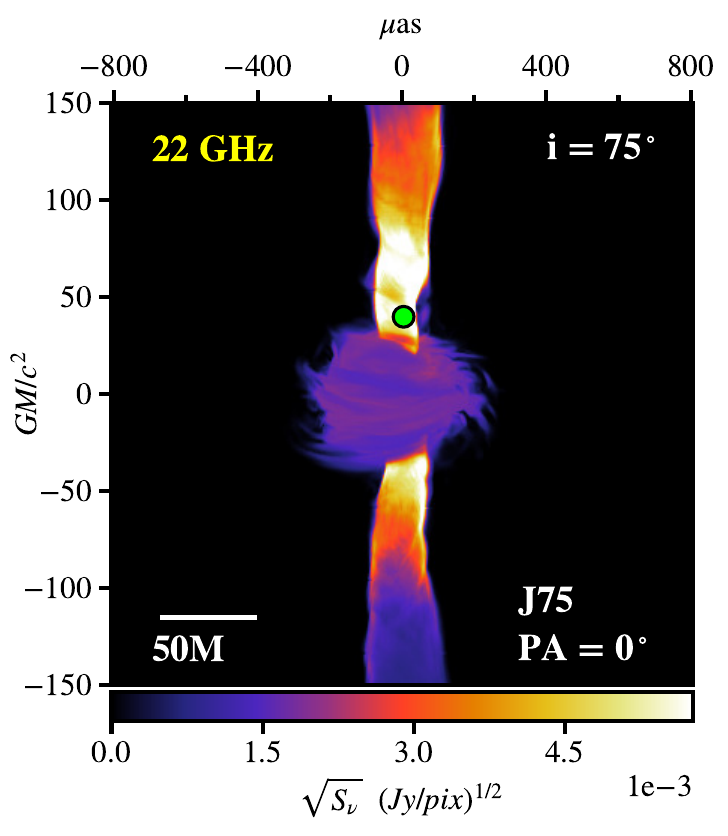}
\includegraphics[width=0.3\textwidth]{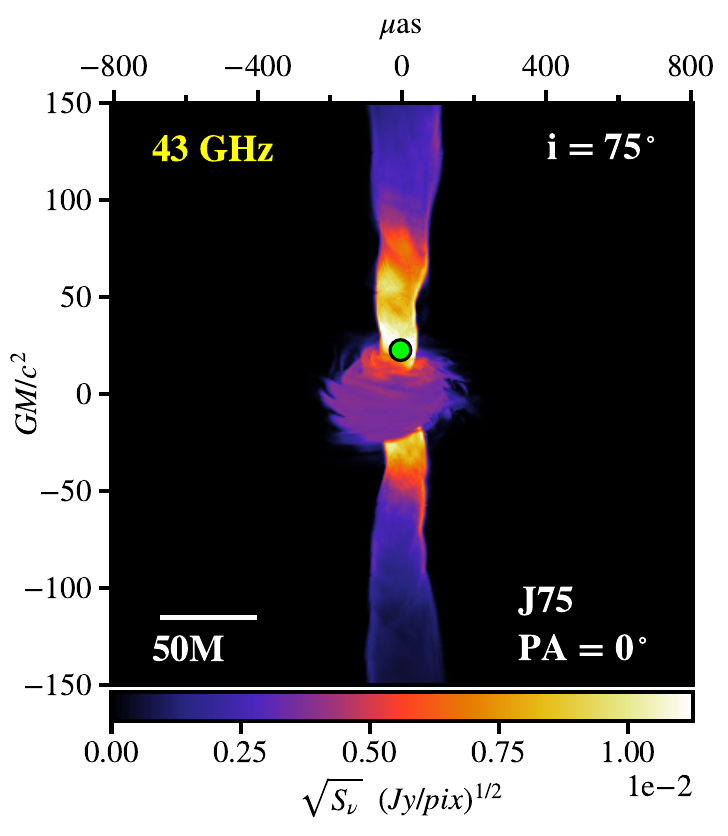}
\includegraphics[width=0.3\textwidth]{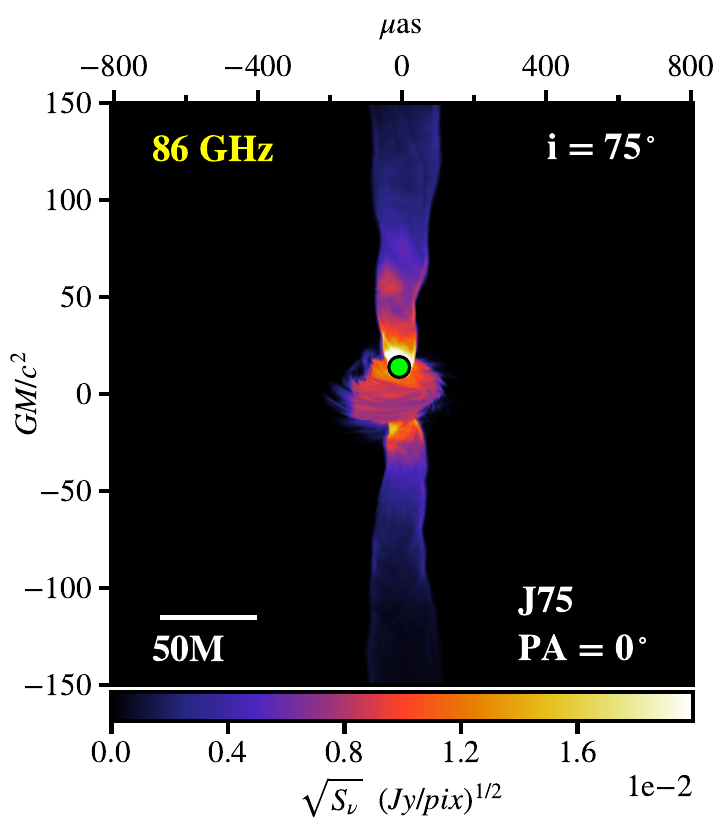}\\
\includegraphics[width=0.3\textwidth]{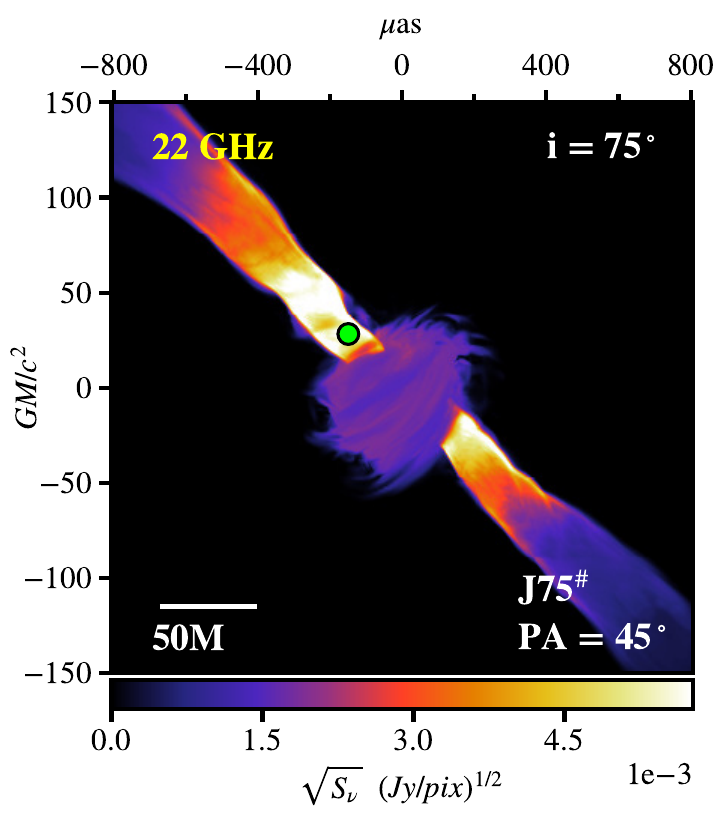}
\includegraphics[width=0.3\textwidth]{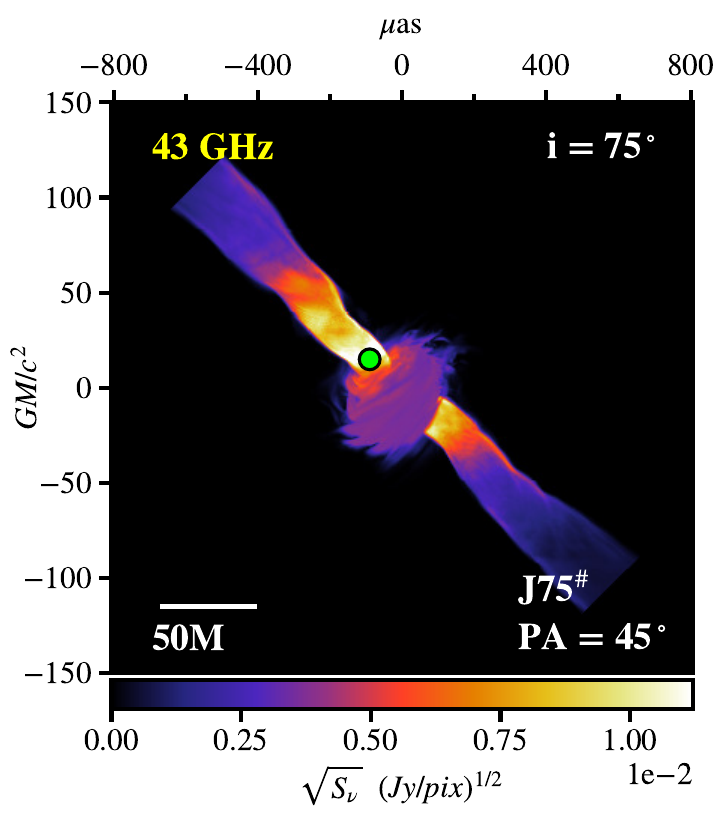}
\includegraphics[width=0.3\textwidth]{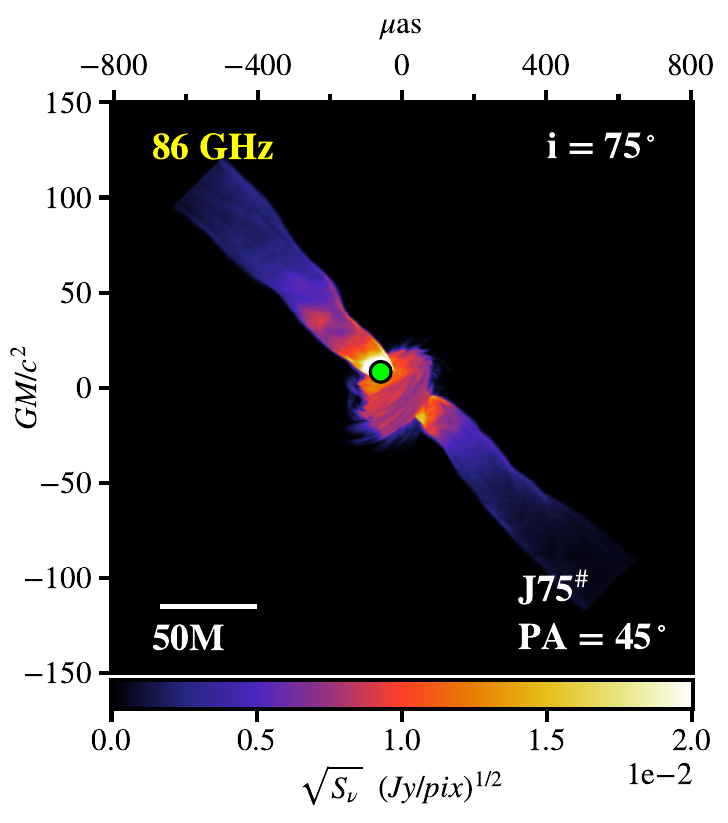}\\
\includegraphics[width=0.3\textwidth]{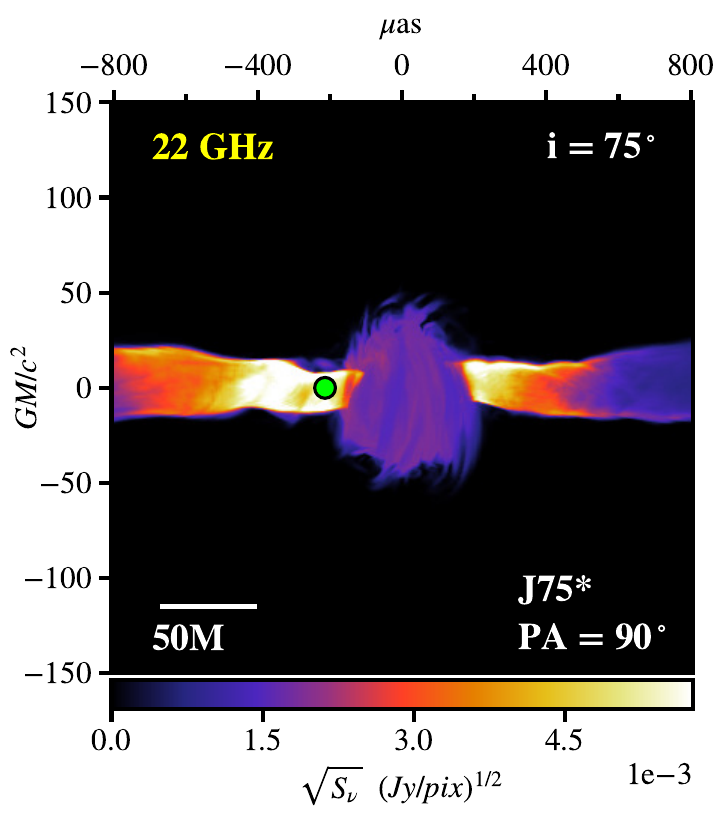}
\includegraphics[width=0.3\textwidth]{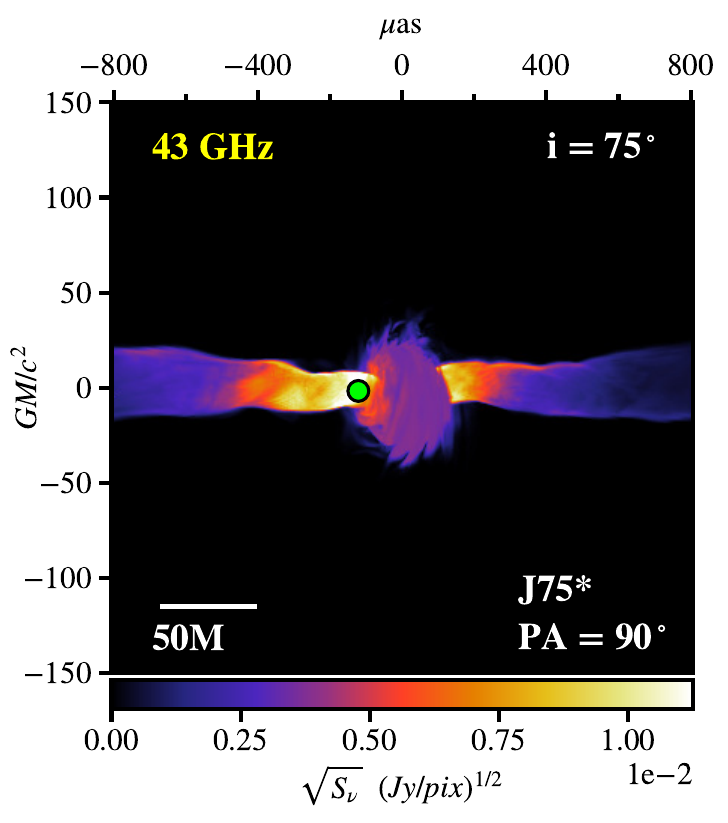}
\includegraphics[width=0.3\textwidth]{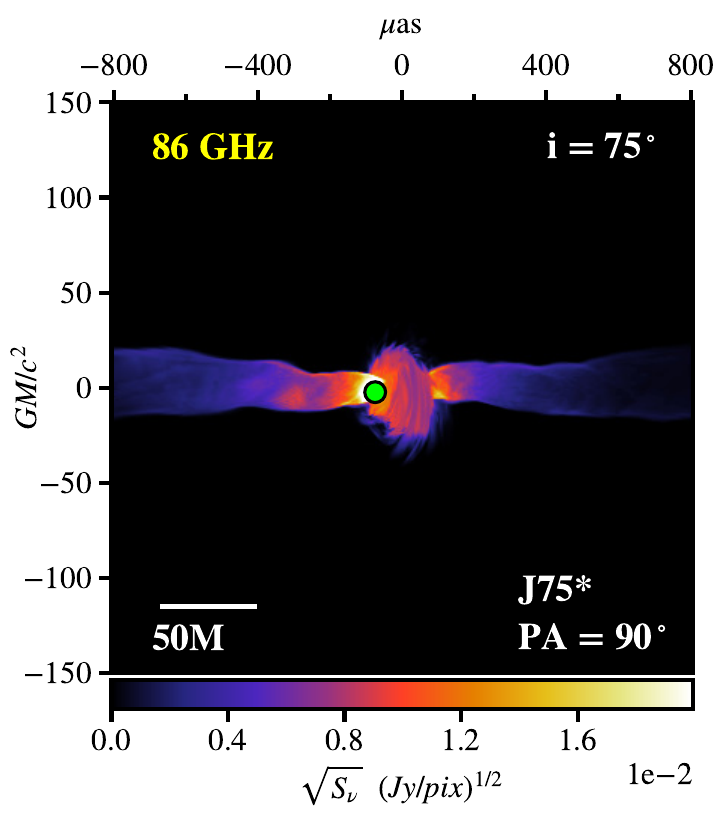}
\caption{\textbf{Unscattered jet model (J75) with different position angle orientations}. Each column shows a jet model 
with inclination $i=75 \degr$ at frequencies of 22~GHz, 43~GHz \& 86~GHz (\textit{left to right}). 
Rows from \textit{top to bottom} show the unscattered jet model with position angles of the BH spin-axis of $PA=0 \degr, 45 \degr, 90 \degr$ (rotated East of North counter-clockwise). Their identifications are J75, J75$^{\#}$, J75$^{*}$, respectively. 
The FOV of the 22~GHz images \textit{(left column)} is doubled compared to the FOV at 43~GHz \& 86~GHz. The green dot indicates the intensity-weighted centroid of each image.}\label{fig:J75_model_PA_rotated}
\end{figure*}

\begin{figure*}
\centering
\includegraphics[width=0.3\textwidth]{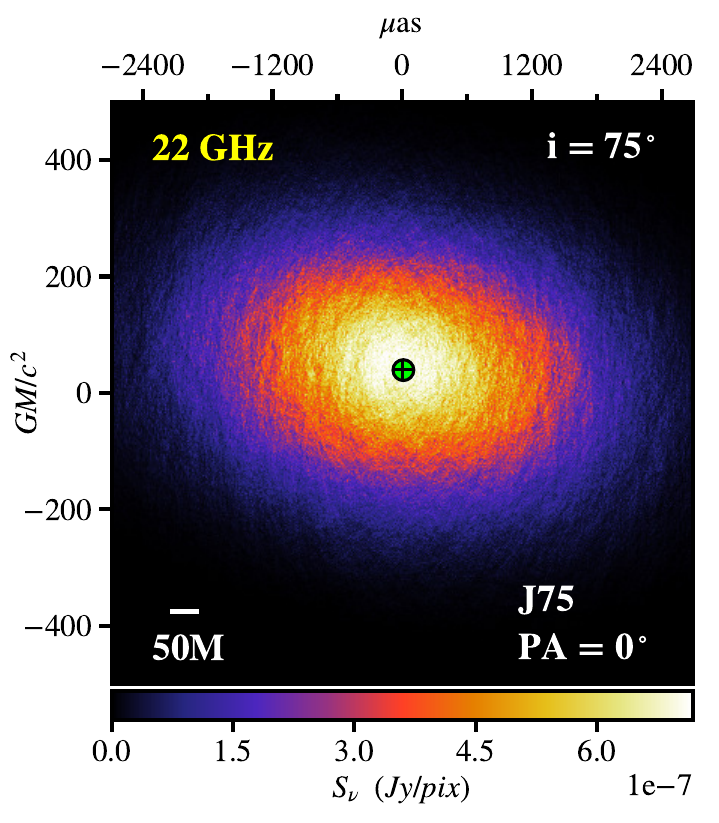}
\includegraphics[width=0.3\textwidth]{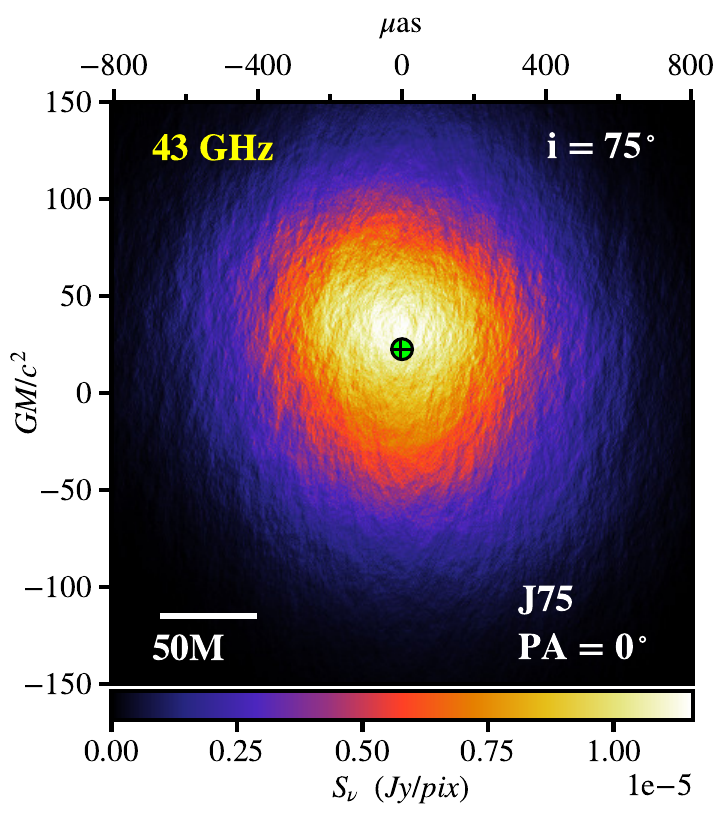}
\includegraphics[width=0.3\textwidth]{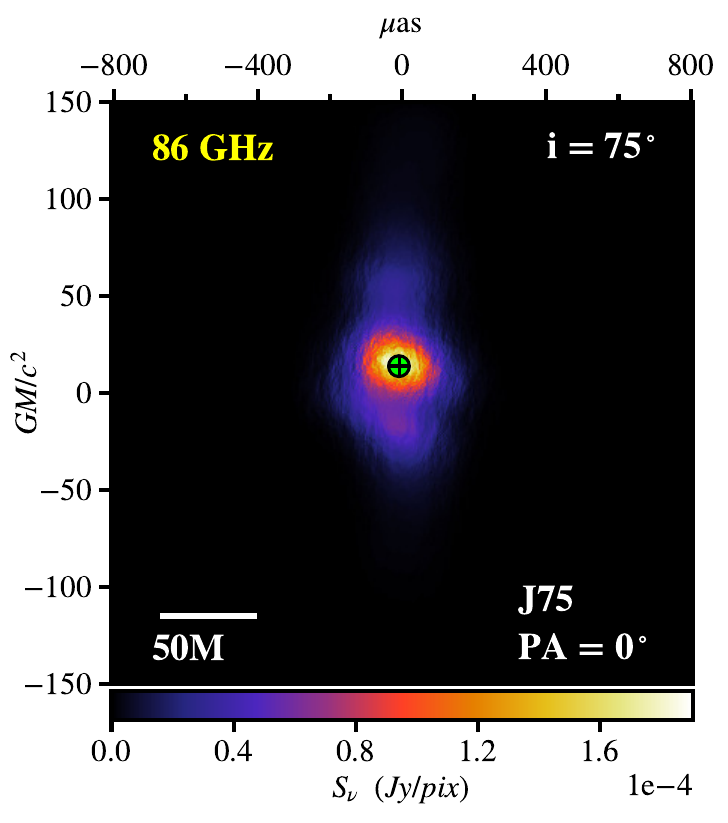}\\
\includegraphics[width=0.3\textwidth]{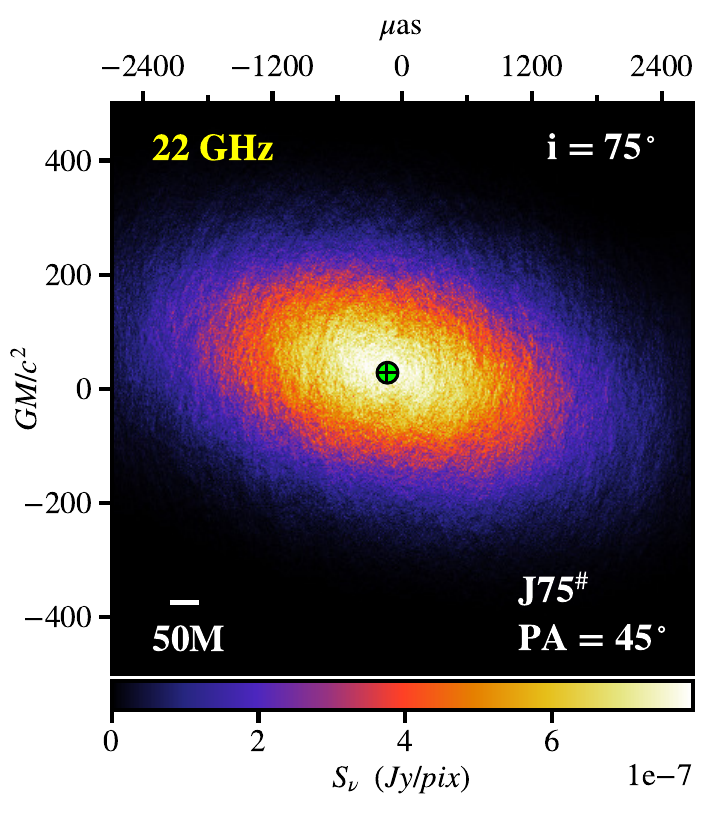}
\includegraphics[width=0.3\textwidth]{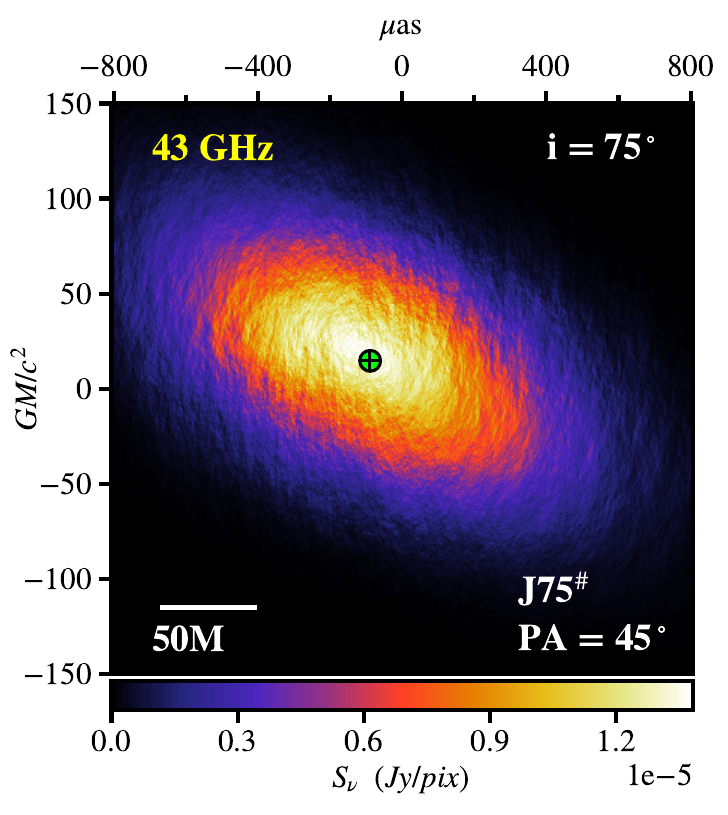}
\includegraphics[width=0.3\textwidth]{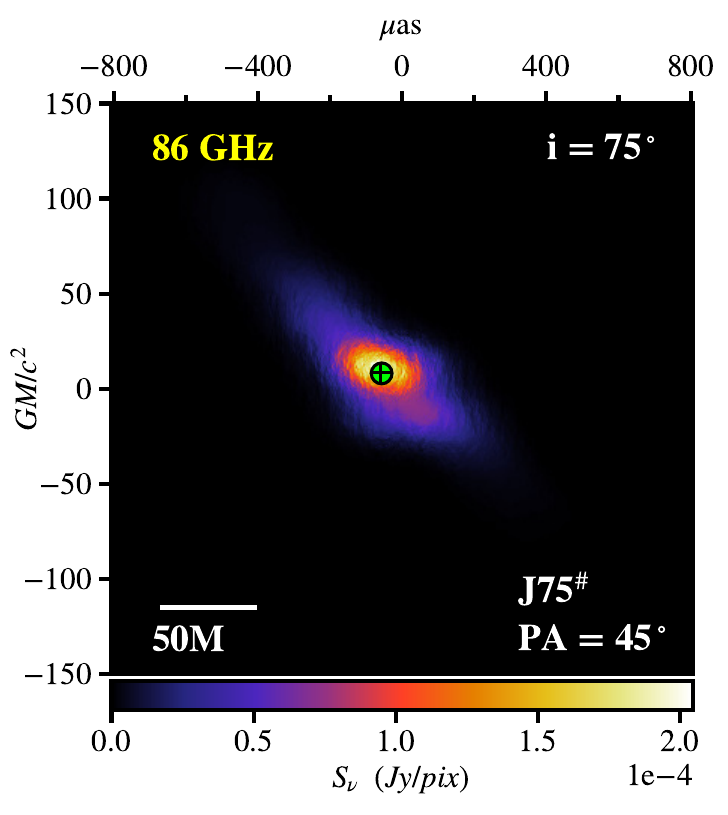}\\
\includegraphics[width=0.3\textwidth]{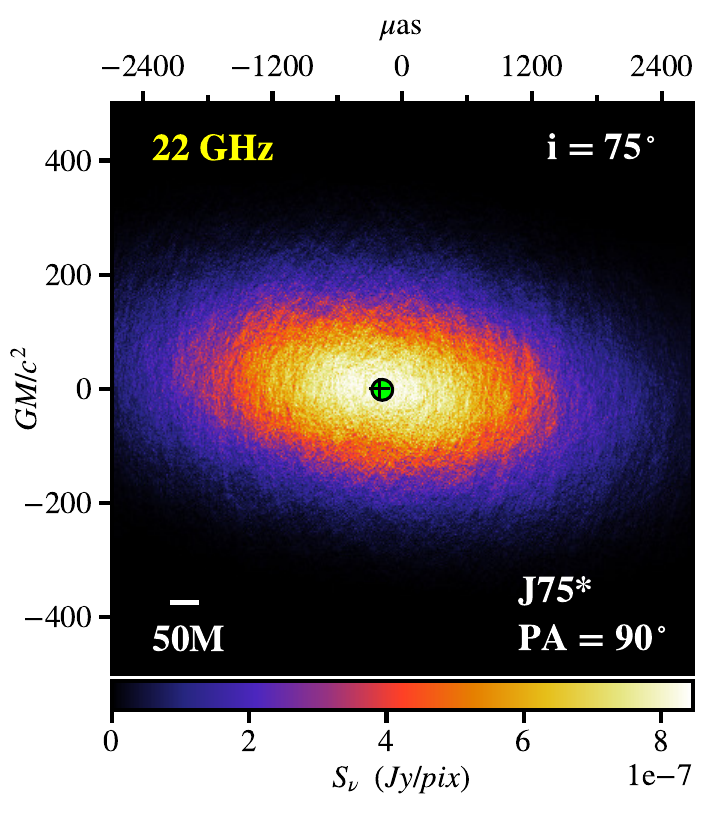}
\includegraphics[width=0.3\textwidth]{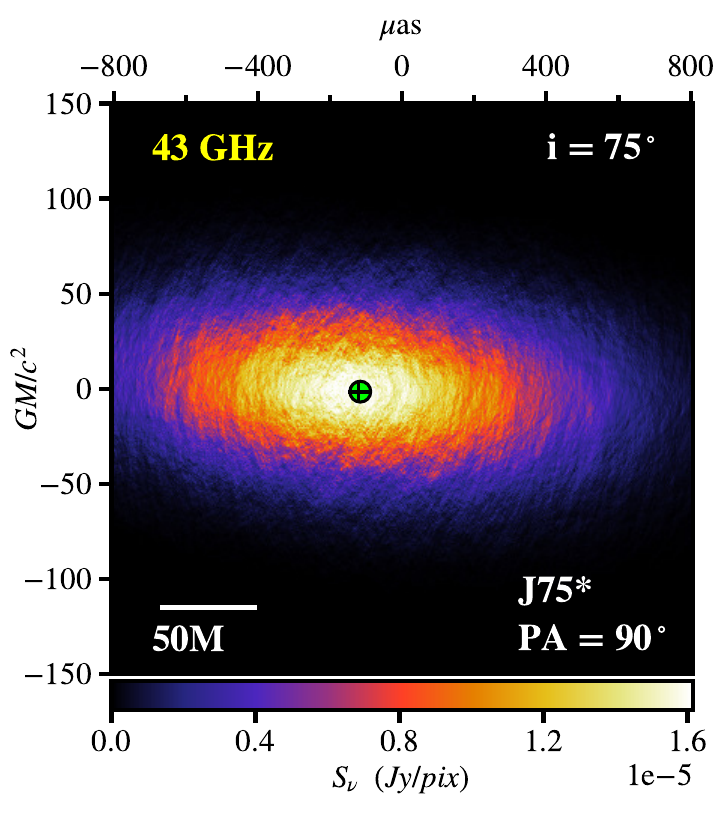}
\includegraphics[width=0.3\textwidth]{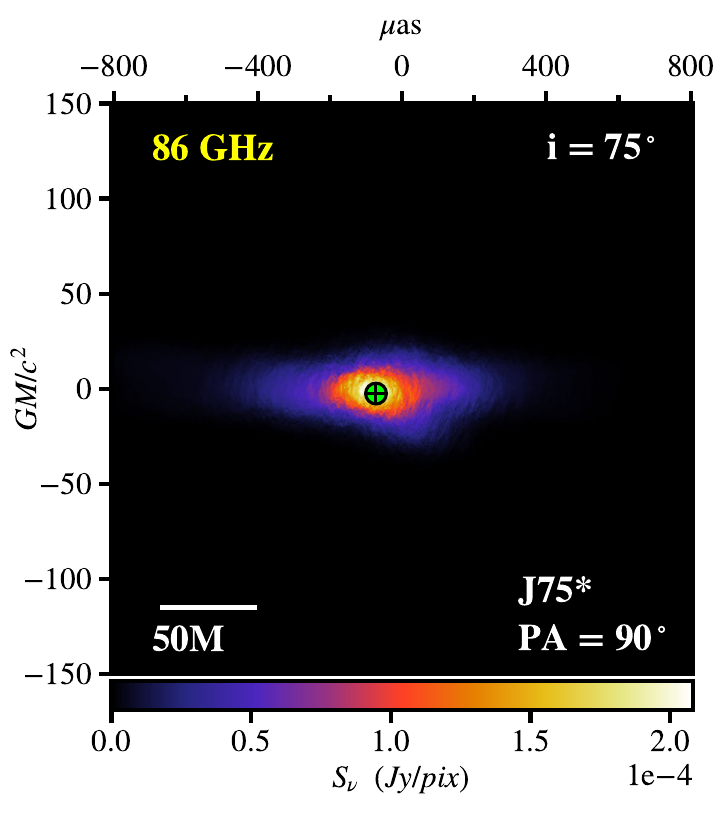}
\caption{\textbf{Scattered jet model (J75) with different position angle orientations}. Each column shows a scattered jet model 
with inclination $i=75 \degr$ at frequencies of 22~GHz, 43~GHz \& 86~GHz (\textit{left to right}). 
Rows from \textit{top to bottom} show the scattered jet model including refractive noise with position angles of the BH spin-axis of $PA=0 \degr, 45 \degr, 90 \degr$ (rotated East of North counter-clockwise). Their identifications are J75, J75$^{\#}$, J75$^{*}$, respectively. The  scattering screen remains fixed 
at $PA_G=81.9\degr$.  
The FOV of the 22~GHz images \textit{(left column)} is doubled compared to the FOV at 43~GHz \& 86~GHz. The green dot indicates the intensity-weighted centroid of each image. Black cross-hairs indicate the location of the intensity-weighted centroid of the \textit{unscattered} jet model.}\label{fig:J75_scattered_PA_rotated}
\end{figure*}

\begin{figure*}
\centering
\includegraphics[width=0.49\textwidth]{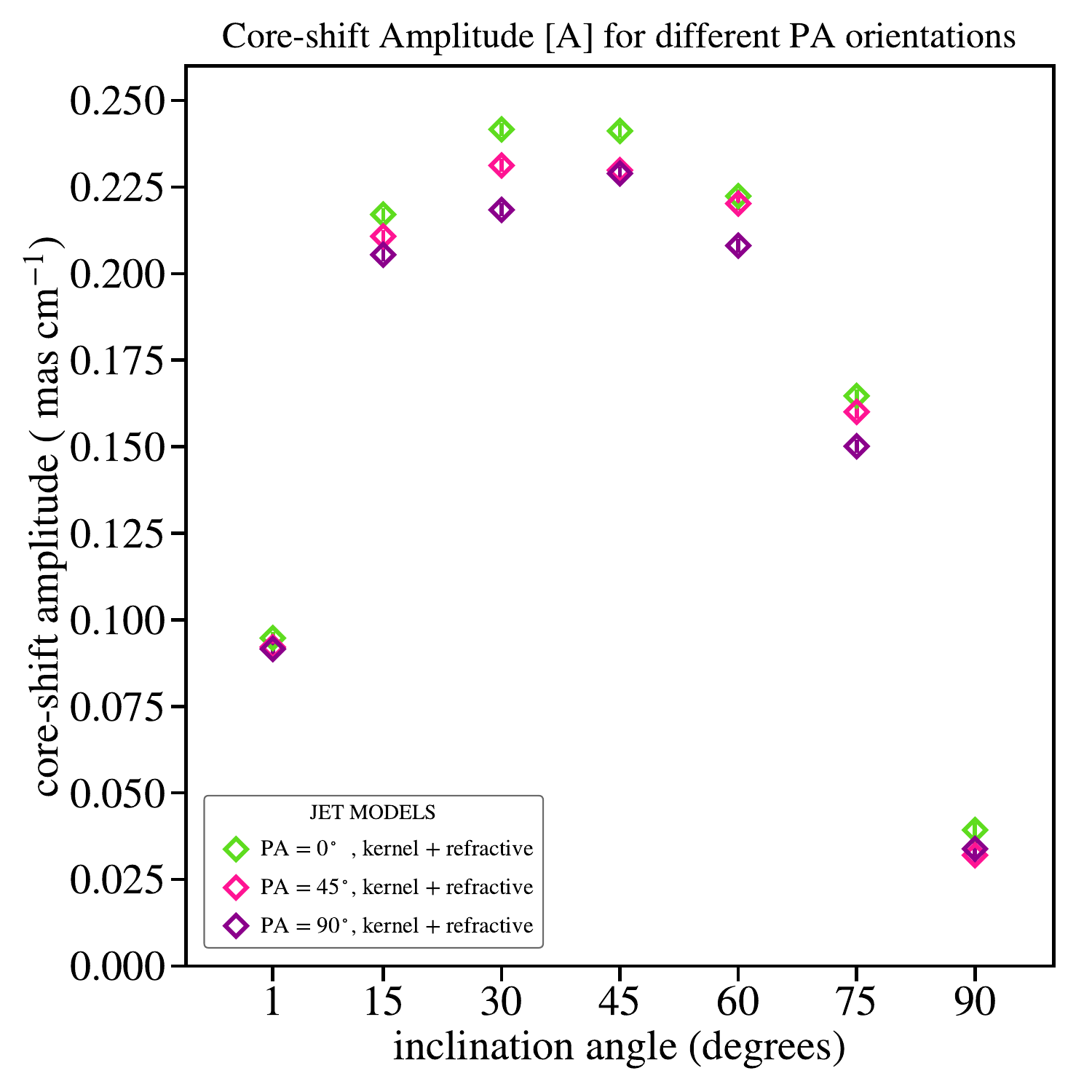}
\includegraphics[width=0.49\textwidth]{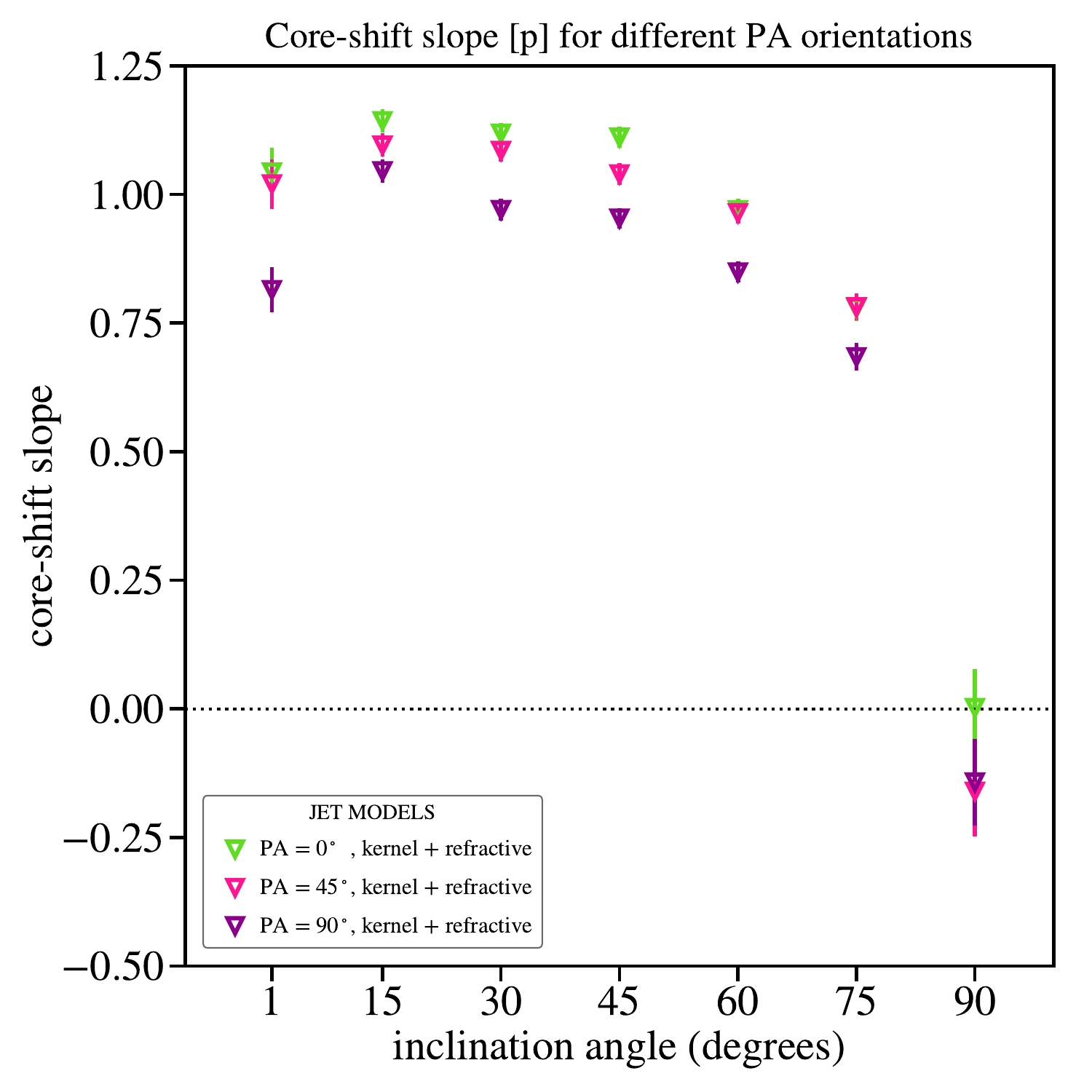}\\
\includegraphics[width=0.49\textwidth]{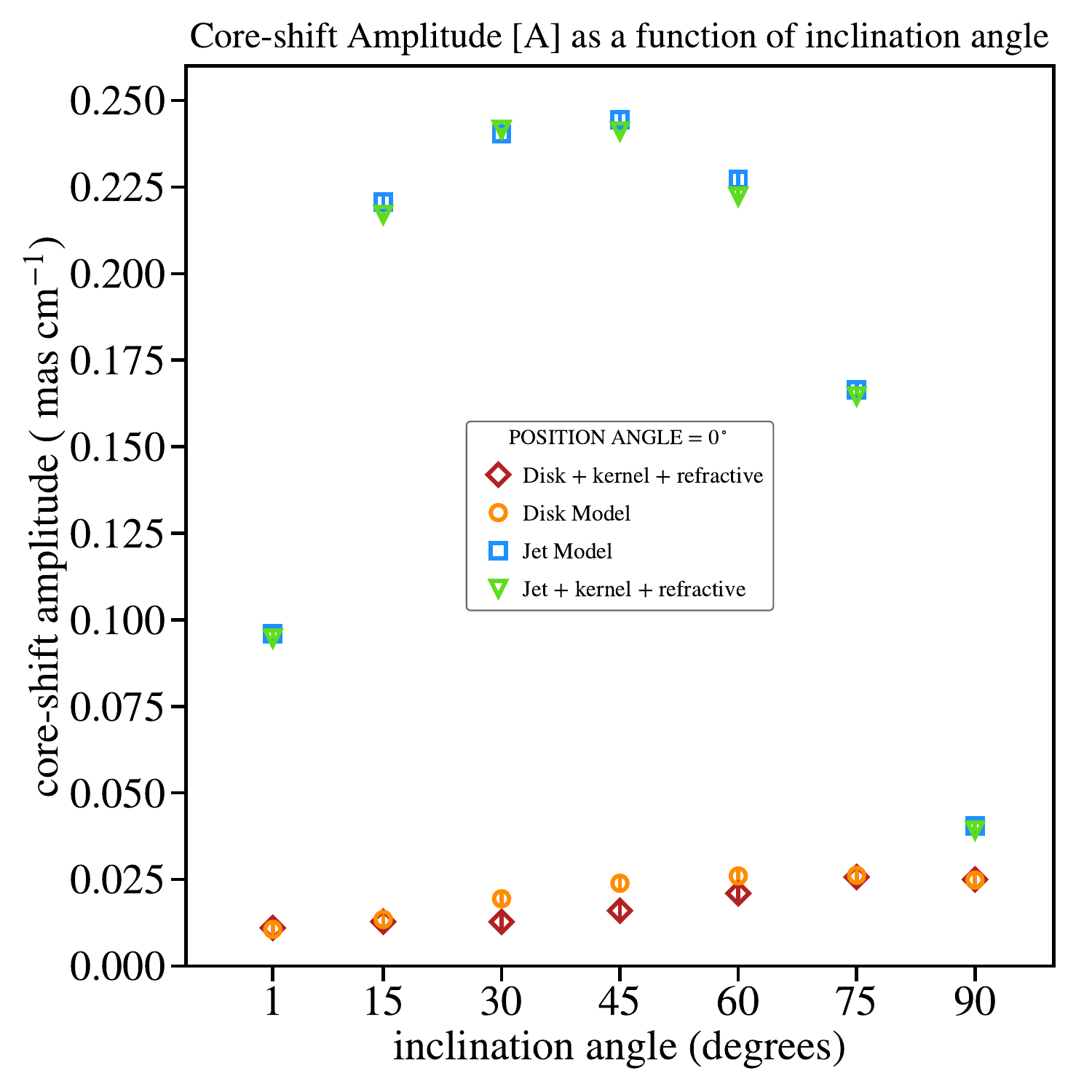}
\includegraphics[width=0.49\textwidth]{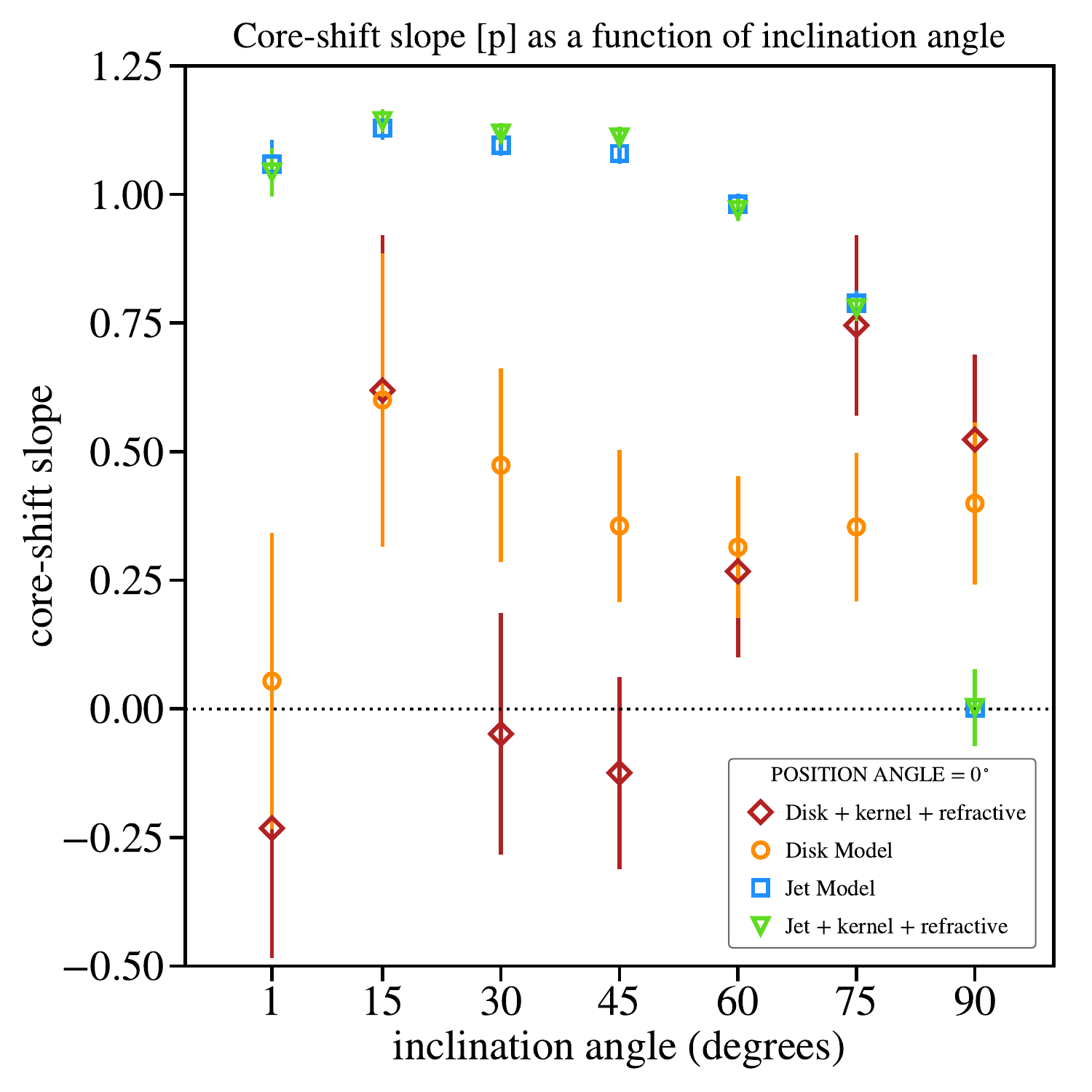}
\caption{\textbf{Core-shift parameters of all models as a function of inclination angle for different spin-axis orientations}. \textit{Top Left:} core-shift amplitude ($A$) versus inclination angle ($i$) for different position angle ($PA$) orientations of the black hole spin-axis. Diamonds represent $A$ values for \textit{jet} models with $PA$= $0\degr$, $45\degr$, $90\degr$ (shown in different colors) which have been scattered by the broadening kernel including refractive noise. \textit{Top Right}: same as the left, but displaying the core-shift slope ($p$), which is the index of the power-law relation $r_{core}=A~ \lambda ^p$ (inverted triangles, scattered). \textit{Bottom Left:} core-shift amplitude ($A$) versus 
inclination angle ($i$) for \textit{disk} models (orange circles), scattered disk models ( brown diamonds), \textit{jet} models (blue squares) and scattered jet models (green inverted triangles) all at a single position angle of $PA$=$0\degr$. \textit{Bottom Right}: same as the left plot, but displaying the core-shift slope ($p$) for the same models. All plots include 1$\sigma$ error bars obtained from the power-law fits. }\label{fig:amp_slope_all_PA_jet_disk}
\end{figure*}

We present model $J75$ as an example.  Figure~\ref{fig:J75_model_PA_rotated} shows jet model images of \sgra for $PA$= $0\degr$, $45\degr$, $90\degr$.   Figure~\ref{fig:J75_scattered_PA_rotated} shows the \textit{scattered} images of \sgra with the corresponding $PA$ orientations. The appearance of \sgra becomes more elongated as the $PA$ increases. In the extreme case where the spin-axis has $PA=90\degr$ (\fig\ref{fig:J75_scattered_PA_rotated}, bottom row), the jets almost line-up completely with the major-axis of the scattering Gaussian ($\Delta PA \sim 8\degr$), as a consequence \sgra is highly broadened in the E-W direction. 

Using centroid calculations, we can investigate the evolution of the core-shift parameters, $A$ and $p$, as a function of inclination angle for the three orientations ($PA=0\degr , 45\degr , 90\degr$) of the jets as presented in Figure~\ref{fig:amp_slope_all_PA_jet_disk}. Comparing \textit{intrinsic} jet models with jet models including angular broadening plus refractive noise for $PA=0\degr$ shows that differences in the core-shift are negligible (blue squares vs. green triangles in Fig.~\ref{fig:amp_slope_all_PA_jet_disk}, bottom left). However, changing the jet orientation with respect to the scattering screen affects core-shift values, with increasing $PA$ the core-shift of scattered jet models starts to deviate from the \textit{intrinsic} model values (see \fig\ref{fig:amp_slope_all_PA_jet_disk}, top left). This suggests that scattering effects become negligible when the source intrinsic $PA$ is close to perpendicular to the $PA$ of the scattering kernel. As \sgra 's jet orientation gets aligned with the scattering kernel, the effect gets larger. However, the overall core-shift amplitude $A$ decreases with increasing $PA$. In addition, larger core-shift amplitudes occur at intermediate inclinations ($i=30\degr,45\degr$) for all jet model orientations.

The core-shift slope $p$ has a value slightly larger than 1 (green triangles) for $PA=0\degr$, about 1 for $PA=45\degr$ (magenta triangles) and lower than 1 for $PA=90\degr$ (purple triangles), so the slope of the power-law (Equation \ref{eq:core_position_lambda}) becomes less steep as the position angle increases (see \fig\ref{fig:amp_slope_all_PA_jet_disk}, top right). In addition, $p$ starts decreasing at inclinations $i \geq 45\degr$. Although there is a downward trend, it is important to note that the fits for models with $i=90\degr$ have much larger error bars (see Fig.~\ref{fig:amp_slope_all_PA_jet_disk}, top right). At $i=90\degr$ we get two negative $p$ values which are caused by two effects in our calculations. First, at this edge-on inclination the polar jets with similar flux densities above and below the faint accretion disk artificially generate centroids very close to the geometrical center of the map, thus $p\sim 0$ (as discussed in Sect.~\ref{sec:coreshift_calculations}). Secondly, as the $PA$ increases the value of $p$ is further brought down, due to the effect of changing the orientation of the jets with respect to the scattering screen $PA$, resulting in negative values (rightmost magenta/purple triangles). Negative $p$ values for the power-law index contradict the physics expected from jet models and observations of AGN.  SSA changes the $\tau _ \nu$ along the jet, at short $\lambda$ the plasma becomes optically-thin, so we can see deeper into the core than at longer $\lambda$. Therefore, we consider these outlier points not physically meaningful.

If we compare jet and disk models it is clear that the former have much larger core-shifts (Fig.~\ref{fig:amp_slope_all_PA_jet_disk}, bottom left), with jet models reaching a peak value of $A \sim 241 \mu  as~cm^{-1}$ while for disk models $A \sim 25 \mu  as~cm^{-1}$. The relation of core position as a function of wavelength does not fit well a power-law for disk models (Fig.~\ref{fig:amp_slope_all_PA_jet_disk}, bottom right). This means that observations of core-shift at mm-wavelengths are a powerful tool to discriminate between jet and disk models for \sgra.

%
\section{Discussion}~\label{sec:discussion}
Here we discuss constraints from observations of \sgra (Sect.~\ref{sec:constraints_observations}) and how future studies can extend the results presented in this work (Sect.~\ref{sec:future}).

%
\subsection{Constraints from Observations}~\label{sec:constraints_observations}
Using VLBI observations the source orientation and type of emission model can be constrained. Recent work by \cite{issaoun:2019_size} presents the first  \textit{intrinsic} image of \sgra at 3.5~mm from observations with GMVA plus ALMA. An intrinsic size of 120$\pm$34~$\mu$as (major-axis) and axial ratio of 1.2$^{+0.3}_{-0.2}$ is reported. The axial ratio indicates a fairly symmetrical source geometry. The authors compare several emission models to the observational constraints, among their models they consider the disk and jet models presented in this work. 

Disk models where $(R_{high}, R_{low})=(3,3)$ do \textit{not} match the size and asymmetry limits from 3.5~mm observations. However, jet-dominated models with $(R_{high}, R_{low})=(20,1)$ fit the constraints but only at inclinations of $i \lesssim 20\degr$. 

In addition, observations of \sgra with the East Asian VLBI Network (EAVN) at 13.6~mm \& 7~mm have obtained an intrinsic size of 704$\pm$102~$\mu$as (major-axis) and axial 
ratio of 1.2$\pm$0.2, and  300$\pm$25~$\mu$as (major-axis) and axial ratio of 1.3$\pm$0.2, respectively \citep{cho:2022}. They choose to compare the intrinsic structure  of \sgra to RIAF models with a thermal plus non-thermal electron distribution function. Their motivation for using a non-thermal component of synchrotron emission is due to the observed excess flux in the spectral energy distribution of \sgra at 13.6~mm \& 7~mm. Emission models with inclination $i$=20$\degr$ and  $i$=$60\degr$ are compared to axial ratio measurements. The results indicate that low inclinations are preferred, somewhere in the range of $i \sim 30\degr$--~$40\degr$.  

From near infrared observations of flaring events in \sgra, orbital parameters of the observed "hot spots" in the innermost accretion region of \sgra yields an inclination of $i \lesssim 25\degr$ \citep{gravity:2018b}. 

Using our Fig.~\ref{fig:amp_slope_all_PA_jet_disk} (left panels) we can set an \textit{upper limit for the core-shift} of jet models including refractive scattering of $A = 241.65 \pm 1.93~ \mu  as~cm^{-1}$, with this upper limit value corresponding to a inclination of $i=30\degr$ (see also Table~\ref{tab:plaw_params_all_PA_jet_scattered}).

\subsection{Future Prospects}~\label{sec:future}
Firstly, the work presented here is not a study on time variability. Our findings on the core-shift are based on snapshots of each model.  Recent work has studied the variability of the centroid position due to refractive scattering and estimated the image wander at 1.3mm to be ~0.53$\mu$as using the same scattering model we use, as well as demonstrated that the effects of refractive image wander can be mitigated by averaging images over multiple observations \cite{zhu:2019}.  In our work we show that the ability to recover core-shift measurements is not impaired by the presence of a scattering screen. The largest uncertainties we obtain in core-shift calculations are of the order of $~5 \mu$as, so we are confident that further studies in which multiple scattered images of \sgra are averaged would still yield measurable core-shifts, and improved power-law fits since inhomogeneities in the emission will be mitigated. On time-scales ranging from intra-day to several months, \sgra displays variable emission at radio, IR and X-ray wavelengths \citep{macquart:2006, yusef-zadeh:2009, akiyama:2013}. The effects of intrinsic source variability (e.g., flaring events or hot spots) could likely impact core-shift measurements because time lags between flare peaks have already been observed at cm/mm wavelengths \citep{yusef-zadeh:2008a, brinkerink:2015}. Thus, the relation between intrinsic source variability and the core-shift of \sgra should be investigated. 

For the scope of this paper we present results for dynamical models that include a combination of an ADAF disk plus a mildly relativistic jet with a thermal electron distribution. These radiatively inefficient accretion flow models are Standard And Normal Evolution models \citep[SANE,][]{narayan:2012} and have low-power jets. It remains to be studied how our core-shift predictions compare to core-shifts from models with a strong disk, strong jet, models where electrons in the jet follow a combination of thermal plus non-thermal power-law distribution \citep{davelaar:2018}, or Magnetically Arrested Disk models \citep[MAD,][]{bisnovatyi:1976,igumenshchev:2003,narayan:2003,mckinney:2012}. Jets of MAD models are more powerful than jets of SANE models and have a wider jet-funnel opening. It is reasonable to think that core-shifts of MAD models could yield different results than our SANE models. However, we would not expect the core-shifts of disk-dominated versus jet-dominated MAD models to be fundamentally different from those presented here.

In the future, this work can be extended by using the core-shift to constrain the $R_{high}$ \& $R_{low}$ values of \sgra emission models and get better parameter estimates from black hole images \citep{EHT:2019_feat} taken by current \& planned EHT arrays, or by prospective space VLBI missions like the Event Horizon Imager \citep{roelofs:2021}. Another avenue of exploration can be to compare the core-shift predictions presented on this study with available archival VLBI observations of \sgra that were specifically designed to measure core-shifts using the so-called \textit{phase-referencing} technique \citep{middelberg:2005} .

%
\section{Summary}\label{sec:conclusions}
In this work we made predictions of the frequency-dependent position of \sgra 's radio core and estimated the core-shift for different jet-plus-disk models. We used 3D GRMHD simulations, ray-tracing and a scattering model to generate flux density maps of \sgra at three frequencies: 22/43/86 GHz. We studied the evolution of core-shift as a function of inclination angle and position angle on the sky, and investigated the effects of interstellar scattering.

We conclude that the core-shift is a clear discriminant between jet-dominated and disk-dominated emission models. It can also be used to constrain the geometry of \sgra such as its inclination and orientation on the sky. Our jet emission models show significantly larger core-shifts - in some cases by a factor of 16 - than disk models at all inclination angles. 

Jet models produced in GRMHD simulations follow well a power-law relation for the frequency-dependent position of \sgra 's radio core and agree with previous predictions of AGN with conical and parabolic jets \citep{blandford_konigl:1979, falcke_biermann:1999a}. Our disk models do not fit well a power-law relation and their core-shifts are fairly insensitive to changes in inclination angle.  
The core-shift is retrievable even in the presence of an interstellar scattering screen located between Earth and \sgra . This scattering kernel is responsible for the angular broadening (blurring) and stochastic changes (refractive noise) observed in \sgra images. 
In jet-dominated emission models, the core-shift amplitude decreases as the orientation of the BH-spin axis increases from $PA=0\degr$ to $90\degr$ (East of North). 

Scattering effects become negligible when the intrinsic $PA$ of the \sgra jet model is close to perpendicular to the $PA$ of the scattering kernel. As \sgra 's jet orientation gets aligned with the scattering kernel, the effect gets larger. Additionally, the largest core-shift amplitudes occur at intermediate inclinations ($i=30\degr,45\degr$) for all jet model orientations.

We also obtain an upper limit for the core-shift of jet models including refractive scattering of $A = 241.65 \pm 1.93~ \mu  as~cm^{-1}$, with this peak value occurring at an inclination of $i=30\degr$. 
Therefore, obtaining core-shifts from multi-wavelength observations gives us the means to further constrain the geometry and emission mechanisms of the plasma surrounding \sgra, and it can be viewed as a great tool to shed light on open-ended debate of disk versus jet models of emission.

%
\begin{acknowledgements}
We thank the anonymous referee(s) for providing comments to improve the quality of our study. We are grateful to the Event Horizon Telescope Publication Committee, Tomohisa Kawashima, Ilje Cho and Michael D. Johnson for their useful comments when reviewing this work . This research is partially supported by the \textit{NWO Spinoza Prize} awarded to H.~Falcke and by
the \textit{ERC Synergy Grant} "BlackHoleCam: Imaging the Event Horizon of Black Holes" awarded to H.~Falcke, M.~Kramer and L.~Rezzolla, Grant 610058. \\

\textit{Software:} eht-imaging python library \citep{chael:2016}, Stochastic Optics \citep{johnson:2018},
IPOLE code \citep{moscibrodzka:2018}, Python, NumPy \citep{numpy:2020}, SciPy \citep{scipy:2020}, Matplotlib \citep{matplotlib}, Astropy \citep{astropy:2013}, Adobe InDesign.
\end{acknowledgements}

%
\bibliographystyle{aa}
\bibliography{local}

\newpage

%
\begin{appendix}

%
\onecolumn
\section{Maps of Disk \& Jet Models (additional inclinations)}~\label{sec:unscattered_models_library}
%
\begin{figure*}[!hp]
\caption{\textbf{Flux Density ($S_\nu$) maps of intrinsic disk models}. Rows of panels show disk models with
 different inclinations ($i=15 \degr$, 45$\degr$ \& 75$\degr$, 
 \textit{top to bottom}). The position angle on the sky of the BH spin-axis is $PA=0 \degr$ for all models. 
 Columns show the maps at different frequencies (22~GHz, 43~GHz \& 86 GHz, \textit{left to right}). 
 The field of view for each panel is 120~$\times$~120~$GM/c^2$ (645~$\times$~645~$\mu$as). The bottom colorbar 
 indicates the flux density in square-root scale, the y-axis shows the distance from the BH  
 (located at 0,0) in units of $GM/c^2$ and the top x-axis shows that same distance in $\mu$as. The green dot represents the position of 
 the intensity weighted centroid.}\label{fig:flux_density_disk_models_more_inclinations}
\centering
\includegraphics[width=0.3\textwidth]{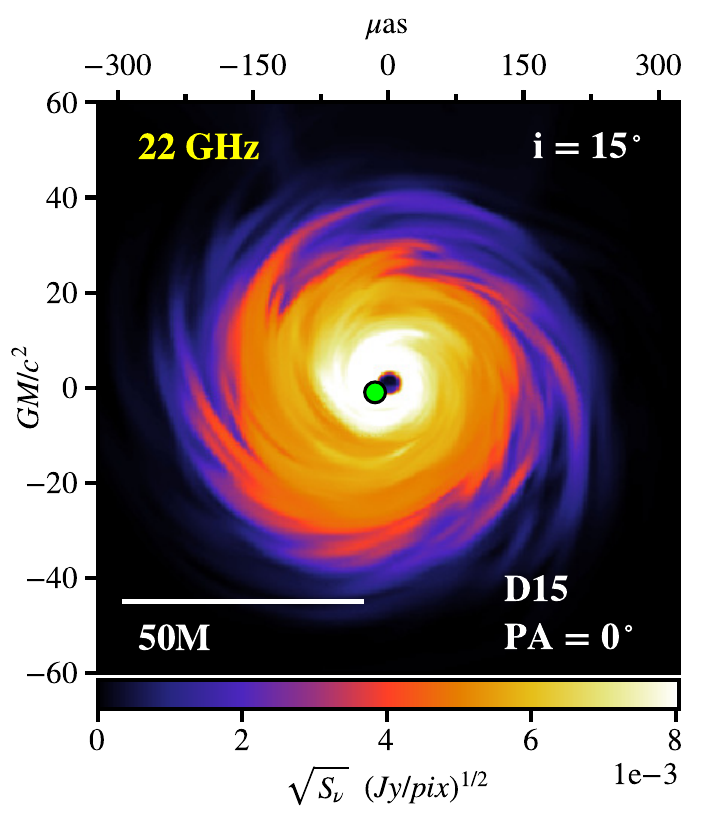}
\includegraphics[width=0.3\textwidth]{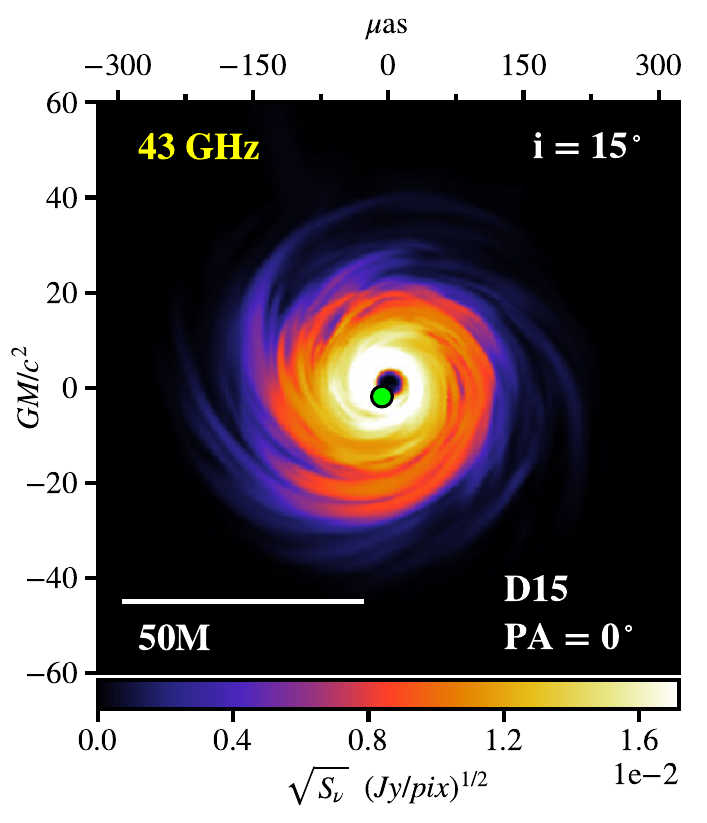}
\includegraphics[width=0.3\textwidth]{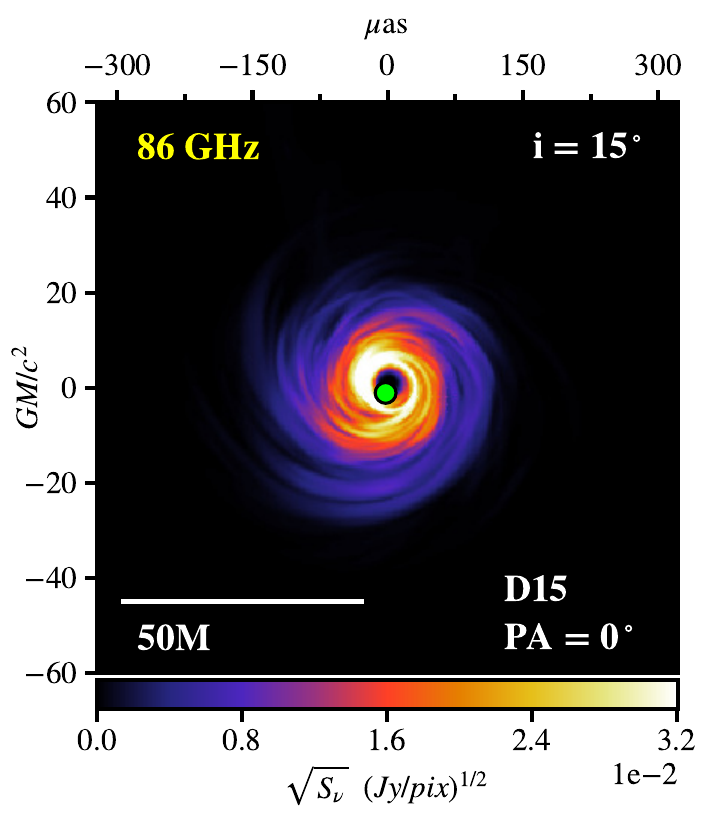}\\
\includegraphics[width=0.3\textwidth]{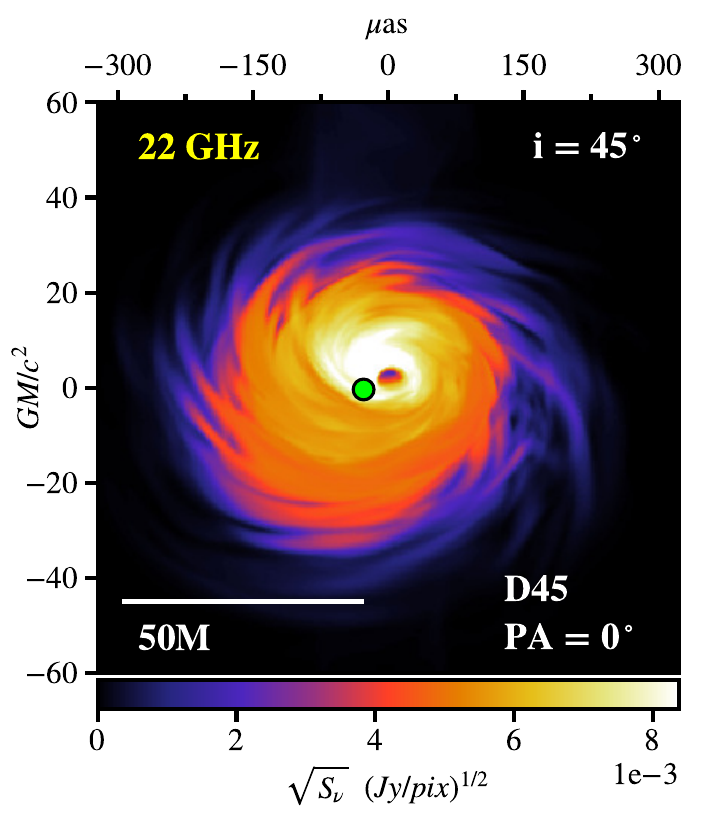}
\includegraphics[width=0.3\textwidth]{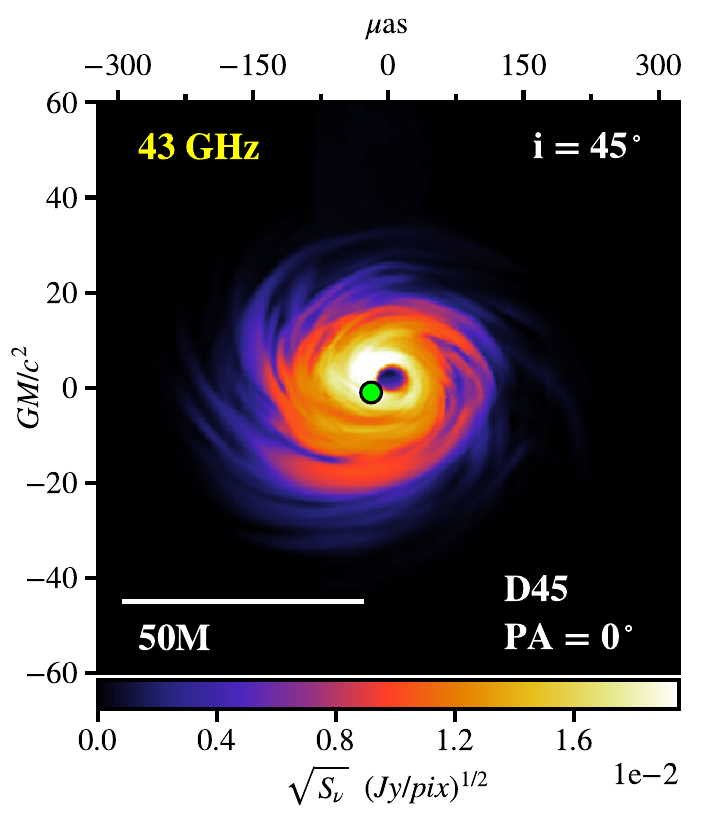}
\includegraphics[width=0.3\textwidth]{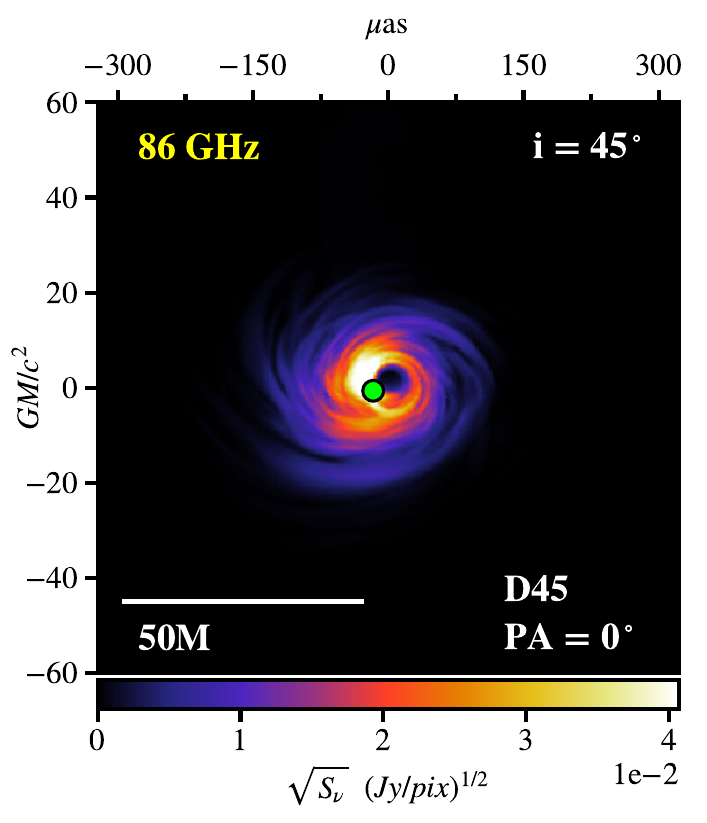}\\
\includegraphics[width=0.3\textwidth]{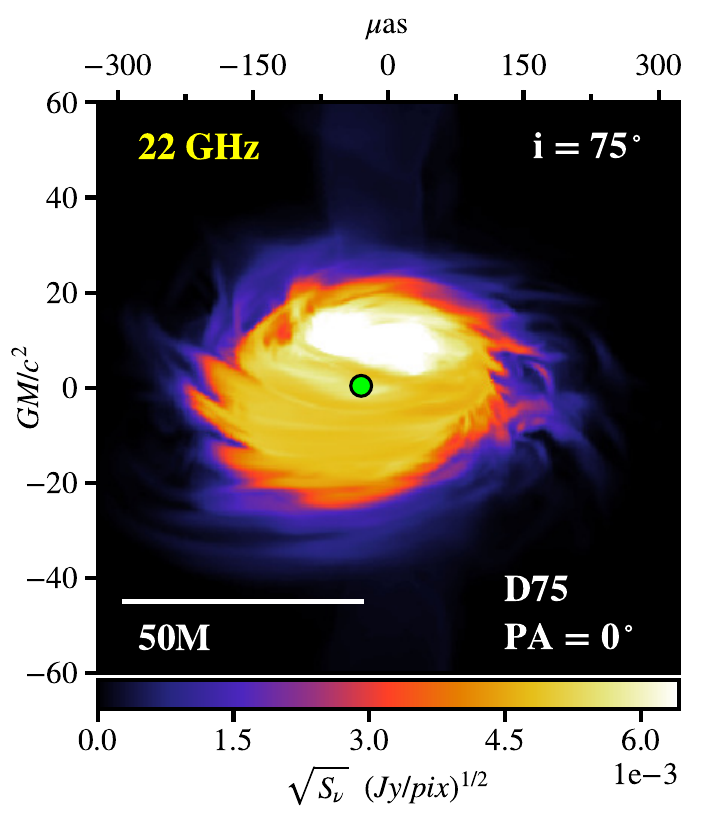}
\includegraphics[width=0.3\textwidth]{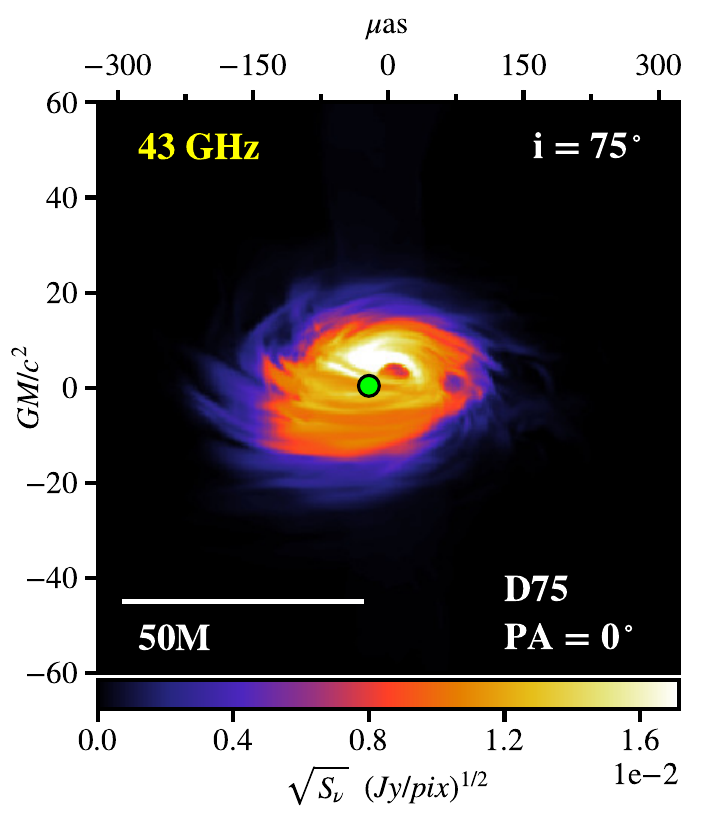}
\includegraphics[width=0.3\textwidth]{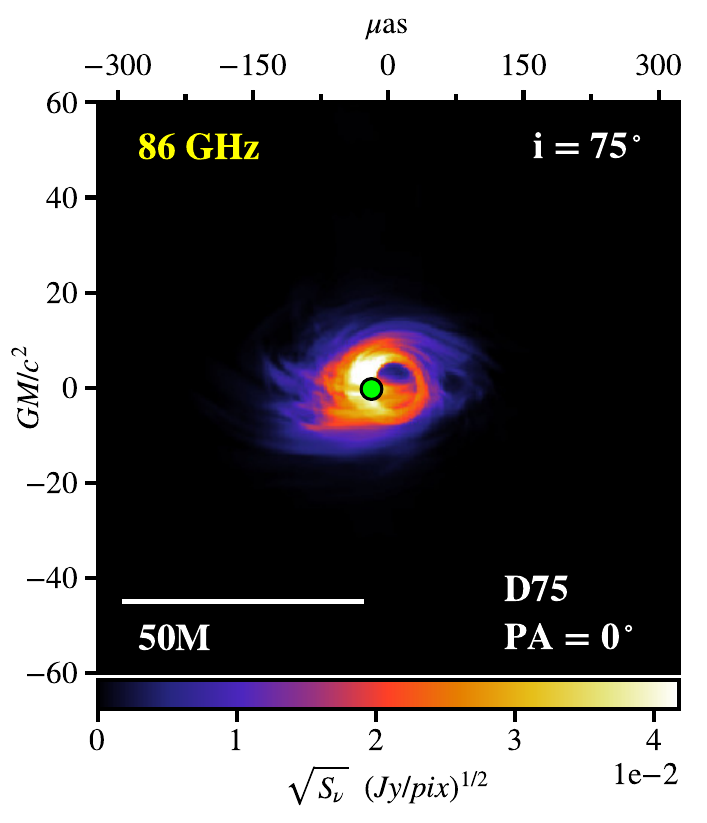}\\
\end{figure*}
\newpage

\begin{figure*}[!hp]
\caption{\textbf{Flux Density ($S_\nu$) maps of intrinsic jet models}. Rows of panels 
show jet models with different inclinations ($i=15 \degr$, 45$\degr$ \& 75$\degr$, 
 \textit{top to bottom}). The position angle on the sky of the BH spin-axis is $PA=0 \degr$ for all models. 
 Columns show the maps at different frequencies (22~GHz, 43~GHz \& 86 GHz, \textit{left to right}). 
 The field of view for each panel is 300~$\times$~300~$GM/c^2$ (1600~$\times$~1600~$\mu$as). The bottom colorbar 
 indicates the flux density in square-root scale, the y-axis shows the distance from the BH  
 (located at 0,0) in units of $GM/c^2$ and the top x-axis shows that same distance in $\mu$as. The green dot represents the position of 
 the intensity weighted centroid.}\label{fig:flux_density_jet_models_more_inclinations}
\centering
\includegraphics[width=0.3\textwidth]{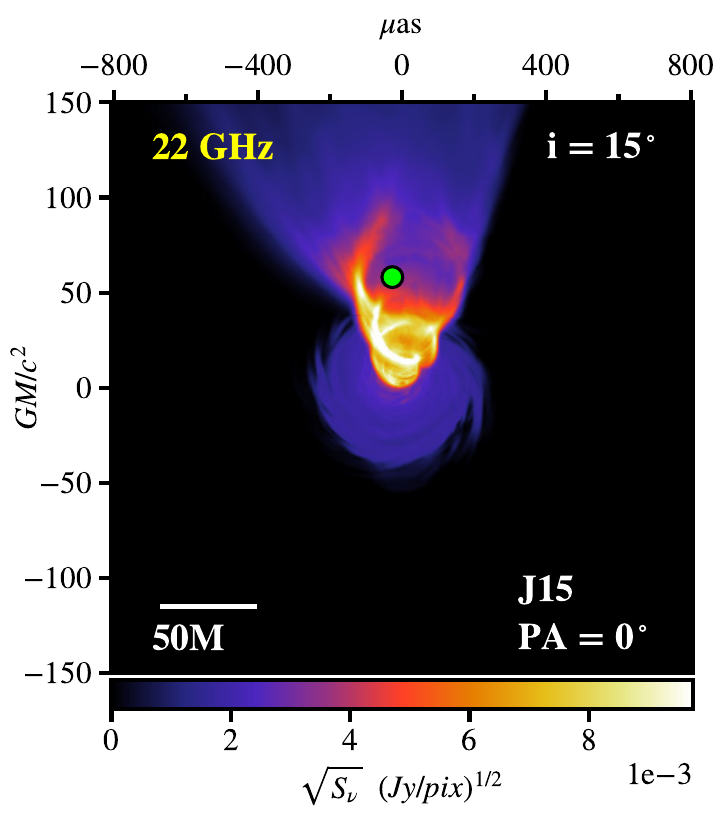}
\includegraphics[width=0.3\textwidth]{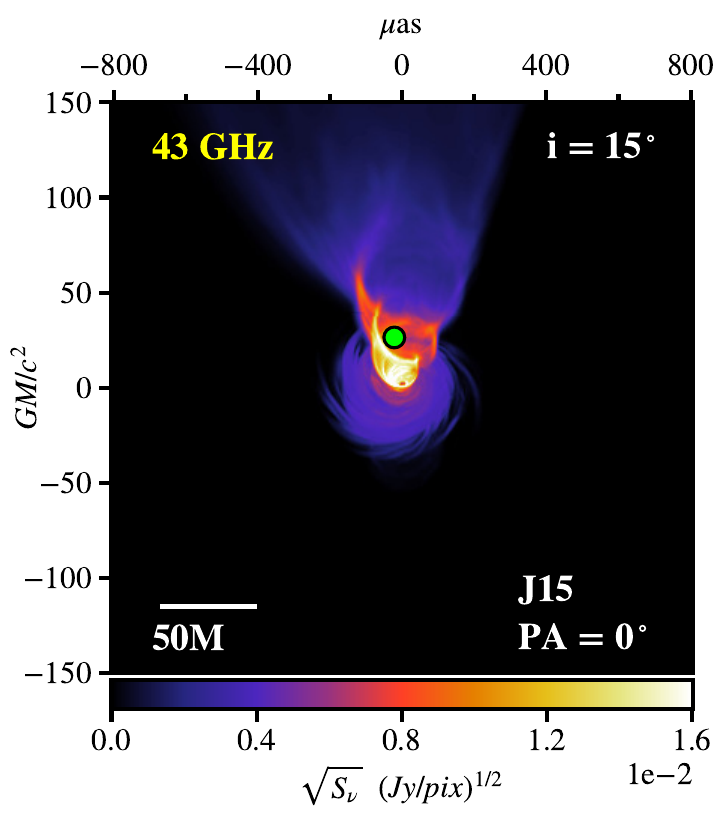}
\includegraphics[width=0.3\textwidth]{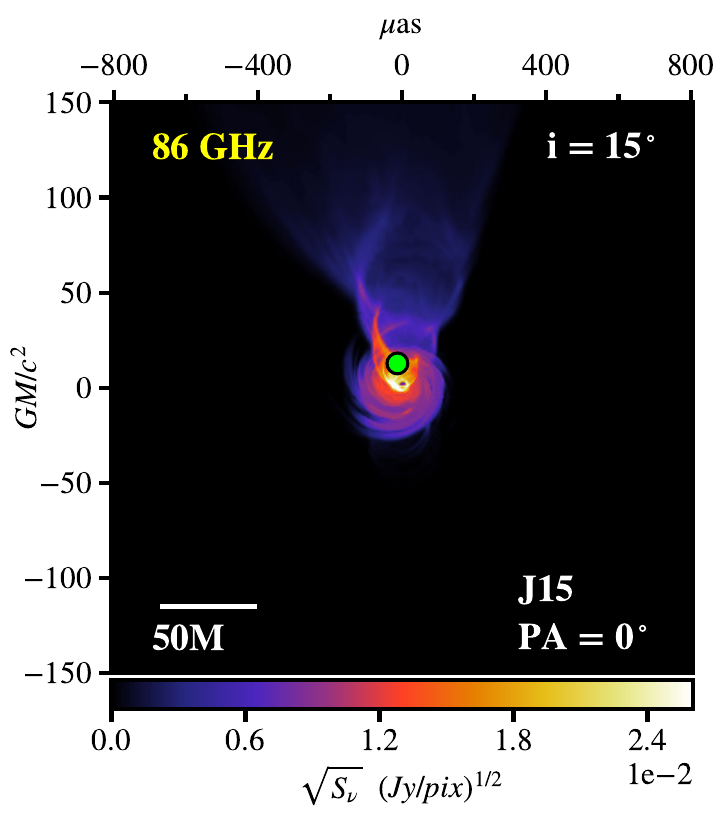}\\
\includegraphics[width=0.3\textwidth]{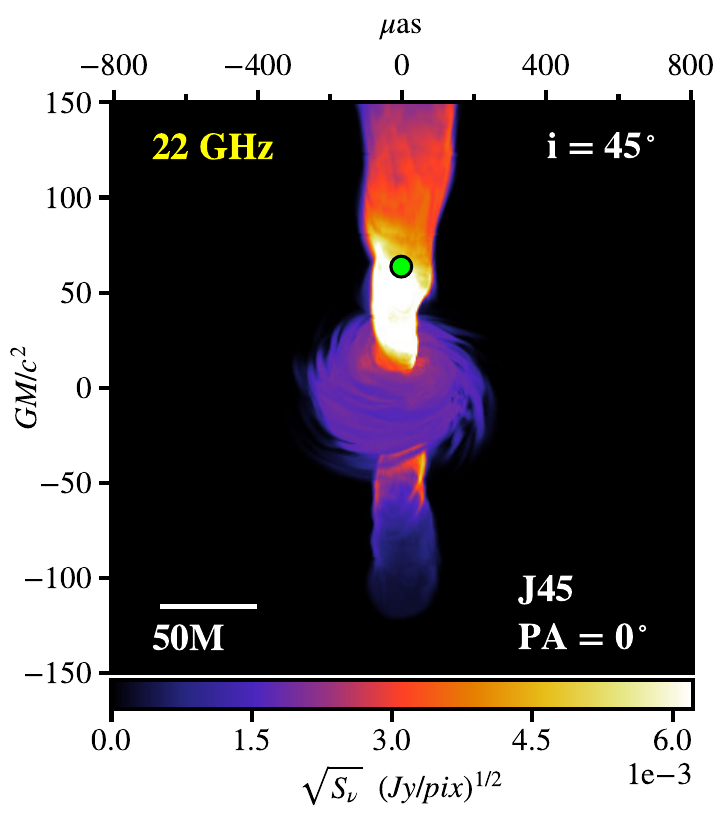}
\includegraphics[width=0.3\textwidth]{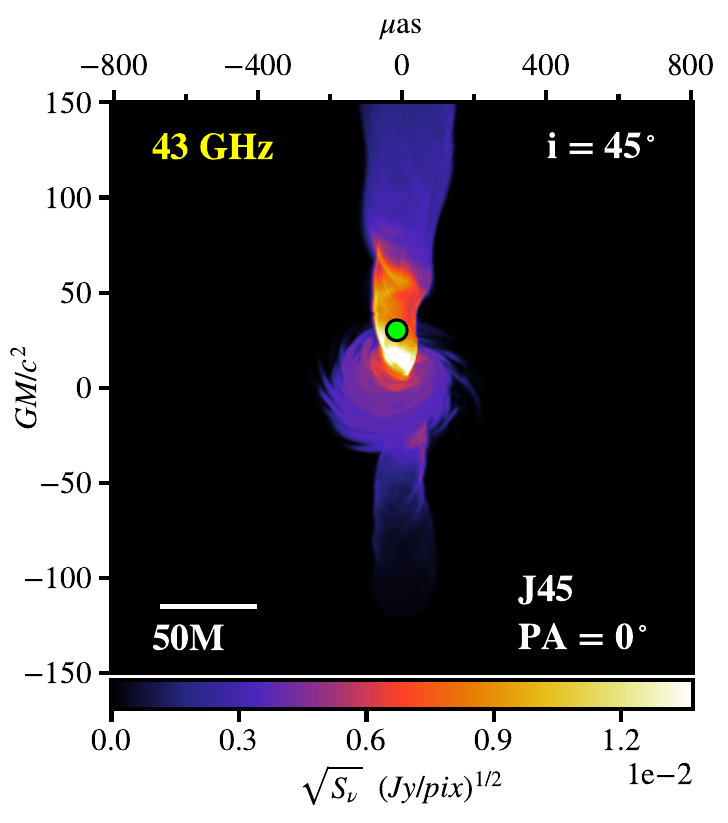}
\includegraphics[width=0.3\textwidth]{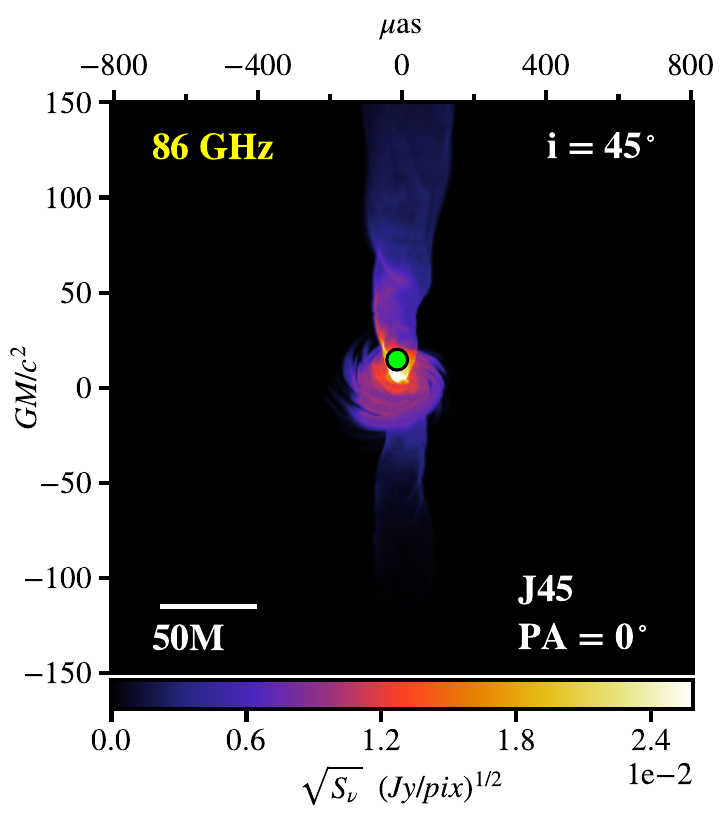}\\
\includegraphics[width=0.3\textwidth]{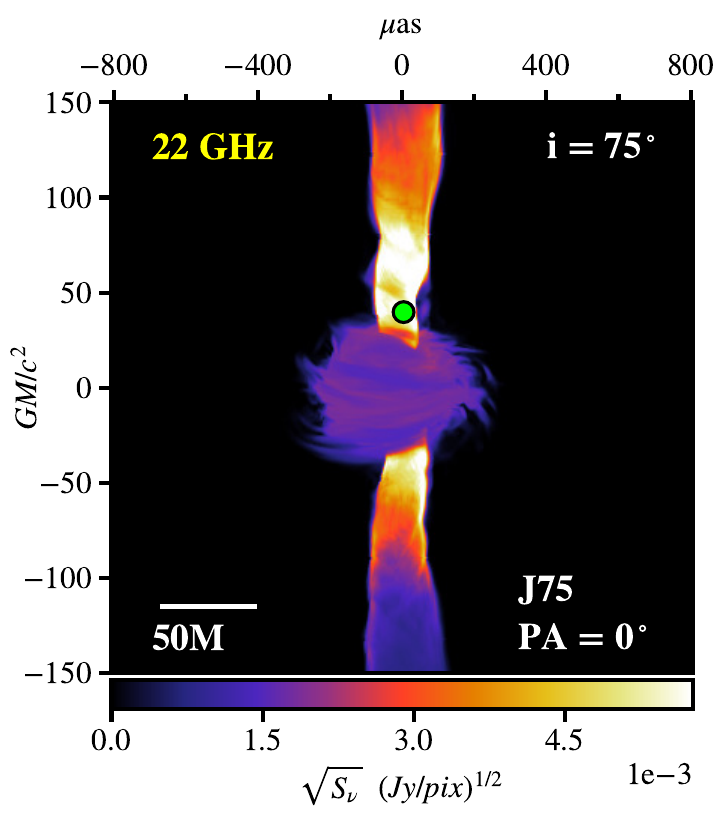}
\includegraphics[width=0.3\textwidth]{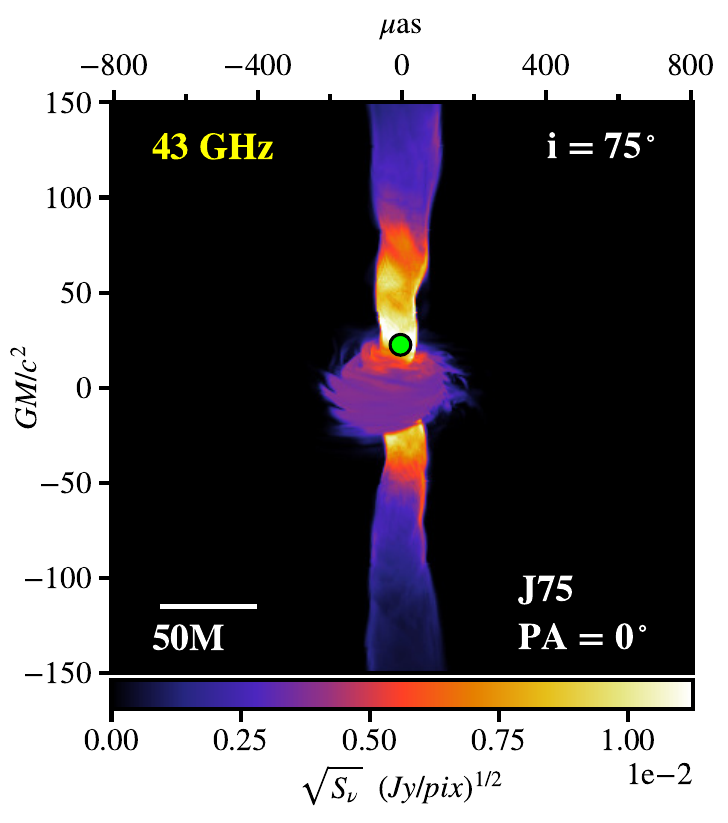}
\includegraphics[width=0.3\textwidth]{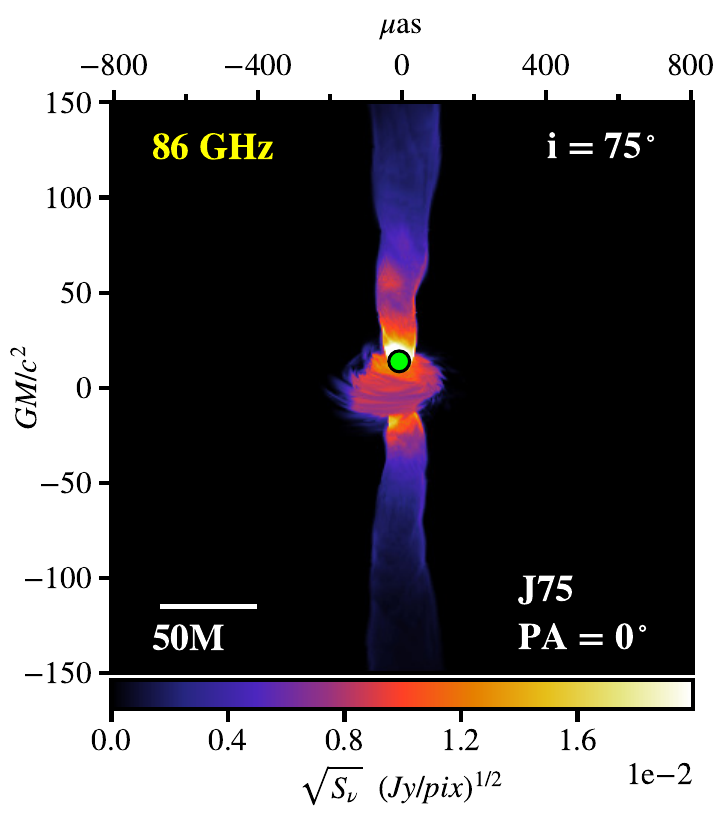}\\
\end{figure*}
\newpage
%
\section{Maps of Jet Models including Scattering (additional inclinations)}~\label{sec:scattered_jet_models_library}
\begin{figure*}[!hp]
\caption{\textbf{Jet model with $i=1 \degr$ including scattering}. Each column shows a jet model with
 inclination $i=1 \degr$ at frequencies of 22~GHz, 43~GHz \& 86~GHz 
 (\textit{left to right}). The position angle on the sky of the BH spin-axis is $PA=0 \degr$. 
 Rows from \textit{top to bottom} show the unscattered 3D-GRMHD jet model, the Gaussian-broadened image, 
 and the average image including refractive scattering. Note that the FOV for images at 22~GHz including 
 scattering \textit{(left column)} is doubled compared to the FOV at 43~GHz \& 86~GHz. The color stretch is 
 also different. The unscattered model panels display the square-root of the flux density (\textit{top row}). 
 The scattered maps are plotted using a linear scale (\textit{second \& third rows}). The green dot indicates
 the intensity-weighted centroid of each image. Black cross-hairs (\textit{second \& third rows}) indicate 
 the location of the intensity-weighted centroid of the model \textit{before} scattering.}\label{fig:scattered_jet_model_i1}
\centering
\includegraphics[width=0.3\textwidth]{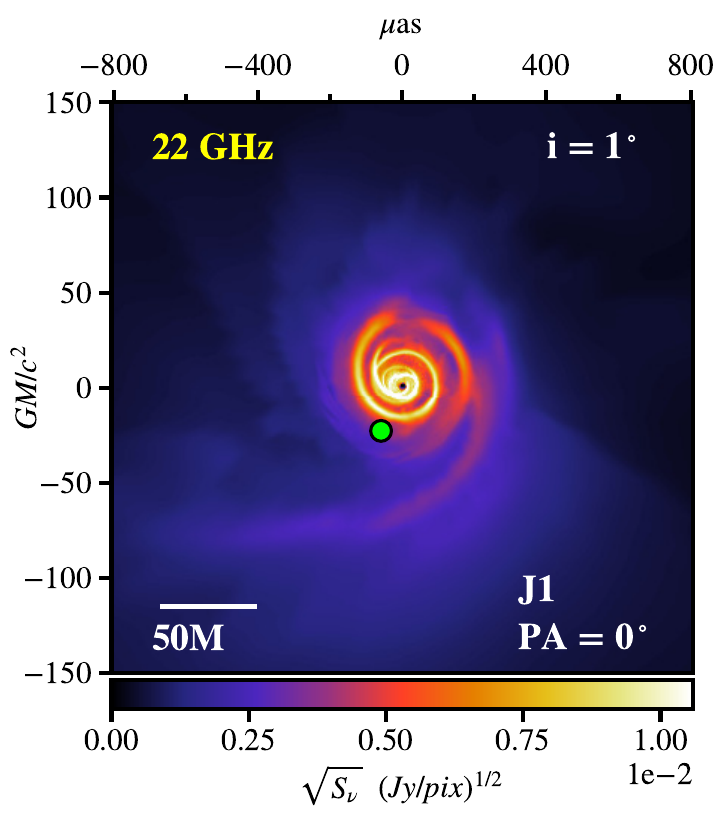}
\includegraphics[width=0.3\textwidth]{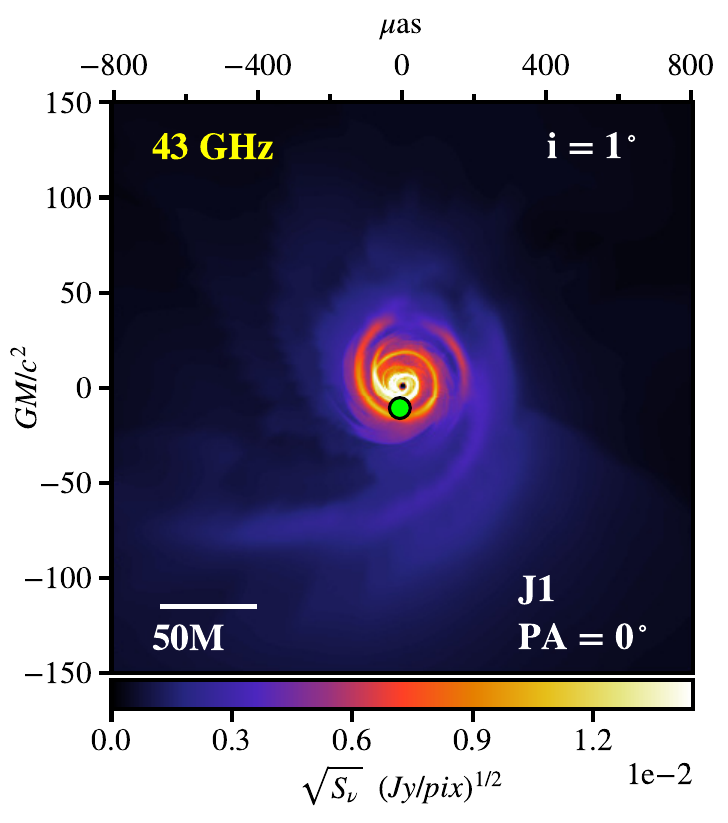}
\includegraphics[width=0.3\textwidth]{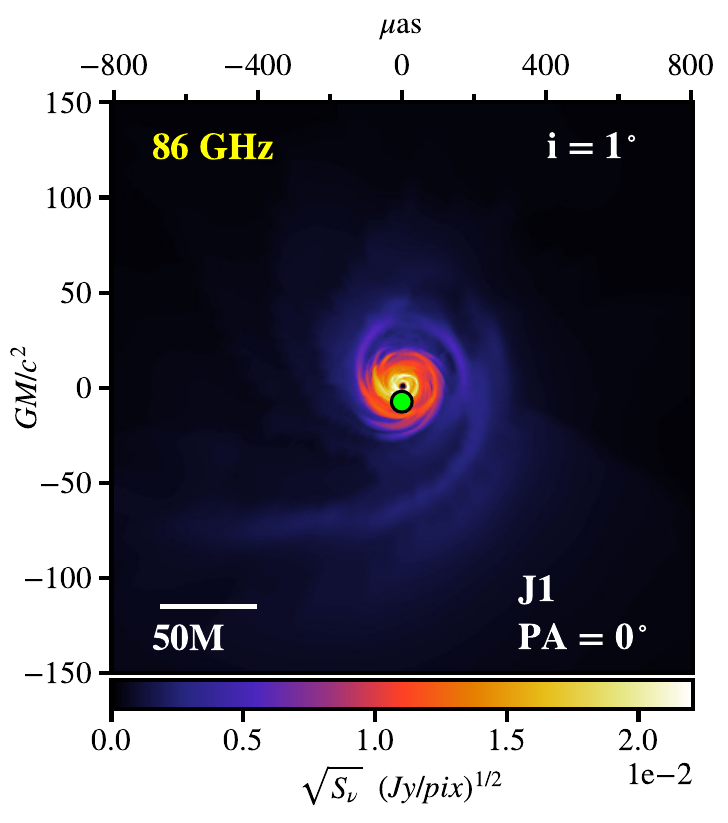}\\
\includegraphics[width=0.3\textwidth]{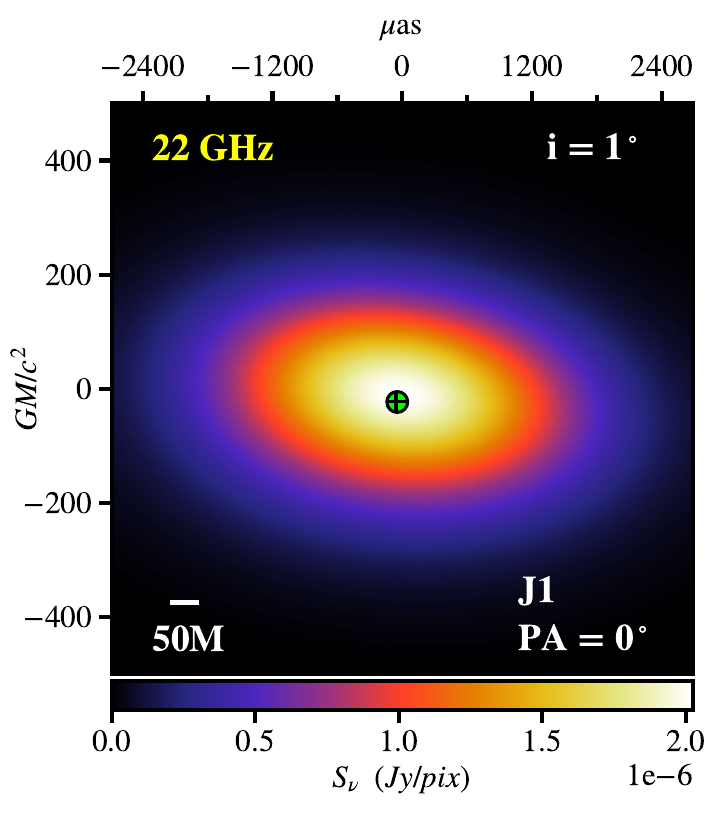}
\includegraphics[width=0.3\textwidth]{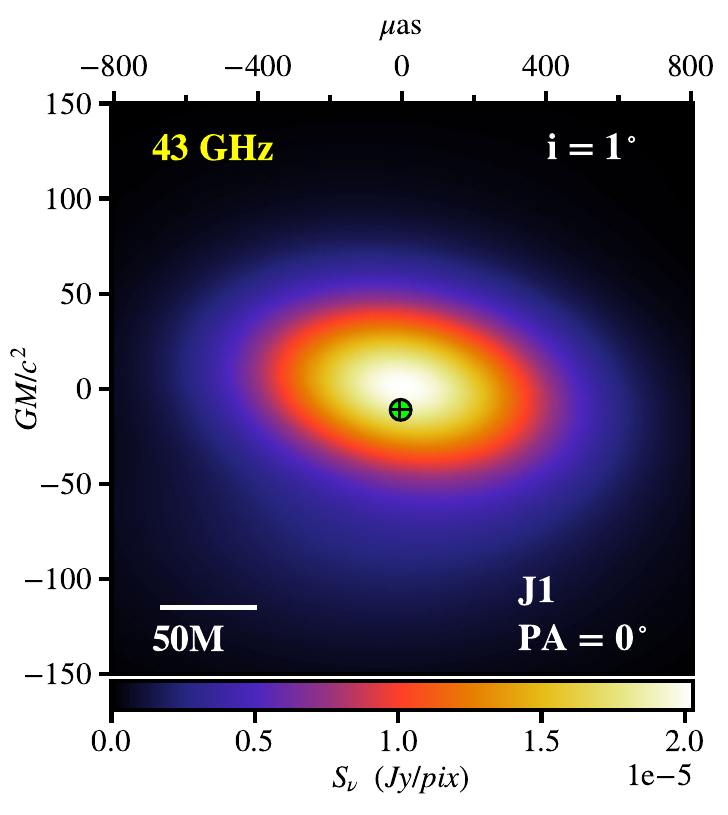}
\includegraphics[width=0.3\textwidth]{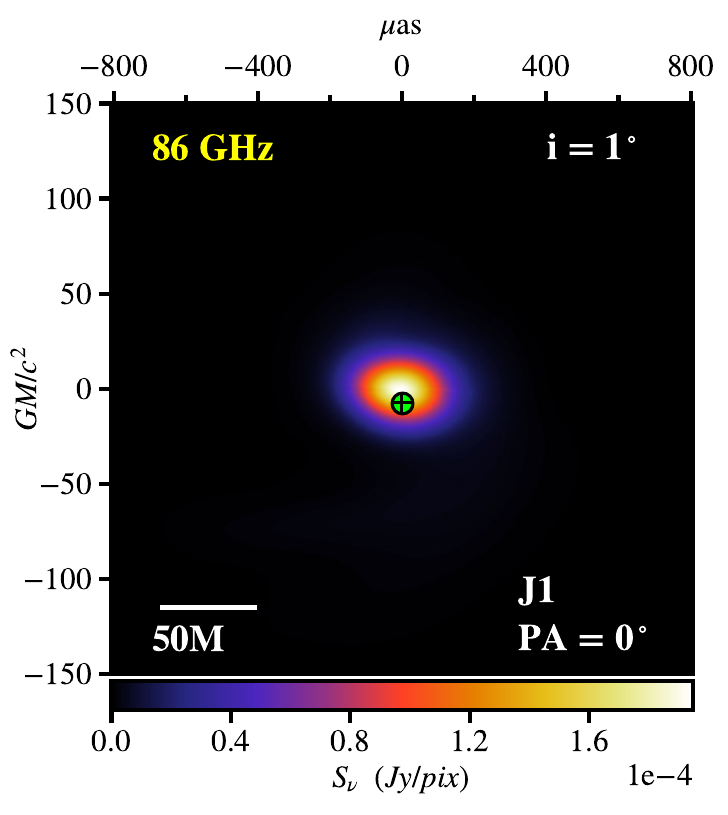}\\
\includegraphics[width=0.3\textwidth]{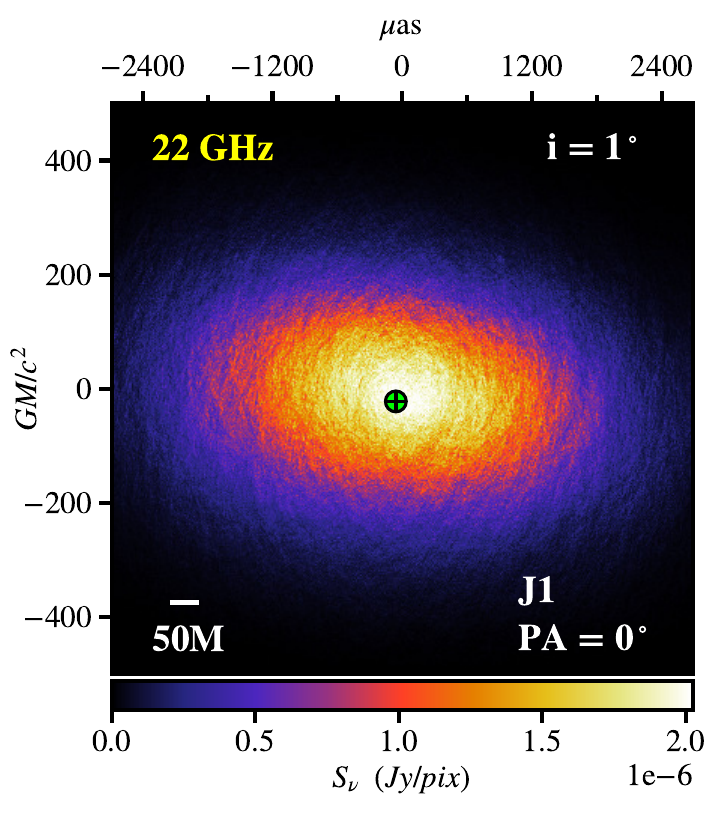}
\includegraphics[width=0.3\textwidth]{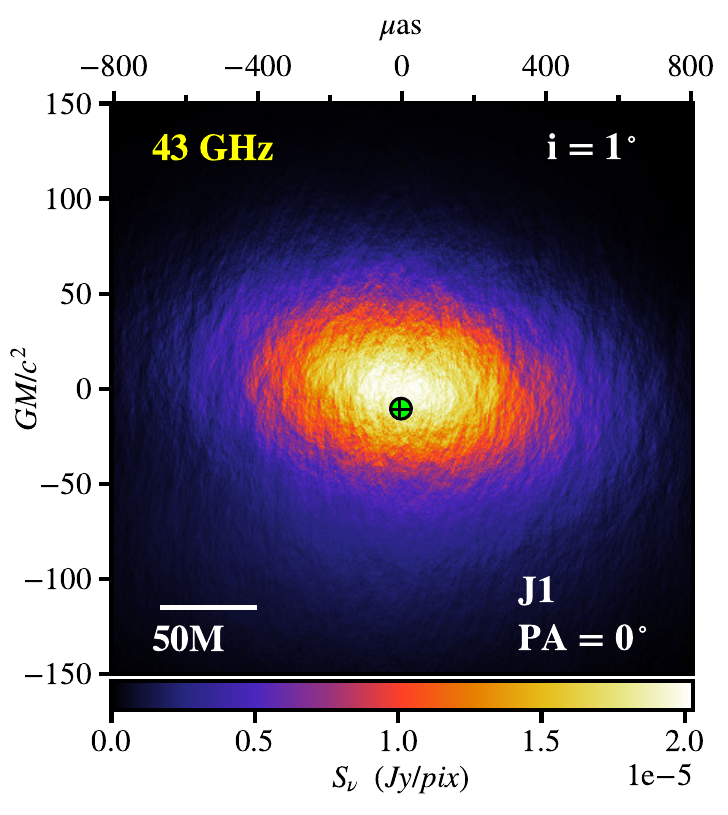}
\includegraphics[width=0.3\textwidth]{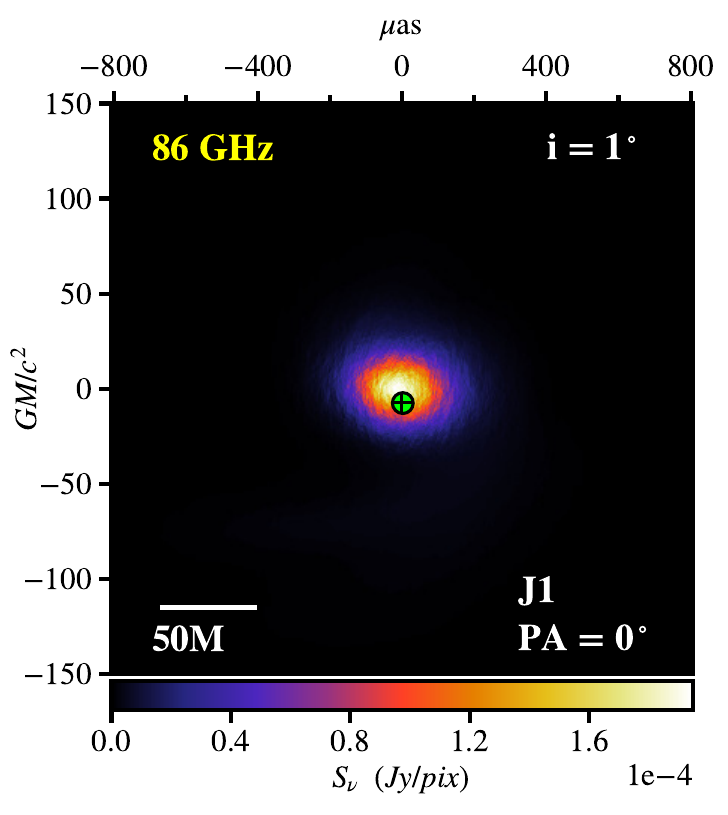}
\end{figure*}
\newpage
\begin{figure*}[!hp]
\caption{\textbf{Jet model with $i=15 \degr$ including scattering}. Each column shows a jet model with
 inclination $i=15 \degr$ at frequencies of 22~GHz, 43~GHz \& 86~GHz 
 (\textit{left to right}). The position angle on the sky of the BH spin-axis is $PA=0 \degr$. 
 Rows from \textit{top to bottom} show the unscattered 3D-GRMHD jet model, the scatter-broadened image, 
 and the average image including refractive scattering. Note that the FOV for images at 22~GHz including 
 scattering \textit{(left column)} is doubled compared to the FOV at 43~GHz \& 86~GHz. The color stretch is 
 also different. The unscattered model panels display the square-root of the flux density (\textit{top row}). 
 The scattered maps are plotted using a linear scale (\textit{second \& third rows}). The green dot indicates
 the intensity-weighted centroid of each image. Black cross-hairs (\textit{second \& third rows}) indicate 
 the location of the intensity-weighted centroid of the model \textit{before} scattering.}\label{fig:scattered_jet_model_i15}
\centering
\includegraphics[width=0.3\textwidth]{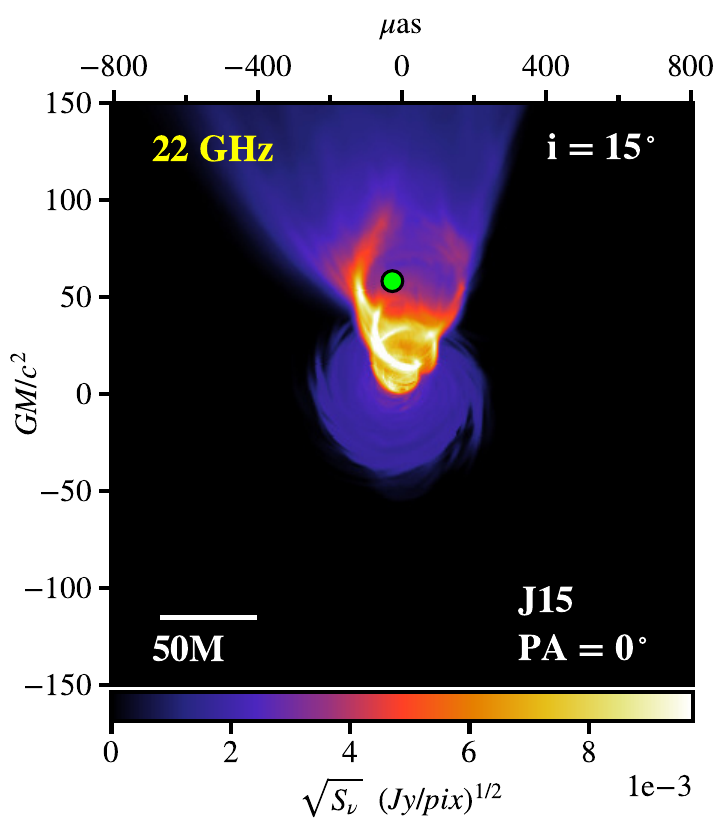}
\includegraphics[width=0.3\textwidth]{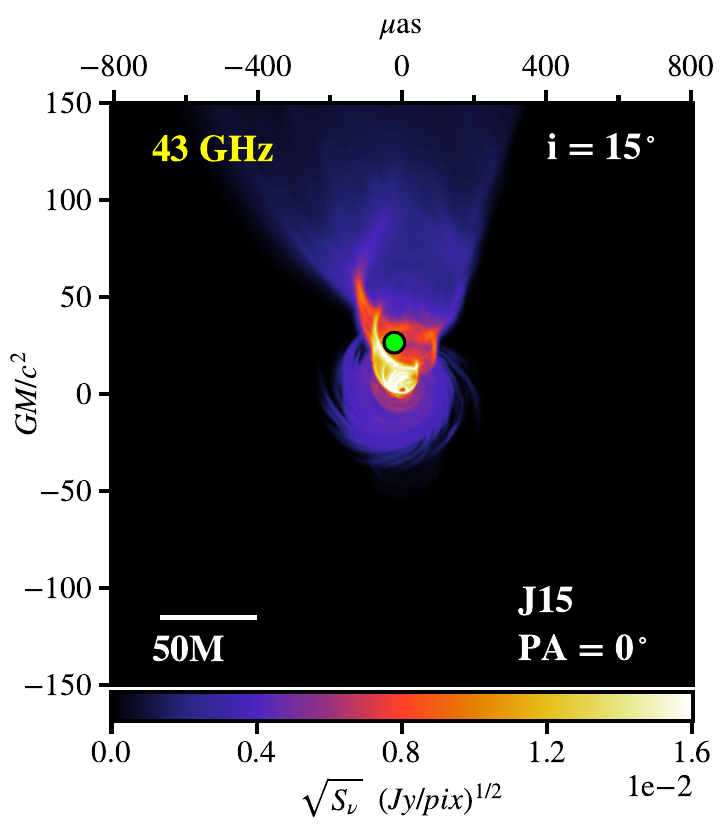}
\includegraphics[width=0.3\textwidth]{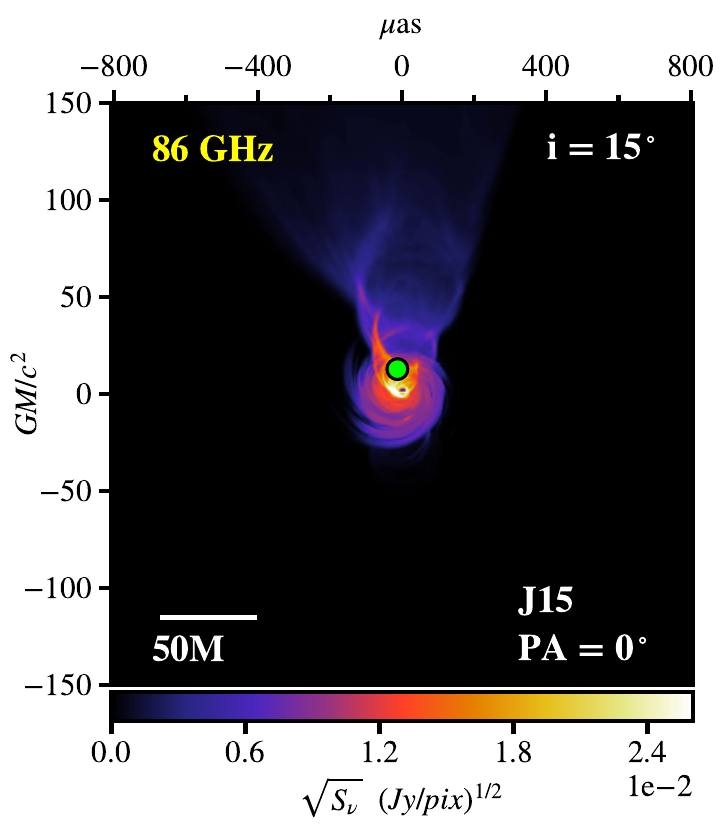}\\
\includegraphics[width=0.3\textwidth]{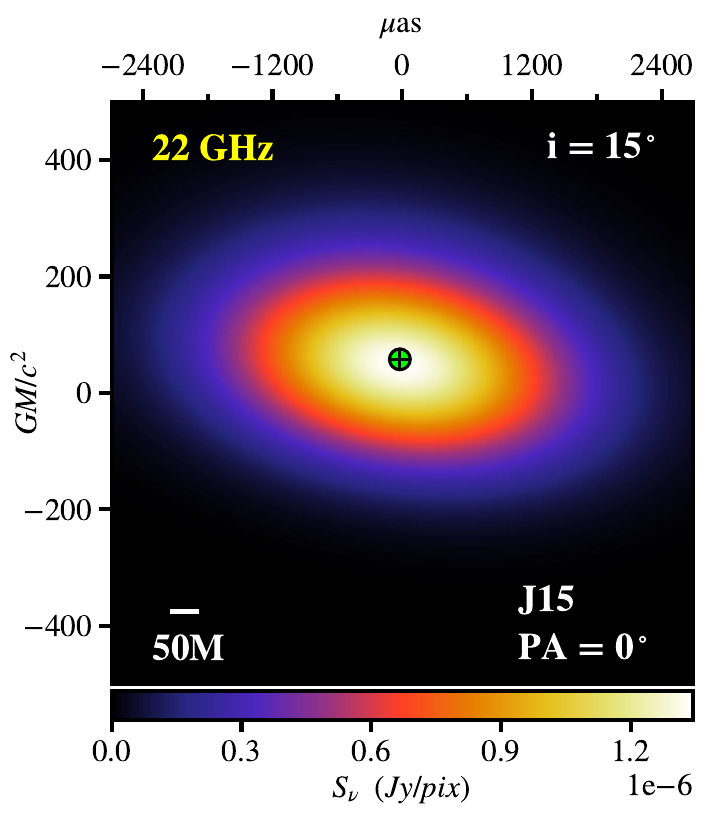}
\includegraphics[width=0.3\textwidth]{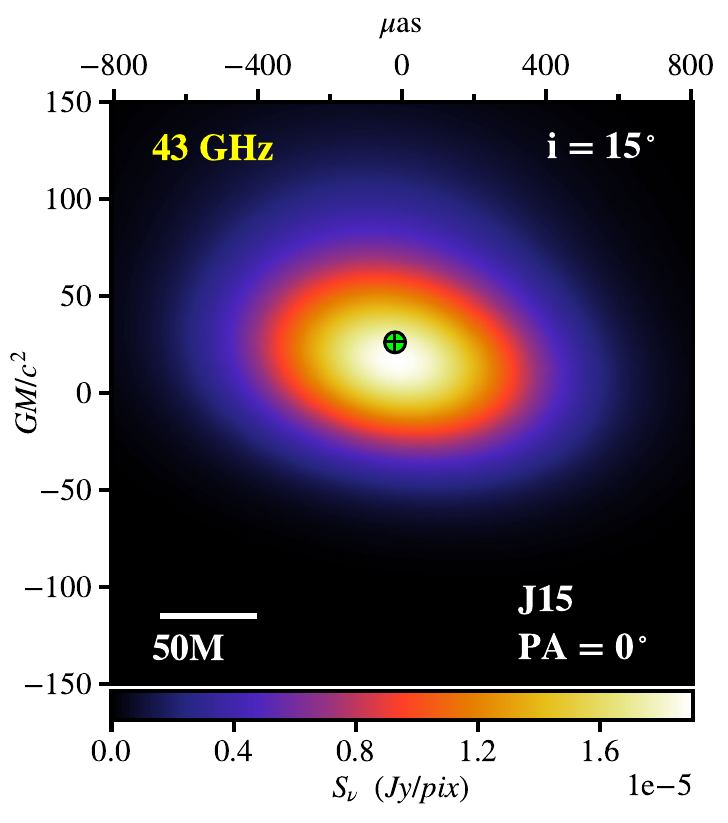}
\includegraphics[width=0.3\textwidth]{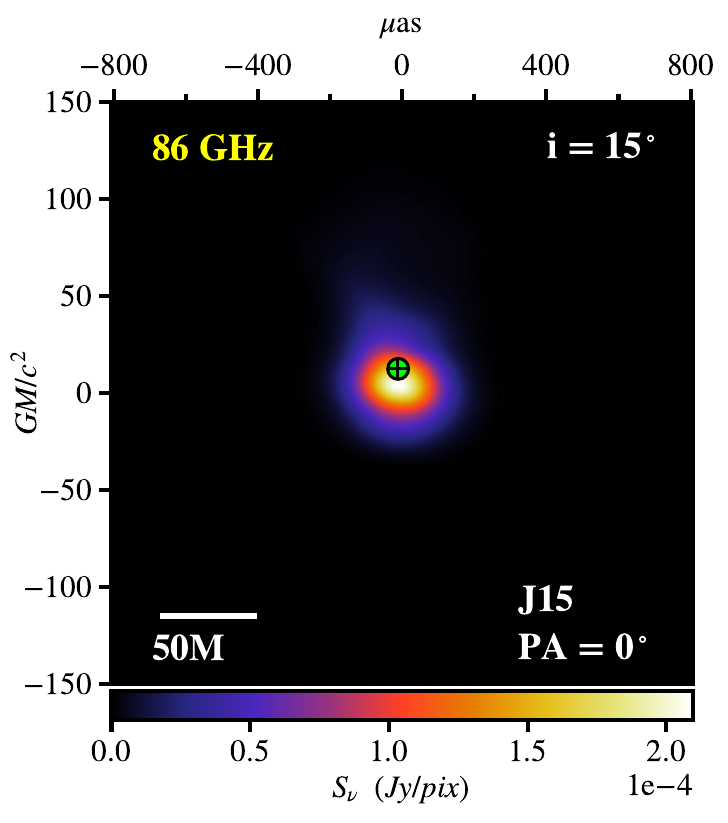}\\
\includegraphics[width=0.3\textwidth]{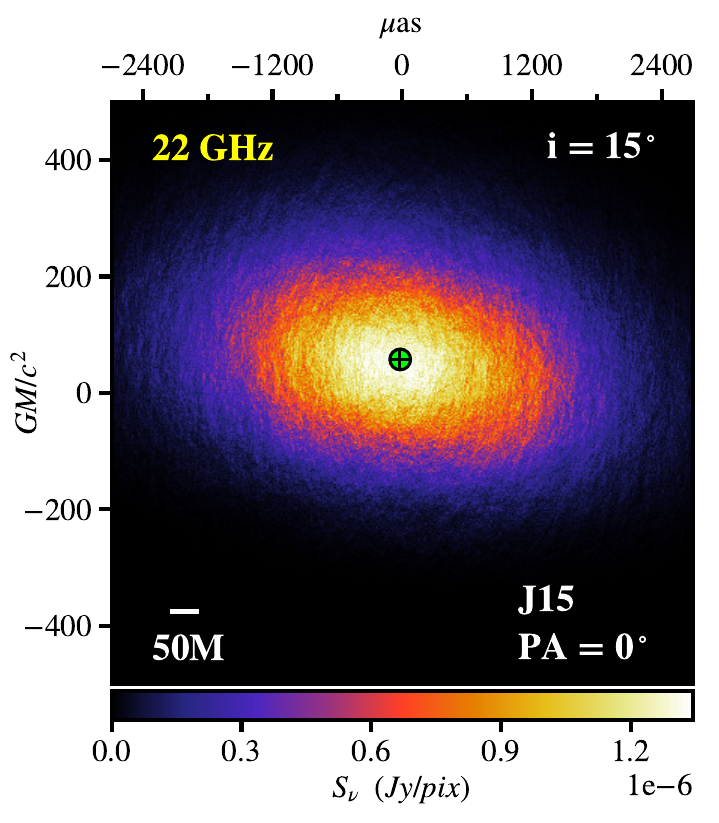}
\includegraphics[width=0.3\textwidth]{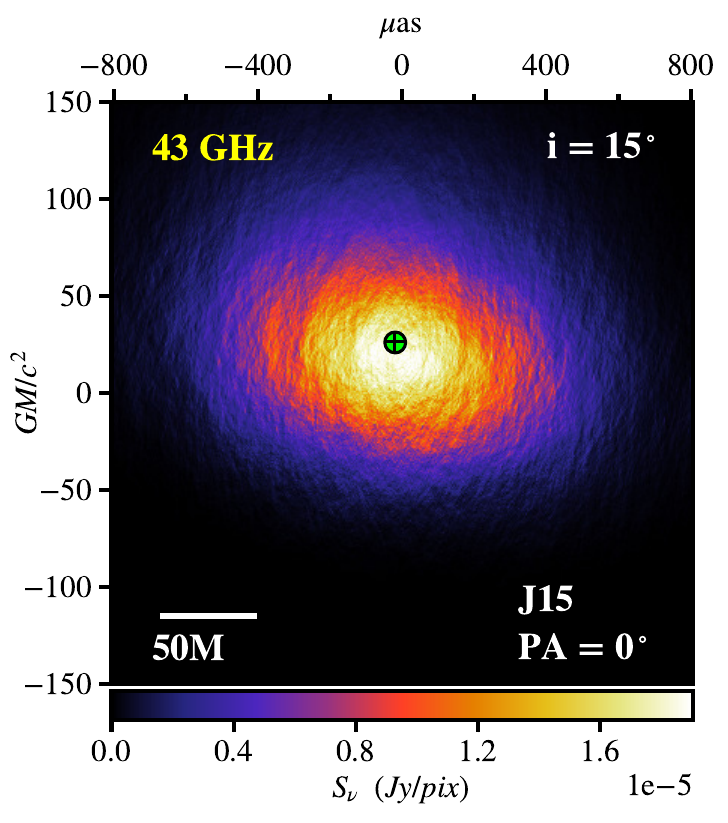}
\includegraphics[width=0.3\textwidth]{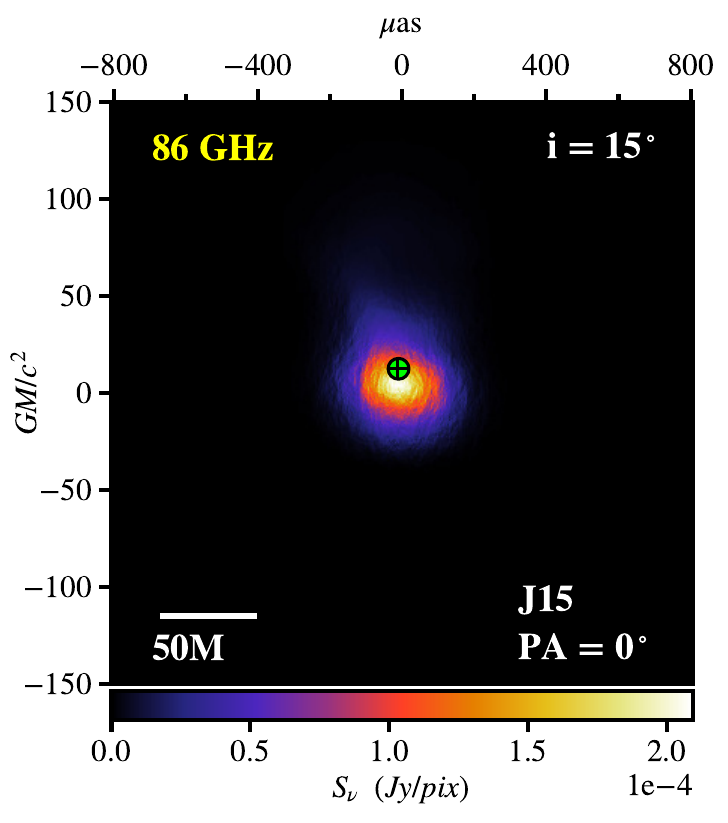}
\end{figure*}
\newpage
\begin{figure*}[!hp]
\caption{\textbf{Jet model with $i=45 \degr$ including scattering}. Each column shows a jet model with
 inclination $i=45 \degr$ at frequencies of 22~GHz, 43~GHz \& 86~GHz 
 (\textit{left to right}). The position angle on the sky of the BH spin-axis is $PA=0 \degr$. 
 Rows from \textit{top to bottom} show the unscattered 3D-GRMHD jet model, the scatter-broadened image, 
 and the average image including refractive scattering. Note that the FOV for images at 22~GHz including 
 scattering \textit{(left column)} is doubled compared to the FOV at 43~GHz \& 86~GHz. The color stretch is 
 also different. The unscattered model panels display the square-root of the flux density (\textit{top row}). 
 The scattered maps are plotted using a linear scale (\textit{second \& third rows}). The green dot indicates
 the intensity-weighted centroid of each image. Black cross-hairs (\textit{second \& third rows}) indicate 
 the location of the intensity-weighted centroid of the model \textit{before} scattering.}\label{fig:scattered_jet_model_i45}
\centering
\includegraphics[width=0.3\textwidth]{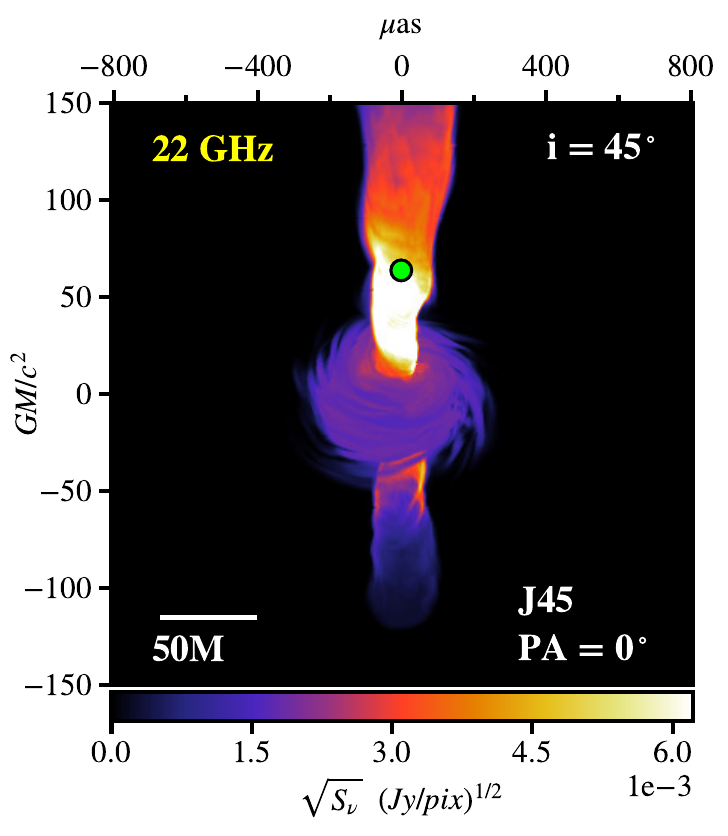}
\includegraphics[width=0.3\textwidth]{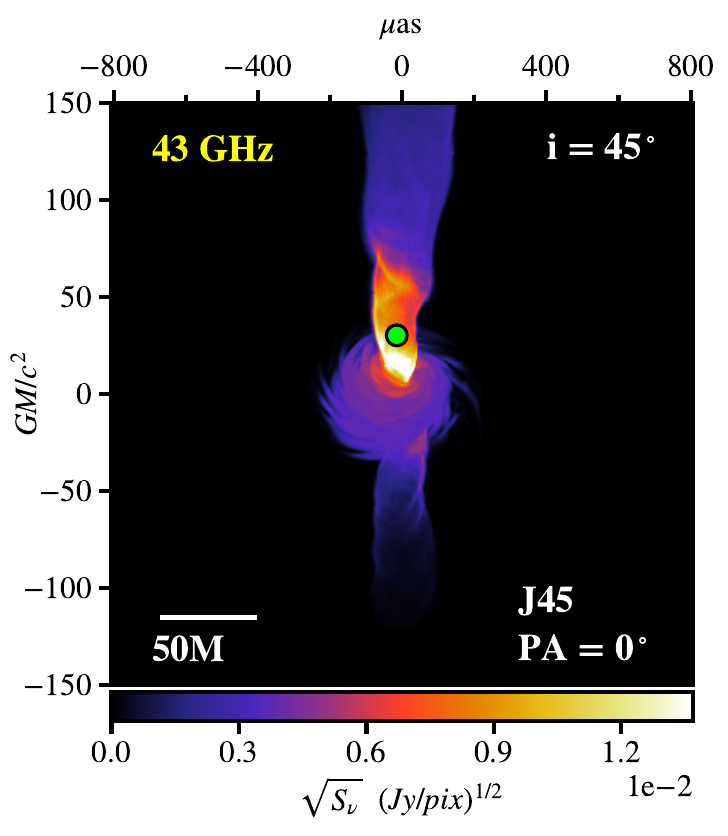}
\includegraphics[width=0.3\textwidth]{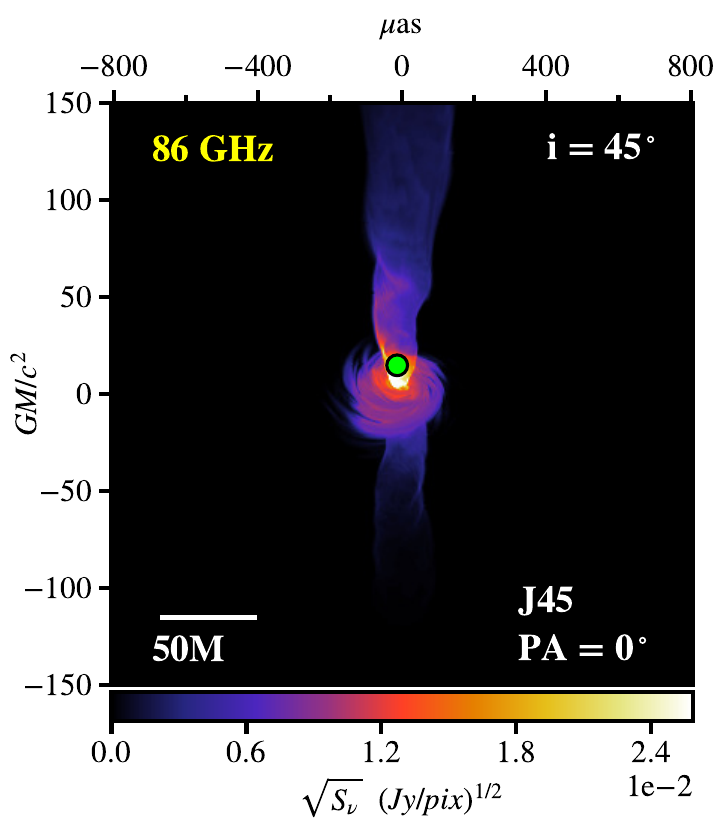}\\
\includegraphics[width=0.3\textwidth]{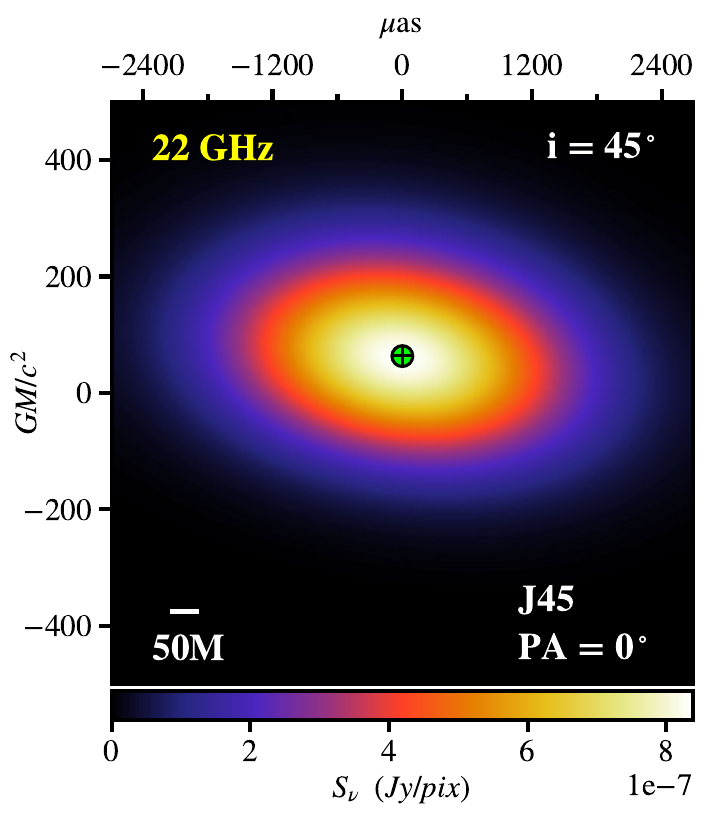}
\includegraphics[width=0.3\textwidth]{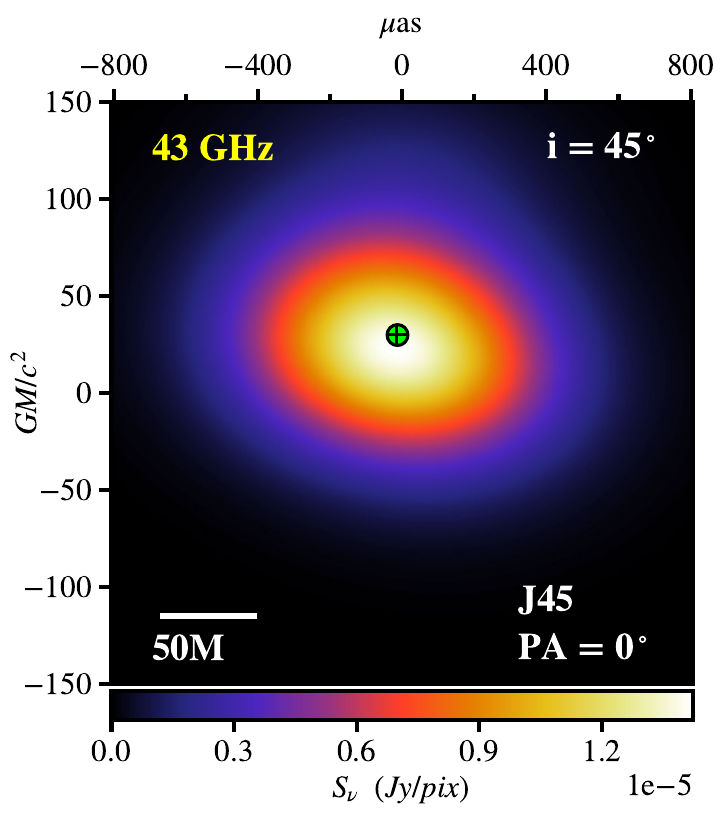}
\includegraphics[width=0.3\textwidth]{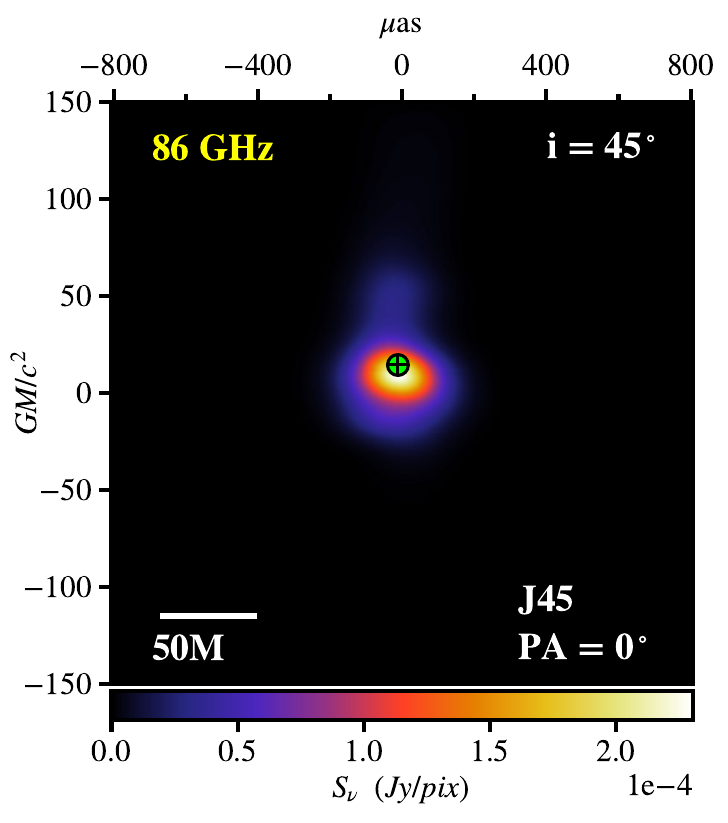}\\
\includegraphics[width=0.3\textwidth]{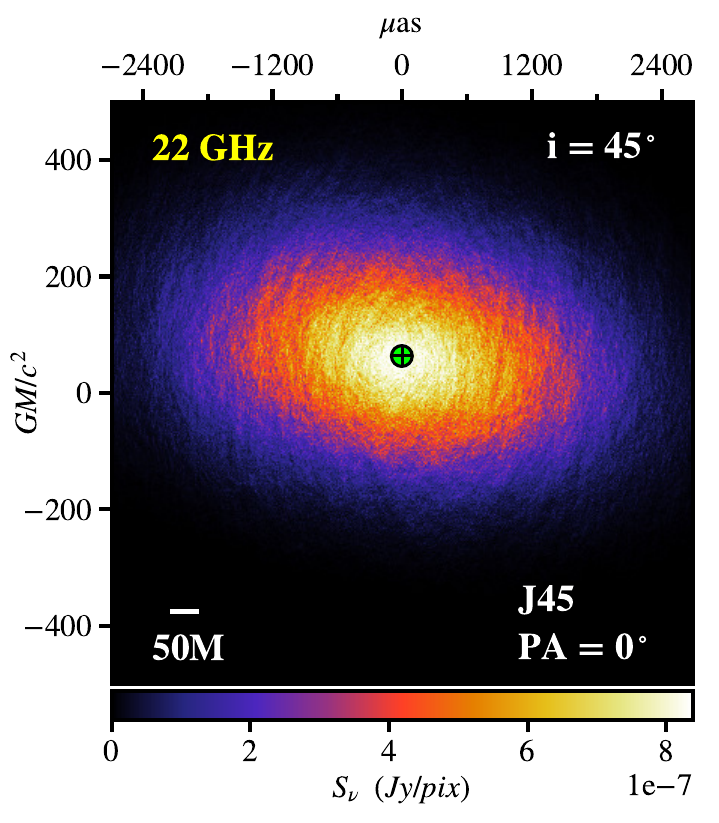}
\includegraphics[width=0.3\textwidth]{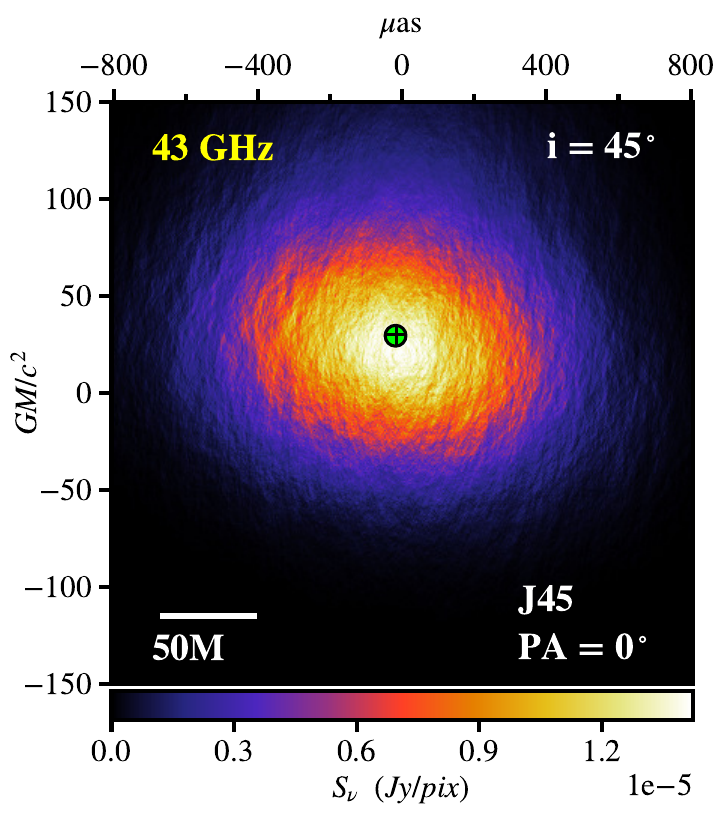}
\includegraphics[width=0.3\textwidth]{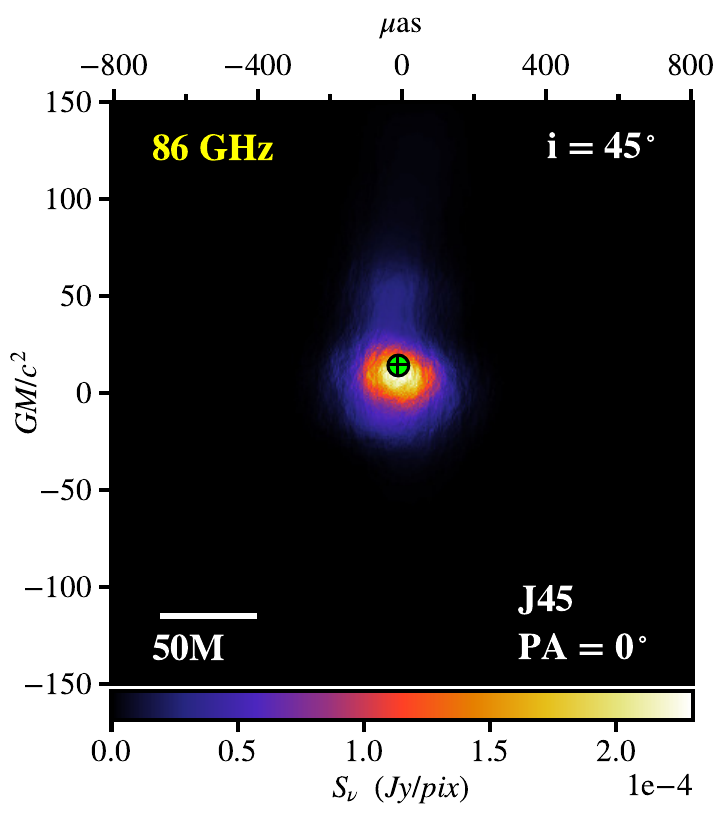}
\end{figure*}
\newpage
\begin{figure*}[!hp]
\caption{\textbf{Jet model with $i=60 \degr$ including scattering}. Each column shows a jet model with
 inclination $i=60 \degr$ at frequencies of 22~GHz, 43~GHz \& 86~GHz 
 (\textit{left to right}). The position angle on the sky of the BH spin-axis is $PA=0 \degr$. 
 Rows from \textit{top to bottom} show the unscattered 3D-GRMHD jet model, the scatter-broadened image, 
 and the average image including refractive scattering. Note that the FOV for images at 22~GHz including 
 scattering \textit{(left column)} is doubled compared to the FOV at 43~GHz \& 86~GHz. The color stretch is 
 also different. The unscattered model panels display the square-root of the flux density (\textit{top row}). 
 The scattered maps are plotted using a linear scale (\textit{second \& third rows}). The green dot indicates
 the intensity-weighted centroid of each image. Black cross-hairs (\textit{second \& third rows}) indicate 
 the location of the intensity-weighted centroid of the model \textit{before} scattering.}\label{fig:scattered_jet_model_i60}
\centering
\includegraphics[width=0.3\textwidth]{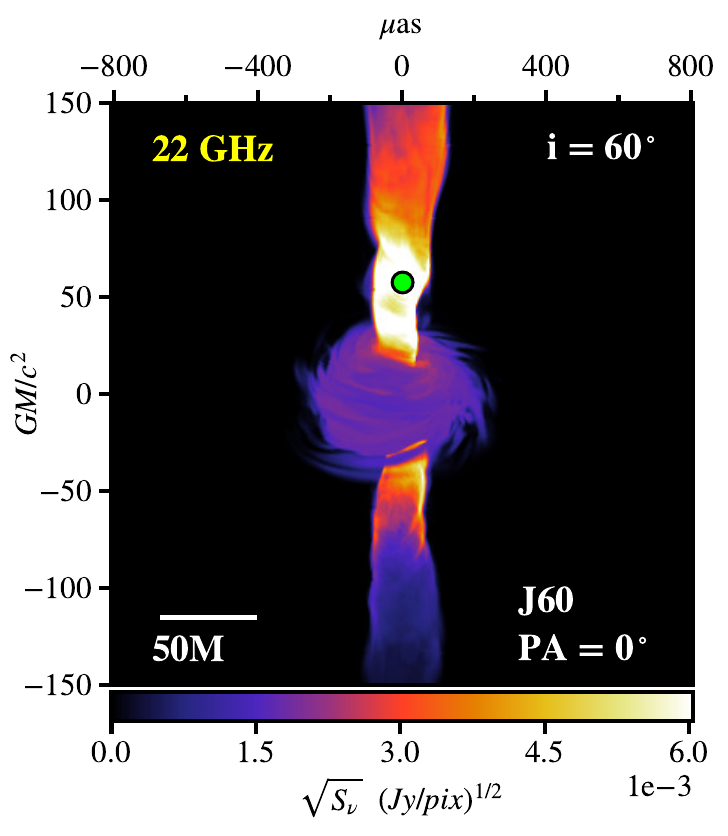}
\includegraphics[width=0.3\textwidth]{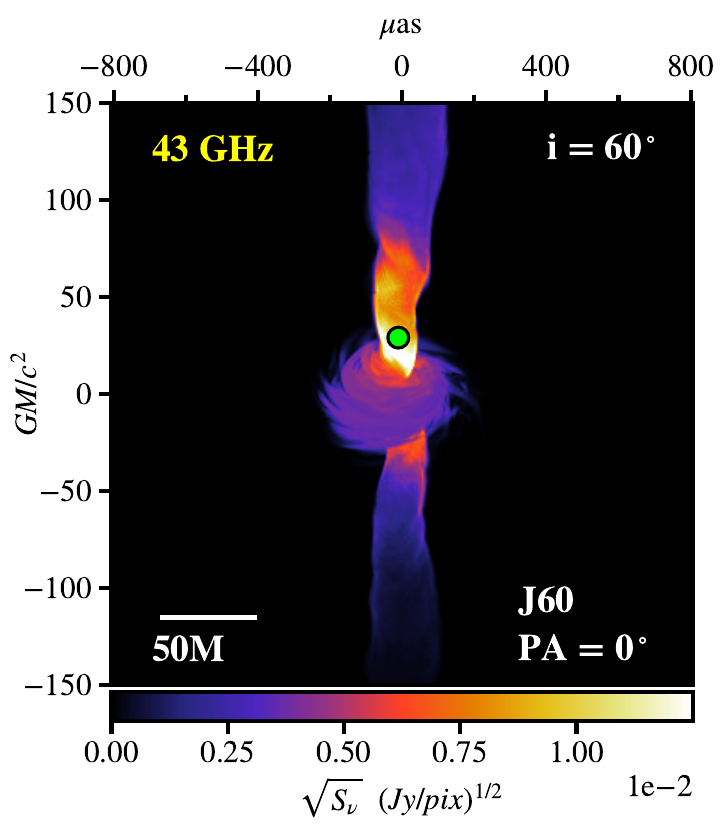}
\includegraphics[width=0.3\textwidth]{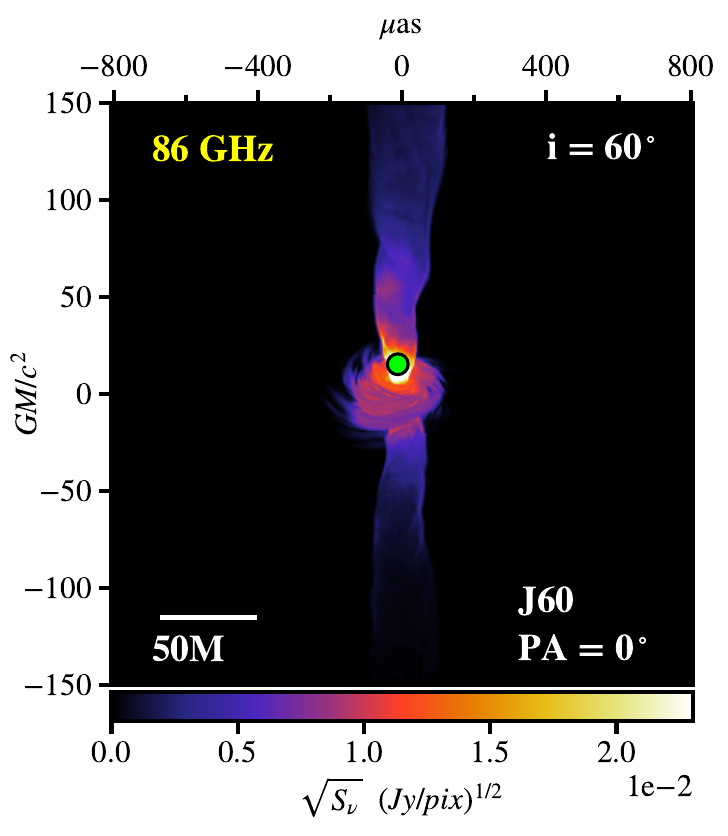}\\
\includegraphics[width=0.3\textwidth]{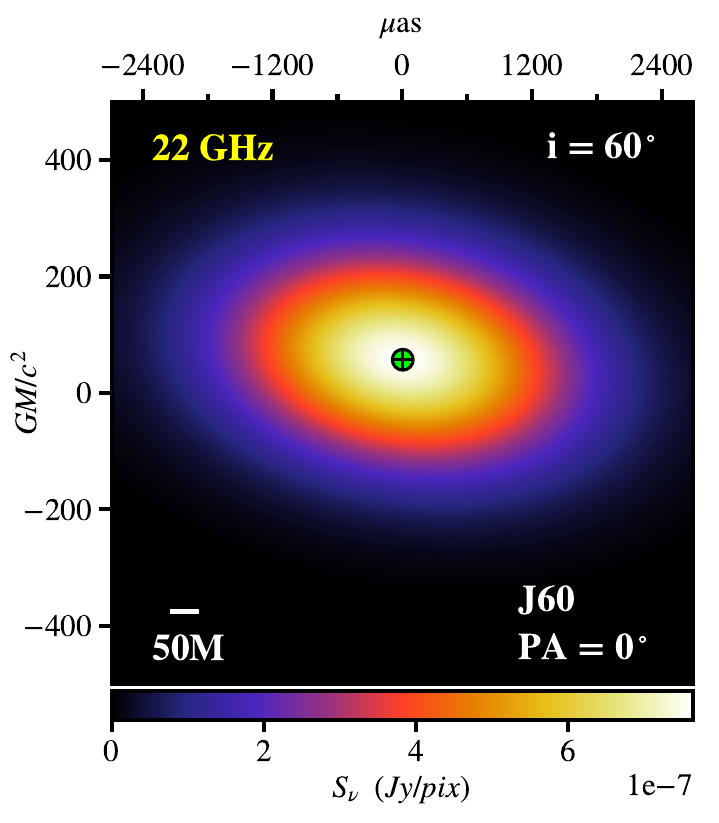}
\includegraphics[width=0.3\textwidth]{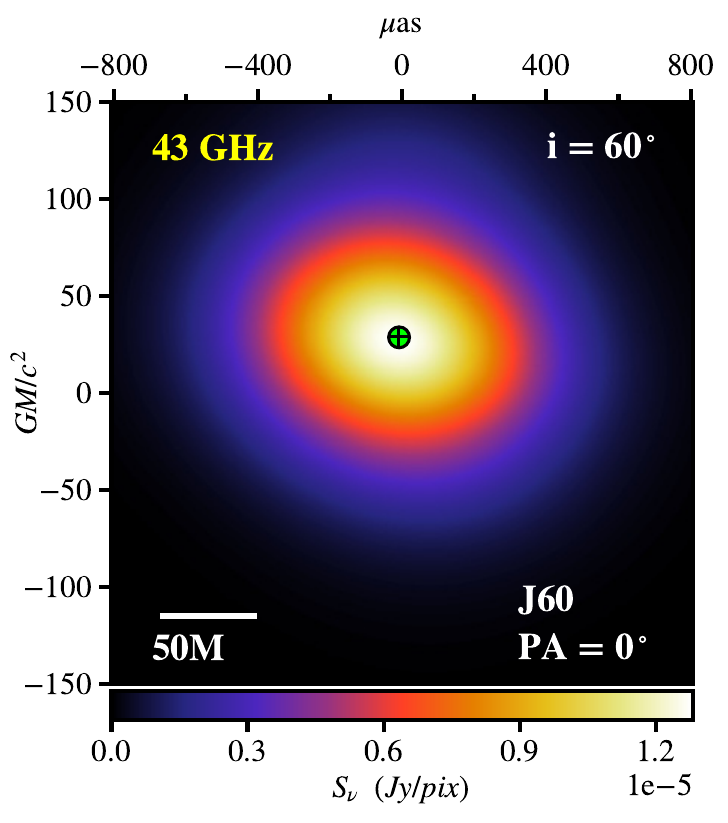}
\includegraphics[width=0.3\textwidth]{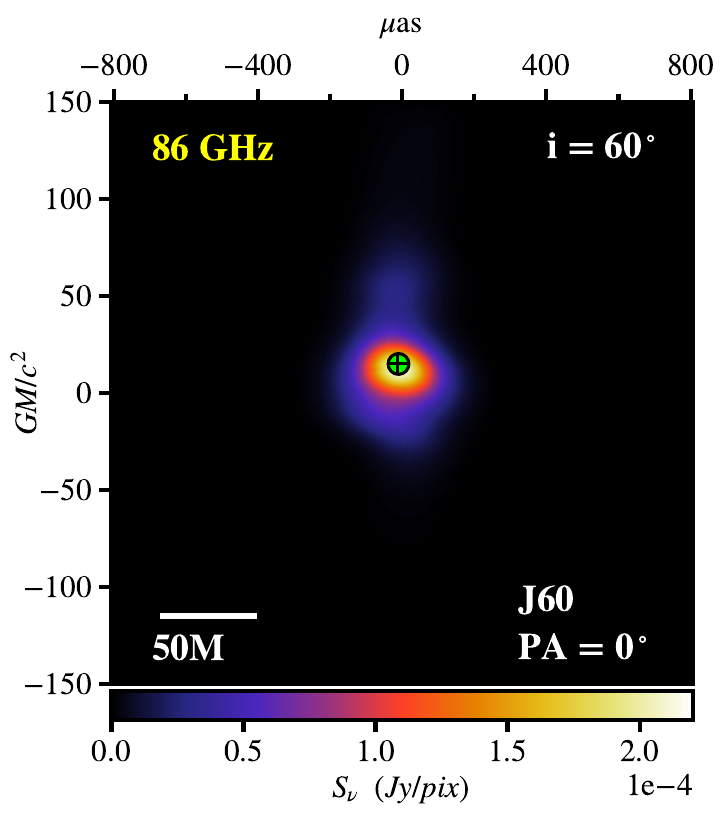}\\
\includegraphics[width=0.3\textwidth]{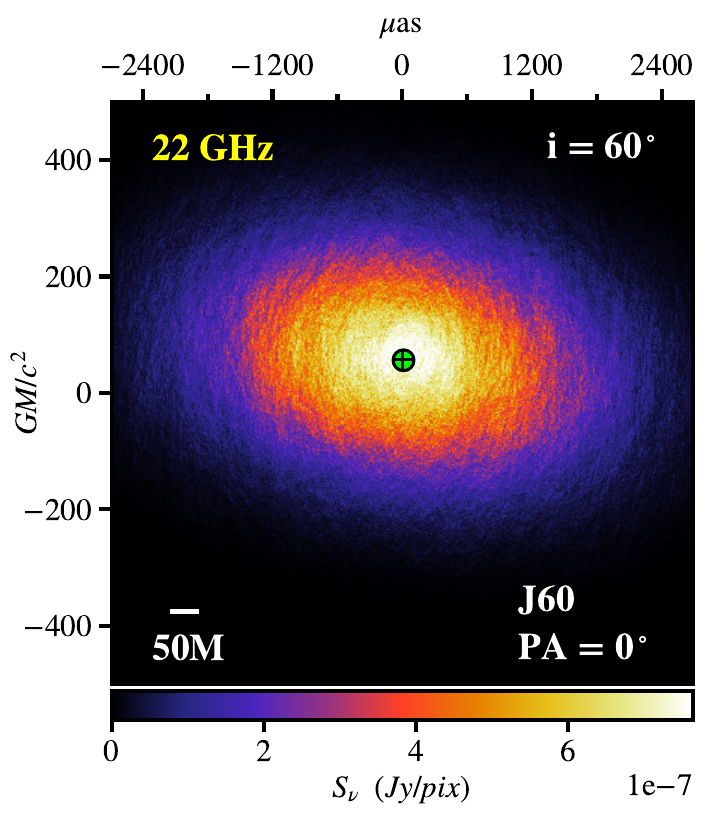}
\includegraphics[width=0.3\textwidth]{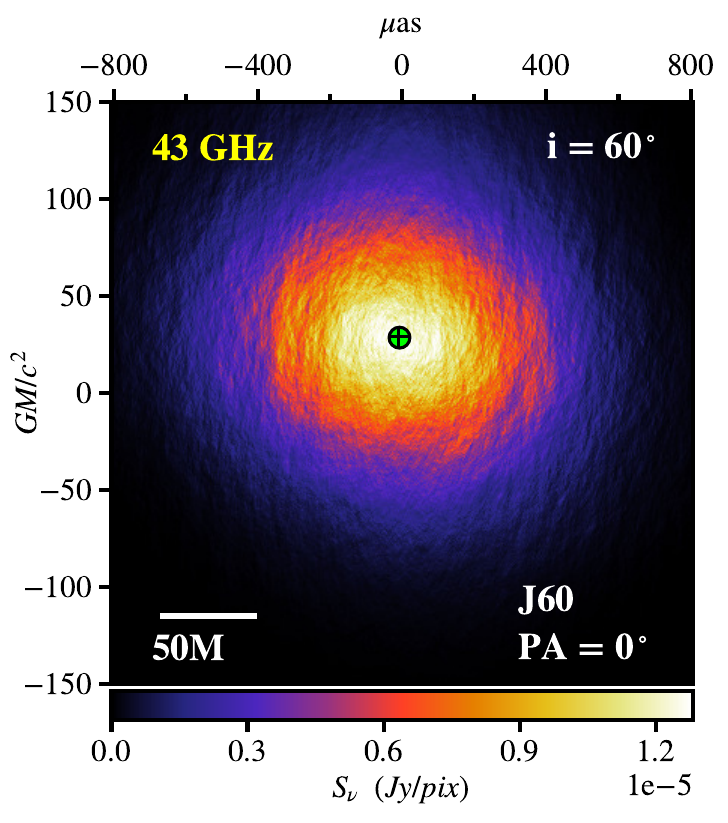}
\includegraphics[width=0.3\textwidth]{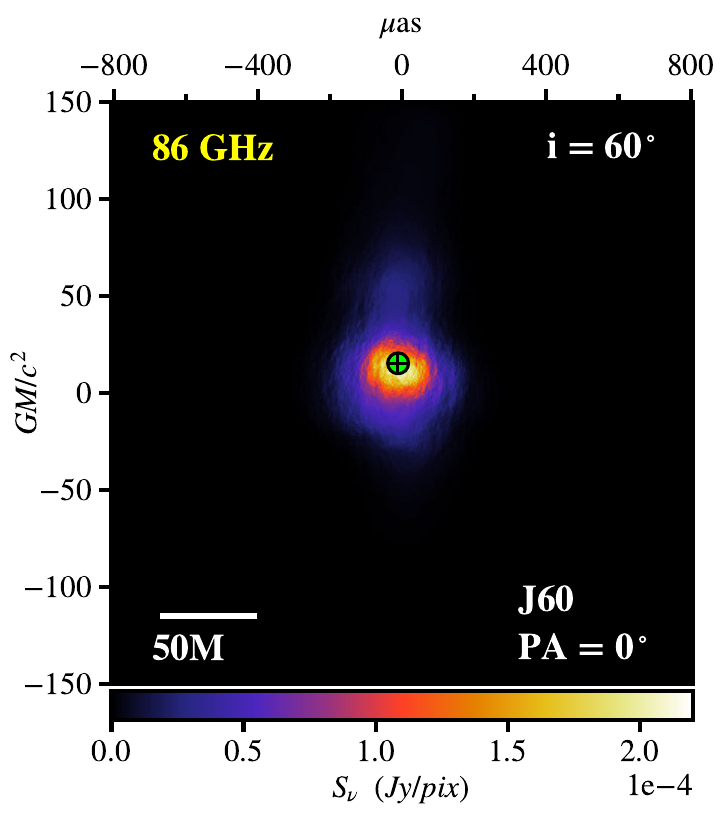}
\end{figure*}
\newpage
\begin{figure*}[!hp]
\caption{\textbf{Jet model with $i=75 \degr$ including scattering}. Each column shows a jet model with
 inclination $i=75 \degr$ at frequencies of 22~GHz, 43~GHz \& 86~GHz 
 (\textit{left to right}). The position angle on the sky of the BH spin-axis is $PA=0 \degr$. 
 Rows from \textit{top to bottom} show the unscattered 3D-GRMHD jet model, the scatter-broadened image, 
 and the average image including refractive scattering. Note that the FOV for images at 22~GHz including 
 scattering \textit{(left column)} is doubled compared to the FOV at 43~GHz \& 86~GHz. The color strecht is 
 also different. The unscattered model panels display the square-root of the flux density (\textit{top row}). 
 The scattered maps are plotted using a linear scale (\textit{second \& third rows}). The green dot indicates
 the intensity-weighted centroid of each image. Black cross-hairs (\textit{second \& third rows}) indicate 
 the location of the intensity-weighted centroid of the model \textit{before} scattering.}\label{fig:scattered_jet_model_i75}
\centering
\includegraphics[width=0.3\textwidth]{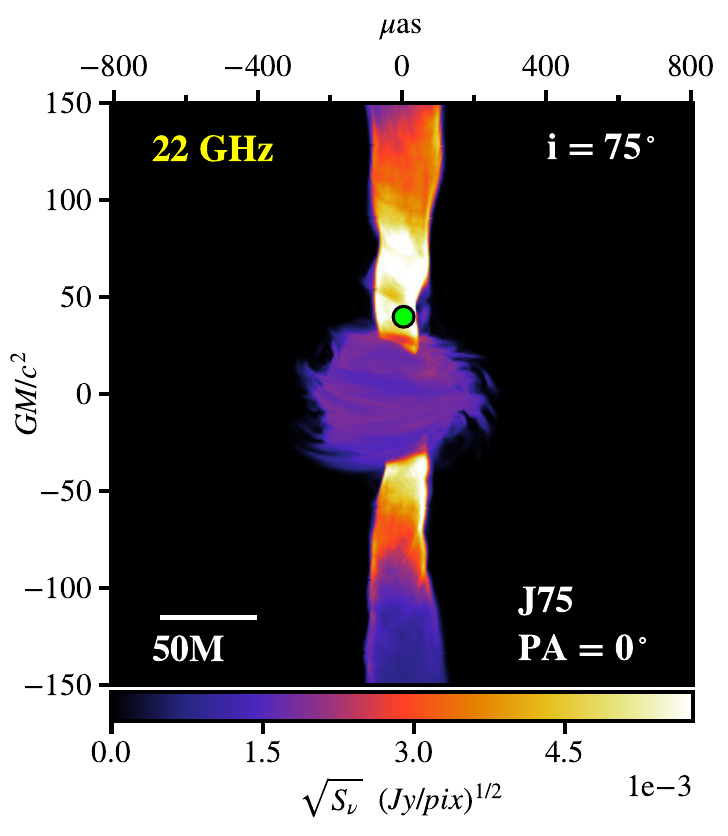}
\includegraphics[width=0.3\textwidth]{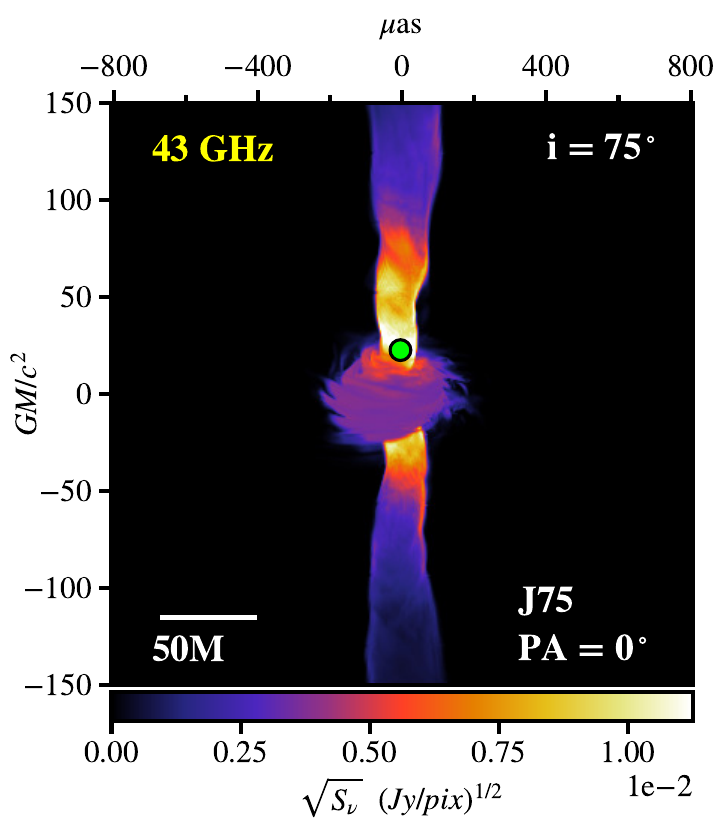}
\includegraphics[width=0.3\textwidth]{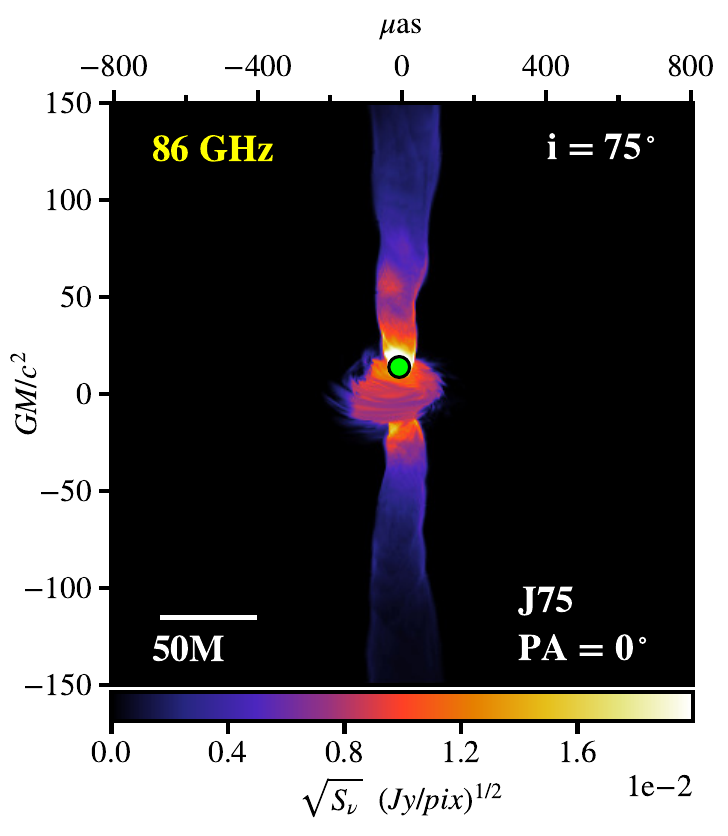}\\
\includegraphics[width=0.3\textwidth]{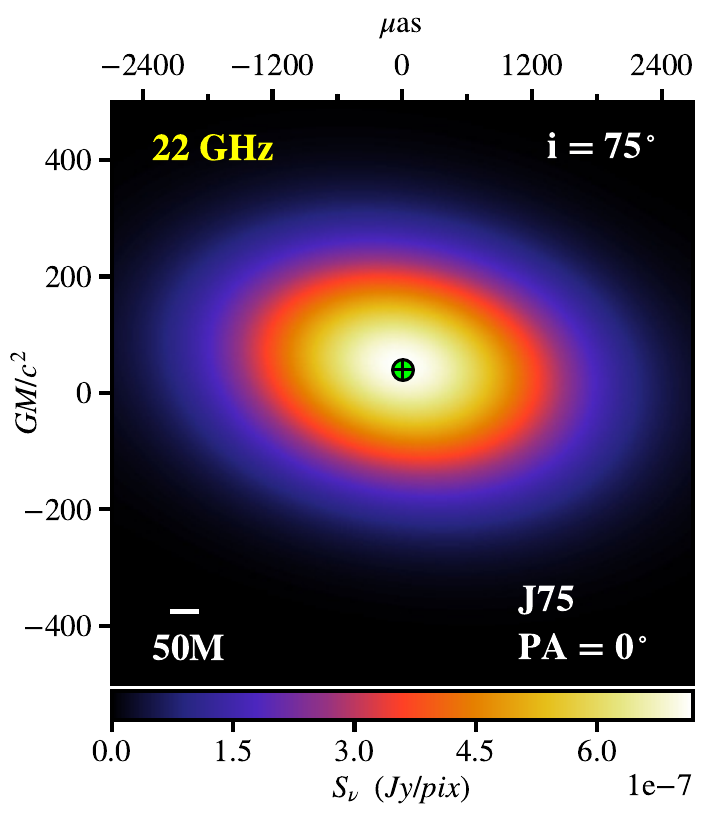}
\includegraphics[width=0.3\textwidth]{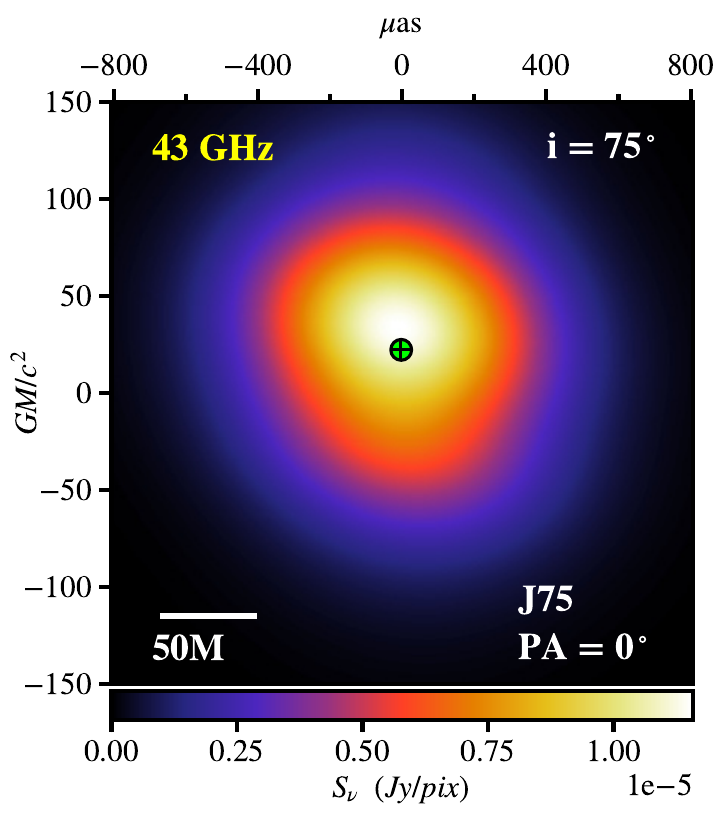}
\includegraphics[width=0.3\textwidth]{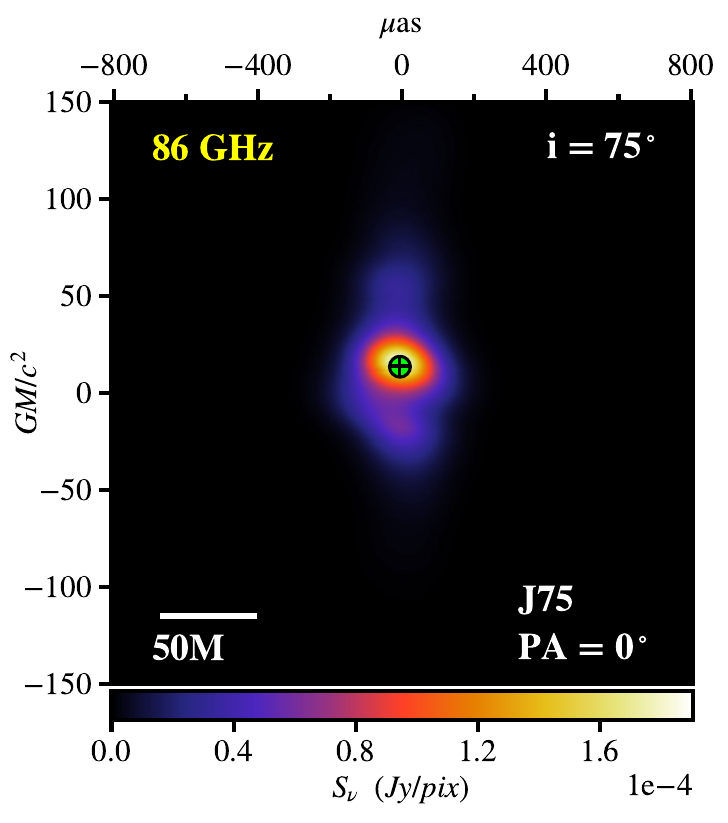}\\
\includegraphics[width=0.3\textwidth]{figures/flux_ref_Jet_i75_f22e9-eps-converted-to.pdf}
\includegraphics[width=0.3\textwidth]{figures/flux_ref_Jet_i75_f43e9-eps-converted-to.pdf}
\includegraphics[width=0.3\textwidth]{figures/flux_ref_Jet_i75_f86e9-eps-converted-to.pdf}
\end{figure*}
\newpage
\begin{figure*}[!hp]
\caption{\textbf{Jet model with $i=90 \degr$ including scattering}. Each column of panels shows a jet model with
 inclination $i=90 \degr$ at frequencies of 22~GHz, 43~GHz \& 86~GHz 
 (\textit{left to right}). The position angle on the sky of the BH spin-axis is $PA=0 \degr$. 
 Rows from \textit{top to bottom} show the unscattered 3D-GRMHD jet model, the scatter-broadened image, and the average image 
 including refractive scattering. Note that the FOV for images at 22~GHz including scattering \textit{(left column)} 
 is doubled compared to the FOV at 43~GHz \& 86~GHz. The color stretch is also different. The unscattered model panels 
 display the square-root of the flux density (\textit{top row}). The scattered maps 
 are plotted using a linear scale (\textit{second \& third rows}). The green dot indicates
 the intensity-weighted centroid of each image. Black cross-hairs (\textit{second \& third rows}) indicate 
 the location of the intensity-weighted centroid of the model \textit{before} scattering.}\label{fig:scattered_jet_model_i90}
\centering
\includegraphics[width=0.3\textwidth]{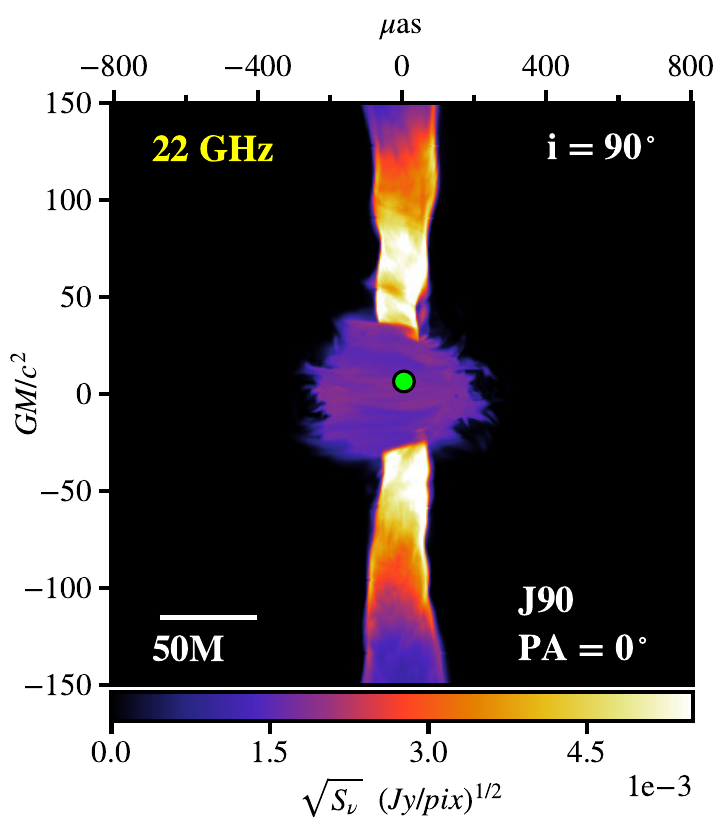}
\includegraphics[width=0.3\textwidth]{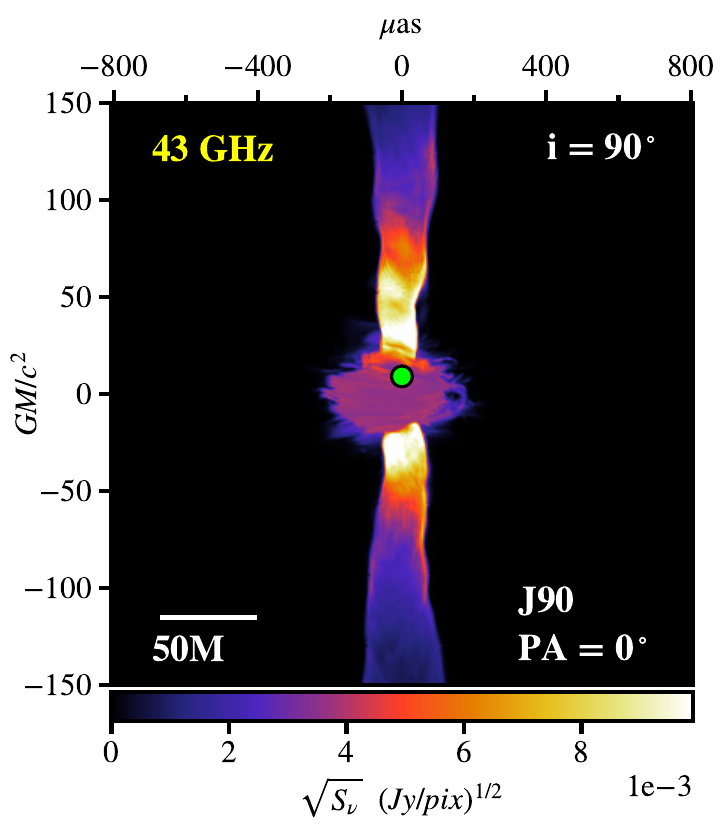}
\includegraphics[width=0.3\textwidth]{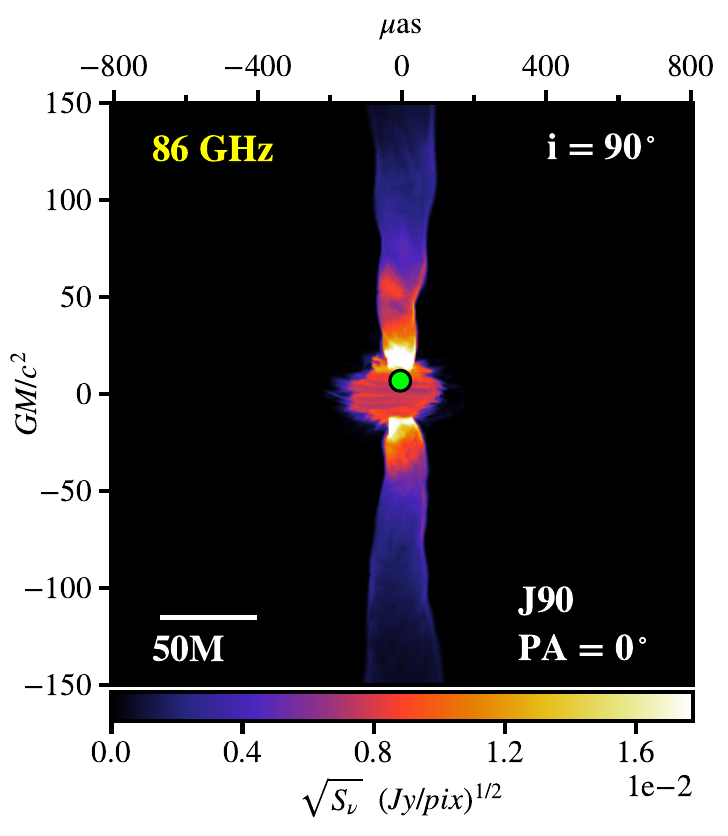}\\
\includegraphics[width=0.3\textwidth]{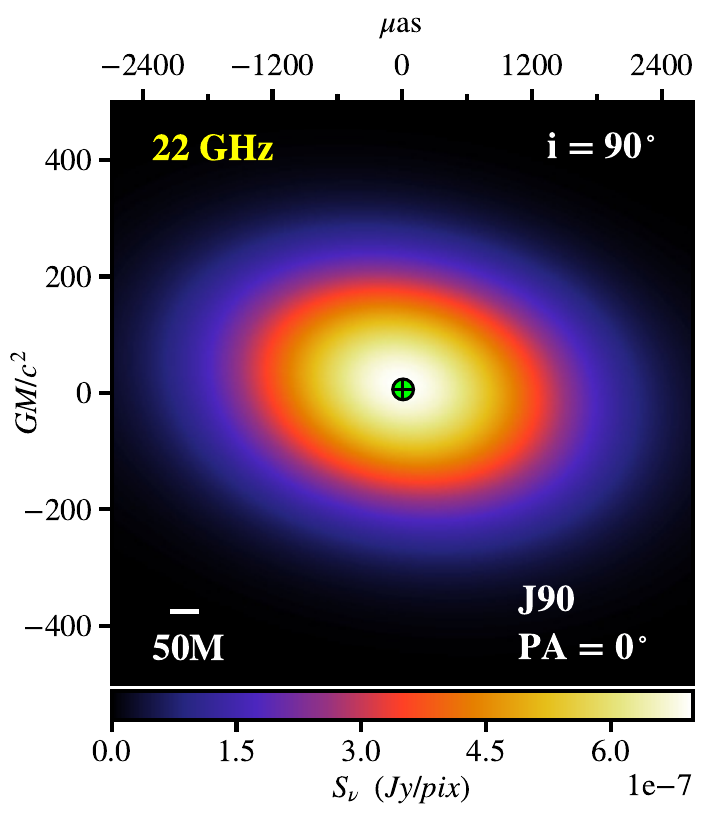}
\includegraphics[width=0.3\textwidth]{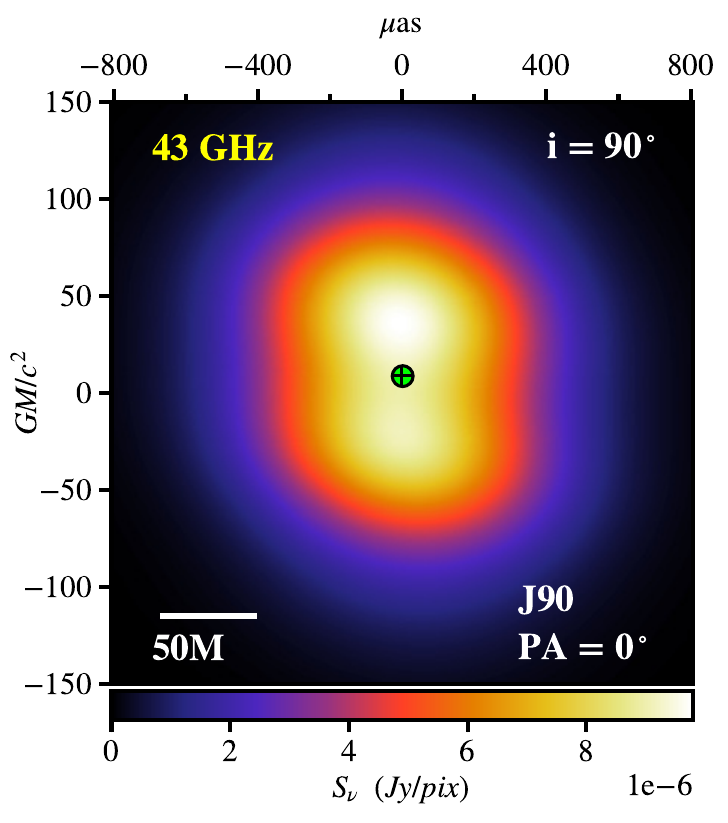}
\includegraphics[width=0.3\textwidth]{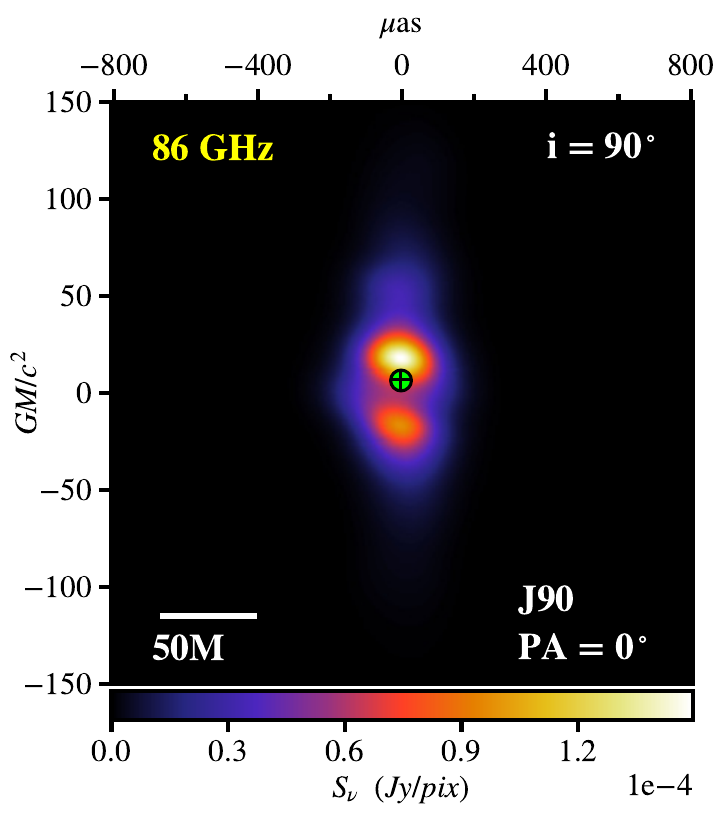}\\
\includegraphics[width=0.3\textwidth]{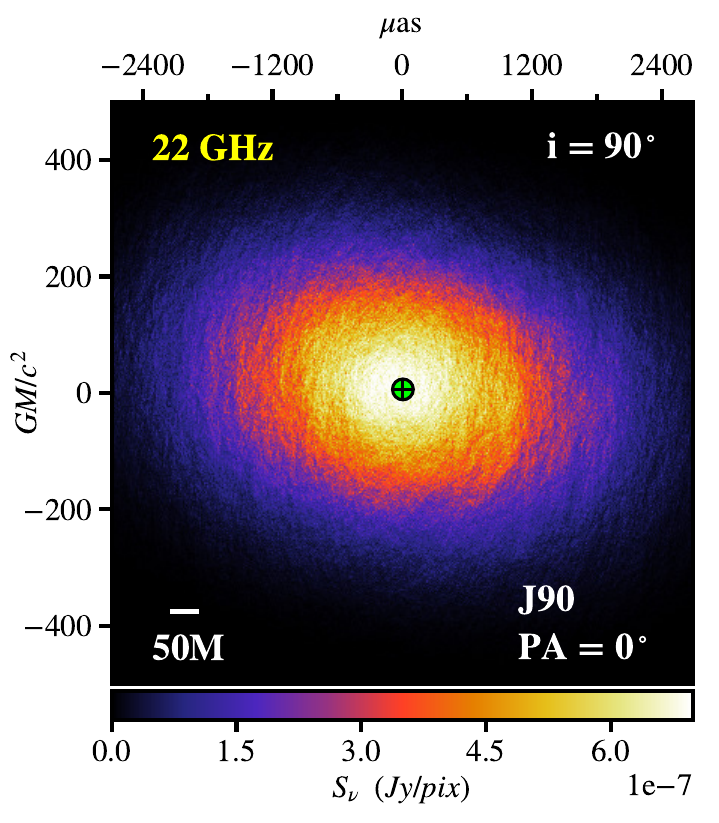}
\includegraphics[width=0.3\textwidth]{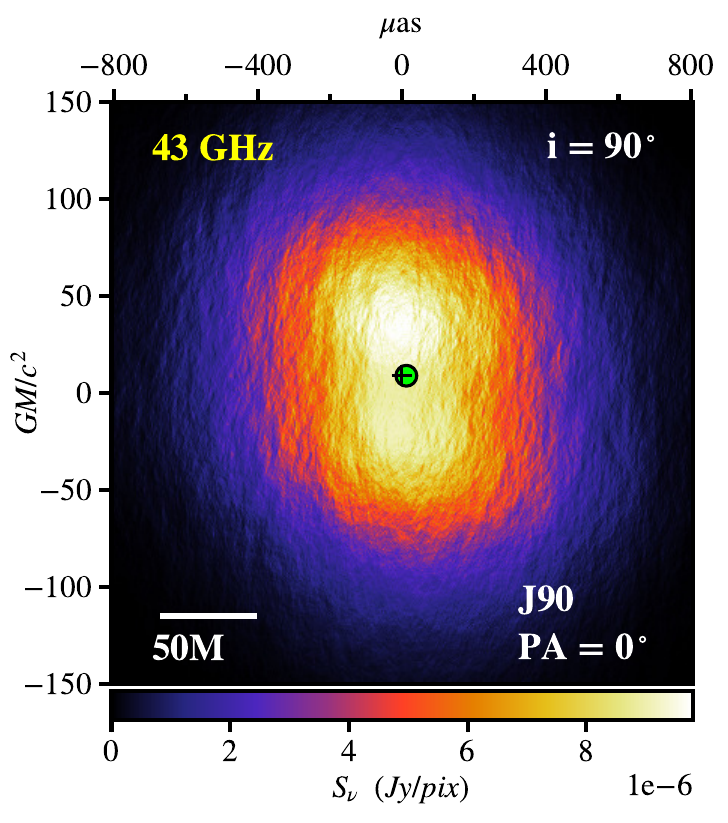}
\includegraphics[width=0.3\textwidth]{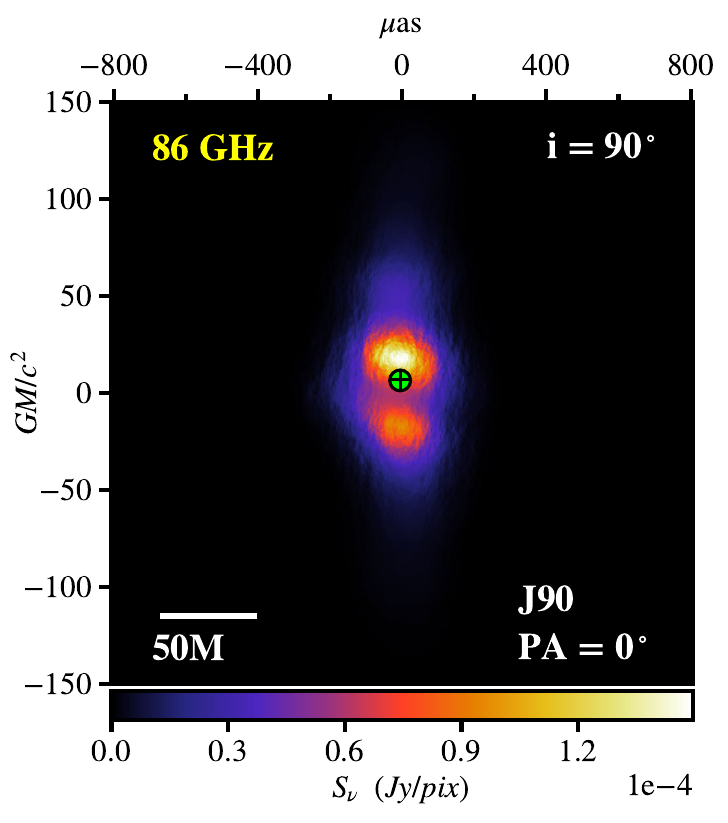}
\end{figure*}

\newpage

\begin{landscape}
\section{Summary Tables of Core-shift Parameters}~\label{sec:coreshift_tables_extra}

\begin{table}[ht!]
\begin{minipage}{.3\linewidth}
\tiny
\centering


\setlength\tabcolsep{3.5pt}
\begin{tabular}[t]{c c c c c c c}
\toprule
\toprule
ID		&$i$				 &index $p$	&$\sigma _p$  	&$A$  				&$\sigma _A$		 &$\chi ^2$          \\
...		&[$\degr$]	& ...				& ...				&[$\mu$as/cm]		&[$\mu$as/cm]		& ...		\\
\midrule
&\multicolumn{6}{c}{Unscattered JET MODELS} \\
\midrule
J1		    & 1			& 1.06		&$\pm$ 0.05  	& 95.99 			&$\pm$ 1.73    			& 2.90 \\ \addlinespace[2pt]
J15		& 15			& 1.13		&$\pm$ 0.02  	& 220.69   		&$\pm$ 1.85  			& 0.19\\ \addlinespace[2pt]
J30		& 30			& 1.10		&$\pm$ 0.02  	& 240.35 			&$\pm$ 1.92 	  		& 0.03\\ \addlinespace[2pt]
J45		& 45			& 1.08		&$\pm$ 0.02  	& 244.51   		&$\pm$ 1.97  			& 0.10\\ \addlinespace[2pt]
J60		& 60			& 0.98		&$\pm$ 0.02 	& 227.20  			&$\pm$ 1.97    			& 0.14\\ \addlinespace[2pt]
J75		& 75			& 0.79		&$\pm$ 0.03  	& 166.37   		&$\pm$ 1.95  			& 0.22\\ \addlinespace[2pt]
J90		& 90			& 0.00		&$\pm$ 0.07  	& 40.45		    	&$\pm$ 2.12    		 	& 2.68\\				
\midrule
&\multicolumn{6}{c}{with Gaussian Broadening} \\
\midrule
J1			& 1			& 0.99		&$\pm$ 0.05  	& 96.09 			&$\pm$ 1.71    			& 2.53 \\ \addlinespace[2pt]
J15		& 15			& 1.15		&$\pm$ 0.02  	& 217.74   		&$\pm$ 1.86  			& 0.15\\ \addlinespace[2pt]
J30		& 30			& 1.11		&$\pm$ 0.02  	& 237.65 			&$\pm$ 1.92 	  		& 0.02\\ \addlinespace[2pt]
J45		& 45			& 1.09		&$\pm$ 0.02  	& 241.86   		&$\pm$ 1.97  			& 0.08\\ \addlinespace[2pt]
J60		& 60			& 0.99		&$\pm$ 0.02 	& 224.66  		&$\pm$ 1.97    			& 0.11\\ \addlinespace[2pt]
J75		& 75			& 0.80		&$\pm$ 0.03  	& 164.03   		&$\pm$ 1.95  			& 0.19\\ \addlinespace[2pt]
J90		& 90			& 0.01		&$\pm$ 0.08  	& 38.55		    	&$\pm$ 2.11    		 	& 2.80\\				
\midrule
&\multicolumn{6}{c}{with Broadening + Refractive Scattering} \\
\midrule
J1			& 1			& 1.04		&$\pm$ 0.05  	& 94.72 			&$\pm$ 1.72    			& 3.67 \\ \addlinespace[2pt]
J15		& 15			& 1.14		&$\pm$ 0.02  	& 217.07   			&$\pm$ 1.86  			& 0.15\\ \addlinespace[2pt]
J30		& 30			& 1.12		&$\pm$ 0.02  	& 241.65			&$\pm$ 1.93 	  		& 0.07\\ \addlinespace[2pt]
J45		& 45			& 1.11		&$\pm$ 0.02  	& 241.21   			&$\pm$ 1.97  			& 0.08\\ \addlinespace[2pt]
J60		& 60			& 0.97		&$\pm$ 0.02 	& 222.39  			&$\pm$ 1.97    			& 0.18\\ \addlinespace[2pt]
J75		& 75			& 0.78		&$\pm$ 0.03  	& 164.68   		&$\pm$ 1.95  			& 0.15\\ \addlinespace[2pt]
J90		& 90			& 0.00		&$\pm$ 0.07  	& 39.29		    	&$\pm$ 2.11    		 	& 3.95\\				
\midrule
&\multicolumn{6}{c}{POSITION ANGLE = $0\degr$} \\
\bottomrule
\end{tabular}
\end{minipage}\hspace*{1em}
\begin{minipage}{.3\linewidth}
%
\tiny
\centering
\setlength\tabcolsep{3.5pt}
\begin{tabular}[t]{c c c c c c c}
\toprule
\toprule
ID		&$i$				  &index $p$	&$\sigma _p$  	&$A$			  	&$\sigma _A$			 &$\chi ^2$          \\
...		&[$\degr$]	& ...				& ...					&[$\mu$as/cm]	  &[$\mu$as/cm]		& ...		\\
\midrule
&\multicolumn{6}{c}{Unscattered JET MODELS} \\
\midrule
J1$\#$	 	 & 1			& 1.12		&$\pm$ 0.05  	& 90.87 			&$\pm$ 1.75    			& 3.62 \\ \addlinespace[2pt]
J15$\#$	 & 15			& 1.14		&$\pm$ 0.02  	& 219.06   		&$\pm$ 1.85  			& 0.22\\ \addlinespace[2pt]
J30$\#$	 & 30		& 1.10		&$\pm$ 0.02  	& 239.64 			&$\pm$ 1.92 	  		& 0.03\\ \addlinespace[2pt]
J45$\#$	 & 45		& 1.08		&$\pm$ 0.02  	& 243.77   		&$\pm$ 1.97  			& 0.10\\ \addlinespace[2pt]
J60$\#$	 & 60		& 0.98		&$\pm$ 0.02 	& 226.42  			&$\pm$ 1.97    			& 0.14\\ \addlinespace[2pt]
J75$\#$	 & 75			& 0.79		&$\pm$ 0.03  	& 165.54   		&$\pm$ 1.95  			& 0.22\\ \addlinespace[2pt]
J90$\#$    & 90			& -0.01		&$\pm$ 0.07  	& 39.50		    	&$\pm$ 2.12    		 	& 2.72\\				
\midrule
&\multicolumn{6}{c}{with Gaussian Broadening} \\
\midrule
J1$\#$	 	 & 1			& 1.05		&$\pm$ 0.05  	& 93.53 			&$\pm$ 1.73    			& 3.44 \\ \addlinespace[2pt]
J15$\#$	 & 15			& 1.10		&$\pm$ 0.02  	& 208.61   		&$\pm$ 1.84  			& 0.02\\ \addlinespace[2pt]
J30$\#$	 & 30		& 1.07		&$\pm$ 0.02  	& 230.04 			&$\pm$ 1.91 	  		& 0.01\\ \addlinespace[2pt]
J45$\#$	 & 45		& 1.06		&$\pm$ 0.02  	& 234.38   		&$\pm$ 1.96  			& 0.01\\ \addlinespace[2pt]
J60$\#$	 & 60		& 0.96		&$\pm$ 0.02 	& 217.60  			&$\pm$ 1.97    			& 0.02\\ \addlinespace[2pt]
J75$\#$	 & 75			& 0.77		&$\pm$ 0.03  	& 158.37   		&$\pm$ 1.95  			& 0.07\\ \addlinespace[2pt]
J90$\#$    & 90			& -0.02		&$\pm$ 0.07  	& 36.16		    		&$\pm$ 2.12    		 	& 3.13\\
\midrule
&\multicolumn{6}{c}{with Broadening + Refractive Scattering} \\
\midrule
J1$\#$			& 1				& 1.02		&$\pm$ 0.05  	& 92.14 			&$\pm$ 1.72    			& 2.77 \\ \addlinespace[2pt]
J15$\#$		& 15				& 1.10		&$\pm$ 0.02  	& 210.77   			&$\pm$ 1.83  			& 0.01\\ \addlinespace[2pt]
J30$\#$		& 30				& 1.09		&$\pm$ 0.02  	& 231.26 			&$\pm$ 1.91 	  		& 0.12\\ \addlinespace[2pt]
J45$\#$		& 45				& 1.04		&$\pm$ 0.02  	& 229.88   		&$\pm$ 1.95  			& 0.02\\ \addlinespace[2pt]
J60$\#$		& 60				& 0.96		&$\pm$ 0.02 	& 220.22  			&$\pm$ 1.97    			& 0.14\\ \addlinespace[2pt]
J75$\#$		& 75				& 0.78		&$\pm$ 0.03  	& 160.07   		&$\pm$ 1.96  			& 0.06\\ \addlinespace[2pt]
J90$\#$		& 90				& -0.16		&$\pm$ 0.09  	& 32.00		    	&$\pm$ 2.16    		 	& 2.04\\				
\midrule
&\multicolumn{6}{c}{POSITION ANGLE = $45\degr$} \\				
\bottomrule
\end{tabular}
\end{minipage}\hspace*{1em}
\begin{minipage}{.3\linewidth}
%
\tiny
\centering
\setlength\tabcolsep{3.5pt}
\begin{tabular}[t]{c c c c c c c}
\toprule
\toprule
ID			&$i$  		 &index $p$	&$\sigma _p$  	&$A$			  	&$\sigma _A$			 &$\chi ^2$          \\
...			&[$\degr$]  & ...			& ...					&[$\mu$as/cm]	  &[$\mu$as/cm]		& ...		\\
\midrule
&\multicolumn{6}{c}{Unscattered JET MODELS} \\
\midrule
J1$*$	 & 1			& 1.06		&$\pm$ 0.05  	& 97.03 				&$\pm$ 1.73    			& 2.95 \\ \addlinespace[2pt]
J15$*$	 & 15			& 1.12		&$\pm$ 0.02  	& 221.40   			&$\pm$ 1.85  			& 0.21\\ \addlinespace[2pt]
J30$*$	 & 30		& 1.09		&$\pm$ 0.02  	& 240.85 				&$\pm$ 1.92 	  		& 0.04\\ \addlinespace[2pt]
J45$*$	 & 45		& 1.07		&$\pm$ 0.02  	& 244.95   			&$\pm$ 1.97  			& 0.12\\ \addlinespace[2pt]
J60$*$	 & 60		& 0.98		&$\pm$ 0.02 	& 227.52  				&$\pm$ 1.97    			& 0.15\\ \addlinespace[2pt]
J75$*$	 & 75			& 0.78		&$\pm$ 0.03  	& 166.57   			&$\pm$ 1.95  			& 0.23\\ \addlinespace[2pt]
J90$*$  & 90		& -0.02		&$\pm$ 0.07  	& 40.43		    		&$\pm$ 2.12    		 	& 2.68\\				
\midrule
&\multicolumn{6}{c}{with Gaussian Broadening} \\
\midrule
J1$*$	 & 1			& 0.93		&$\pm$ 0.04  	& 94.33 				&$\pm$ 1.69    			& 1.90 \\ \addlinespace[2pt]
J15$*$	 & 15			& 1.01		&$\pm$ 0.02  	& 204.71   			&$\pm$ 1.81  			& 0.01\\ \addlinespace[2pt]
J30$*$	 & 30		& 0.99		&$\pm$ 0.02  	& 223.61 				&$\pm$ 1.90 	  		& 0.11\\ \addlinespace[2pt]
J45$*$	 & 45		& 0.98		&$\pm$ 0.02  	& 227.42   			&$\pm$ 1.95  			& 0.04\\ \addlinespace[2pt]
J60$*$	 & 60		& 0.88		&$\pm$ 0.02 	& 210.86  				&$\pm$ 1.96    			& 0.02\\ \addlinespace[2pt]
J75$*$	 & 75			& 0.69		&$\pm$ 0.03  	& 153.62   			&$\pm$ 1.96  			& 0.01\\ \addlinespace[2pt]
J90$*$   & 90		& -0.06		&$\pm$ 0.08  	& 37.40		    		&$\pm$ 2.13    		 	& 3.32\\				
\midrule
&\multicolumn{6}{c}{with Broadening + Refractive Scattering} \\
\midrule
J1$*$		& 1			& 0.81		&$\pm$ 0.04  	& 91.56 				&$\pm$ 1.67    			& 0.75 \\ \addlinespace[2pt]
J15$*$		& 15			& 1.05		&$\pm$ 0.02  	& 205.46   			&$\pm$ 1.83  			& 0.11\\ \addlinespace[2pt]
J30$*$		& 30			& 0.97		&$\pm$ 0.02  	& 218.43 				&$\pm$ 1.89 	  		& 0.80\\ \addlinespace[2pt]
J45$*$		& 45			& 0.95		&$\pm$ 0.02  	& 228.92   			&$\pm$ 1.95  			& 0.18\\ \addlinespace[2pt]
J60$*$		& 60			& 0.84		&$\pm$ 0.02 	& 208.08 				&$\pm$ 1.96    			& 0.01\\ \addlinespace[2pt]
J75$*$		& 75			& 0.68		&$\pm$ 0.03  	& 150.14	   			&$\pm$ 1.96  			& 0.02\\ \addlinespace[2pt]
J90$*$		& 90			& -0.14		&$\pm$ 0.08  	& 33.84		    		&$\pm$ 2.16    		 	& 5.77\\				
\midrule
&\multicolumn{6}{c}{POSITION ANGLE = $90\degr$} \\		
\bottomrule
\end{tabular}
\end{minipage}
\vspace{1cm}
\caption{ \textbf{Power-law parameters for jet models with spin-axis orientations of $PA=0\degr , 45\degr , 90\degr$}. 
These are results from power-law fits using a least-squares fitting method. Values include the power-law index ($p$), 
its error ($\sigma _p$), the power-law coefficient ($A$) which indicates the amount of core-shift, its error ($\sigma _A$), 
and the chi-squared value of the fit (~$\chi ^2$). Models are labeled with an ID according to their inclination angle ($i$) and
orientation.}\label{tab:plaw_params_all_PA_jet_scattered}
\end{table}
\end{landscape}

%
\begin{landscape}
\begin{table}[ht!]
\begin{minipage}{.3\linewidth}
\tiny
\centering

\setlength\tabcolsep{3.5pt}
\begin{tabular}[t]{c c c c c c c}
\toprule
\toprule
ID		&$i$				 &index $p$	&$\sigma _p$  	&$A$   				&$\sigma _A$  			&$\chi ^2$          \\
...		&[$\degr$]	& ...			& ...					&[$\mu$as/cm]		&[$\mu$as/cm]			& ...		\\
\midrule
&\multicolumn{6}{c}{Unscattered DISK MODELS} \\
\midrule
D1			& 1		& 0.05		&$\pm$ 0.29  	& 10.60 				&$\pm$ 2.11    			& 0.13 \\ \addlinespace[2pt]
D15		& 15			& 0.60		&$\pm$ 0.29  	& 13.46   				&$\pm$ 1.99  			& 0.30\\ \addlinespace[2pt]
D30		& 30			& 0.47		&$\pm$ 0.19  	& 19.43 				&$\pm$ 2.05 	  		& 0.01\\ \addlinespace[2pt]
D45		& 45			& 0.36		&$\pm$ 0.15  	& 23.90   				&$\pm$ 2.15  			& 0.13\\ \addlinespace[2pt]
D60		& 60			& 0.31		&$\pm$ 0.14 	& 26.02  				&$\pm$ 2.28    			& 0.14\\ \addlinespace[2pt]
D75		& 75			& 0.35		&$\pm$ 0.14  	& 26.19   				&$\pm$ 2.44  			& 0.14\\ \addlinespace[2pt]
D90		& 90			& 0.40		&$\pm$ 0.15  	& 24.82		    		&$\pm$ 2.51    		 	& 0.01\\				
\midrule
&\multicolumn{6}{c}{with Gaussian Broadening} \\
\midrule
D1			& 1			& -0.11		&$\pm$ 0.25  	& 11.68 				&$\pm$ 2.15    			& 0.32 \\ \addlinespace[2pt]
D15		& 15			& 0.38		&$\pm$ 0.27  	& 13.09   				&$\pm$ 2.04  			& 0.36\\ \addlinespace[2pt]
D30		& 30			& 0.42		&$\pm$ 0.20  	& 17.69 				&$\pm$ 2.06 	  		& 0.01\\ \addlinespace[2pt]
D45		& 45			& 0.32		&$\pm$ 0.16  	& 21.44   				&$\pm$ 2.16  			& 0.04\\ \addlinespace[2pt]
D60		& 60			& 0.28		&$\pm$ 0.15 	& 23.13 				&$\pm$ 2.29    			& 0.06\\ \addlinespace[2pt]
D75		& 75			& 0.32		&$\pm$ 0.16  	& 23.06   				&$\pm$ 2.44  			& 0.10\\ \addlinespace[2pt]
D90		& 90			& 0.38		&$\pm$ 0.18  	& 22.04		    		&$\pm$ 2.51   		 	& 0.02\\				
\midrule
&\multicolumn{6}{c}{with Broadening + Refractive Scattering} \\
\midrule
D1			& 1			& -0.23		&$\pm$ 0.05  	& 11.02 				&$\pm$ 2.18    			& 0.76 \\ \addlinespace[2pt]
D15		& 15			& 0.62		&$\pm$ 0.02  	& 12.81   				&$\pm$ 1.98  			& 0.04\\ \addlinespace[2pt]
D30		& 30			& -0.05		&$\pm$ 0.02  	& 12.78 				&$\pm$ 2.19 	  		& 1.08\\ \addlinespace[2pt]
D45		& 45			& -0.12		&$\pm$ 0.02  	& 15.99   				&$\pm$ 2.28  			& 0.67\\ \addlinespace[2pt]
D60		& 60			& 0.26		&$\pm$ 0.02 	& 20.97  				&$\pm$ 2.30   			& 1.37\\ \addlinespace[2pt]
D75		& 75			& 0.74		&$\pm$ 0.03  	& 25.71   				&$\pm$ 2.34 			& 3.26\\ \addlinespace[2pt]
D90		& 90			& 0.52		&$\pm$ 0.07  	& 25.00		    		&$\pm$ 2.48    		 	& 0.05\\				
\midrule
&\multicolumn{6}{c}{POSITION ANGLE = $0\degr$} \\
\bottomrule
\end{tabular}
\end{minipage}\hspace*{1em}
\begin{minipage}{.3\linewidth}
%
\tiny
\centering
\setlength\tabcolsep{3.5pt}
\begin{tabular}[t]{c c c c c c c}
\toprule
\toprule
ID		&$i$				  &index $p$	&$\sigma _p$  	&$A$ 		  	&$\sigma _A$				 &$\chi ^2$          \\
...		&[$\degr$]	& ...			& ...					&[$\mu$as/cm]		&[$\mu$as/cm]			& ...		\\
\midrule
&\multicolumn{6}{c}{Unscattered DISK MODELS} \\
\midrule
D1$\#$	 	 & 1			& 0.23		&$\pm$ 0.27  	& 12.08 			&$\pm$ 2.07    			& 0.19 \\ \addlinespace[2pt]
D15$\#$	 & 15			& 0.54		&$\pm$ 0.24  	& 15.63 	  		&$\pm$ 2.00  			& 0.23\\ \addlinespace[2pt]
D30$\#$	 & 30		& 0.41		&$\pm$ 0.16  	& 21.72 			&$\pm$ 2.06 	  		& 0.01\\ \addlinespace[2pt]
D45$\#$	 & 45		& 0.32		&$\pm$ 0.13  	& 26.12   			&$\pm$ 2.16  			& 0.12\\ \addlinespace[2pt]
D60$\#$	 & 60		& 0.28		&$\pm$ 0.12 	& 28.15  			&$\pm$ 2.29    			& 0.13\\ \addlinespace[2pt]
D75$\#$	 & 75			& 0.32		&$\pm$ 0.13  	& 28.25   			&$\pm$ 2.45  			& 0.22\\ \addlinespace[2pt]
D90$\#$   & 90			& 0.36		&$\pm$ 0.14  	& 26.99		    	&$\pm$ 2.52   		 	& 0.01\\				
\midrule
&\multicolumn{6}{c}{with Gaussian Broadening} \\
\midrule
D1$\#$	 	 & 1			& 0.16		&$\pm$ 0.22  	& 14.50 			&$\pm$ 2.09    			& 0.28 \\ \addlinespace[2pt]
D15$\#$	 & 15			& 0.40		&$\pm$ 0.20  	& 17.43   			&$\pm$ 2.03  			& 0.22\\ \addlinespace[2pt]
D30$\#$	 & 30		& 0.35		&$\pm$ 0.15  	& 22.69 			&$\pm$ 2.08 	  		& 0.01\\ \addlinespace[2pt]
D45$\#$	 & 45		& 0.28		&$\pm$ 0.13  	& 26.48   			&$\pm$ 2.17  			& 0.06\\ \addlinespace[2pt]
D60$\#$	 & 60		& 0.24		&$\pm$ 0.12 	& 28.02  			&$\pm$ 2.30    			& 0.08\\ \addlinespace[2pt]
D75$\#$	 & 75			& 0.28		&$\pm$ 0.13  	& 27.80   			&$\pm$ 2.46 			& 0.14\\ \addlinespace[2pt]
D90$\#$  & 90			& 0.33		&$\pm$ 0.14  	& 27.03		    		&$\pm$ 2.53    		 	& 0.01\\				
\midrule
&\multicolumn{6}{c}{with Broadening + Refractive Scattering} \\
\midrule
D1	$\#$		& 1			& 0.33		&$\pm$ 0.21  	& 16.17 					&$\pm$ 2.05    			& 0.05 \\ \addlinespace[2pt]
D15$\#$	& 15			& 0.42		&$\pm$ 0.19  	& 18.48   				&$\pm$ 2.02  			& 1.03\\ \addlinespace[2pt]
D30$\#$	& 30			& 0.64		&$\pm$ 0.14  	& 27.73 				&$\pm$ 2.00 	  		& 0.15\\ \addlinespace[2pt]
D45$\#$	& 45			& 0.28		&$\pm$ 0.13  	& 25.48   				&$\pm$ 2.18 			& 2.12\\ \addlinespace[2pt]
D60$\#$	& 60			& 0.26		&$\pm$ 0.12 	& 29.41  				&$\pm$ 2.31    			& 1.28\\ \addlinespace[2pt]
D75$\#$	& 75			& 0.25		&$\pm$ 0.13  	& 27.40   				&$\pm$ 2.47  			& 0.11\\ \addlinespace[2pt]
D90$\#$	& 90			& 0.38		&$\pm$ 0.13  	& 28.70		    		&$\pm$ 2.50    		 	& 0.02\\				
\midrule
&\multicolumn{6}{c}{POSITION ANGLE = $45\degr$} \\
\bottomrule
\end{tabular}
\end{minipage}\hspace*{1em}
\begin{minipage}{.3\linewidth}
%
\tiny
\centering
\setlength\tabcolsep{3.5pt}
\begin{tabular}[t]{c c c c c c c}
\toprule
\toprule
ID			&$i$  		&index $p$	&$\sigma _p$  	&$A$  				&$\sigma _A$			 &$\chi ^2$          \\
..		&[$\degr$]	& ...				& ...					&[$\mu$as/cm]		&[$\mu$as/cm]		& ...		\\
\midrule
&\multicolumn{6}{c}{Unscattered DISK MODELS} \\
\midrule
D1$*$	 	 & 1			& 0.44		&$\pm$ 0.29  	& 12.48 			&$\pm$ 2.01    			& 0.24 \\ \addlinespace[2pt]
D15$*$	 	 & 15			& 0.55		&$\pm$ 0.22  	& 16.96   			&$\pm$ 1.99  			& 0.15\\ \addlinespace[2pt]
D30$*$	 & 30		& 0.39		&$\pm$ 0.15  	& 23.47 			&$\pm$ 2.07 	  		& 0.03\\ \addlinespace[2pt]
D45$*$	 & 45		& 0.30		&$\pm$ 0.12  	& 28.06   			&$\pm$ 2.17  			& 0.13\\ \addlinespace[2pt]
D60$*$	 & 60		& 0.26		&$\pm$ 0.12 	& 30.22  			&$\pm$ 2.30    			& 0.13\\ \addlinespace[2pt]
D75$*$	 	 & 75			& 0.29		&$\pm$ 0.12  	& 30.39   			&$\pm$ 2.45 			& 0.13\\ \addlinespace[2pt]
D90$*$   	 & 90		& 0.33		&$\pm$ 0.13  	& 29.06		    	&$\pm$ 2.53   		 	& 0.01\\				
\midrule
&\multicolumn{6}{c}{with Gaussian Broadening} \\
\midrule
D1$*$	 	 & 1			& 0.42		&$\pm$ 0.23  	& 15.06 			&$\pm$ 2.02    			& 0.36 \\ \addlinespace[2pt]
D15$*$	 	 & 15			& 0.44		&$\pm$ 0.18  	& 19.77   			&$\pm$ 2.02  			& 0.12\\ \addlinespace[2pt]
D30$*$	 & 30		& 0.33		&$\pm$ 0.13  	& 26.06 			&$\pm$ 2.09 	  		& 0.02\\ \addlinespace[2pt]
D45$*$	 & 45		& 0.26		&$\pm$ 0.11  	& 30.46   			&$\pm$ 2.18 			& 0.10\\ \addlinespace[2pt]
D60$*$	 & 60		& 0.24		&$\pm$ 0.11 	& 32.42  			&$\pm$ 2.30    			& 0.11\\ \addlinespace[2pt]
D75$*$	 	 & 75			& 0.27		&$\pm$ 0.11  	& 32.43   			&$\pm$ 2.46 			& 0.15\\ \addlinespace[2pt]
D90$*$   	& 90			& 0.30		&$\pm$ 0.12  	& 31.31		    		&$\pm$ 2.54    		 	& 0.01\\				
\midrule
&\multicolumn{6}{c}{with Broadening + Refractive Scattering} \\
\midrule
D1	$*$		& 1			& 0.60		&$\pm$ 0.25  	& 15.48 			&$\pm$ 1.98    			& 0.06 \\ \addlinespace[2pt]
D15$*$		& 15			& 0.51		&$\pm$ 0.17  	& 22.06   			&$\pm$ 2.01  			& 0.01\\ \addlinespace[2pt]
D30$*$	& 30			& 0.21		&$\pm$ 0.14  	& 23.01 			&$\pm$ 2.12 	  		& 0.11\\ \addlinespace[2pt]
D45$*$	& 45			& 0.21		&$\pm$ 0.11  	& 30.31   			&$\pm$ 2.18  			& 0.11\\ \addlinespace[2pt]
D60$*$	& 60			& 0.22		&$\pm$ 0.12 	& 29.71  			&$\pm$ 2.31    			& 0.01\\ \addlinespace[2pt]
D75$*$		& 75			& 0.19		&$\pm$ 0.12  	& 29.23   			&$\pm$ 2.50  			& 0.17\\ \addlinespace[2pt]
D90$*$	& 90			& 0.33		&$\pm$ 0.11  	& 34.29		    	&$\pm$ 2.53    		 	& 0.86\\				
\midrule
&\multicolumn{6}{c}{POSITION ANGLE = $90\degr$} \\
\bottomrule
\end{tabular}
\end{minipage}
\vspace{1cm}
\caption{ \textbf{Power-law parameters for disk models with spin-axis orientations of $PA=0\degr , 45\degr , 90\degr$}. 
These are results from power-law fits using a least-squares fitting method. Values include the power-law index ($p$), 
its error ($\sigma _p$), the power-law coefficient ($A$) which indicates the amount of core-shift, its error ($\sigma _A$), 
and the chi-squared value of the fit (~$\chi ^2$). Models are labeled with an ID according to their inclination angle ($i$) and
orientation.}\label{tab:plaw_params_all_PA_disk_scattered}
\end{table}
\end{landscape}
\newpage

%
\section{Maps of Disk Models including Scattering (additional inclinations)}~\label{sec:scattered_disk_models_library}
\begin{figure*}[!hp]
\caption{\textbf{Disk model with $i=1 \degr$ including scattering}. Each column shows a disk model with
 inclination $i=1 \degr$ at frequencies of 22~GHz, 43~GHz \& 86~GHz 
 (\textit{left to right}). The position angle on the sky of the BH spin-axis is $PA=0 \degr$. 
 Rows from \textit{top to bottom} show the unscattered 3D-GRMHD jet model, the scatter-broadened image, 
 and the average image including refractive scattering. Note that the FOV for images at 22~GHz including 
 scattering \textit{(left column)} is doubled compared to the FOV at 43~GHz \& 86~GHz. The color stretch is 
 also different. The unscattered model panels display the square-root of the flux density (\textit{top row}). 
 The scattered maps are plotted using a linear scale (\textit{second \& third rows}). The green dot indicates
 the intensity-weighted centroid of each image. Black cross-hairs (\textit{second \& third rows}) indicate 
 the location of the intensity-weighted centroid of the model \textit{before} scattering.}\label{fig:scattered_disk_model_i1}
\centering
\includegraphics[width=0.3\textwidth]{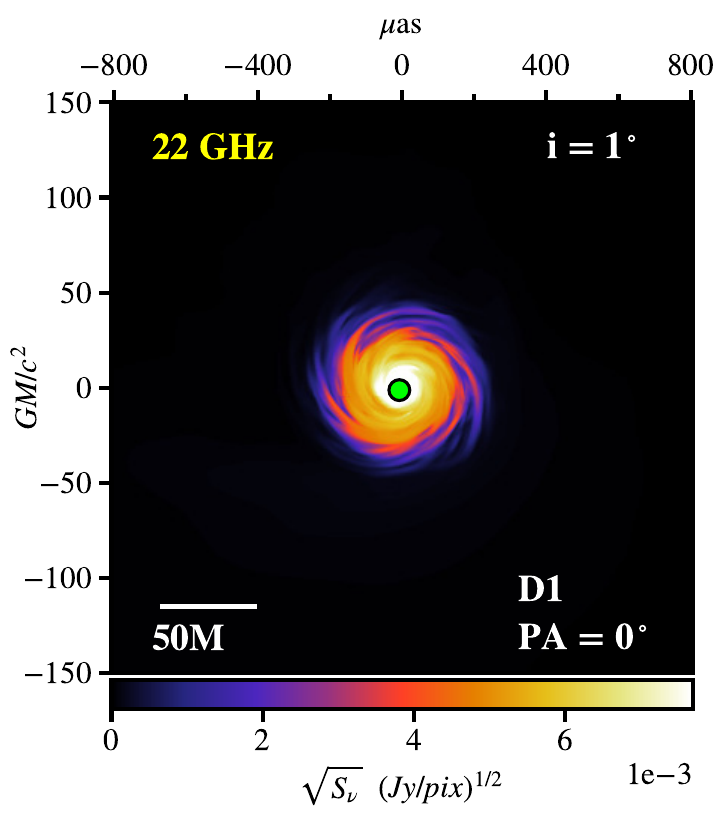}
\includegraphics[width=0.3\textwidth]{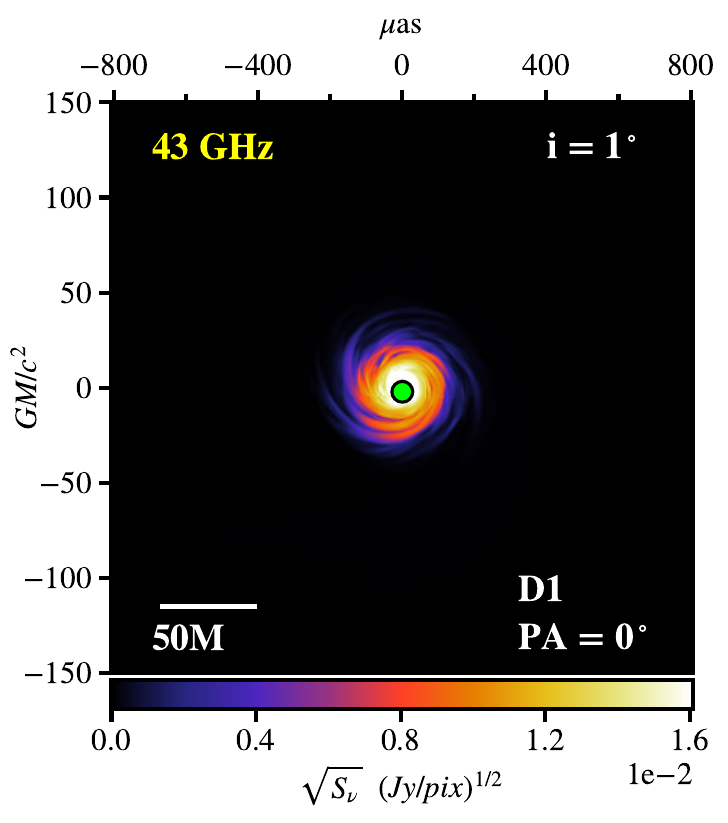}
\includegraphics[width=0.3\textwidth]{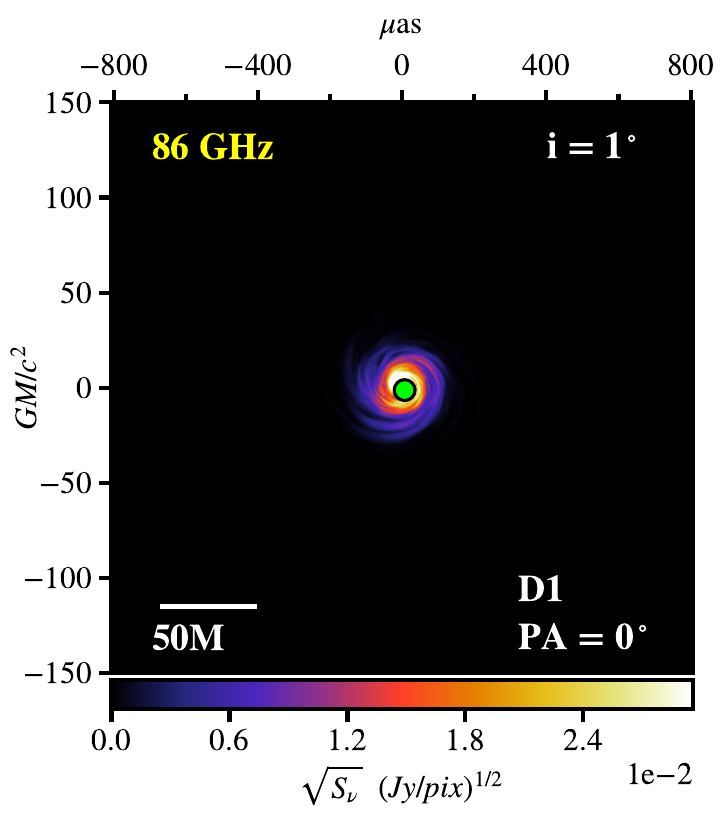}\\
\includegraphics[width=0.3\textwidth]{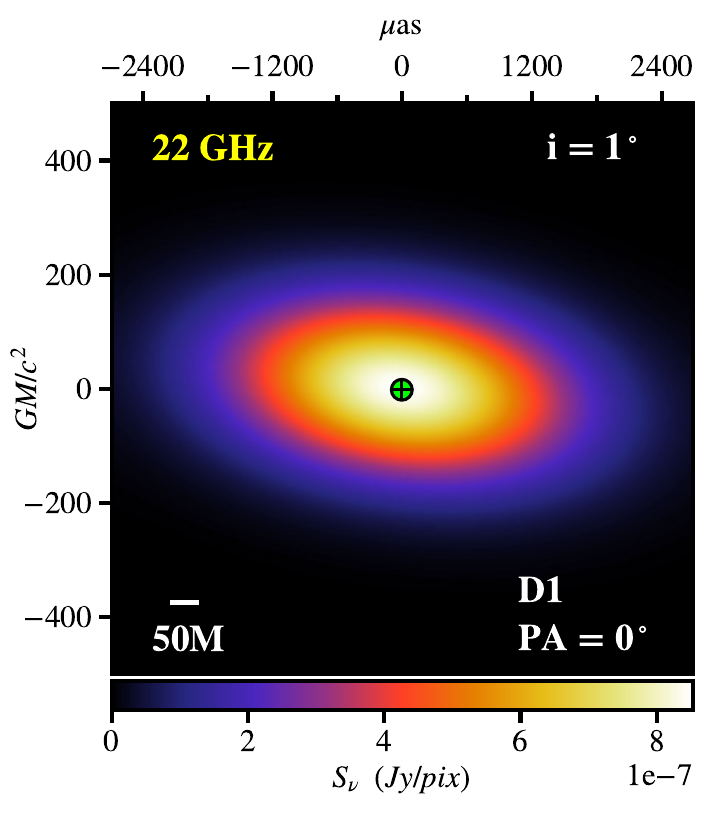}
\includegraphics[width=0.3\textwidth]{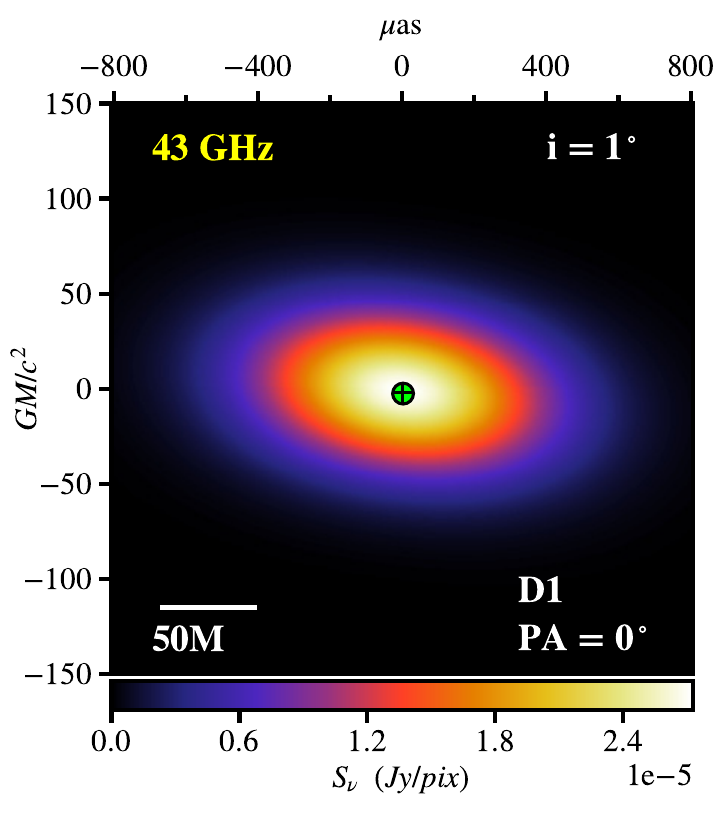}
\includegraphics[width=0.3\textwidth]{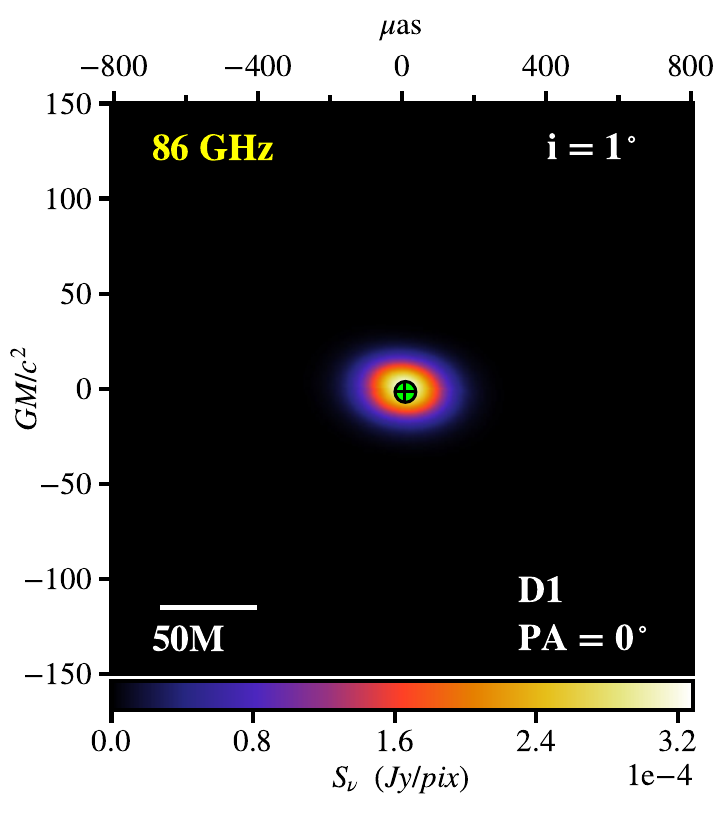}\\
\includegraphics[width=0.3\textwidth]{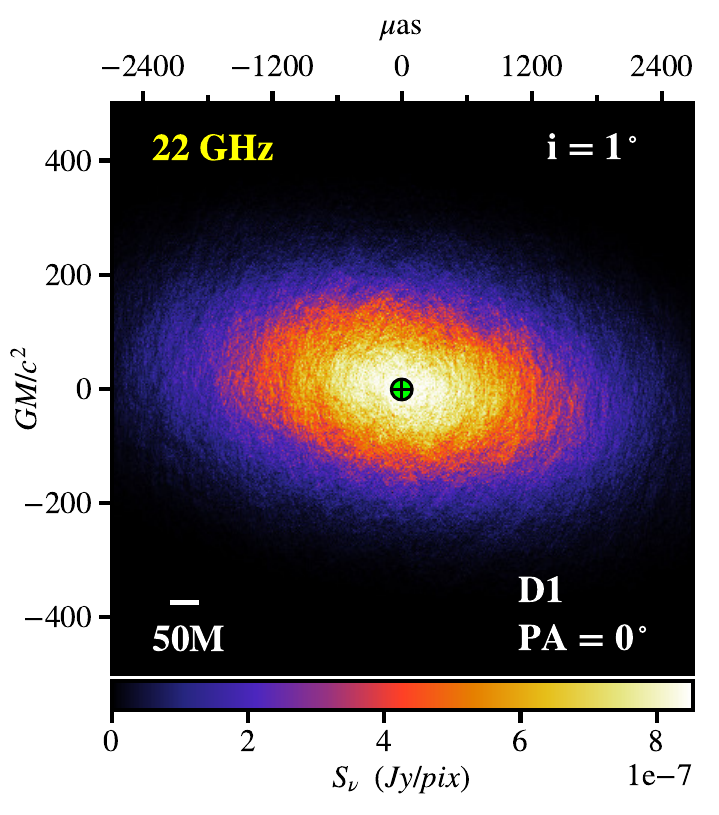}
\includegraphics[width=0.3\textwidth]{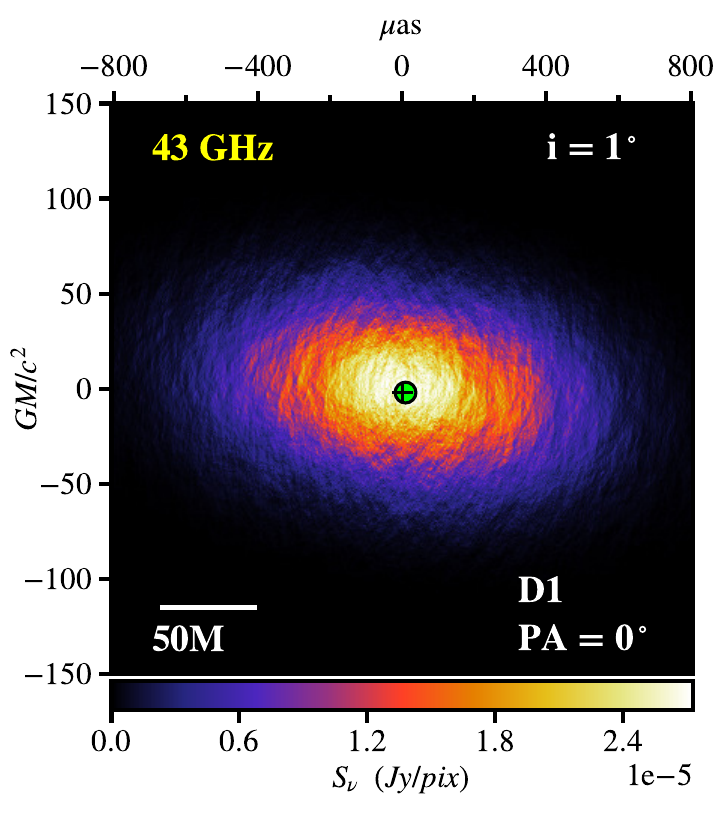}
\includegraphics[width=0.3\textwidth]{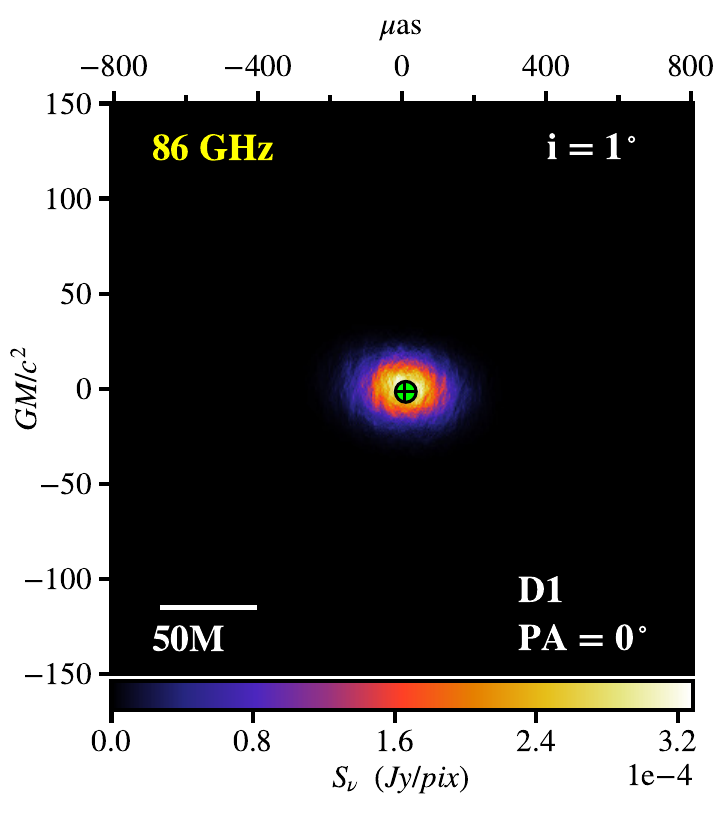}
\end{figure*}
\newpage
\begin{figure*}[!hp]
\caption{\textbf{Disk model with $i=15 \degr$ including scattering}. Each column shows a disk model with
 inclination $i=15 \degr$ at frequencies of 22~GHz, 43~GHz \& 86~GHz 
 (\textit{left to right}). The position angle on the sky of the BH spin-axis is $PA=0 \degr$. 
 Rows from \textit{top to bottom} show the unscattered 3D-GRMHD jet model, the scatter-broadened image, 
 and the average image including refractive scattering. Note that the FOV for images at 22~GHz including 
 scattering \textit{(left column)} is doubled compared to the FOV at 43~GHz \& 86~GHz. The color stretch is 
 also different. The unscattered model panels display the square-root of the flux density (\textit{top row}). 
 The scattered maps are plotted using a linear scale (\textit{second \& third rows}). The green dot indicates
 the intensity-weighted centroid of each image. Black cross-hairs (\textit{second \& third rows}) indicate 
 the location of the intensity-weighted centroid of the model \textit{before} scattering.}\label{fig:scattered_disk_model_i15}
\centering
\includegraphics[width=0.3\textwidth]{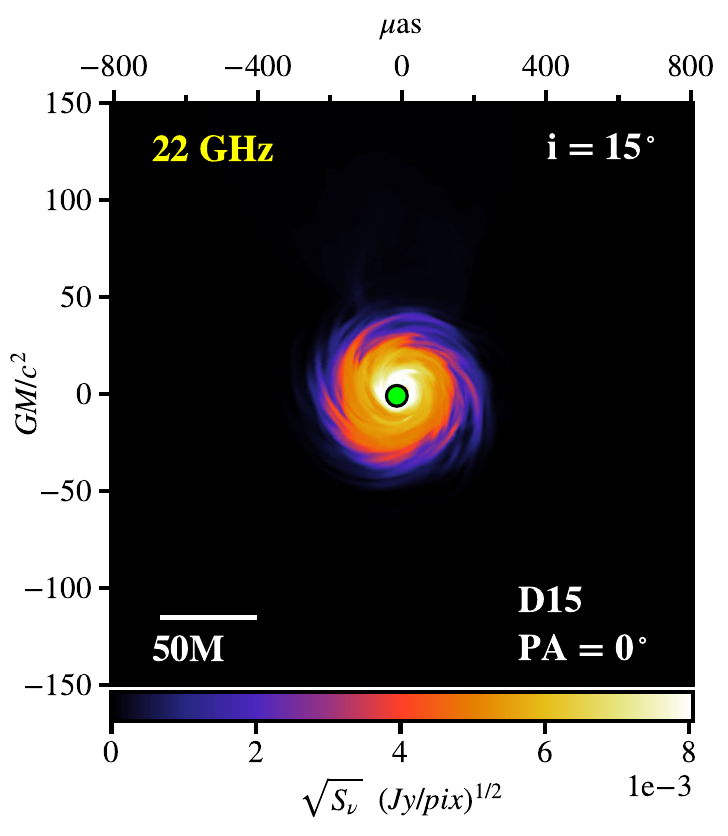}
\includegraphics[width=0.3\textwidth]{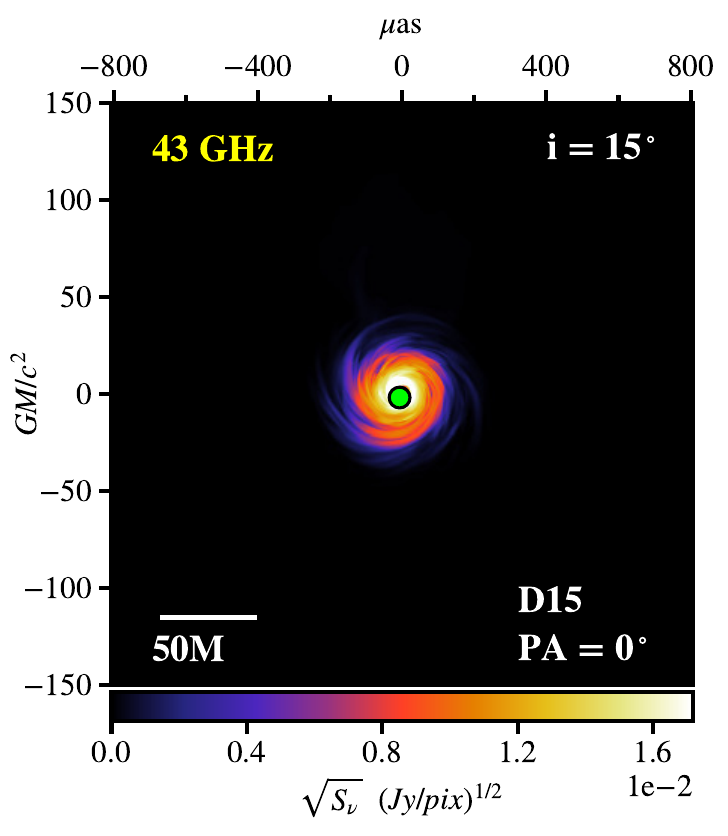}
\includegraphics[width=0.3\textwidth]{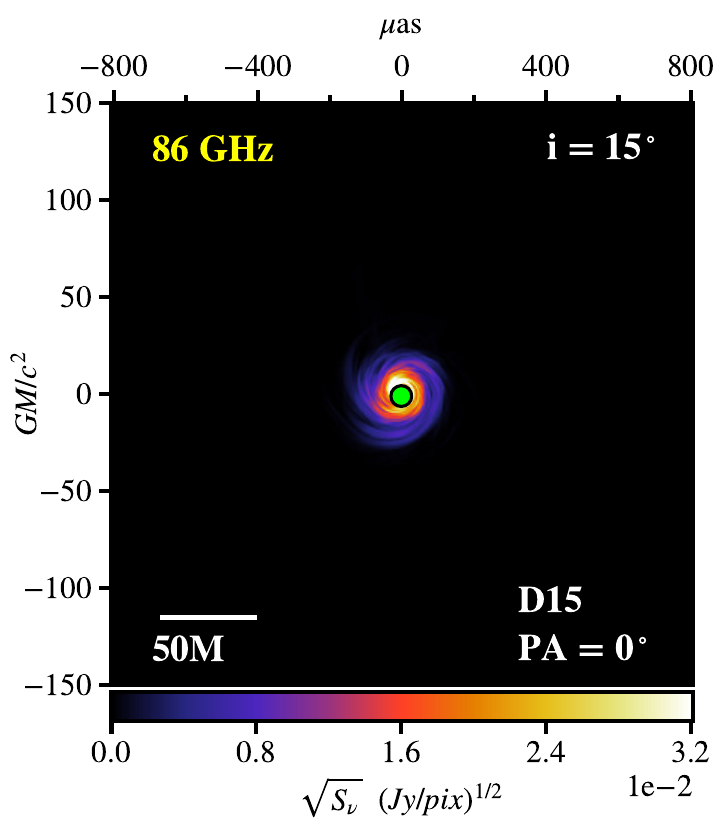}\\
\includegraphics[width=0.3\textwidth]{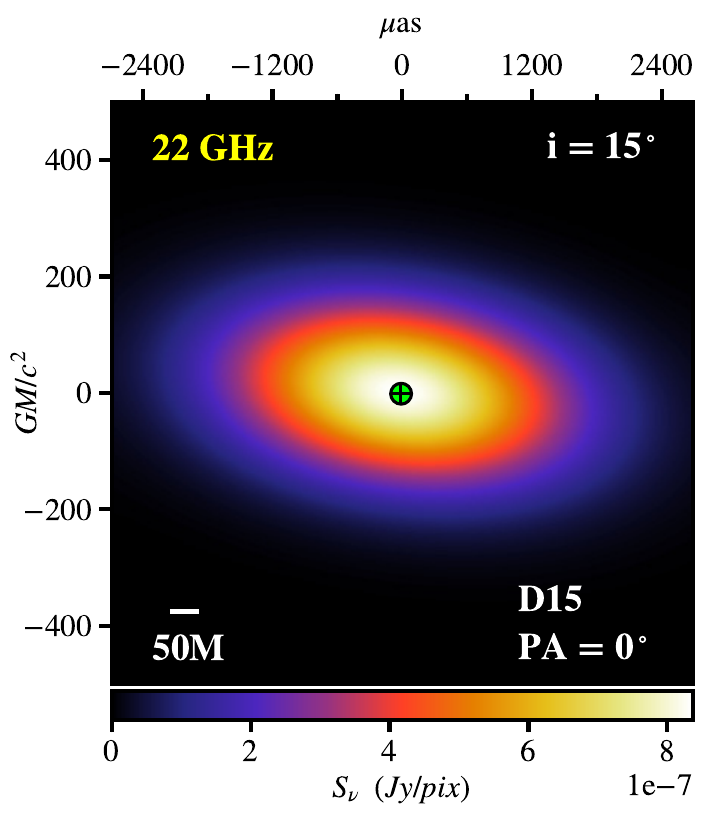}
\includegraphics[width=0.3\textwidth]{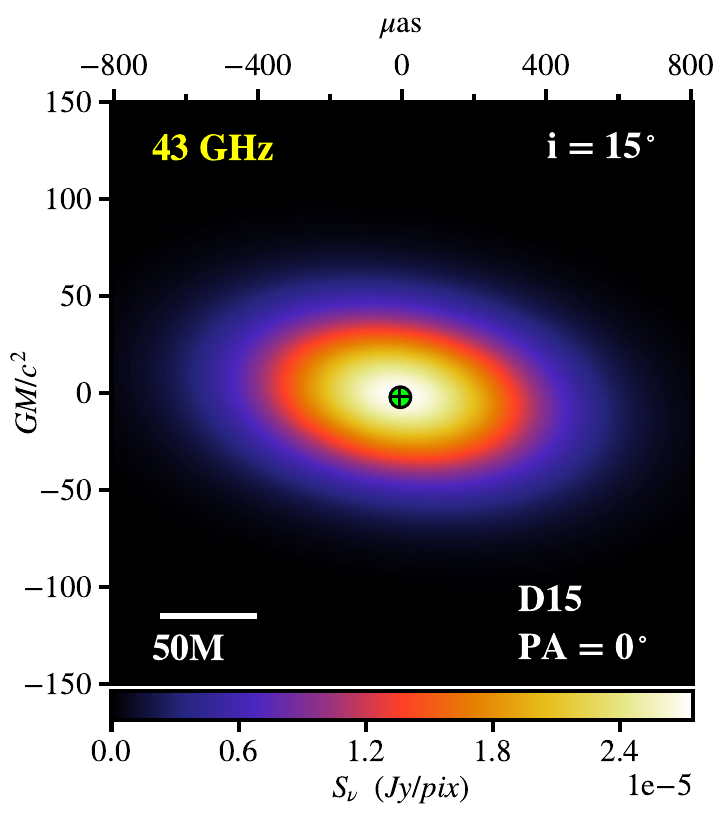}
\includegraphics[width=0.3\textwidth]{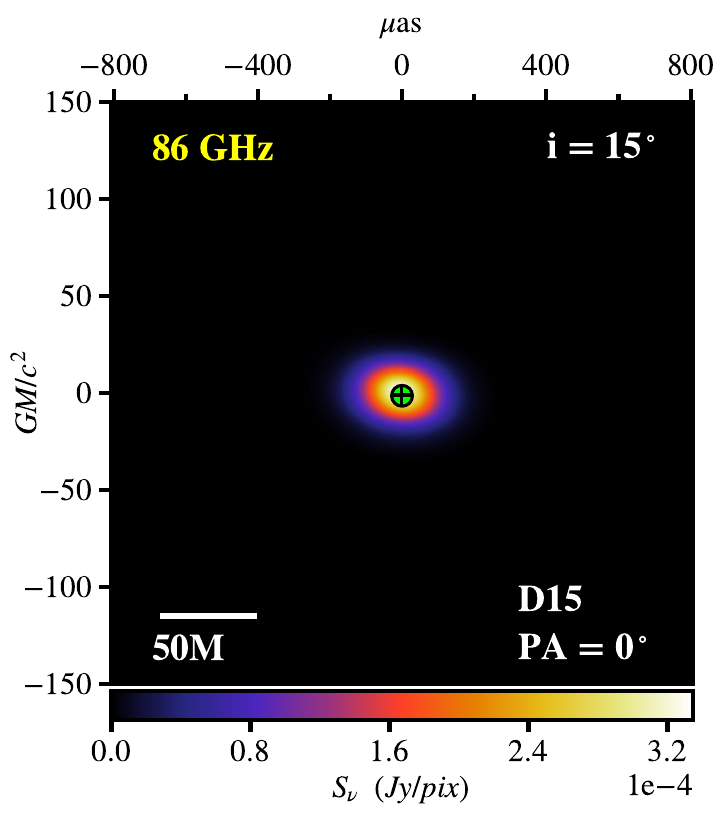}\\
\includegraphics[width=0.3\textwidth]{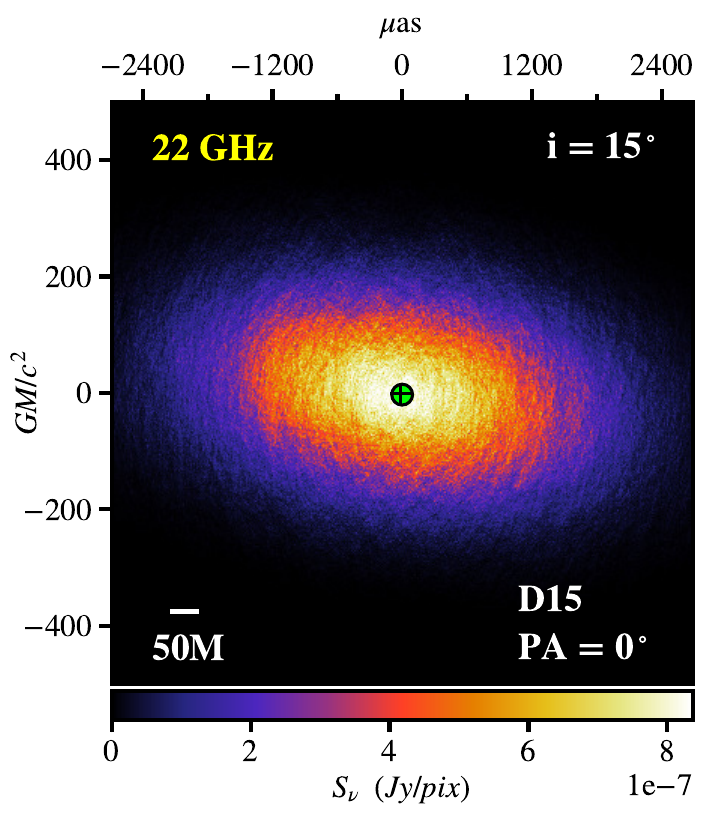}
\includegraphics[width=0.3\textwidth]{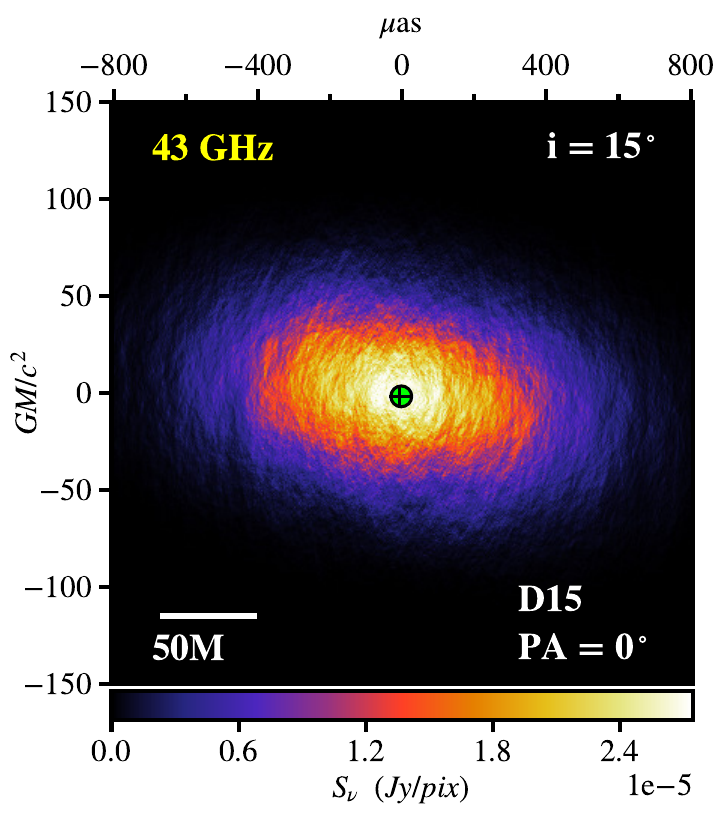}
\includegraphics[width=0.3\textwidth]{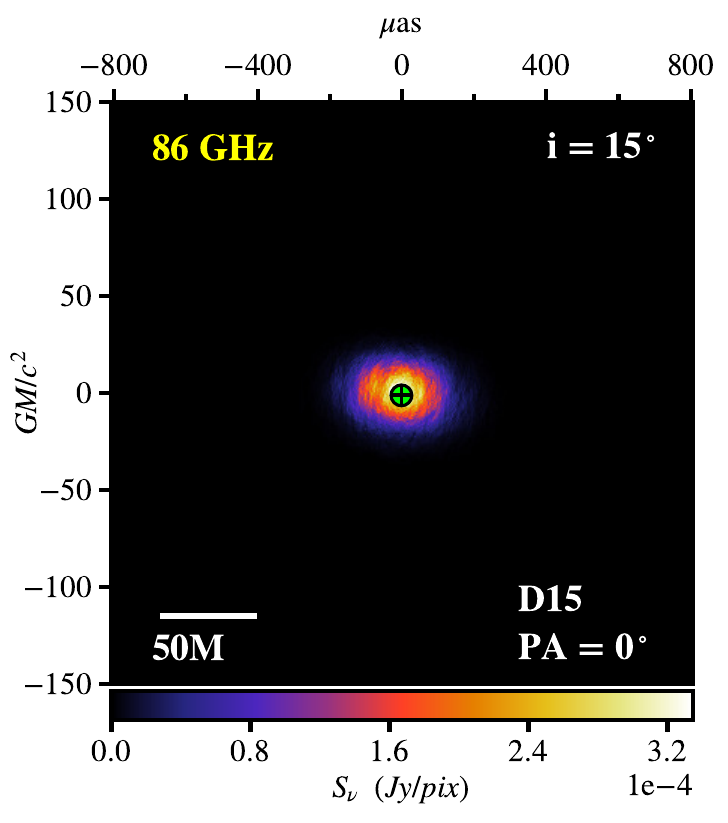}
\end{figure*}
\begin{figure*}[!hp]
\caption{\textbf{Disk model with $i=30 \degr$ including scattering}. Each column shows a disk model with
 inclination $i=30 \degr$ at frequencies of 22~GHz, 43~GHz \& 86~GHz 
 (\textit{left to right}). The position angle on the sky of the BH spin-axis is $PA=0 \degr$. 
 Rows from \textit{top to bottom} show the unscattered 3D-GRMHD jet model, the scatter-broadened image, 
 and the average image including refractive scattering. Note that the FOV for images at 22~GHz including 
 scattering \textit{(left column)} is doubled compared to the FOV at 43~GHz \& 86~GHz. The color stretch is 
 also different. The unscattered model panels display the square-root of the flux density (\textit{top row}). 
 The scattered maps are plotted using a linear scale (\textit{second \& third rows}). The green dot indicates
 the intensity-weighted centroid of each image. Black cross-hairs (\textit{second \& third rows}) indicate 
 the location of the intensity-weighted centroid of the model \textit{before} scattering.}\label{fig:scattered_disk_model_i30}
\centering
\includegraphics[width=0.3\textwidth]{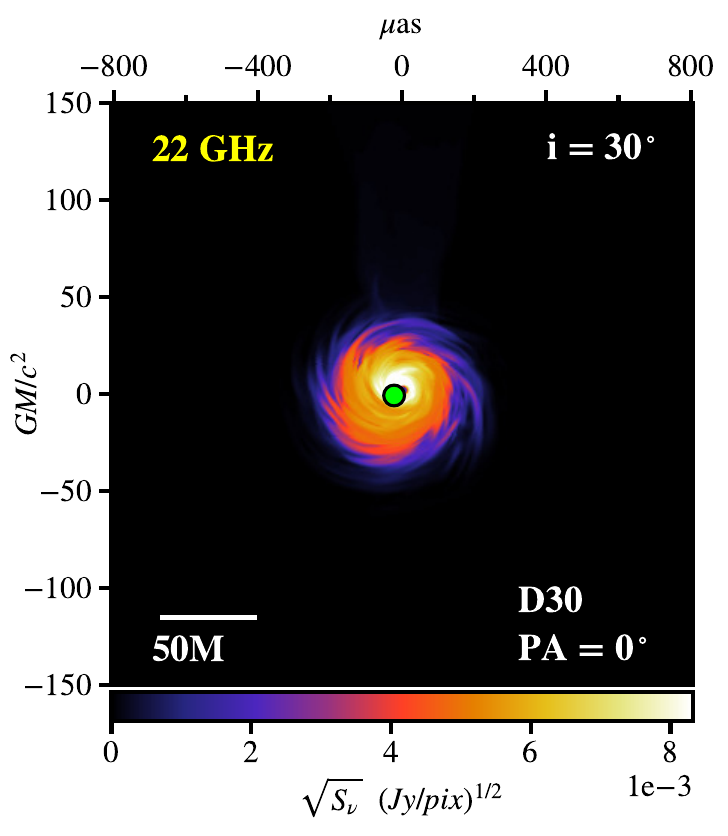}
\includegraphics[width=0.3\textwidth]{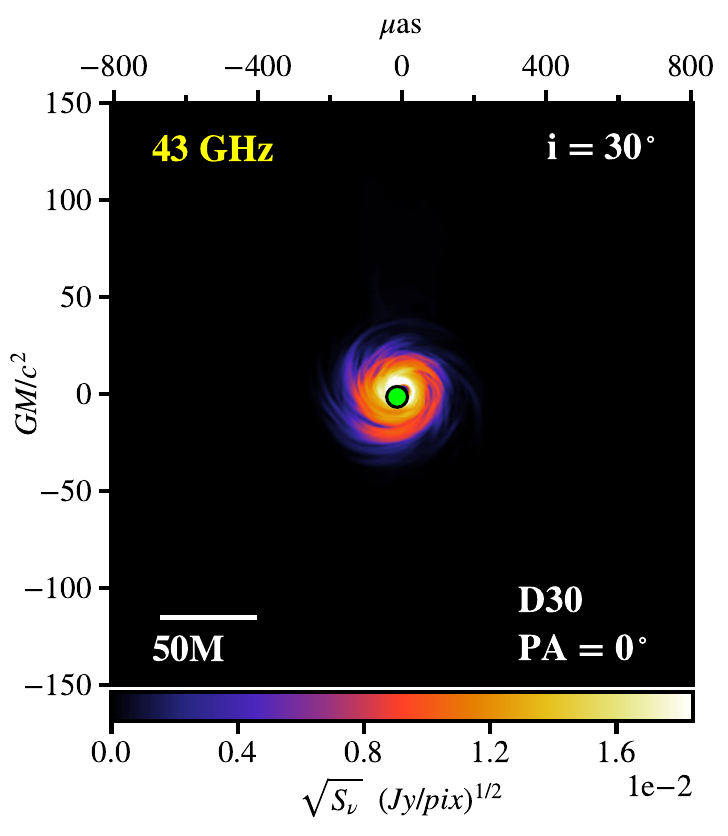}
\includegraphics[width=0.3\textwidth]{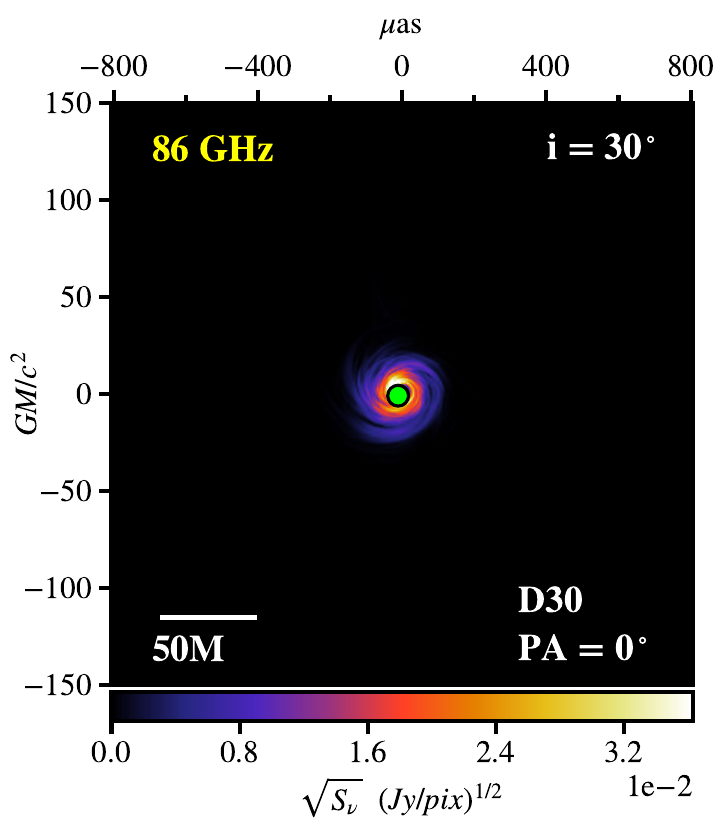}\\
\includegraphics[width=0.3\textwidth]{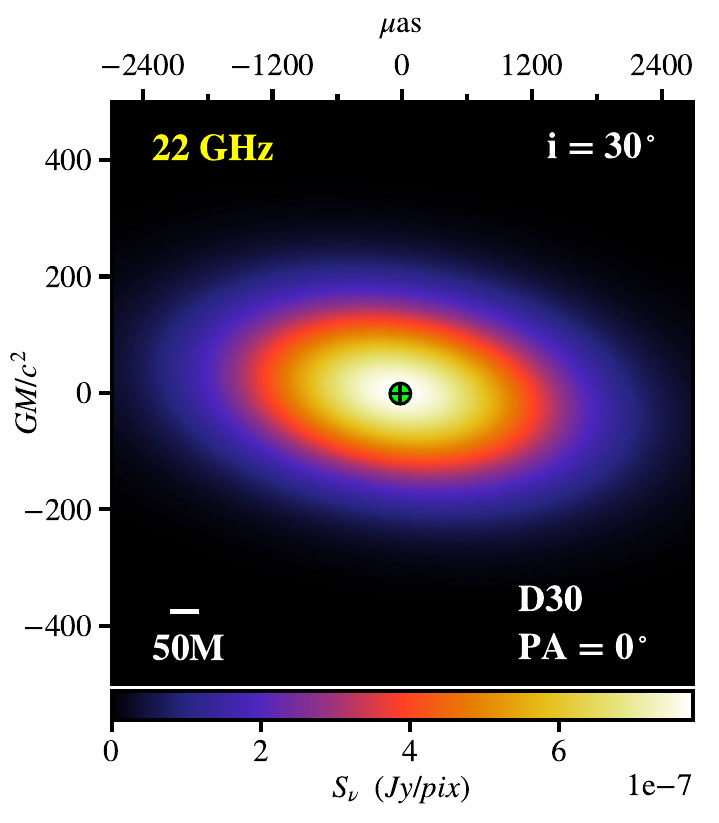}
\includegraphics[width=0.3\textwidth]{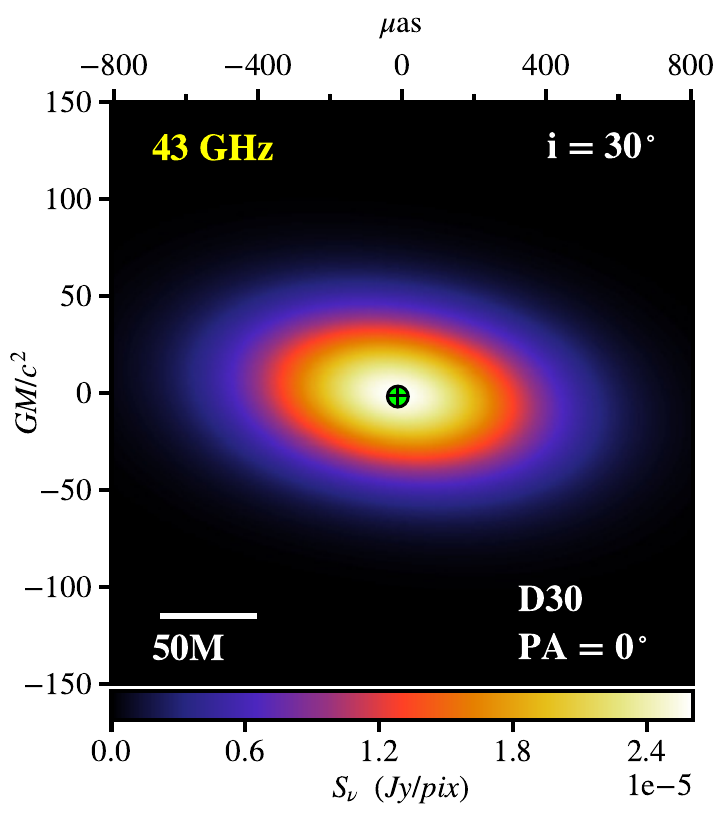}
\includegraphics[width=0.3\textwidth]{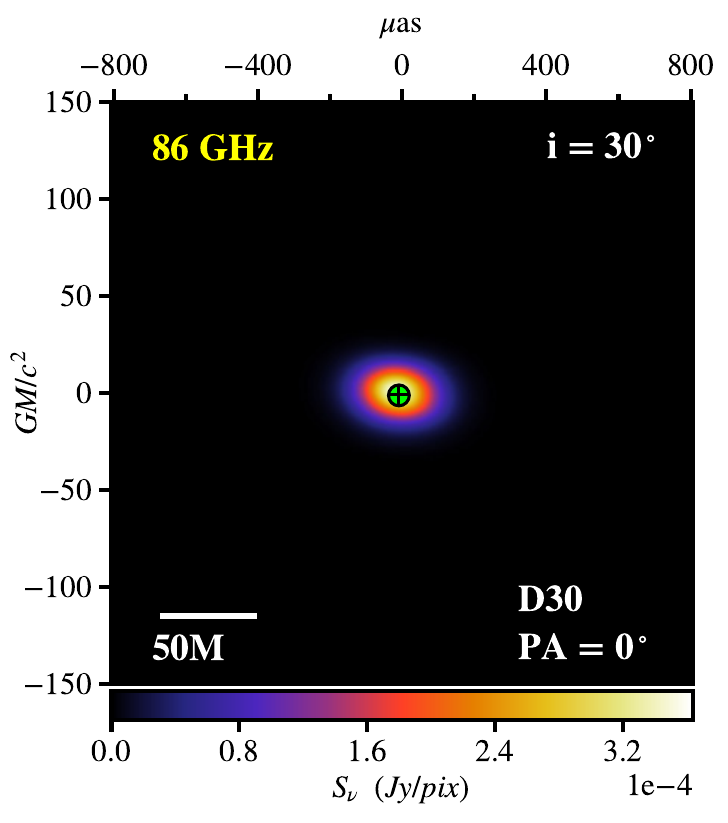}\\
\includegraphics[width=0.3\textwidth]{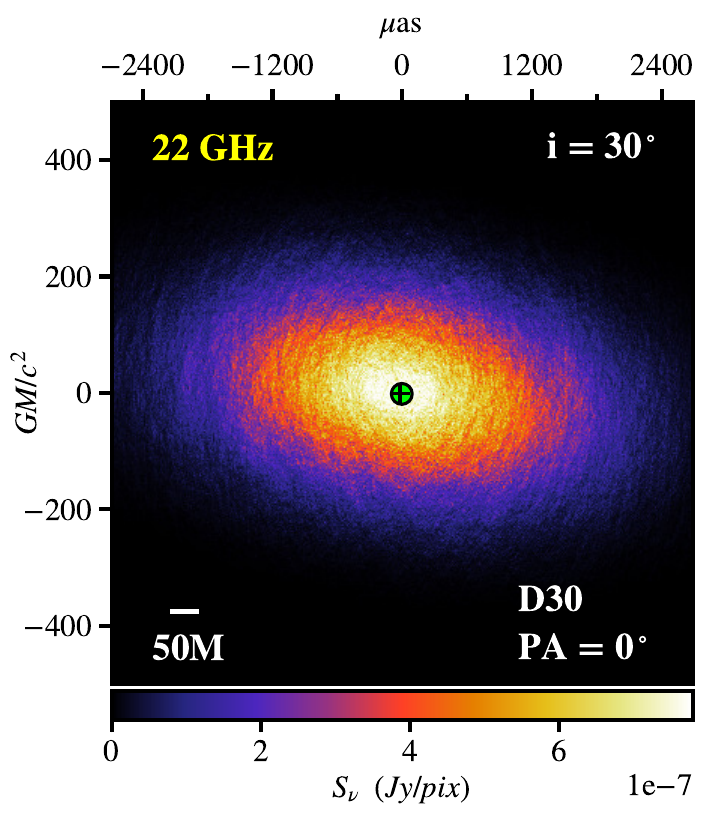}
\includegraphics[width=0.3\textwidth]{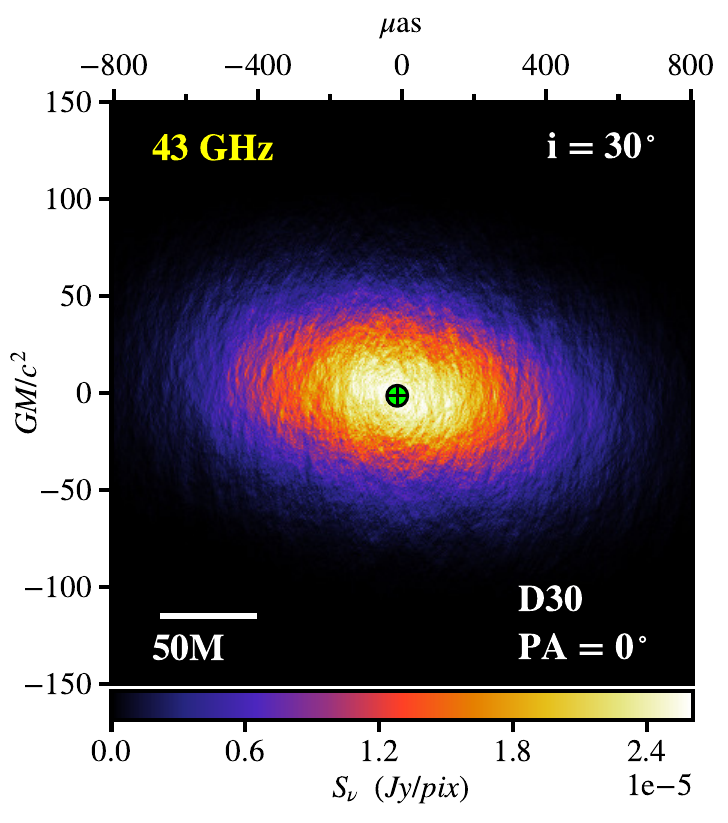}
\includegraphics[width=0.3\textwidth]{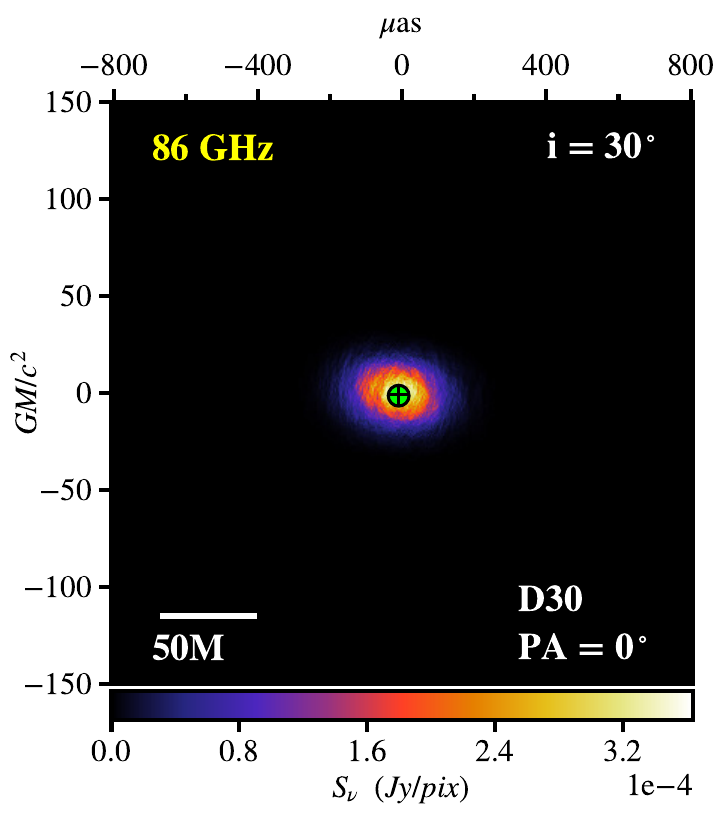}
\end{figure*}
\begin{figure*}[!hp]
\caption{\textbf{Disk model with $i=45 \degr$ including scattering}. Each column shows a disk model with
 inclination $i=45 \degr$ at frequencies of 22~GHz, 43~GHz \& 86~GHz 
 (\textit{left to right}). The position angle on the sky of the BH spin-axis is $PA=0 \degr$. 
 Rows from \textit{top to bottom} show the unscattered 3D-GRMHD jet model, the scatter-broadened image, 
 and the average image including refractive scattering. Note that the FOV for images at 22~GHz including 
 scattering \textit{(left column)} is doubled compared to the FOV at 43~GHz \& 86~GHz. The color stretch is 
 also different. The unscattered model panels display the square-root of the flux density (\textit{top row}). 
 The scattered maps are plotted using a linear scale (\textit{second \& third rows}). The green dot indicates
 the intensity-weighted centroid of each image. Black cross-hairs (\textit{second \& third rows}) indicate 
 the location of the intensity-weighted centroid of the model \textit{before} scattering.}\label{fig:scattered_disk_model_i45}
\centering
\includegraphics[width=0.3\textwidth]{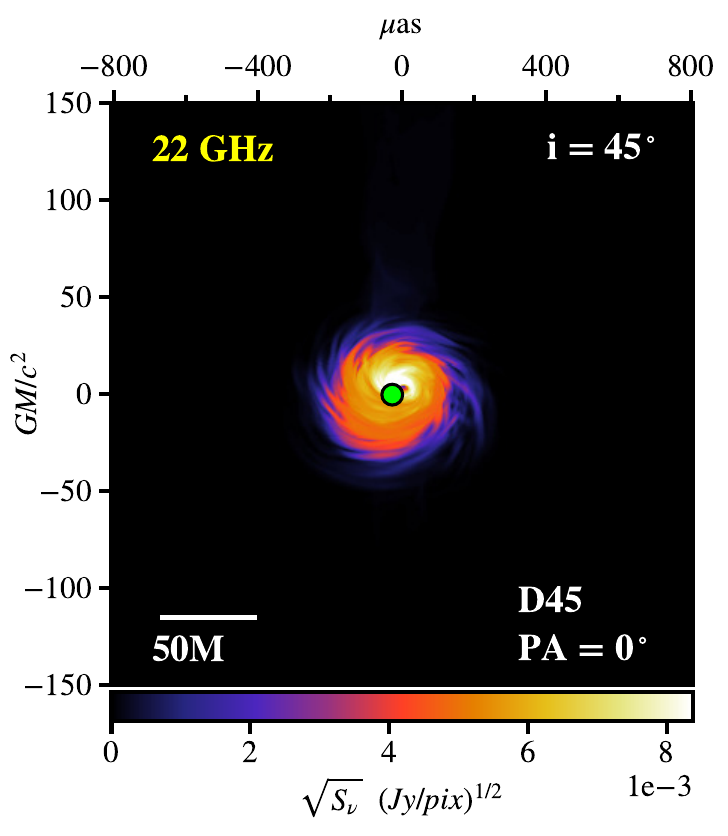}
\includegraphics[width=0.3\textwidth]{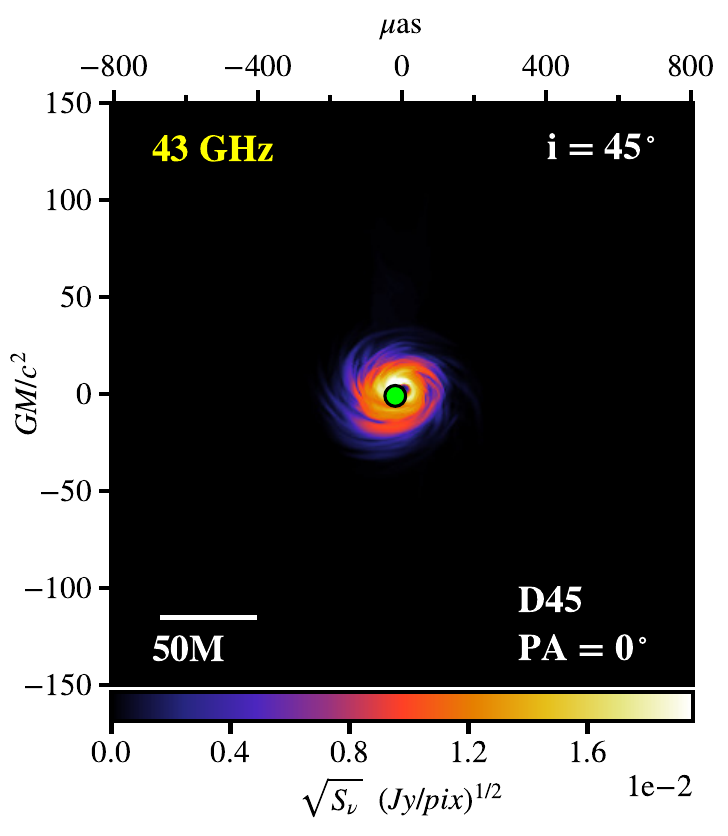}
\includegraphics[width=0.3\textwidth]{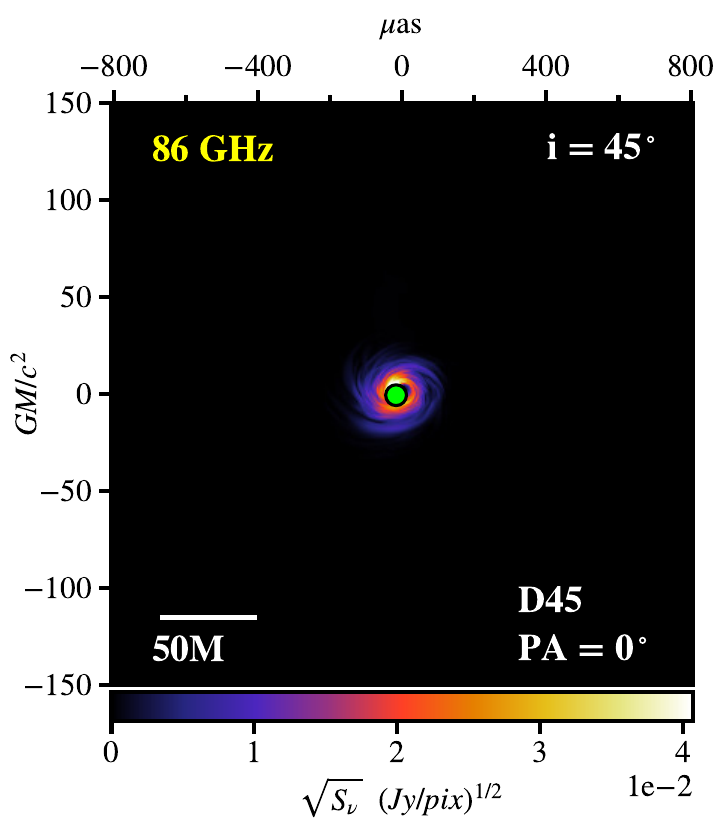}\\
\includegraphics[width=0.3\textwidth]{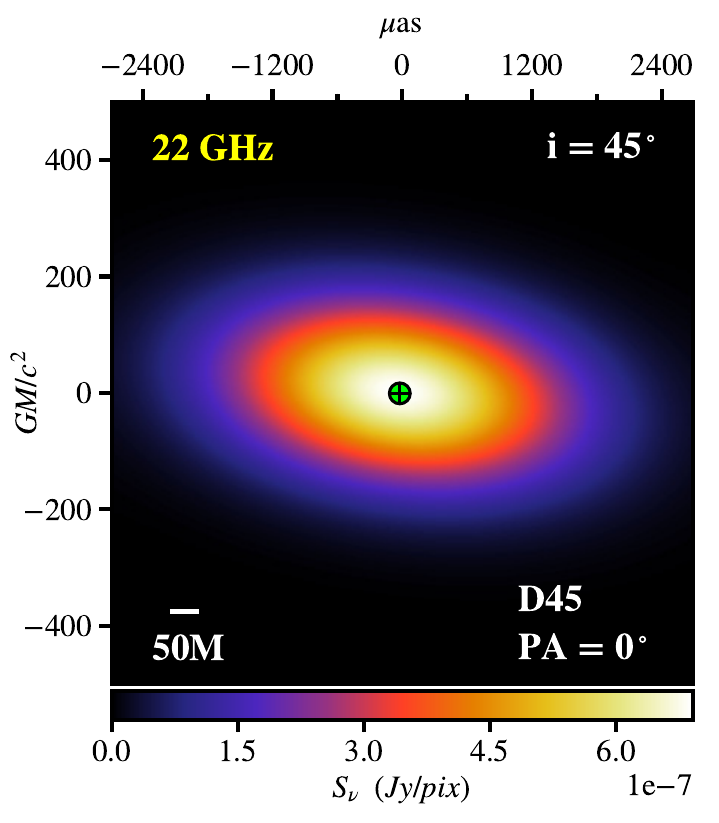}
\includegraphics[width=0.3\textwidth]{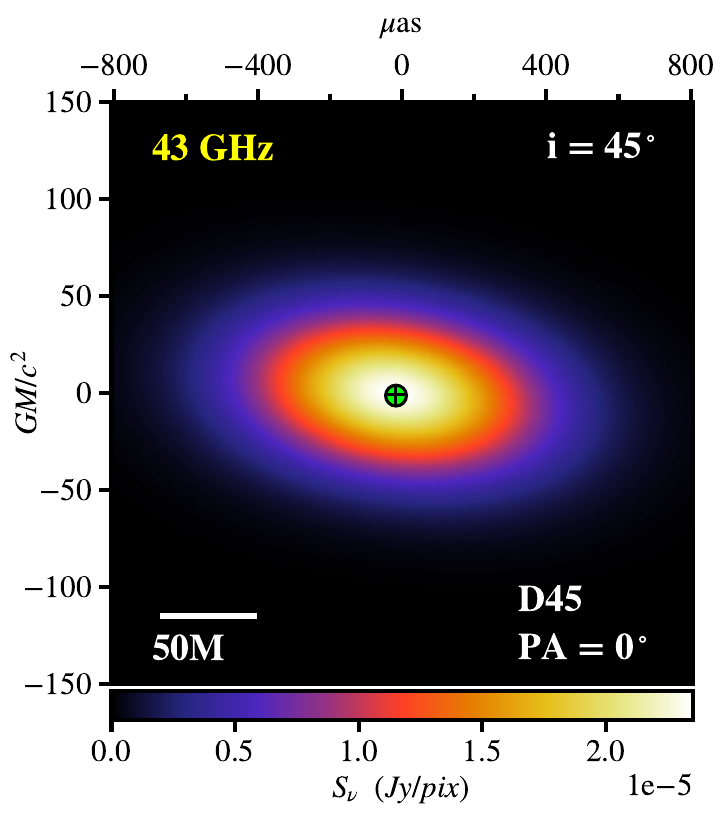}
\includegraphics[width=0.3\textwidth]{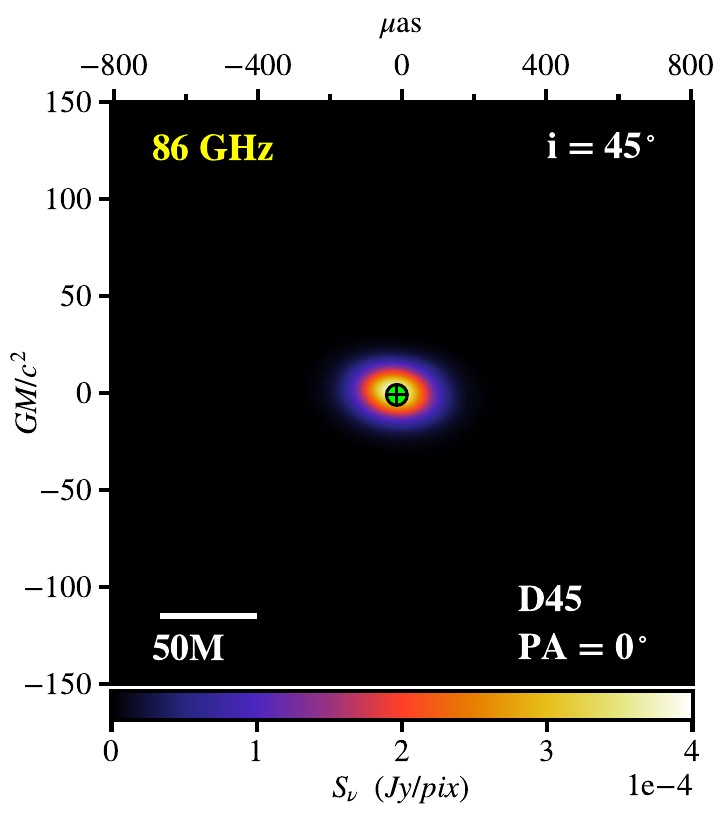}\\
\includegraphics[width=0.3\textwidth]{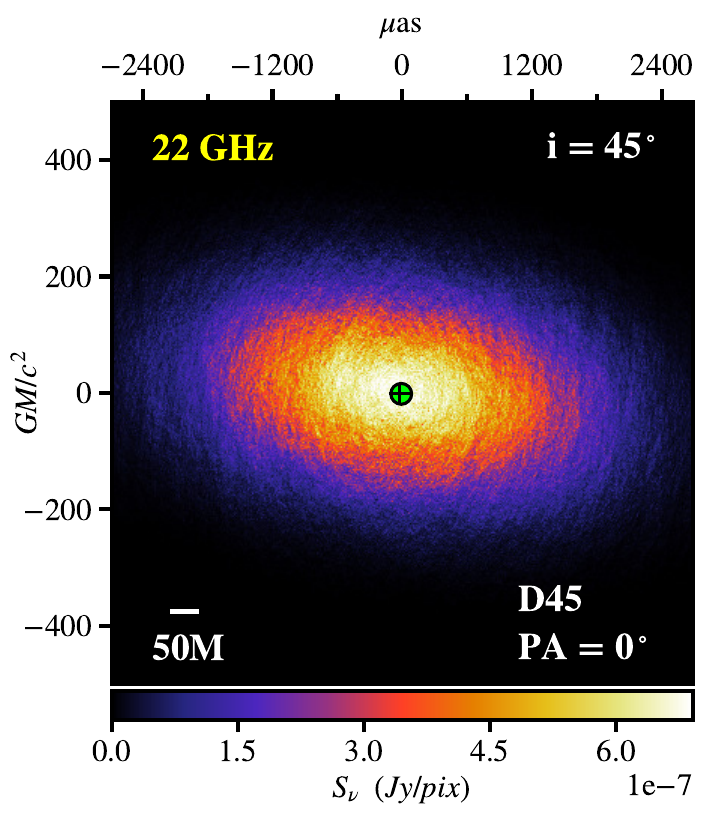}
\includegraphics[width=0.3\textwidth]{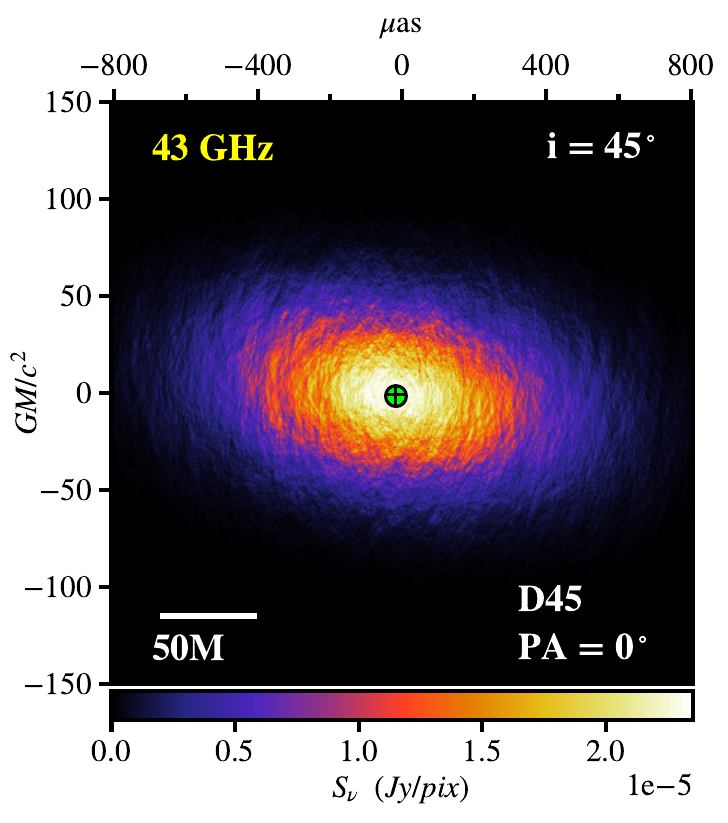}
\includegraphics[width=0.3\textwidth]{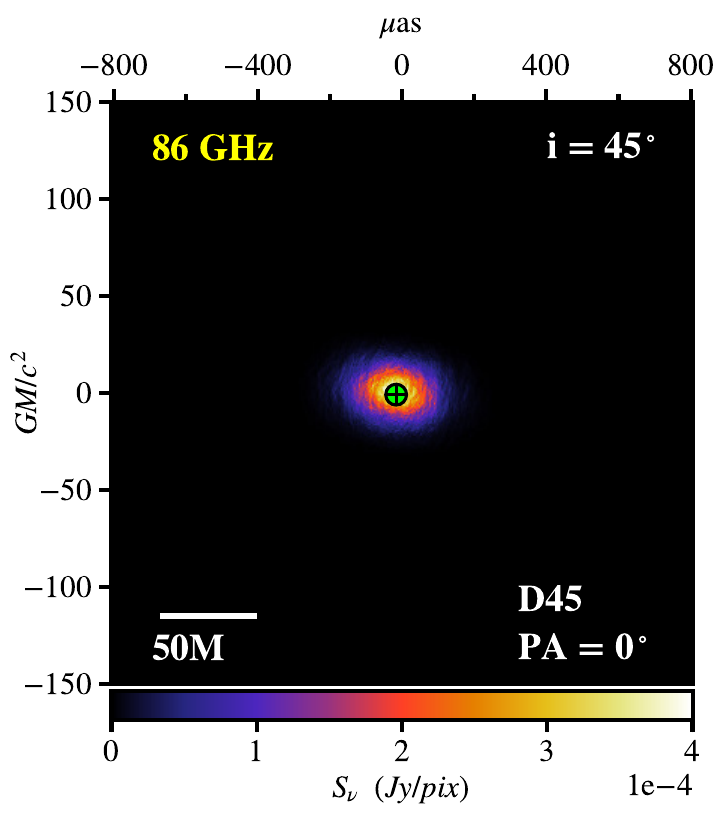}
\end{figure*}
\begin{figure*}[!hp]
\caption{\textbf{Disk model with $i=60 \degr$ including scattering}. Each column shows a disk model with
 inclination $i=60 \degr$ at frequencies of 22~GHz, 43~GHz \& 86~GHz 
 (\textit{left to right}). The position angle on the sky of the BH spin-axis is $PA=0 \degr$. 
 Rows from \textit{top to botton} show the unscattered 3D-GRMHD jet model, the scatter-broadened image, 
 and the average image including refractive scattering. Note that the FOV for images at 22~GHz including 
 scattering \textit{(left column)} is doubled compared to the FOV at 43~GHz \& 86~GHz. The color strecht is 
 also different. The unscattered model panels display the square-root of the flux density (\textit{top row}). 
 The scattered maps are plotted using a linear scale (\textit{second \& third rows}). The green dot indicates
 the intensity-weighted centroid of each image. Black cross-hairs (\textit{second \& third rows}) indicate 
 the location of the intensity-weighted centroid of the model \textit{before} scattering.}\label{fig:scattered_disk_model_i60}
\centering
\includegraphics[width=0.3\textwidth]{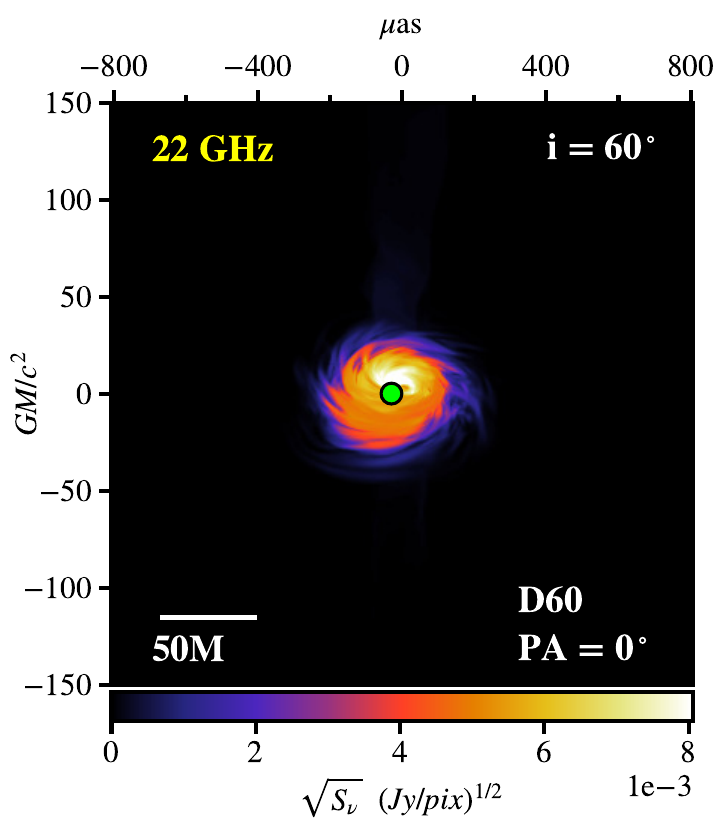}
\includegraphics[width=0.3\textwidth]{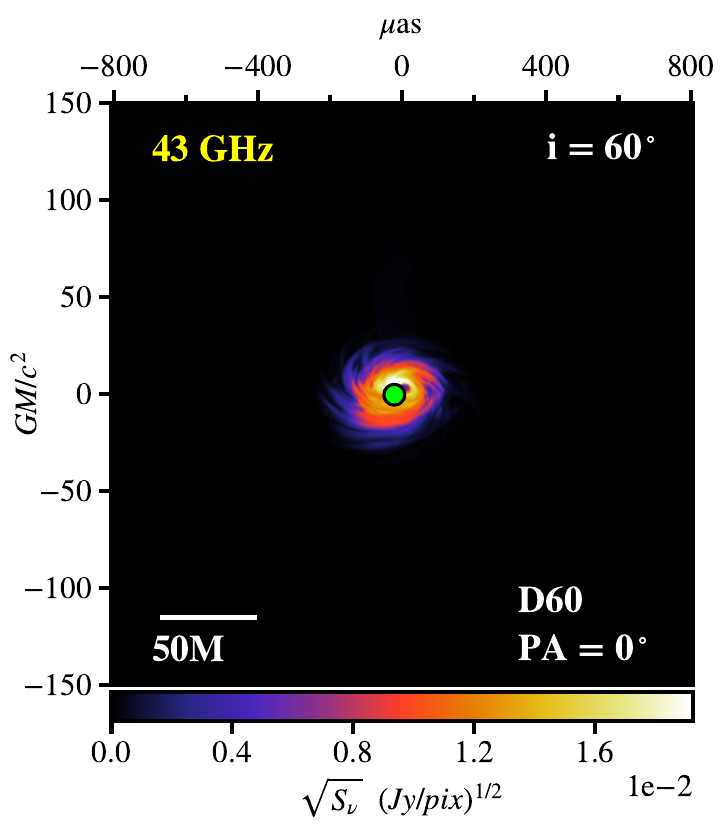}
\includegraphics[width=0.3\textwidth]{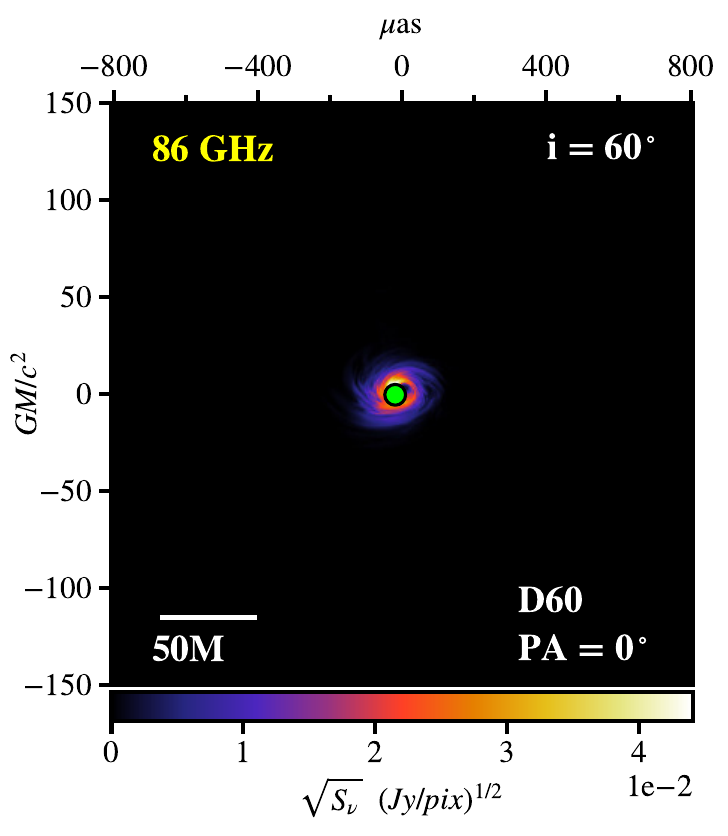}\\
\includegraphics[width=0.3\textwidth]{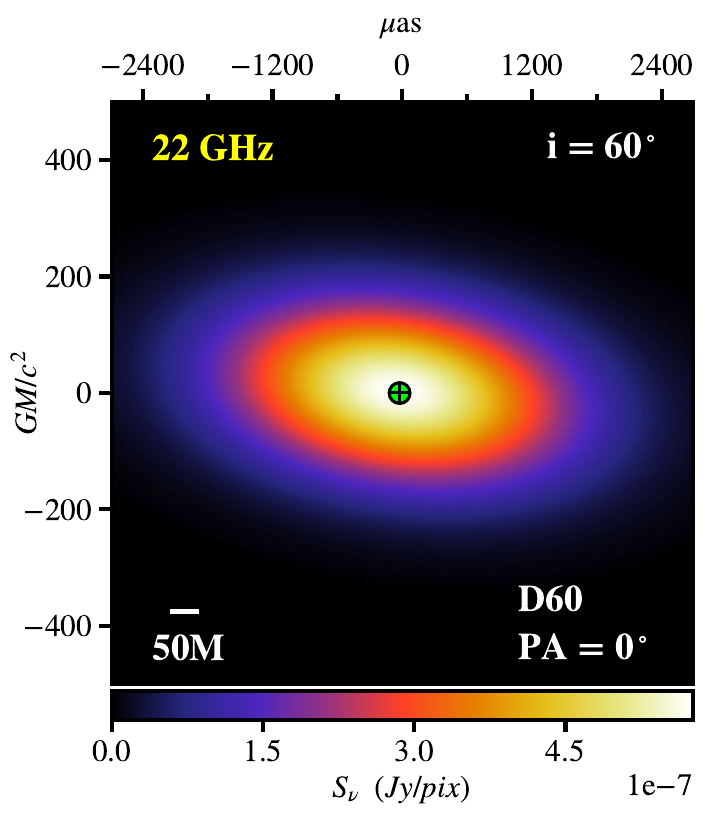}
\includegraphics[width=0.3\textwidth]{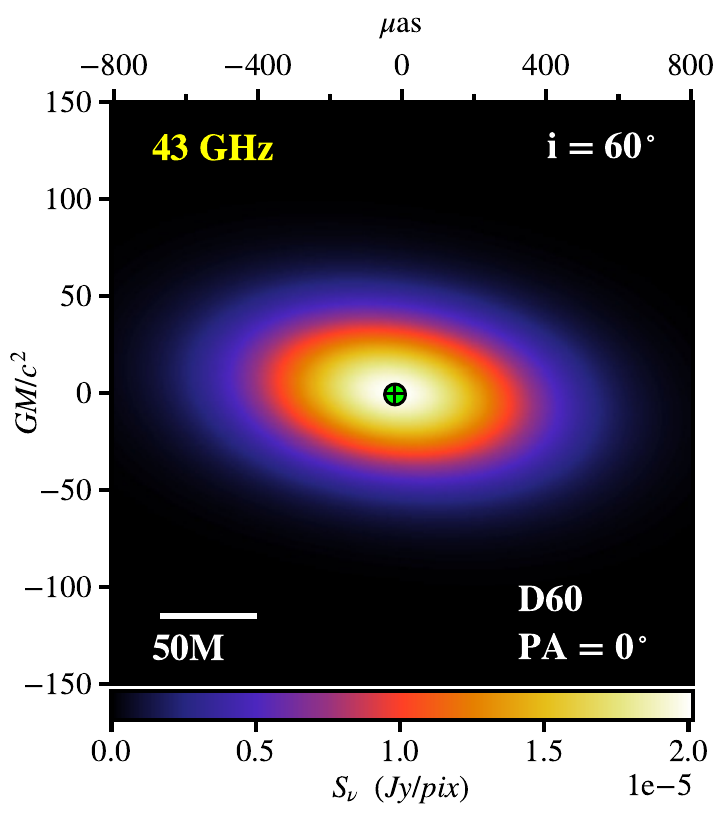}
\includegraphics[width=0.3\textwidth]{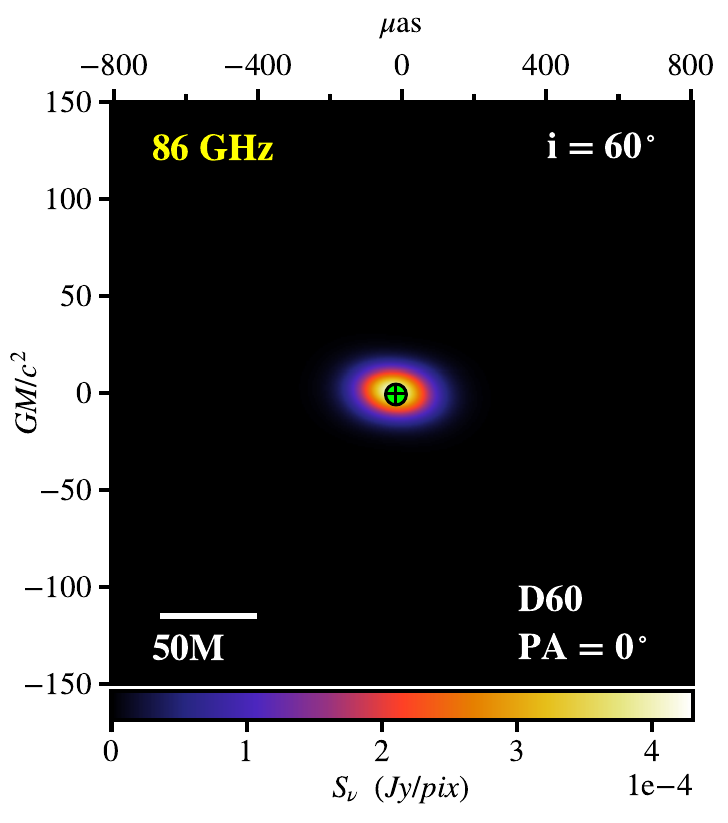}\\
\includegraphics[width=0.3\textwidth]{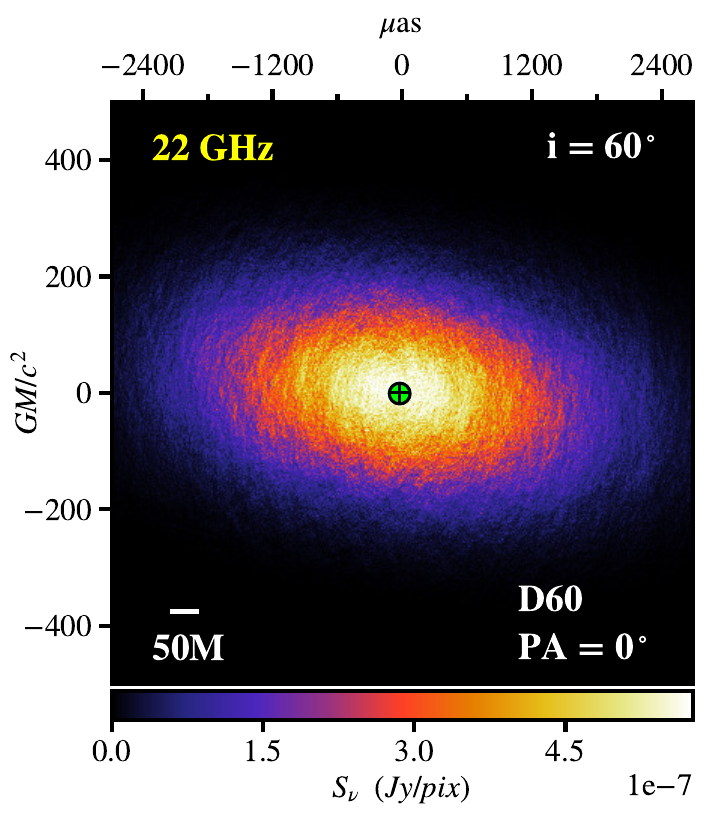}
\includegraphics[width=0.3\textwidth]{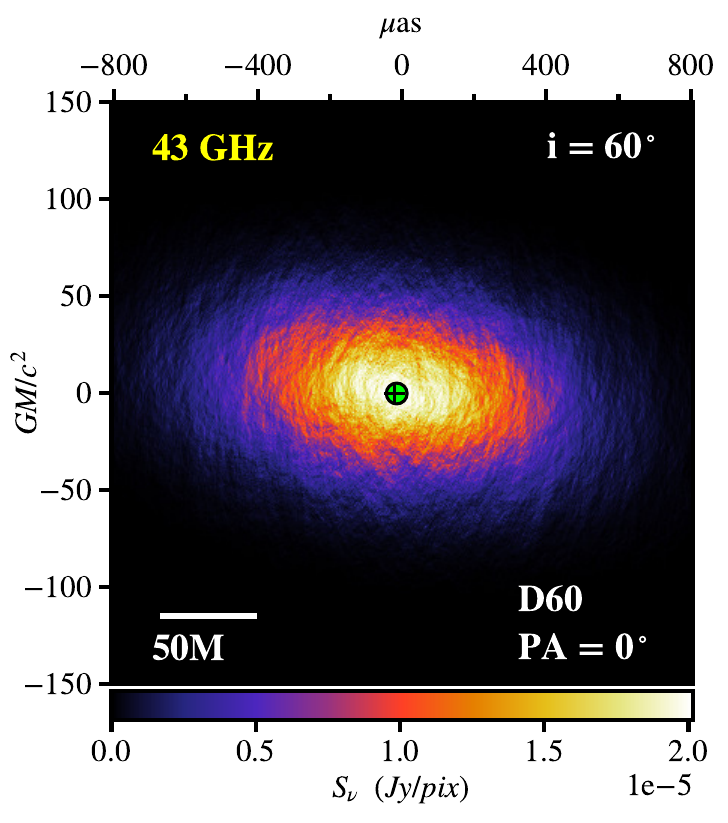}
\includegraphics[width=0.3\textwidth]{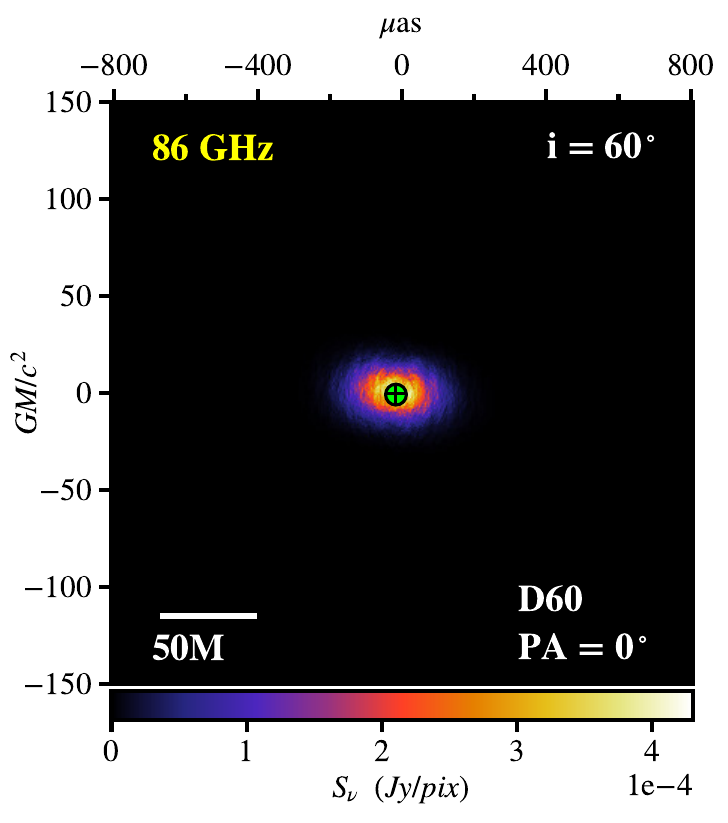}
\end{figure*}
\begin{figure*}[!hp]
\caption{\textbf{Disk model with $i=75 \degr$ including scattering}. Each column shows a disk model with
 inclination $i=75 \degr$ at frequencies of 22~GHz, 43~GHz \& 86~GHz 
 (\textit{left to right}). The position angle on the sky of the BH spin-axis is $PA=0 \degr$. 
 Rows from \textit{top to botton} show the unscattered 3D-GRMHD jet model, the scatter-broadened image, 
 and the average image including refractive scattering. Note that the FOV for images at 22~GHz including 
 scattering \textit{(left column)} is doubled compared to the FOV at 43~GHz \& 86~GHz. The color strecht is 
 also different. The unscattered model panels display the square-root of the flux density (\textit{top row}). 
 The scattered maps are plotted using a linear scale (\textit{second \& third rows}). The green dot indicates
 the intensity-weighted centroid of each image. Black cross-hairs (\textit{second \& third rows}) indicate 
 the location of the intensity-weighted centroid of the model \textit{before} scattering.}\label{fig:scattered_disk_model_i75}
\centering
\includegraphics[width=0.3\textwidth]{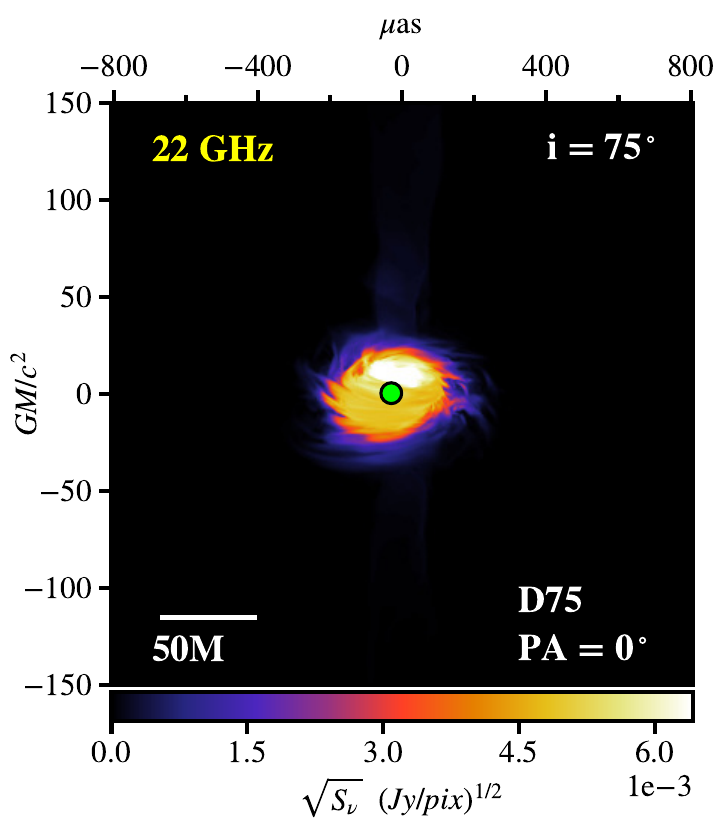}
\includegraphics[width=0.3\textwidth]{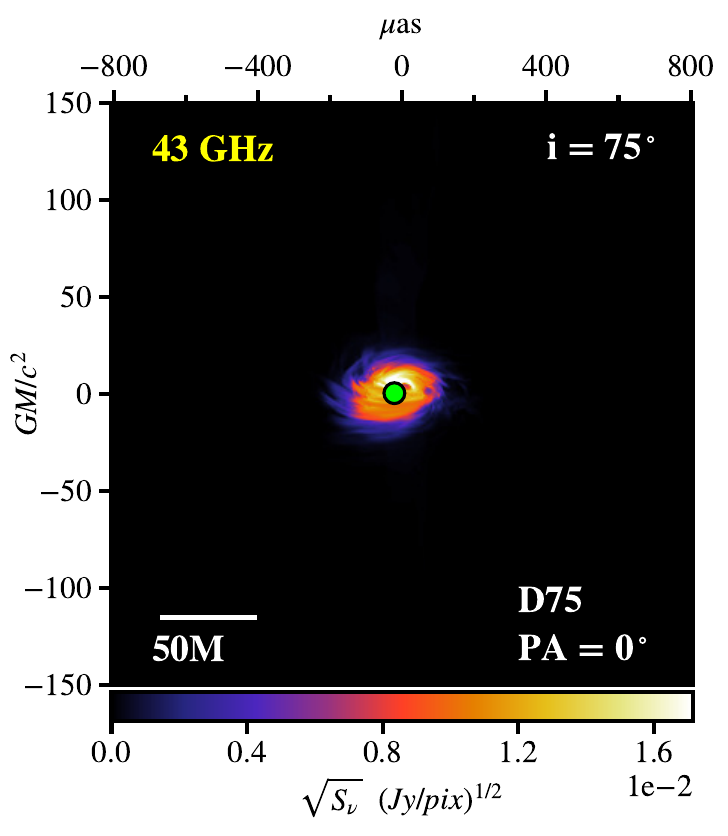}
\includegraphics[width=0.3\textwidth]{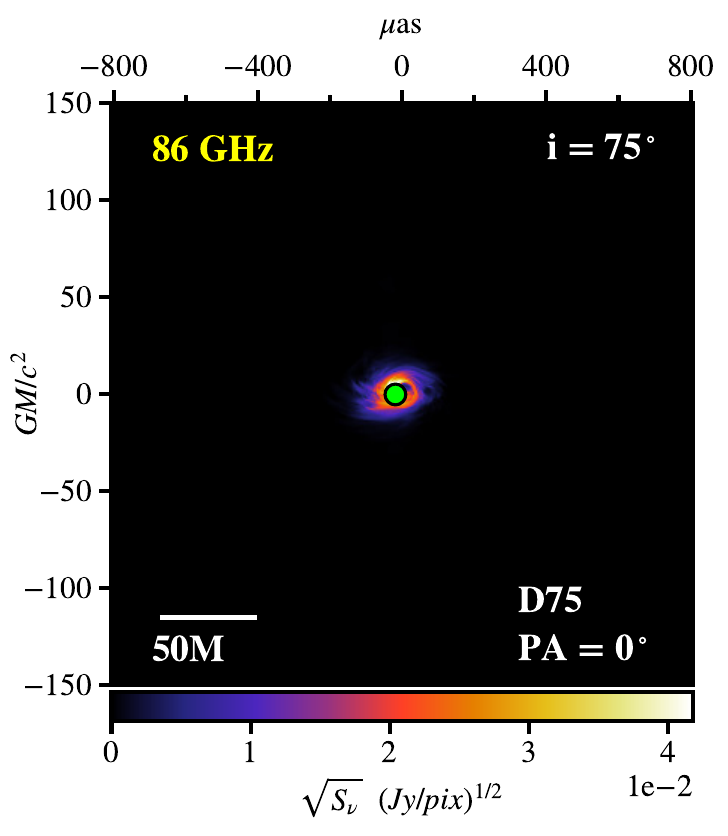}\\
\includegraphics[width=0.3\textwidth]{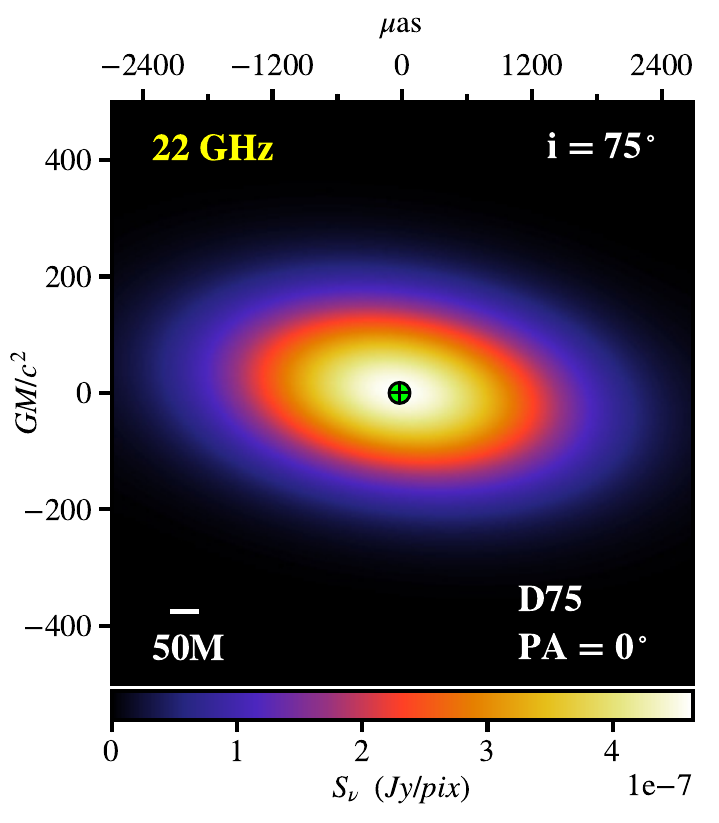}
\includegraphics[width=0.3\textwidth]{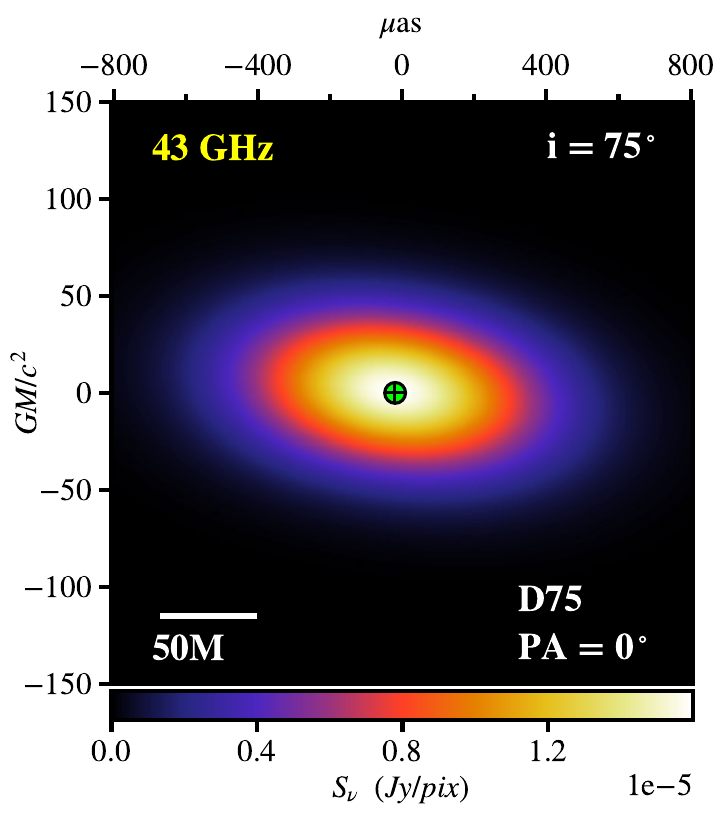}
\includegraphics[width=0.3\textwidth]{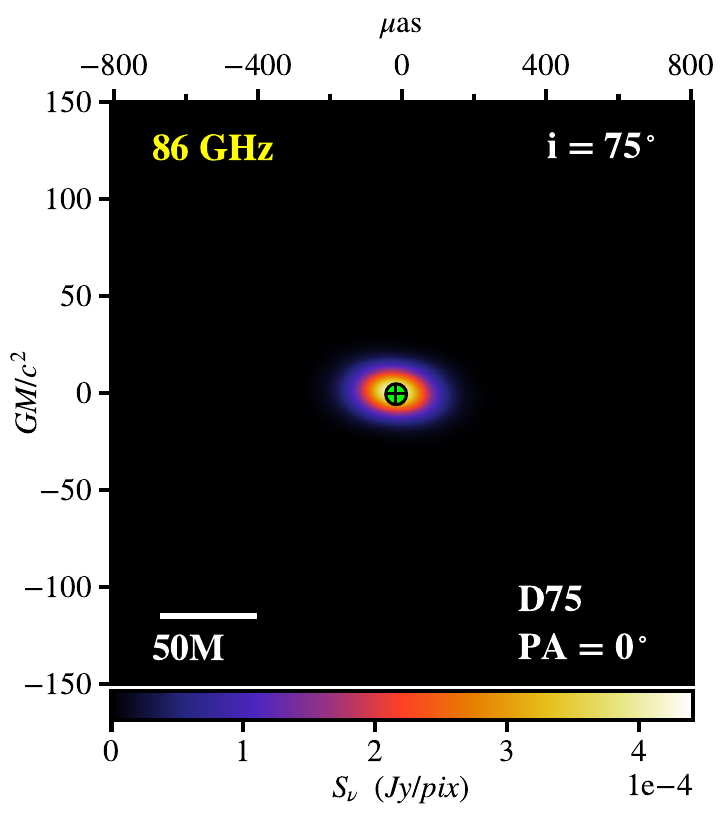}\\
\includegraphics[width=0.3\textwidth]{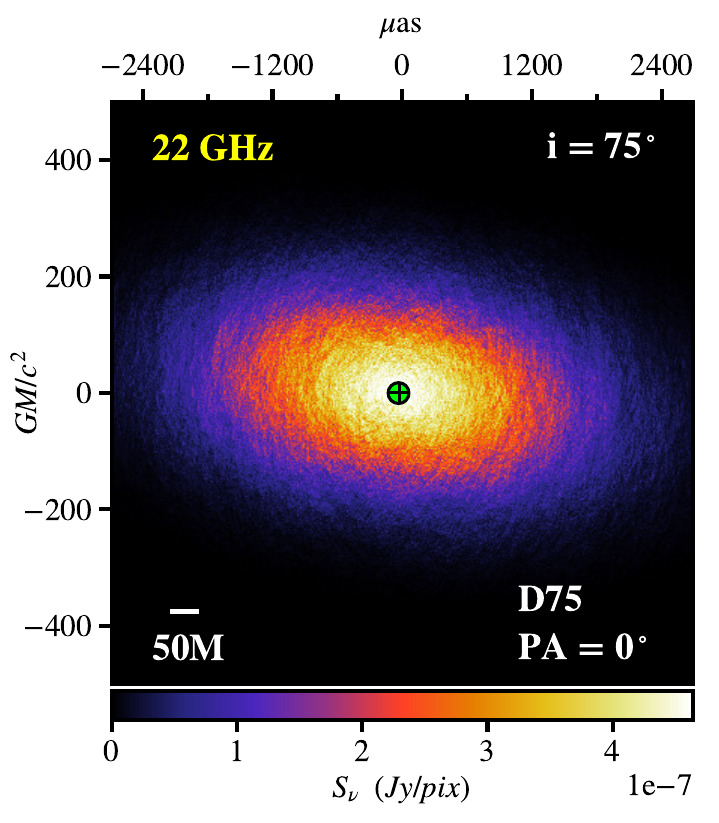}
\includegraphics[width=0.3\textwidth]{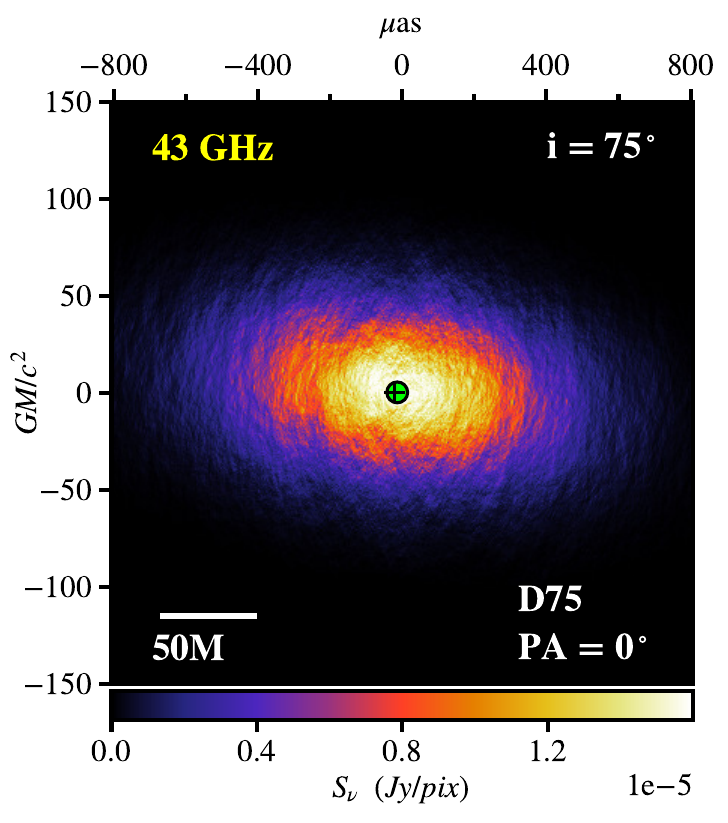}
\includegraphics[width=0.3\textwidth]{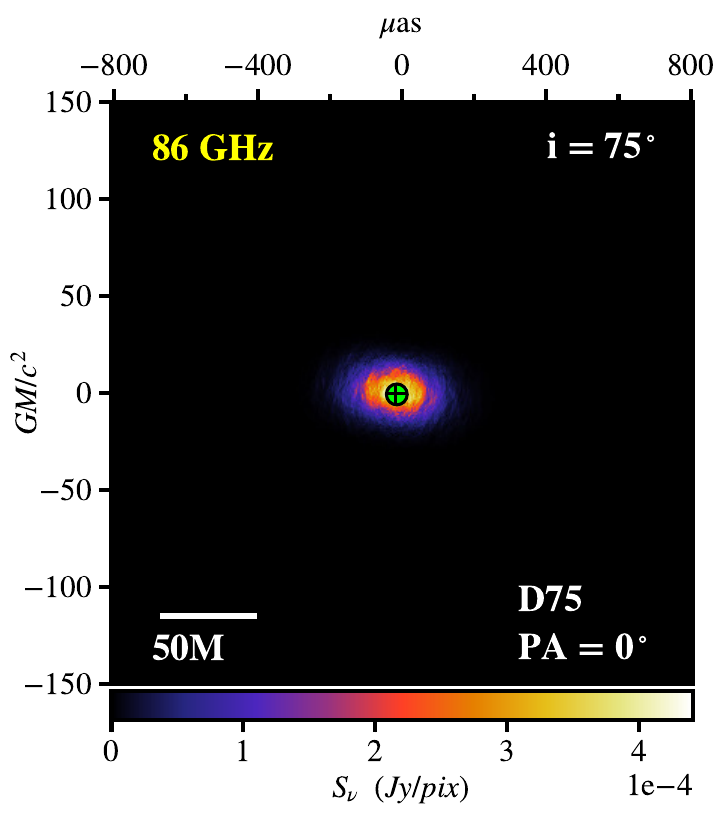}
\end{figure*}
\begin{figure*}[!hp]
\caption{\textbf{Disk model with $i=90 \degr$ including scattering}. Each column shows a disk model with
 inclination $i=90 \degr$ at frequencies of 22~GHz, 43~GHz \& 86~GHz 
 (\textit{left to right}). The position angle on the sky of the BH spin-axis is $PA=0 \degr$. 
 Rows from \textit{top to botton} show the unscattered 3D-GRMHD jet model, the scatter-broadened image, 
 and the average image including refractive scattering. Note that the FOV for images at 22~GHz including 
 scattering \textit{(left column)} is doubled compared to the FOV at 43~GHz \& 86~GHz. The color strecht is 
 also different. The unscattered model panels display the square-root of the flux density (\textit{top row}). 
 The scattered maps are plotted using a linear scale (\textit{second \& third rows}). The green dot indicates
 the intensity-weighted centroid of each image. Black cross-hairs (\textit{second \& third rows}) indicate 
 the location of the intensity-weighted centroid of the model \textit{before} scattering.}\label{fig:scattered_disk_model_i90}
\centering
\includegraphics[width=0.3\textwidth]{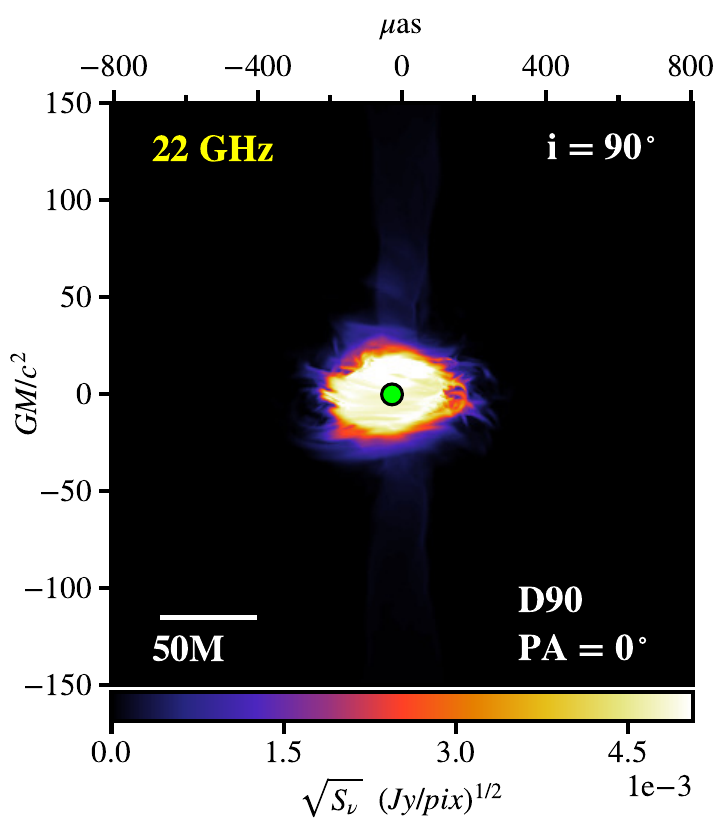}
\includegraphics[width=0.3\textwidth]{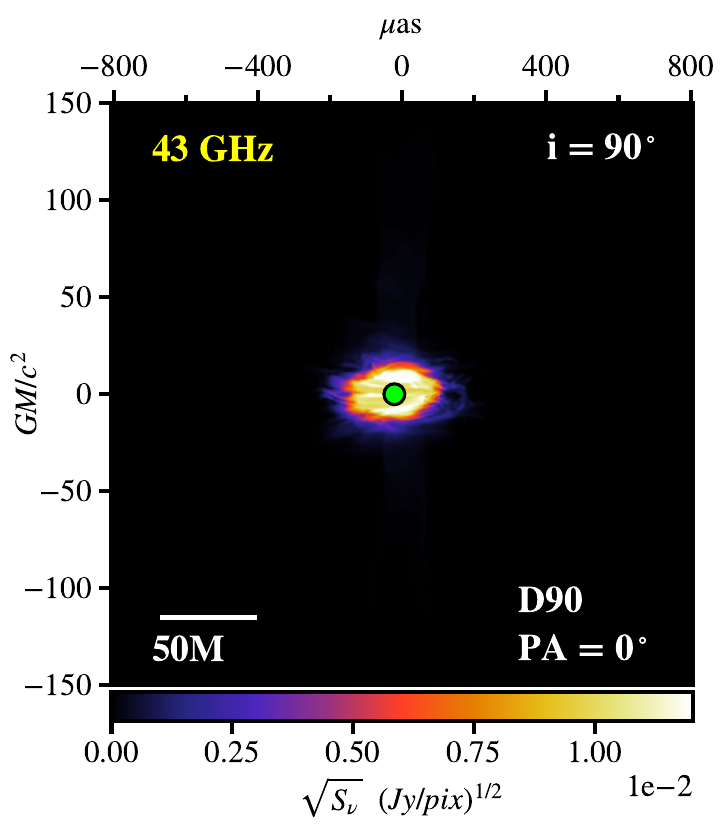}
\includegraphics[width=0.3\textwidth]{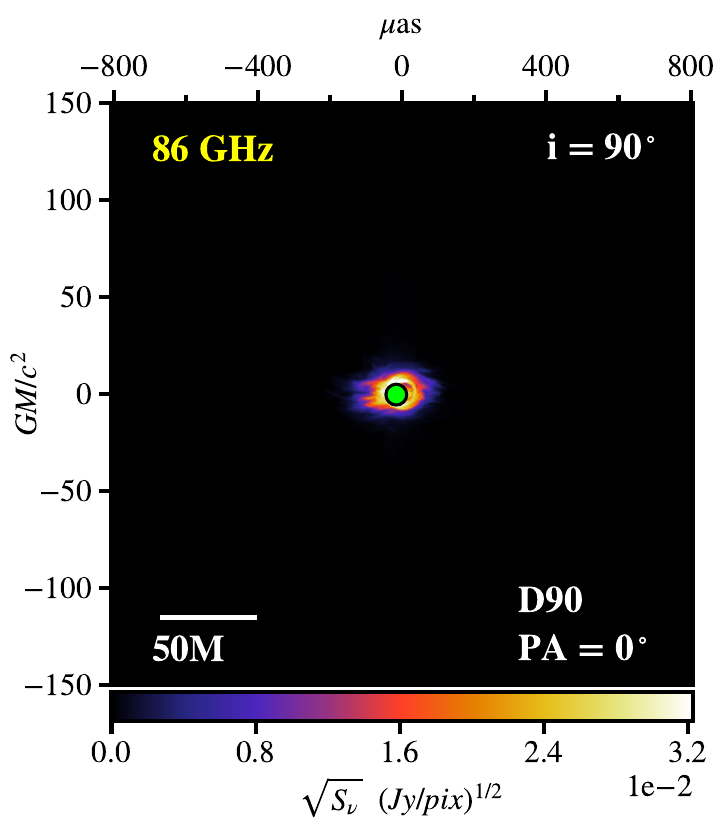}\\
\includegraphics[width=0.3\textwidth]{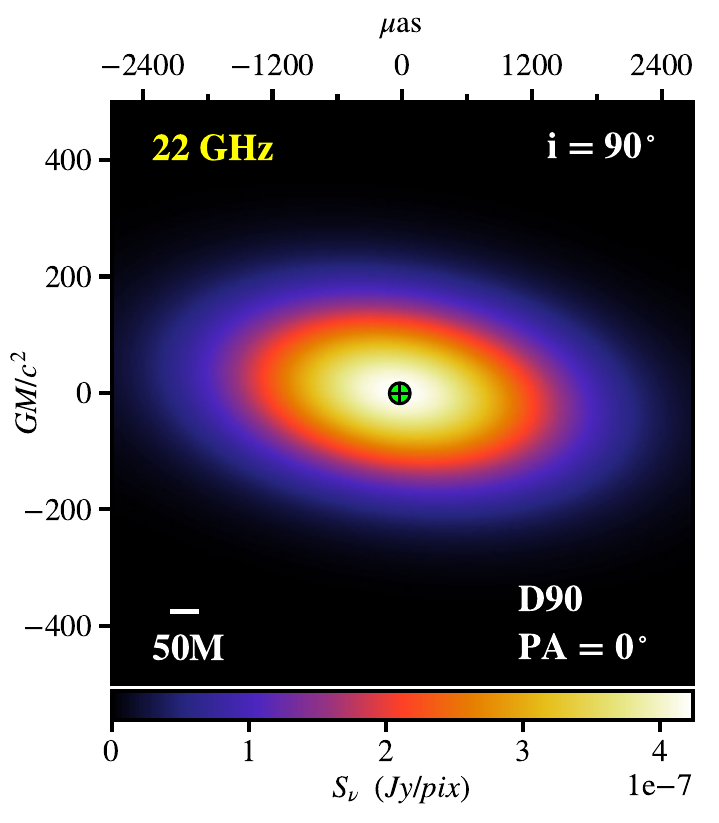}
\includegraphics[width=0.3\textwidth]{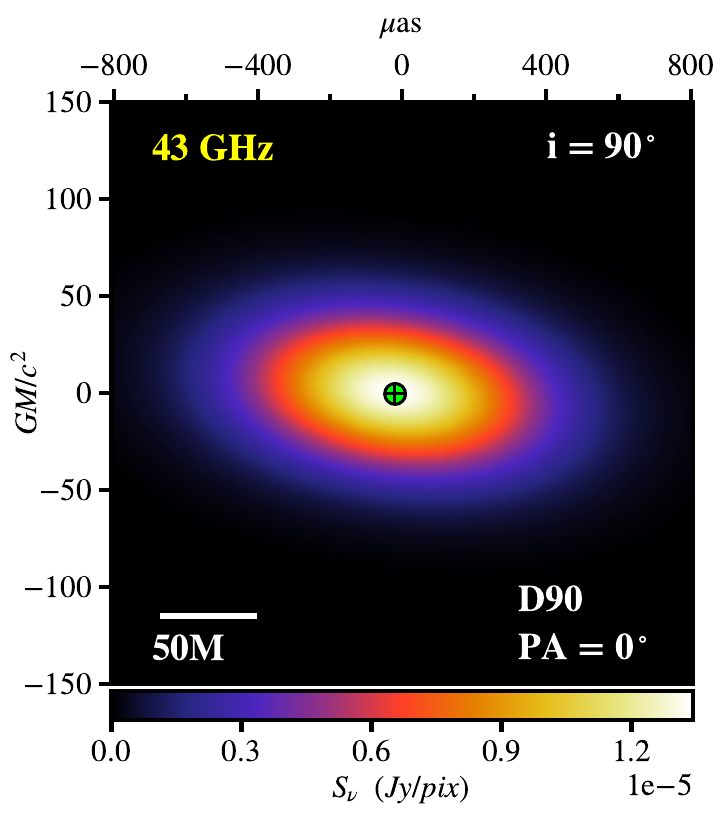}
\includegraphics[width=0.3\textwidth]{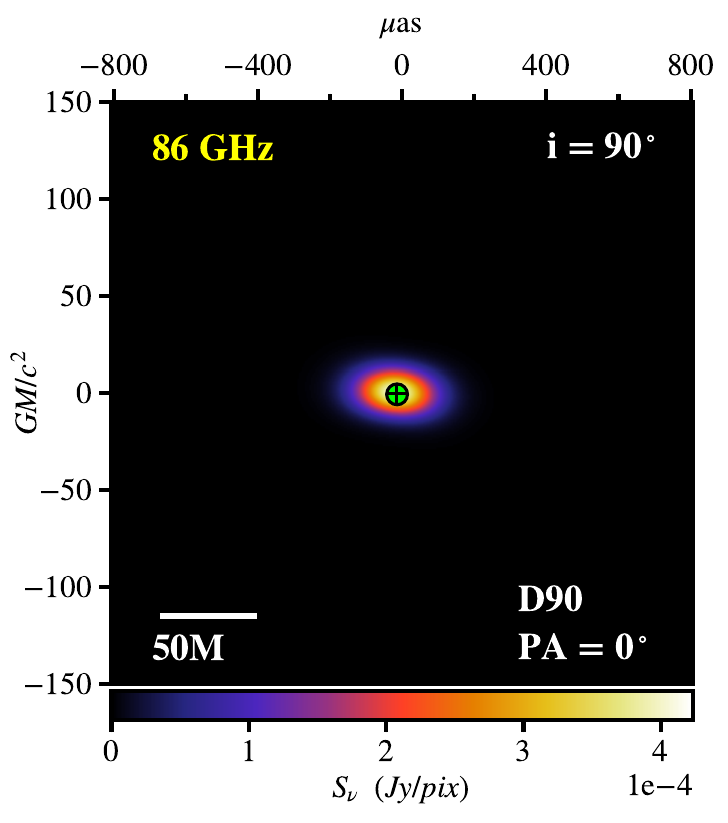}\\
\includegraphics[width=0.3\textwidth]{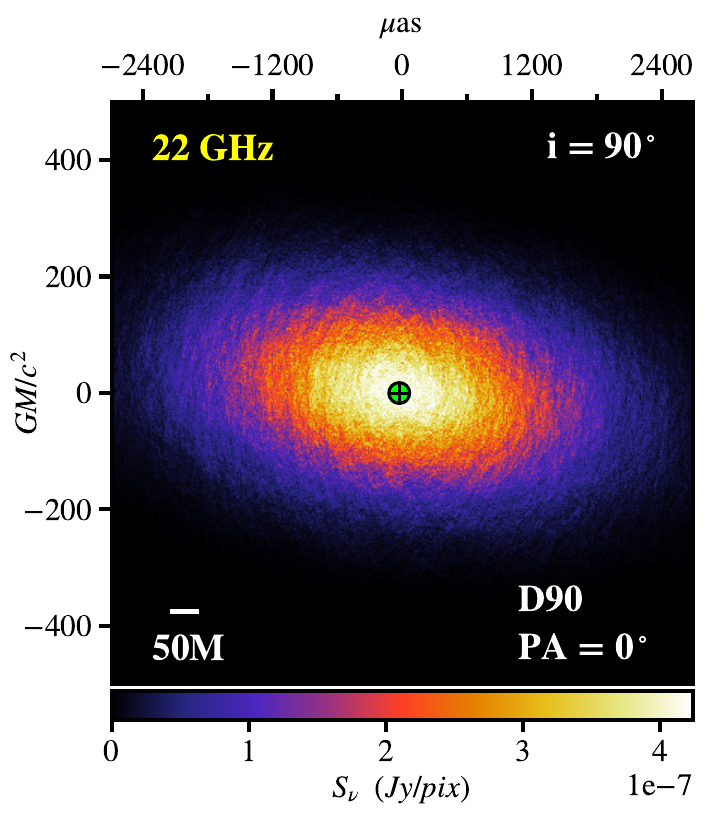}
\includegraphics[width=0.3\textwidth]{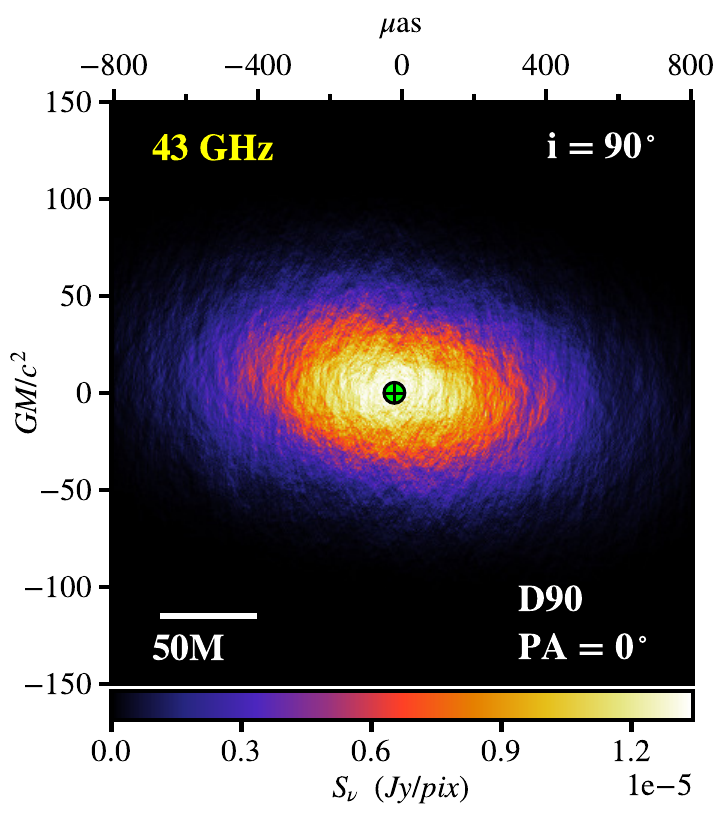}
\includegraphics[width=0.3\textwidth]{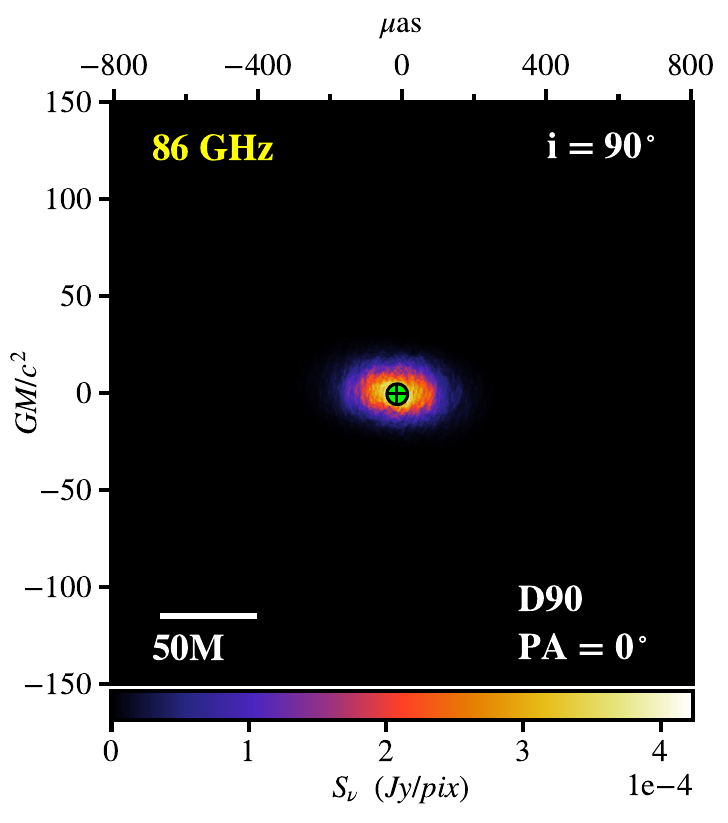}
\end{figure*}

\end{appendix}


\end{document}